\documentclass{aa}

\usepackage{graphicx}

\usepackage[utf8]{inputenc}
\usepackage{txfonts}
\usepackage[encapsulated]{CJK}

\usepackage{amssymb,amsmath,wasysym}
\usepackage{color}
\usepackage{comment}

\usepackage{siunitx}

\usepackage{url}
\usepackage[dvipsnames]{xcolor}
\usepackage[colorlinks=true, allcolors=blue, breaklinks=true]{hyperref}

\newcommand*\cthanks[1][\value{footnote}]{\footnotemark[#1]}

\DeclareRobustCommand\cjkwong{\begin{CJK*}{Bg5}{bsmi}\CJKchar[Bg5]{"B6}{"C0}\CJKchar[Bg5]{"B9}{"C5}\CJKchar[Bg5]{"B9}{"46}\end{CJK*}}

\newcommand*\micron{$\mu$m}
\newcommand*\kms{\ensuremath{\,\text{km\,s}^{-1}}}
\newcommand*\nkms{\ensuremath{\text{km\,s}^{-1}}}
\newcommand*\percm{\ensuremath{\text{cm}^{-1}}}
\newcommand*\percc{\ensuremath{\text{cm}^{-3}}}
\newcommand*\fluxunits{\ensuremath{\,\text{mJy\,beam}^{-1}}}
\newcommand*\intfluxunits{\ensuremath{\,\text{mJy\,beam}^{-1}\,\text{km\,s}^{-1}}}
\newcommand*\velgradunits{\ensuremath{\,\text{km\,s}^{-1}\,\text{arcsec}^{-1}}}

\newcommand*\ammabun{\ensuremath{n_{\text{NH}_3}/n_{\text{H}_2}}}
\newcommand*\bdopp{\ensuremath{b_\text{Dopp}}}
\newcommand*\lsun{\ensuremath{L_{\odot}}}
\newcommand*\lstar{\ensuremath{L_{\star}}}
\newcommand*\msun{\ensuremath{M_{\odot}}}
\newcommand*\mdot{\ensuremath{\dot M}}
\newcommand*\mspy{\ensuremath{\,M_{\odot}\,\text{yr}^{-1}}}
\newcommand*\rsun{\ensuremath{R_{\odot}}}
\newcommand*\rstar{\ensuremath{R_{\star}}}
\newcommand*\rin{\ensuremath{R_\text{in}}}
\newcommand*\refd{\ensuremath{R_e}}
\newcommand*\tstar{\ensuremath{T_{\star}}}
\newcommand*\vlsr{\ensuremath{V_\text{LSR}}}
\newcommand*\vin{\ensuremath{V_\text{in}}}
\newcommand*\vexp{\ensuremath{V_\text{exp}}}

\newcommand*\ratran{\mbox{\large\textsc{ratran}}}

\newcommand*\herschel{\mbox{\textit{Herschel}}}
\newcommand*\hifi{\mbox{\textit{Herschel}/HIFI}}
\newcommand*\pacs{\mbox{\textit{Herschel}/PACS}}

\newcommand*\hubble{\mbox{\textit{Hubble}}}
\newcommand*\planck{\mbox{\textit{Planck}}}

\newcommand*\ik{\mbox{IK Tau}}
\newcommand*\vy{\mbox{VY CMa}}
\newcommand*\vx{\mbox{VX Sgr}}
\newcommand*\nml{\mbox{NML Cyg}}
\newcommand*\oh{\mbox{OH\,231.8+4.2}}
\newcommand*\irc{\mbox{IRC\,+10420}}
\newcommand*\cw{\mbox{IRC\,+10216}}
\newcommand*\omi{\mbox{$o$ Cet}}
\newcommand*\rleo{\mbox{R Leo}}
\newcommand*\egg{\mbox{CRL\,2688}}
\newcommand*\crl{\mbox{CRL\,618}}
\newcommand*\ck{\mbox{CK Vulpeculae}}

\newlength{\elvlwidth}
\newlength{\specwidth}
\newlength{\jvlaheight}
\newlength{\aipswidth}
\newlength{\irtfwidth}
\newlength{\profheight}
\newlength{\ircheight}
\newlength{\ircparamh}
\newcommand*{\tm}[1]{\tablefootmark{#1}}

\newcommand*\aasperm{{\textcopyright} AAS. Reproduced with permission.}

\begin{document}

\title{Circumstellar ammonia in oxygen-rich evolved stars
\thanks{{\herschel} is an ESA space observatory with science instruments provided by European-led Principal Investigator consortia and with important participation from NASA.}\fnmsep
\thanks{Based on observations carried out under project numbers 216-09, 212-10, and 052-15 with the IRAM 30m Telescope. IRAM is supported by INSU/CNRS (France), MPG (Germany) and IGN (Spain).}\fnmsep
\thanks{Based on observations with ISO, an ESA project with instruments funded by ESA Member States (especially the PI countries: France, Germany, the Netherlands and the United Kingdom) and with the participation of ISAS and NASA.}
}

\author{K.~T.~Wong ({\cjkwong})\inst{1,2}\fnmsep
\thanks{Member of the International Max Planck Research School (IMPRS) for Astronomy and Astrophysics at the Universities of Bonn and Cologne.}\fnmsep
\thanks{Permanent email address: \texttt{tat612@connect.hku.hk}}
\and K.~M.~Menten\inst{1}
\and T.~Kami\'{n}ski\inst{3}
\and F.~Wyrowski\inst{1}
\and J.~H.~Lacy\inst{4}\fnmsep
\thanks{Visiting Astronomer at the Infrared Telescope Facility, which is operated by the University of Hawaii under contract NNH14CK55B with the National Aeronautics and Space Administration.}
\and T.~K.~Greathouse\inst{5}\fnmsep
\cthanks[6]}

\institute{Max-Planck-Institut f\"{u}r Radioastronomie, Auf dem H\"{u}gel 69, 53121 Bonn, Germany \\
\email{[ktwong; kmenten; wyrowski]@mpifr-bonn.mpg.de}
\and Institut de Radioastronomie Millim\'{e}trique, 300 rue de la Piscine, 38406 Saint-Martin-d'H\`{e}res, France \\
\email{wong@iram.fr} (present address)
\and Harvard-Smithsonian Center for Astrophysics, 60 Garden Street, Cambridge, MA 02138, USA \\
\email{tomasz.kaminski@cfa.harvard.edu}
\and Department of Astronomy, University of Texas, Austin, TX 78712, USA \\
\email{lacy@astro.as.utexas.edu}
\and Southwest Research Institute, 6220 Culebra Road, San Antonio, TX 78238, USA \\
\email{tgreathouse@swri.edu}}

\titlerunning{Ammonia in O-rich evolved stars}
\authorrunning{K.~T.~Wong et al.}

\date{{{\aap} accepted}}

 
\abstract
{The circumstellar ammonia (NH$_3$) chemistry in evolved stars is poorly understood. Previous observations and modelling showed that NH$_3$ abundance in oxygen-rich stars is several orders of magnitude above that predicted by equilibrium chemistry.}
{We would like to characterise the spatial distribution and excitation of NH$_3$ in the oxygen-rich circumstellar envelopes (CSEs) of four diverse targets: IK Tau, VY CMa, OH 231.8+4.2, and IRC +10420.}
{We observed NH$_3$ emission from the ground state in the inversion transitions near 1.3 cm with the Very Large Array (VLA) and submillimetre rotational transitions with the Heterodyne Instrument for the Far-Infrared (HIFI) aboard \textit{Herschel} Space Observatory from all four targets. For IK Tau and VY CMa, we observed NH$_3$ rovibrational absorption lines in the $\nu_2$ band near 10.5 $\mu$m with the Texas Echelon Cross Echelle Spectrograph (TEXES) at the NASA Infrared Telescope Facility (IRTF). We also attempted to search for the rotational transition within the excited vibrational state ($\varv_2=1$) near 2 mm with the IRAM 30m Telescope. Non-LTE radiative transfer modelling, including radiative pumping to the vibrational state, was carried out to derive the radial distribution of NH$_3$ in the CSEs of these targets.}
{We detected NH$_3$ inversion and rotational emission in all four targets. IK Tau and VY CMa show blueshifted absorption in the rovibrational spectra. We did not detect vibrationally excited rotational transition from IK Tau. Spatially resolved VLA images of IK Tau and IRC +10420 show clumpy emission structures; unresolved images of VY CMa and OH 231.8+4.2 indicate that the spatial-kinematic distribution of NH$_3$ is similar to that of assorted molecules, such as SO and SO$_2$, that exhibit localised and clumpy emission. Our modelling shows that the NH$_3$ abundance relative to molecular hydrogen is generally of the order of $10^{-7}$, which is a few times lower than previous estimates that were made without considering radiative pumping and is at least 10 times higher than that in the carbon-rich CSE of IRC +10216. NH$_3$ in OH 231.8+4.2 and IRC +10420 is found to emit in gas denser than the ambient medium. Incidentally, we also derived a new period of IK Tau from its $V$-band light curve.}
{NH$_3$ is again detected in very high abundance in evolved stars, especially the oxygen-rich ones. Its emission mainly arises from localised spatial-kinematic structures that are probably denser than the ambient gas. Circumstellar shocks in the accelerated wind may contribute to the production of NH$_3$. Future mid-infrared spectroscopy and radio imaging studies are necessary to constrain the radii and physical conditions of the formation regions of NH$_3$.}


\maketitle


\section{Introduction}
\label{sec:intro}

Ammonia (NH$_3$) was the first discovered interstellar polyatomic molecule \citep{cheung1968} and is a ubiquitous component of the interstellar medium. NH$_3$ has been detected in a wide variety of astronomical environments such as cold and infrared dark clouds \citep[IRDCs; e.g.][]{cheung1973,pillai2006}, molecular clouds \citep[e.g.][]{harju1993,mills2013,arai2016}, massive star-forming regions \citep[e.g.][]{urquhart2011,wienen2012}, and, recently, in an interacting region between an IRDC and the nebula of a luminous blue variable candidate \citep{rizzo2014}, a stellar merger remnant candidate \citep{kaminski2015}, and a protoplanetary disc \citep{salinas2016}. It is also an important species in the atmosphere of gas-giant (exo)planets \citep[e.g.][]{betz1996,madhusudhan2016} and is considered to be one of the possible reactants in the synthesis of prebiotic organic molecules such as amino acids in interstellar space \citep[][and references therein]{mccollom2013}.

NH$_3$ has also been detected in the circumstellar envelopes (CSEs) around a variety of evolved stars. The first detection of the molecule was based on mid-infrared (MIR; 10\,{\micron}) absorption spectra in the rovibrational (hereafter, $\nu_2$) band from the high mass-loss carbon-rich asymptotic giant branch (AGB) star {\cw} \citep{betz1979} and several oxygen-rich (super)giants belonging to very different mass and luminosity regimes, including $o$ Ceti, VY Canis Majoris, VX Sagittarii, and {\irc} \citep{mclaren1980,betz1985}. Vibrational ground-state inversion transitions near 1.3\,cm (the radio $K$ band) have also been detected from the AGB stars {\cw} and IK Tauri \citep[e.g.][]{bell1980,menten1995}, several (pre-)planetary nebulae ({\egg}, {\oh}, {\crl}) \citep{nqr1984,morris1987,tb1988,tb1996,mp1992a,mp1992b,mp1993,mp1995}, and the hypergiant {\irc} \citep{menten1995}. More recently, submillimetre observations with space observatories have detected rotational transitions of NH$_3$ from evolved stars of all chemical types, including the C-rich star {\cw} \citep[$\text{C/O} > 1$;][]{hasegawa2006,schmidt2016}, several O-rich stars \citep[$\text{C/O} < 1$;][]{menten2010,teyssier2012}, and the S-type star W Aquilae \citep[$\text{C/O} \approx 1$;][]{danilovich2014}.

Ever since the initial discoveries of NH$_3$ in evolved stars, it has been known that its abundance in circumstellar environments is much higher, by \textit{several orders} of magnitude, than that predicted by chemical models assuming dissociative and local thermodynamical equilibria (hereafter, equilibrium chemistry). Under equilibrium chemistry, most of the nitrogen nuclei in the photosphere of cool giants and supergiants ($T_\text{eff} \lesssim 3000\,\text{K}$) are locked in molecular nitrogen (N$_2$) and few are left to form NH$_3$. As a result, the photospheric NH$_3$ abundance of cool (super)giants is typically estimated to be about $10^{-12}$--$10^{-10}$ relative to molecular hydrogen (H$_2$) \citep[e.g.][]{tsuji1964,dolan1965,fujita1966,tsuji1973,mccabe1979,johnson1982}. On the other hand, the abundance derived from observations ranges from $10^{-8}$ to $10^{-6}$ \citep[e.g.][]{betz1979,kwok1981,mp1992a,menten1995}. Furthermore, circumstellar chemical models have to `inject' NH$_3$ at a high abundance (${\sim}10^{-6}$--$10^{-5}$; as derived from observations) in the inner CSE ($r<10^{16}\,\text{cm}$) in order to explain the observed abundances of NH$_3$ and other nitrogen-bearing molecules \citep[e.g.][]{nercessian1989,willacy1997,gobrecht2016,li2016}.

There are a few possible explanations for the overabundance problem of circumstellar ammonia. Pulsation-driven shocks in the inner wind may dissociate molecular N$_2$, thereby releasing free N atoms to form NH$_3$. However, theoretical models including shock-induced chemistry within the dust condensation zone have failed to produce a significantly higher NH$_3$ abundance than yielding from equilibrium chemistry. In the carbon-rich chemical model of \citet{willacy1998}, even strong shocks with velocity as high as $20{\kms}$ are not able to enhance the NH$_3$ abundance by even an order of magnitude; in the inner wind of O-rich stars, NH$_3$ can only be produced at an abundance of ${\sim}10^{-10}$--$10^{-9}$ under shock-induced chemistry \citep{duari1999,gobrecht2016}. In the recent model of \citet{gobrecht2016}, NH$_3$ may form at an abundance ${\sim}10^{-9}$ only if the shock radius is very close to the star and the shock velocity is very high ($r_\text{shock} = 1\,\rstar$ and $V_\text{shock} = 32{\kms}$ in their adopted model). A shock at an outer radius or with a lower velocity would result in lower post-shock gas densities, under which the formation of other N-bearing molecules, such as HCN and PN, are favoured over NH$_3$ (Gobrecht 2017, priv. comm.). The chemical models predict that the NH$_3$ abundance drops rapidly at radii outside of the shock \citep{duari1999,gobrecht2016}, because the molecule is converted to species such as HCN and NO (Gobrecht 2017, priv. comm.). The amount of shock-generated NH$_3$ remaining beyond the dust condensation zone is vanishingly small \citep{gobrecht2016}.

Alternatively, ultraviolet (UV) photons from the interstellar radiation field may dissociate N$_2$ molecules in the inner CSE. \citet{decin2010nat} have suggested that a clumpy CSE may allow deep penetration of UV radiation into the inner radii (${<}10^{14}\,\text{cm}$), thus triggering photochemistry that would have only occurred in the outer envelopes. Detailed chemical models have shown that the NH$_3$ abundance could reach a few times $10^{-8}$ in the clumpy CSE of high mass-loss AGB stars \citep[${\gtrsim}10^{-6}\mspy$;][]{decin2010nat,agundez2010}, which is consistent with the refined value of the observed NH$_3$ abundance in the carbon star {\cw} \citep{schmidt2016}. On the other hand, if the CSE is smooth, nitrogen molecules at the inner radii may be shielded from dissociative photons by dust, H, H$_2$, CO, and N$_2$ itself \citep{li2013}. Recent calculations by \citet{li2016} have shown that self-shielding of N$_2$ may shift its photodissociation region by about seven times further, to $r{\sim}10^{17}\,\text{cm}$ in O-rich CSEs. The CO imaging survey by \citet{cc2010} has indicated a certain degree of clumpiness in AGB stellar winds.

In addition to NH$_3$, several N-bearing molecules (such as NO, NS, PN) have been detected in O-rich objects, starting with HCN \citep{deguchi1985,deguchi1986,nercessian1989}. These O-rich sources towards which N-bearing molecules other than HCN have been detected include {\oh} \citep{vp2015,sc2015}, {\irc} \citep{ql2013,ql2016}, and {\ik} \citep{vp2017}. This may suggest that there could be a general enhancement of the production of $^{14}$N from stellar nucleosynthesis \citep[e.g.][]{sc2015,ql2016}. For intermediate-mass AGB stars between 4 and 8 solar masses, this could happen due to hot bottom burning \citep[e.g.][]{karakas2014}; in A- to F-type supergiants, nitrogen may be enriched by a factor of a few due to rotationally induced mixing \citep[e.g.][]{lyubimkov2011}. However, chemical models assuming a high $^{14}$N-enrichment by a factor of 40 in the pre-planetary nebula {\oh} do not seem to reproduce the observed abundances of N-bearing molecules \citep{vp2015,sc2015}.

The NH$_3$ emission in the radio inversion transitions from {\ik} and {\irc} has been modelled by \citet{menten1995}. In their models, which assumed local thermodynamic equilibrium (LTE) excitation of the molecule, a high NH$_3$ abundance\footnote{All abundances discussed in this paper are in fractions of the abundance of molecular hydrogen (H$_2$).} of $4 \times 10^{-6}$ is required for both stars to explain the observed inversion line intensities. This NH$_3$ abundance pertains over the accelerated circumstellar wind from an inner radius of approximately $10$--$20\,\rstar$ to ${\sim}1300\,\rstar$ from {\ik} and ${\sim}8600\,\rstar$ from {\irc}. Statistical equilibrium calculations by \citet{menten2010} have also shown that a similarly high NH$_3$ abundance of ${\sim}3 \times 10^{-7}$--$3 \times 10^{-6}$ is required to reproduce the strong submillimetre ground-state rotational transition $J_K=1_0$--$0_0$ from ortho-NH$_3$. In their models, however, the radio inversion transitions from para-NH$_3$ in {\ik} and {\irc} can only be explained by an abundance lower by the factor of ${\sim}2.5$--$3.0$ \citep{menten2010}. Hence, the derived ortho-to-para NH$_3$ ratio seems to deviate from the equilibrium value of unity expected for warm circumstellar gas (see Sect. \ref{sec:mol}). They noted that their non-LTE models only included energy levels within the vibrational ground state of NH$_3$ and hence infrared radiative excitation, which may be important in populating the rotational energy levels, was not considered. The recent radiative transfer modelling by \citet{schmidt2016} on submillimetre NH$_3$ rotational emission from {\cw} has confirmed that MIR pumping near 10\,{\micron} of the molecule from the ground state ($\varv=0$) to the lowest vibrationally excited state, $\varv_2=1$, is necessary to reconcile the discrepancy between the derived ortho- and para-NH$_3$ abundances. In {\cw}, the inclusion of MIR excitation also reduces the derived NH$_3$ abundance for the rotational transitions from a high value of $1 \times 10^{-6}$ \citep{hasegawa2006} to ${\sim}3 \times 10^{-8}$ \citep{schmidt2016}.

In this study, we present multi-wavelength observations and radiative transfer modelling of NH$_3$ from four oxygen-rich stars of different masses that are at different stages of post-main sequence stellar evolution in order to improve our knowledge of the abundance and spatial distribution of the NH$_3$ molecule in O-rich circumstellar environments. The lists of observed NH$_3$ transition lines can be found in Tables \ref{tab:obs_lowfreq} and \ref{tab:obs_mir}. Basic information on our target stars, along with the modelling results, can be found in Table \ref{tab:models}. In Sect. \ref{sec:mol}, we discuss the molecular physics of ammonia and observable transitions at various wavelengths. In Sect. \ref{sec:obs}, we describe our observations at radio, (sub)millimetre, and infrared wavelengths, whereas details of the radio and submillimetre observations are also summarised in Appendices \ref{sec:app_vla} and \ref{sec:app_hifi}. In Sect. \ref{sec:modelling}, we explain the general set-up of our radiative transfer modelling. Since the four targets in our study differ from each other in terms of the structure of their CSEs and the amount of available NH$_3$ data, and hence the considerations in the input physical models for modelling, we shall present and discuss the observational results and radiative transfer models for each target separately in Sect. \ref{sec:results}. In addition, we also derive the new period and phase-zero date of {\ik}'s pulsation in Appendix \ref{sec:app_iktau} because the pulsation phase of {\ik} is important for our discussion. We summarise our results in Sect. \ref{sec:conclsions}. 


\section{The ammonia molecule}
\label{sec:mol}

In its equilibrium configuration, NH$_3$ is a symmetric top molecule and the nitrogen atom at one of the equilibrium positions can tunnel through the plane formed by the three hydrogen atoms to the other equilibrium position. The inverting (non-rigid) NH$_3$ molecule corresponds to the molecular symmetry point group $\boldsymbol{D}_{3h}$ \citep{longuethiggins1963}. There are six fundamental modes of vibration for this type of molecule, namely, the symmetric stretch ($\nu_1$: near 3.0\,{\micron}), the symmetric bend or the `umbrella' mode ($\nu_2$: near 10.3\,{\micron} and 10.7\,{\micron}), the doubly degenerate asymmetric stretch ($\nu_{3}^{\ell_3}$: near 2.9\,{\micron}), and the doubly degenerate asymmetric bend ($\nu_{4}^{\ell_4}$: near 6.1\,{\micron}) \citep{howard1934,herzbergbook1945}. \citet{schmidt2016} have recently shown that, in AGB circumstellar envelopes, only the ground vibrational state ($\varv=0$) and the lowest vibrationally excited state ($\varv_2=1$) play dominant roles in populating the most important energy levels in the ground vibrational state. In this study, therefore, we only consider the energy levels in the ground and $\varv_2=1$ states of the molecule as done by \citet{schmidt2016}.

The rotational energy levels of NH$_3$ are characterised by several quantum numbers, including $J$ and $K$, which represent the total angular momentum and its projection along the axis of rotational symmetry, respectively. The existence of two equilibria leads to a symmetric (s, $+$) and an asymmetric (a, $-$) combination of the wave functions \citep[e.g. Sect. II, 5(d) of][]{herzbergbook1945}. Therefore, each of the rotational $(J,K)$ levels is split into two inversion levels, except for the levels of the $K=0$ ladder in which half of the energy levels are missing due to the Pauli exclusion principle \citep*[e.g. Sect. 3-4 of][]{townesbook1955}. Dipole selection rules dictate a change in symmetric/asymmetric combination, i.e. $+ \leftrightarrow -$ \citep[e.g. Chap. 12 of][]{bunkerbook1979}, in any transition of the NH$_3$ molecule. In the ground state, transition between the two inversion levels within a $(J,K)$ state (where $K \ne 0$) corresponds to a wavelength of about 1.3\,cm (a frequency of around 24\,GHz) and many $(J,K)$ inversion lines can be observed in the radio $K$ band. Modern digital technology allows a large instantaneous bandwidth, so that the newly upgraded Karl G. Jansky Very Large Array (VLA) and the Effelsberg 100-m Radio Telescope \citep{kbandrx} allow simultaneous observations of a multitude of NH$_3$ inversion lines \citep[see, e.g., Fig. A1 of][]{gong2015}.

Rotational transitions follow the selection rule $\Delta J=0, \pm 1$, except for $K=0$ for which $\Delta J = \pm 1$ only. In each $K$-ladder, energy levels with $J>K$ are known as non-metastable levels because they decay rapidly to the metastable level ($J=K$) via $\Delta J = -1$ transitions in the submillimetre and far-infrared wavelengths. In the ground state, important low-$J$ rotational transitions include $J_K=1_0$--$0_0$, $2_K$--$1_K$, and $3_K$--$2_K$ near the frequencies of 572\,GHz, 1.2\,THz, and 1.8\,THz, respectively, frequencies that are inaccessible to ground-based telescopes primarily due to tropospheric water vapour absorption \citep[e.g.][]{yang2011}. Hence these lines can only be observed with space telescopes or airborne observatories. In the $\varv_2=1$ state, the two important transitions at (sub)millimetre wavelengths are $J_K=0_0$--$1_0$ at 466.246\,GHz and $2_1$--$1_1$ at 140.142\,GHz. These two vibrationally excited lines have previously been detected from a few star-forming regions in the Galaxy \citep{schilke1990,schilke1992}, but searches had been unsuccessful in the carbon-rich AGB star {\cw} \citep{mauersberger1988,schilke1990}.

Since the axis of rotational symmetry is also along the direction of the molecule's electric dipole moment, no torque exists along this axis and the projected angular momentum does not change upon rotations. In other words, rotational transitions follow the selection rule $\Delta K=0$ \citep[e.g.][]{oka1976} and radiative transitions between different $K$-ladders of NH$_3$ are generally forbidden\footnote{Radiative transitions with $\Delta K = \pm 3$ are weakly allowed due to centrifugal distortion, in which the molecular rotation induces a small dipole moment in the direction perpendicular to the principal axis of NH$_3$ \citep{watson1971,oka1976}. However, these transitions are important only under very low gas density \citep[${\sim}200\,{\percc}$;][]{oka1971}. In circumstellar environments, $\Delta K = \pm 3$ transitions are dominated by collisions.}. Transitions between the metastable levels between $K$-ladders are only possible via collisions \citep{ho1983}.

Absorption of an infrared photon leads to radiative vibrational excitation of NH$_3$ ($\varv_2=1 \leftarrow 0$). The strongest transitions are in the MIR $N$ band near 10\,{\micron}. The rovibrational transitions are classified as $P$, $Q$, or $R$ branches if $J(\varv_2=1)-J(\text{ground}) = -1$, $0$, or $+1$, respectively \citep[e.g.][]{herzbergbook1945}. Individual transitions are labelled by the lower (ground-state) level and the upper level can be inferred through selection rules \citep[e.g.][]{tsuboi1958}. For examples, $aR(0,0)$ represents the transition between $\varv=0$ $J_K=0_0(\text{a})$ and $\varv_2=1$ $J_K=1_0(\text{s})$, while $sP(1,0)$ represents the transition between $\varv=0$ $J_K=1_0(\text{s})$ and $\varv_2=1$ $J_K=0_0(\text{a})$. The letters a and s represent the asymmetric and symmetric states, respectively.

\begin{figure*}[!htbp]
\centering
\includegraphics[width=\elvlwidth]{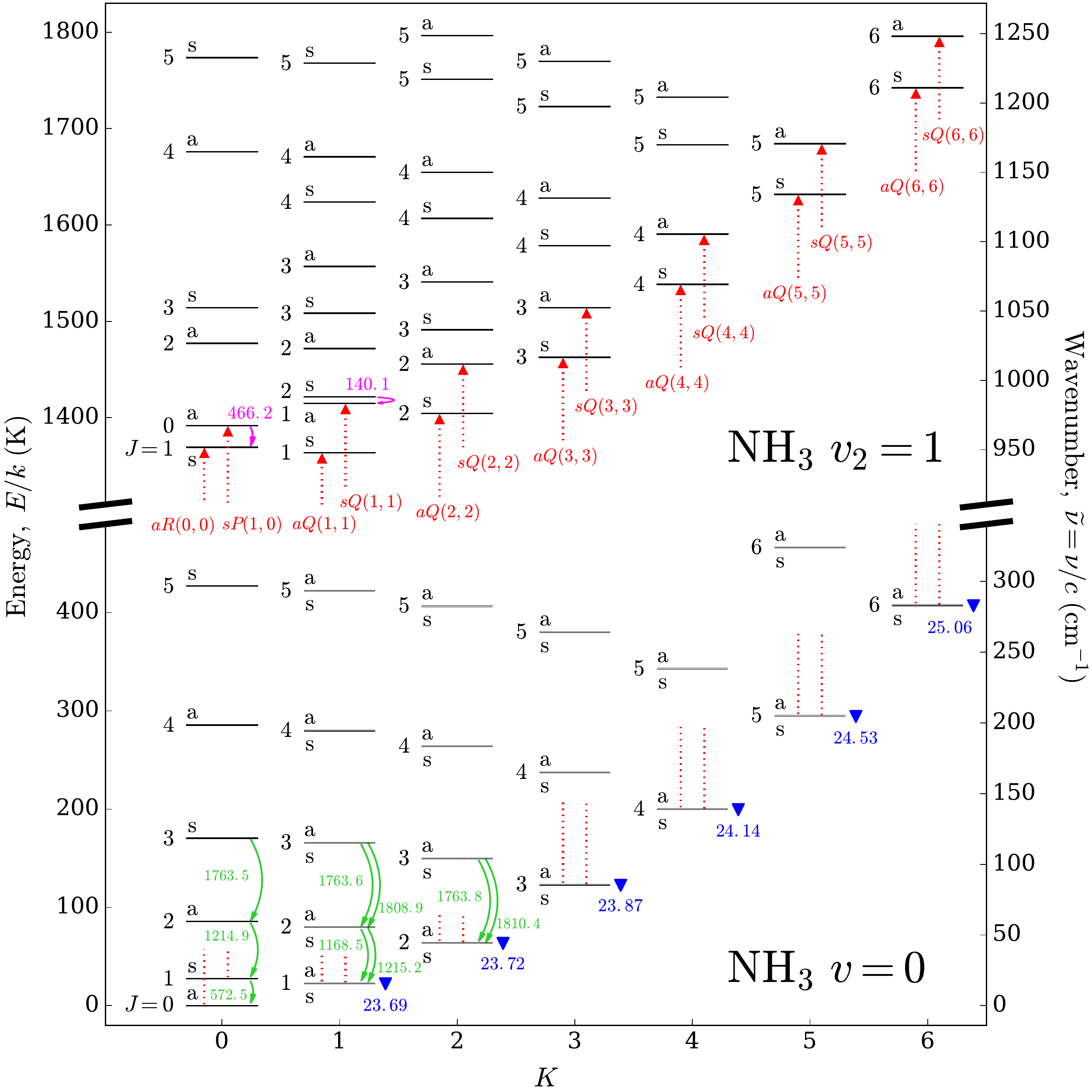}
\caption[]{Energy level diagram of NH$_3$ showing the ground ($\varv=0$; \textit{lower half}) and the first excited vibrational ($\varv_2=1$; \textit{upper half}) states. The \textit{truncated} vertical axes on the left and right show the energy from the ground ($E/k$; K) and the corresponding wavenumber ($\tilde{\nu}=\nu/c$; $\text{cm}^{-1}$), respectively; the horizontal axis shows the quantum number $K$. The integral value next to each level or inversion doublet shows the quantum number $J$; the letter `a' or `s' represents the symmetry of the level. Ground-state inversion doublets ($K>0$) can only be resolved when the figure is zoomed in due to the small energy differences. Important transitions as discussed in this article are labelled: curved green and magenta arrows indicate the (sub)millimetre rotational line emission within the ground and excited states, respectively; blue triangles mark the radio metastable inversion line emission in the ground state; and upward red dotted arrows indicate the absorption in the MIR $\nu_2$ band. These transitions are labelled by their rest frequencies in GHz (see Table \ref{tab:obs_lowfreq}), except the MIR transitions which are labelled by their transition names (see Table \ref{tab:obs_mir}).}
\label{fig:levels}
\end{figure*}

Each hydrogen atom in the NH$_3$ molecule can take the nuclear spin of $+1/2$ or $-1/2$, and hence the total nuclear spin can either be $I=1/2$ or $3/2$. Since radiation and collisions in interstellar gas in general do not change the spin orientation, transitions between ortho-NH$_3$ ($I=3/2$; $K=0,3,6,\ldots$) and para-NH$_3$ ($I=1/2$; $K=1,2,4,5,\ldots$) are highly forbidden \citep{cheung1969,ho1983}. Therefore, the energy levels in ortho- and para-NH$_3$ should be treated separately in excitation analyses. \citet{faure2013} found that the ortho-to-para abundance ratio of NH$_3$ deviates from unity if the molecule was formed in cold interstellar gas of temperature below 30\,K. In circumstellar environments where the gas temperatures are generally much higher than 100\,K, the ortho- and para-NH$_3$ abundances are expected to be identical.

Figure \ref{fig:levels} shows the energy level diagram of NH$_3$ in the ground ($\varv=0$) and first excited vibrational state ($\varv_2=1$). Important transitions in this article are annotated with their frequencies or transition names. Observations of these transitions will be discussed in the next section.


\section{Observations}
\label{sec:obs}

We observed various NH$_3$ transitions in four oxygen-rich stars of different evolutionary stages: {\ik}, a long-period variable (LPV) with a high mass-loss rate; {\oh}, a bipolar pre-planetary nebula (PPN) around an OH/IR star; {\vy}, a red supergiant (RSG); and {\irc}, a yellow hypergiant (YHG). These targets sample both the low-/intermediate-mass (LPV/PPN) and the high-mass (RSG/YHG) regimes of post-main sequence stellar evolution. They are also among the most chemically rich and the most studied objects of their classes. All four targets exhibited strong NH$_3$ emission in previous observations \citep[e.g.][]{menten1995,menten2010}.

In the following subsections, we discuss the multi-wavelength observations of NH$_3$ emission and absorption in the CSEs of these four targets. For an overview, Table \ref{tab:obs_lowfreq} lists all the radio inversion and (sub)millimetre rotational lines and Table \ref{tab:obs_mir} lists all the MIR vibrational transitions that are included in our study.


\subsection{Radio inversion transitions}
\label{sec:vla}

Table \ref{tab:obs_vla} summarises all the relevant VLA observations and the corresponding spectral setups in this study. The two ground-state metastable inversion transitions of NH$_3$, $(J,K)=(1,1)$ and $(2,2)$, were observed with the NRAO\footnote{The National Radio Astronomy Observatory is a facility of the National Science Foundation operated under cooperative agreement by Associated Universities, Inc.} VLA in 2004 for {\ik} and {\irc} (legacy ID: AM797). The observations were performed in the D configuration, which is the most compact one, with the correlator in the 2AD mode. The two inversion lines were observed at their sky frequencies simultaneously in individual IF channels of different circular polarisations. Polarisations are irrelevant in our analysis. The total bandwidth of each IF channel is $12.5\,\text{MHz}$ (${\sim}160{\kms}$) with a spectral resolution of $390.625\,\text{kHz}$ (${\sim}5{\kms}$). 

The absolute flux calibrators during the observations of {\ik} and {\irc} were 3C48 (J0137+3309) and 3C286 (J1331+3030), respectively. Based on the time-dependent coefficients determined by \citet{perley2013}, the flux densities of 3C48 and 3C286 during the observations were $1.10\,\text{Jy}$ and $2.41\,\text{Jy}$, respectively. The interferometer phase of {\ik} was referenced to the gain calibrator J0409+1217 and that of {\irc} was referenced to J1922+1530. Both calibrators are about $4\degr$ from the respective targets.

The inversion spectra of both stars have been published by \citet{menten2010}. In this study, we repeated the data reduction and radiative transfer modelling of the same VLA dataset. The data were reduced by the Astronomical Image Processing System (AIPS; version 31DEC14). Part of the observations of {\ik} suffered from poor phase stability, probably due to bad weather, so the later half of the observation (time ranging between 16:15:00 and 19:16:00) had to be discarded. The rest of the calibration and imaging were done in the standard manner. We imaged the data with the CLEAN algorithm under natural weighting to minimise the rms noise and restored the final images with a circular beam of $3{\farcs}7$ in full width at half maximum (FWHM), which is approximately the geometric mean of the FWHMs of the major and minor axes of the raw beam.

With the new wideband correlator, WIDAR\footnote{Wideband Interferometer Digital Architecture.}, the upgraded VLA can cover multiple NH$_3$ inversion lines simultaneously in the radio $K$ band, particularly around 22--25\,GHz. We performed three 2-GHz $K$-band surveys towards {\vy}, {\oh}, and {\ik} with different array configurations and frequency coverages. {\vy} and {\oh} were observed on 2013 April 05 in the D configuration under the project VLA/13A-325. The correlator was tuned to cover the six lowest metastable transitions from $(1,1)$ to $(6,6)$. 3C147 (J0542+498) was observed as the absolute flux calibrator, J0731-2341 and J0748-1639 were observed as the gain (relative amplitude and phase) calibrators of {\vy} and {\oh}, respectively. We used CASA\footnote{Common Astronomy Software Applications.} version 4.2.2 to manually calibrate and image the dataset \citep{casa}. Spectral line calibration and continuum subtraction were performed in the standard manner similar to that presented in the CASA tutorial\footnote{\url{https://casaguides.nrao.edu/index.php?title=EVLA_high_frequency_Spectral_Line_tutorial_-_IRC\%2B10216_-_CASA_4.2\&oldid=15543}}. For both targets, we used natural weighting to image all six inversion lines. The rms noise in each channel map ($\Delta V = 3{\kms}$) is roughly $0.8\,\fluxunits$ and the restoring beam is about $4{\arcsec}$ in FWHM. We also tried imaging under uniform weighting with a beam of ${\sim}3{\arcsec}$, but the NH$_3$ emission was still not clearly resolved. We restored all the images with a common circular beam of $4{\farcs}0$, which is the upper limit of the geometric means of the FWHMs of the major and minor axes of the raw beam of the six transitions. 

{\ik} was also observed in 2015 and 2016 with D, C, and B configurations with baselines ranging from 35\,m to ${\sim}11\,\text{km}$ under the projects VLA/15B-167 and VLA/16A-011. The absolute flux (3C48) and gain calibrators (J0409+1217) were the same as in the 2004 observations. We conducted the line surveys at lower frequencies to search for non-metastable NH$_3$ emission and to cover the strong water masers near 22.235\,GHz. The highest covered metastable line is NH$_3$ $(4,4)$ (see Table \ref{tab:obs_vla}). All data were processed in CASA version 4.6.0. For each day of observations, we first performed standard phase-referenced calibration, then used the strong H$_2$O maser to self-calibrate the science target\footnote{See the CASA tutorial at \url{https://casaguides.nrao.edu/index.php?title=VLA_high_frequency_Spectral_Line_tutorial_-_IRC\%2B10216-CASA4.6\&oldid=20567}}. The last observation of {\ik} on 2016 July 01 suffered from very poor phase stability and the rms noises of the visibility data were at least ten times higher than for any other observations. This dataset is therefore excluded from our imaging and analysis. All other visibility data of the metastable lines (from $(1,1)$ to $(4,4)$) were then continuum subtracted, concatenated, and imaged using natural weighting. The images were restored with a common, circular beam with a FWHM of $0{\farcs}45$. For better visualisation, we further smoothed the images with a $0{\farcs}7$-FWHM Gaussian kernel and the resultant beam has a FWHM of $0{\farcs}83$.

The continuum data of the ${\sim}2$-GHz windows of {\ik}, {\vy}, and {\oh} were also imaged in the similar manner. The images were fitted with a single Gaussian component. Table \ref{tab:cont} presents the results of the continuum measurements of the three targets.


\subsection{Submillimetre rotational transitions}
\label{sec:hifi}

NH$_3$ rotational transitions were observed with the HIFI\footnote{Heterodyne Instrument for the Far-Infrared \citep{deGraauw2010}.} instrument aboard the ESA {\herschel} Space Observatory \citep{pilbratt2010}. The detection of the lowest rotational line, $J_K=1_0$--$0_0$, towards three O-rich targets (except {\oh}) in our study has already been reported and analysed by \citet{menten2010} using spectra exclusively from the {\herschel} Guaranteed Time Key Programme HIFISTARS \citep{bujarrabal2011}. After the successful detection of the $1_0$--$0_0$ line, follow-up observations were carried out to cover higher transitions, i.e. $J=2$--$1$ and $J=3$--$2$. In addition to the HIFISTARS data, we also include the data of all available NH$_3$ rotational transitions of our targets in the {\herschel} Science Archive (HSA; v8.0)\footnote{\url{http://archives.esac.esa.int/hsa/whsa/}} to extend the coverage of NH$_3$ transitions and to improve the signal-to-noise (S/N) ratio of the transitions covered by HIFISTARS.

Table \ref{tab:obs_hifi} summarises the HIFI spectra, which are all double sideband (DSB) spectra, used in this study. All but one observations were carried out in dual beam switch (DBS) mode. In DBS mode, ON-OFF subtraction is done twice by first chopping the HIFI beam between the science target and an emission-free reference with one being the ON position and the other being OFF, and then swapping the ON/OFF roles of the target and reference positions \citep[Sect. 8.1.1 of][]{deGraauw2010}. This method helps reducing the baseline ripple caused by standing waves. The spectrum with the Observation ID (OBSID) of 1342190840 was taken in frequency switch mode, which results in significant baseline ripple. However, the frequency range of the NH$_3$ $J_K=1_0$--$0_0$ line is small compared to that of the ripple and the S/N ratio of the line is quite high, allowing decent baseline removal.

Since there are no User Provided Data Products (UPDPs) for the observations other than those of HIFISTARS, we use the pipeline-processed products of all data, including those from HIFISTARS observations, to ensure uniform data processing among the spectra. We downloaded the Level 2 data products from the HSA via the {\herschel} Interactive Processing Environment (HIPE; v15.0.0) software \citep{ott2010}. All products have been processed with the {\herschel} Standard Product Generation (SPG) pipeline (v14.1.0) during the final bulk reprocessing of the HSA. Apart from the flux differences caused by the adopted main-beam efficiencies as discussed below, there are no significant differences between the UPDPs and pipeline products of the HIFISTARS observations.

For each data product, we averaged the signals from the H and V polarisation channels and stitched the spectrometer subbands together. The spectra in antenna temperature ($T_\text{A}^{\ast}$) scale were then exported to FITS format by the \texttt{hiClass} task and loaded in CLASS\footnote{Continuum and Line Analysis Single-dish Software.} of the GILDAS package\footnote{Grenoble Image and Line Data Analysis Software; \url{http://www.iram.fr/IRAMFR/GILDAS/}} \citep{gildas}, where further processing including polynomial baseline removal, main-beam efficiency correction, and noise-weighted averaging of spectra from different observations were carried out.

The antenna temperature scale of the spectra was converted to main-beam temperature ($T_\text{mb}$) by the relation 
\begin{equation}
T_\text{mb} = T_\text{A}^{\ast} \frac{\eta_\text{l}}{\eta_\text{mb}},
\end{equation}
where $\eta_\text{l} = 0.96$ is the forward efficiency \citep{roelfsema2012} and $\eta_\text{mb}$ is the main-beam efficiency. The main-beam efficiency for each NH$_3$ line frequency is the average of $\eta_\text{mb, H}$ and $\eta_\text{mb, V}$ for both polarisations, which are evaluated from the parameters reported by \citet{mueller2014} in October 2014. Their $\eta_\text{mb}$ values are consistently smaller (by 10\%--20\%) than and supersede those reported by \citet{roelfsema2012} assuming Gaussian beam shape. Previous publications that include some of our NH$_3$ spectra were based on the old $\eta_\text{mb}$ values \citep[e.g.][]{menten2010,bujarrabal2012,justtanont2012,teyssier2012,alcolea2013}.

By combining the calibration uncertainty budget in \citet{roelfsema2012} in quadrature, we estimate the calibration uncertainties to be about 10\% for NH$_3$ $J=1$--$0$ (band 1) and 15\% for $J=2$--$1$ and $J=3$--$2$ (bands 5 \& 7). The sizes of the NH$_3$-emitting regions of our targets as constrained by VLA observations are ${\lesssim}4\arcsec$, much smaller than the FWHM of the telescope beam, which ranges from about $12\arcsec$ ($J=3$--$2$) to $37\arcsec$ ($J=1$--$0$).


\subsection{Millimetre rotational transition within the \texorpdfstring{$\varv_2=1$}{v2=1} state}
\label{sec:iram}

We searched for the $J_K=2_1$--$1_1$ rotational transition within the vibrationally excited state $\varv_2=1$ at 140.142\,GHz (2.14\,mm) from {\ik} with the IRAM\footnote{Institut de Radioastronomie Millim{\'e}trique.} 30m Telescope on Pico Veleta. The frequency range of interest was covered by an earlier line survey (IRAM projects 216-09 and 212-10; PI: C. S{\'a}nchez Contreras), which has subsequently been published by \citet{vp2016,vp2017}, and no NH$_3$ line was detected at the rms noise in $T_\text{mb}$ scale of 1.8\,mK at the velocity resolution of $4.3{\kms}$. In order to obtain a tighter constraint, we conducted a deep integration at the frequency of the 2-mm rotational line in August 2015 under the project 052-15.

We observed {\ik} with the Eight MIxer Receiver (EMIR) in the 2\,mm (E150) band \citep{carter2012}. ON-OFF subtraction was done with the wobbler switching mode with a throw of ${\pm}120\arcsec$. Pointing was checked regularly every ${\sim}1$ hour with Uranus or a nearby quasar (J0224+0659 or J0339-0146, selected by a compromise between angular distance and 2-mm flux density), and focus was checked with Uranus every ${\sim}3$ hours and after sunrise. The total ON-source integration time is about 10.1 hours. The data were calibrated with the MIRA\footnote{Multichannel Imaging and Calibration Software for Receiver Arrays.} software in GILDAS. We adopt the main-beam and forward efficiencies to be $\eta_\text{mb}=0.74$ and $\eta_\text{l}=0.93$, respectively, which are the characteristic values at 145\,GHz \citep{kramer2013}. The rms noise near 140\,GHz in $T_\text{mb}$ scale is 0.61\,mK at the velocity resolution of 5.0{\kms}. By comparing the line detection from other molecules, our spectrum is consistent with that presented by \citet{vp2017} within the noise.


\begin{table*}[!htbp]
\caption{Observations of $^{14}$NH$_3$ lines in the radio and submillimetre wavelengths.}
\label{tab:obs_lowfreq}
\centering
\begin{tabular}{c S[table-format=8.6,table-align-text-post=false] S[table-format=4.3] ccc}
\hline\hline
Transition & {Rest frequency\tablefootmark{a}} & {Energy\tablefootmark{a,b}} & Telescope/instrument & Band & Ref. for \\
$J_K(\text{sym})$ & {(MHz)} & {(K)} & & & line freq. \\ \hline
$1_1(\text{a})$--$1_1(\text{s})$ &   23694.4955(2) &  23.26 & VLA & $K$ (radio) & 1 \\ 
$2_2(\text{a})$--$2_2(\text{s})$ &   23722.6333(2) &  64.45 & VLA & $K$ (radio) & 1 \\ 
$3_3(\text{a})$--$3_3(\text{s})$ &   23870.1292(2) & 123.54 & VLA & $K$ (radio) & 1 \\
$4_4(\text{a})$--$4_4(\text{s})$ &   24139.4163(2) & 200.52 & VLA & $K$ (radio) & 2 \\
$5_5(\text{a})$--$5_5(\text{s})$ &   24532.9887(2) & 295.37 & VLA & $K$ (radio) & 2 \\
$6_6(\text{a})$--$6_6(\text{s})$ &   25056.03(1)   & 408.06 & VLA & $K$ (radio) & 3 \\
\hline
$\varv_2=1$ $2_1(\text{s})$--$1_1(\text{a})$ & 140141.807(23) & 1421.59 & IRAM/EMIR & E150 & 4 \\
\hline
$1_0(\text{s})$--$0_0(\text{a})$ &   572498.160(2) & 27.48 & {\hifi} & 1b & 5 \\
$2_1(\text{s})$--$1_1(\text{a})$ & 1168452.39(2) & 79.34 & {\hifi} & 5a & 6 \\ 
$2_0(\text{a})$--$1_0(\text{s})$ & 1214852.94(2) & 85.78 & {\hifi} & 5a & 6 \\
$2_1(\text{a})$--$1_1(\text{s})$ & 1215245.71(2) & 80.45 & {\hifi} & 5a & 6 \\
$3_0(\text{s})$--$2_0(\text{a})$ & 1763524.4(1) & 170.42 & {\hifi} & 7a & 7 \\
$3_1(\text{s})$--$2_1(\text{a})$ & 1763601.2(1) & 165.09 & {\hifi} & 7a & 7 \\
$3_2(\text{s})$--$2_2(\text{a})$ & 1763823.2(1) & 149.10 & {\hifi} & 7a & 7 \\
$3_1(\text{a})$--$2_1(\text{s})$ & 1808934.6(1) & 166.16 & {\hifi} & 7b & 7 \\
$3_2(\text{a})$--$2_2(\text{s})$ & 1810380.0(1) & 150.19 & {\hifi} & 7b & 7 \\
\hline
\end{tabular}
\tablebib{
(1) \citet{kukolich1967}; 
(2) \citet{kukolich1970}; 
(3) \citet{poynter1975}; 
(4) \citet{belov1998}; 
(5) \citet{cazzoli2009}; 
(6) \citet{winnewisser1996};
(7) \citet{yu2010}.}
\tablefoot{
\tablefoottext{a}{The NH$_3$ experimental line frequencies and upper-level energies are retrieved from the calculations and compilations by \citet{yu2010}. Code for original references are given in the last column. These values are the same as those in the Jet Propulsion Laboratory (JPL) catalogue \citep{jpl1998}. The numbers shown in parentheses are the uncertainties corresponding to the last digit(s) of the quoted frequencies. The line frequencies are generally consistent with previous calculations by \citet{poynter1983}, except for $2_0(\text{a})$--$1_0(\text{s})$ near 1214.9\,GHz where the difference is about 6\,MHz (${\sim}1.4{\kms}$).}
\tablefoottext{b}{$E_\text{up}/k$, the upper energy level in K.}
}
\end{table*}


\subsection{Mid-infrared rovibrational transitions}
\label{sec:irtf}

The $\nu_2$ band ($\varv_2=1 \leftarrow 0$) of ammonia absorbs MIR $N$-band radiation from the central star and its circumstellar dust shells. Previous observations have detected multiple NH$_3$ absorption lines from the carbon star {\cw} and a few O-rich (super)giants (see Table \ref{tab:obs_mir}). The highest spectral resolutions were achieved by the Berkeley infrared heterodyne receiver at the then McMath Solar Telescope of Kitt Peak National Observatory\footnote{Currently as the McMath-Pierce Solar Telescope of National Solar Observatory, operated by the Association of Universities for Research in Astronomy, Inc. (AURA), under cooperative agreement with the National Science Foundation.} \citep{betz1975} and the MPE\footnote{Max Planck Institute for Extraterrestrial Physics.} 10-micron heterodyne receivers at various telescopes \citep{schrey1985,rothermel1986} with a resolving power of $R = \lambda/{\Delta\lambda} \approx 1\text{--}6 \times 10^6$. Other old MIR observations made use of the Berkeley heterodyne receiver at the Infrared Spatial Interferometer \citep{monnier2000i} and the Fourier Transform Spectrometer at the Kitt Peak Mayall Telescope \citep{keady1993} with $R \sim 10^5$.

In order to obtain a broader coverage of $\nu_2$ absorption lines from evolved stars, we performed new MIR observations of {\ik}, {\vy}, and the archetypal eponymous Mira variable {\omi} with the Texas Echelon Cross Echelle Spectrograph \citep[TEXES;][]{lacy2002} at the 3.0-metre NASA IRTF\footnote{The Infrared Telescope Facility is operated by the University of Hawaii under contract NNH14CK55B with the National Aeronautics and Space Administration.} under the programme 2016B064. The targets were observed on (UT) 2016 December 22 and 23 in high-resolution mode with a resolving power of $R \sim 10^5$ \citep{lacy2002}. Each resolution element is sampled by 4 pixels and the width of each spectral pixel is about 0.003\,{\percm}. In this article, we present the spectra of {\ik} and {\vy} in three wavelength (wavenumber) settings near 10.3 (966), 10.5 (950) and 10.7\,{\micron} (930\,{\percm}), covering in total 14 rovibrational transitions as listed in Table \ref{tab:obs_mir}. The width of the slit is $1{\farcs}4$. The on-source time for each target in each setting was about 10 minutes and the total observing time per setting was about 3 hours. The dwarf planet Ceres was observed as the telluric calibrator for the setting near 10.7\,{\micron} (930\,{\percm}) and the asteroid Vesta was observed for the rest. 

The FWHM of the IRTF's point-spread function in these settings is about $0{\farcs}9$. The visual seeing during our observations was ${\lesssim}1{\farcs}0$\footnote{From the Mauna Kea Atmospheric Monitor, available at \url{http://mkwc.ifa.hawaii.edu/current/seeing/index.cgi}}, which corresponds to about $0{\farcs}6$ at 10\,{\micron} \citep[e.g.][]{vdancker2016}. Hence, our observations are diffraction-limited.

The data were processed with a procedure similar to the TEXES data reduction pipeline \citep{lacy2002}, except that telluric lines were removed by a separate fitting routine. In the pipeline, the calibration frame (flat field) was the difference between the black chopper blade and sky measurements, $S_\nu(\text{black}-\text{sky})$, which partially removes telluric lines \citep[see Eq. (3) of][]{lacy2002}. In our processing, the data spectra were flat-fielded with the black measurements and the intermediate source intensity is given by
\begin{equation}
I_\nu(\text{target}) \approx \frac{S_\nu(\text{target})}{S_\nu(\text{black})} B_\nu(T_{\text{black}}),
\label{eq:texes}
\end{equation}
where $S_\nu$ and $B_\nu$ are the measured signal and the Planck's law at the frequency $\nu$. Each intermediate spectrum was then normalised and divided by a modelled telluric transmission spectrum. The parameters of the telluric transmission model were chosen to make the divided spectrum as flat as possible and to fit the observed $S_\nu(\text{black}-\text{sky})$, with the atmospheric temperatures and humidities constrained by weather balloon data. Finally, each target spectrum was divided by a residual spectrum smoothed with a Gaussian kernel of $60{\kms}$ in FWHM. The last division was necessary to remove broad wiggles due to the echelon blaze function and interference fringes, but may also have removed or created broad wings on the observed circumstellar lines.

We verified the wavenumber scale of the calibrated data by comparing the minima of atmospheric transmission with the rest wavenumbers of strong telluric lines in the GEISA spectroscopic database\footnote{\textit{Gestion et Etude des Informations Spectroscopiques Atmosph{\'e}riques} [Management and Study of Atmospheric Spectroscopic Information], \url{http://ara.abct.lmd.polytechnique.fr/}} \citep[2015 edition;][]{geisa2016}. Almost all strong telluric lines are identified as due to CO$_2$, except for the H$_2$O line at 948.262881\,{\percm}. The differences between the transmission minima and telluric lines do not exceed $4{\kms}$ in radio local standard of rest (LSR) velocity. The uncertainties of the rest wavenumbers of our targeted NH$_3$ lines (Table \ref{tab:obs_mir}) correspond to velocities of ${<}0.002{\kms}$.

We converted the wavenumber (frequency) axis from topocentric to LSR frame of reference using the ephemeris software GILDAS/ASTRO\footnote{A Software To pRepare Observations.}. In order to remove local wiggles in the normalised spectra of the targets, for {\ik} in particular, we performed polynomial baseline subtraction for some NH$_3$ spectra in GILDAS/CLASS. The removal of telluric line contamination was not perfect. Hence, the residual baseline was not always stable within the possible width of the lines and for some spectra it was difficult to identify the continuum. Although the strong absorption features were not severely affected, any much weaker emission features (if they exist at all) may be lost. 

We also compare the old $aR(0,0)$ and $aQ(2,2)$ spectra of {\vy} and the $aR(0,0)$ spectrum of {\irc} as detected with the Berkeley heterodyne receiver at the McMath Solar Telescope. These spectra were digitised from the scanned paper of \citet[][{\aasperm}]{mclaren1980} with the online application \texttt{WebPlotDigitizer}\footnote{\url{http://arohatgi.info/WebPlotDigitizer/}} \citep[version 3.10;][]{webplotdigitizer}. We did not detect NH$_3$ absorption from {\omi}; therefore we only briefly discuss its non-detection in Sect. \ref{sec:results_chem}.


\begin{table*}[!htbp]
\caption{List of MIR NH$_3$ $\nu_2$ transitions searched with the TEXES instrument at the NASA IRTF under the programme 2016B064 and their previous detection in evolved stars.}
\label{tab:obs_mir}
\centering
\begin{tabular}{crrrrcc}
\hline\hline
Transition & Wavelength \hspace{-1.5ex} & Wavenumber\tablefootmark{a} \hspace{-1.2ex} & $E_\text{up}/k$ \hspace{0.5ex} & $A_\text{ul}$\tablefootmark{b} \hspace{0.3ex} & Previous detection & Ref. for \\
           & ({\micron}) \hspace{1.5ex} & ({\percm}) \hspace{1.8ex} & (K) \hspace{1.5ex} & ($\text{s}^{-1}$) \hspace{0.0ex} & in evolved stars\tablefootmark{c} & detection \\ \hline
$sQ(1,1)$  & 10.33060   & 967.997763  & 1414.86 &  7.94 &        & \\
$sQ(2,2)$\tablefootmark{d} & 10.33337 & 967.738421 & 1455.67 & 10.57 &  &  \\
$sQ(3,3)$  & 10.33756   & 967.346340  & 1514.19 & 11.60 &        & \\
$sQ(4,4)$  & 10.34324   & 966.814717  & 1590.39 & 12.67 &        & \\
$sQ(5,5)$  & 10.35035   & 966.151127  & 1684.27 & 13.20 &        & \\
$sQ(6,6)$  & 10.35890   & 965.353911  & 1795.79 & 13.58 &        & \\
\hline
$aR(0,0)$  & 10.50667   & 951.776244  & 1369.39 &  5.49 & {\irc}, {\vx} & 1 \\
           &            &             &         &       & {\vy}  & 1, 2 \\
           &            &             &         &       & {\nml} & 2 \\
           &            &             &         &       & {\cw}  & 2, 3, 4, 5 \\
$sP(1,0)$\tablefootmark{e} & 10.54594 & 948.232047 & 1391.77 & 14.88 & {\nml} & 2 \\
\hline
$aQ(1,1)$  & 10.73390   & 931.627754  & 1363.67 &  7.74 &        & \\
$aQ(2,2)$  & 10.73730   & 931.333249  & 1404.43 & 10.32 & {\cw}  & 2, 4, 6, 7 \\
           &            &             &         &       & {\vy}  & 1, 2, 8 \\
           &            &             &         &       & {\omi} & 9 \\
$aQ(3,3)$  & 10.74394   & 930.757028  & 1462.69 & 11.60 & {\vy}  & 2 \\
           &            &             &         &       & {\cw}  & 2, 4, 8 \\
$aQ(4,4)$  & 10.75387   & 929.898108  & 1538.44 & 12.36 & {\cw}  & 4 \\
$aQ(5,5)$  & 10.76711   & 928.754479  & 1631.64 & 12.86 &        &   \\
$aQ(6,6)$  & 10.78373   & 927.323036  & 1742.27 & 13.22 & {\cw}  & 2, 4, 6 \\
           &            &             &         &       & {\vy}  & 2, 8 \\
\hline
\end{tabular}
\tablebib{
(1) \citet{mclaren1980}; 
(2) \citet{goldhaber1988}; 
(3) \citet{betz1980}; 
(4) \citet{keady1993}; 
(5) \citet{fonfria2017}; 
(6) \citet{betz1979}; 
(7) \citet{monnier2000i}; 
(8) \citet{monnier2000s}; 
(9) \citet{betz1985}.
}
\tablefoot{
\tablefoottext{a}{The uncertainties of the wavenumbers are smaller than 0.000004\,{\percm} (${\sim}0.0013{\kms}$). The wavenumbers and uncertainties are computed from the data in supplementary material 6 of \citet{chen2006}, which is also publicly available at \url{https://library.osu.edu/projects/msa/jmsa_06.htm}.}
\tablefoottext{b}{The Einstein $A$ coefficient. The values are retrieved from Table S1 of \citet[][supplementary material]{yu2010}.}
\tablefoottext{c}{Additional $\nu_2$ transitions have been detected towards the red hypergiants {\vy} and {\nml} by \citet{goldhaber1988} and towards {\cw} by \citet[][quoted in \citealp{monnier2000s}]{holler1999}. \citet{schrey1986} also reported detection of $sQ(2,1)$ or $sQ(2,2)$ absorption towards the Mira-type AGB star R Leonis; however, the purported detection had very low S/N ratios in the spectra and the position of the more probable $sQ(2,2)$ line was too far away from the suspected absorption features.}
\tablefoottext{d}{Potentially blended with telluric CO$_2$ $R(8)$ absorption at 967.707233\,{\percm}.}
\tablefoottext{e}{Potentially blended with telluric o-H$_2$O $J_{K_a,K_c}=12_{5,8}$--$11_{0,11}$ absorption at 948.262881\,{\percm}.}
}
\end{table*}


\section{Radiative transfer modelling}
\label{sec:modelling}

We modelled the NH$_3$ spectra with the 1-D radiative transfer code {\ratran}\footnote{\url{http://www.sron.rug.nl/~vdtak/ratran/}} \citep[version 1.97; ][]{ratran}, assuming that the distribution of NH$_3$ is spherically symmetric. {\ratran} solves the coupled level population and radiative transfer equations with the Monte Carlo method and produces an output image cube for each modelled transition. We then convolved the model image cubes with the same telescope point-spread function or restoring beam as that of the observed data. We processed the modelled images and spectra with the Miriad software package\footnote{\url{http://www.atnf.csiro.au/computing/software/miriad/}} \citep{miriad}. We explain the molecular data files and input dust temperature profiles in the following subsections, and discuss the input physical models for each target in Sect. \ref{sec:results}.


\subsection{\texorpdfstring{NH$_3$}{NH3} molecular data}
\label{sec:moldata}

\subsubsection{Energy levels and radiative transitions}
\label{sec:moldata_levels}

As discussed in Sect. \ref{sec:mol}, the energy levels of ortho- ($K=0,3,6,\ldots$) and para-NH$_3$ ($K=1,2,4,5,\ldots$) are essentially independent and the two species have to be modelled separately. We obtained the molecular data files of ortho- and para-NH$_3$ from \citet{schmidt2016}. The energy levels and transition line strengths were taken from the `BYTe' hot NH$_3$ line list\footnote{\url{http://www.exomol.com/data/molecules/NH3/14N-1H3/BYTe/}} \citep{yurchenko2011}. The data files for our modelling consist of all the energy levels up to the energy corresponding to 3300\,{\percm} (or 4750\,K) in the ground vibrational state and the $\varv_2=1$ state. There are 172 energy levels (up to $J=15$) and 1579 radiative transitions for ortho-NH$_3$, and 340 levels and 4434 transitions for para-NH$_3$.


\subsubsection{Collisional rate coefficients}
\label{sec:moldata_coeffs}

In the model of \citet{schmidt2016}, the collisional rate coefficients used for the ground vibrational state of NH$_3$ were calculated by \citet{danby1988} for the collisions with para-H$_2$ ($j=0$) at temperatures $15\,\text{K} \le T_\text{kin} \le 300\,\text{K}$ and were extrapolated to higher temperature assuming that they are proportional to $\sqrt{T_\text{kin}}$, where $T_\text{kin}$ is the kinetic temperature of para-H$_2$. These rate coefficients were computed in the transitions involving energy levels up to $J_\text{ortho}=6$ ($E_\text{up}/k < 600\,\text{K}$) for ortho-NH$_3$ and $J_\text{para}=5$ ($E_\text{up}/k < 450\,\text{K}$) for para-NH$_3$. Hence, higher energy levels are populated via radiative transitions only. By adopting some constant rate coefficients for other transitions, \citet{schmidt2016} found that the collisional transitions involving higher ground-state levels or vibrationally excited levels were not important.

\citet{bouhafs2017} reported new calculations on the rate coefficients of the collisions between NH$_3$ and atomic hydrogen (H), ortho-H$_2$, and para-H$_2$ at temperatures up to 200\,K and in transitions involving energy levels up to $E_\text{up}/k = 600\,\text{K}$ ($J_\text{ortho} \le 6$ and $J_\text{para} \le 7$). Their calculations were based on the full-dimensional (9-D) NH$_3$--H potential energy surface (PES) of \citet{liguo2014} and the five-dimensional NH$_3$--H$_2$ PES of \citet{maret2009} with an ad hoc approximation of the inversion wave functions using the method of \citet{green1976}. The rate coefficients of NH$_3$--para-H$_2$ are thermalised over $j=0$ and $2$ and they are consistent with those of \citet{danby1988} within a factor of 2. We do not consider the collisions with H in our models. The coefficients of NH$_3$ with ortho- and para-H$_2$ at 200\,K are consistent within a factor of ${\sim}2$ (up to 4). We weight the coefficients of the two H$_2$ species by assuming a thermal H$_2$ ortho-to-para ratio of 3 \citep{flower1984}. Since the coefficients for most low-$J$ and small-$\Delta J$ transitions do not vary much with temperature \citep[e.g. Fig. 2 of][]{bouhafs2017}, we apply constant extrapolation of the coefficients from 200\,K to higher temperatures. Vibrations were not included in the 5-D PES of NH$_3$--H$_2$. Quantum dynamics calculations have shown that vibrationally elastic ($\Delta \varv=0$; purely rotational) excitation of linear molecules by collisions with noble gases (e.g. helium) is almost independent of the vibrational state $\varv$ \citep[][and references therein]{roueff2013,balanca2017}. This also appears to be valid in the elastic ($\Delta \varv_2 = 0$) transitions of the H$_2$O--H$_2$ collision system \citep{faure2008}; therefore, we assume that the elastic (rotation-inversion) transitions within the $\varv_2=1$ state to have the same rate coefficients as the `corresponding elastic transitions'\footnote{\label{fn:trans}The transition of the same $\Delta J$, $\Delta K$, and change in symmetry.} within the ground state. For vibrationally inelastic transitions ($\varv_2=1 \rightarrow 0$), we scale the coefficients of the `corresponding elastic transitions' within the ground state by a constant factor of 0.05 \citep[Faure 2017, priv. comm.;][]{faure2008}, taking detailed balance into account. For all other transitions, i.e. vibrationally elastic (within $\varv_2=1$) or inelastic ($\varv_2=1 \rightarrow 0$) transitions that do not have a `corresponding elastic transition' within the ground vibrational state, we adopt a constant rate coefficient of $2 \times 10^{-11}\,\text{cm}^3\text{s}^{-1}$ for vibrationally elastic transitions and $1 \times 10^{-12}\,\text{cm}^3\text{s}^{-1}$ for vibrationally inelastic transitions.

Our extrapolations in temperature and rovibrational transition are admittedly quite crude but they are necessary to make. Similar to the analysis of \citet[][their Sect. 3]{schmidt2016}, we compare the synthesised spectra of the transitions discussed in this article using the above-mentioned extended coefficients with those using only the coefficients of either \citet{danby1988} or \citet{bouhafs2017}. Neglecting collisional transitions of high-$J$ and $\varv_2$-excited levels, the coefficients of \citet{danby1988} and \citet{bouhafs2017} do not produce qualitatively different spectra. While the spectra of {\ik} are insensitive to the presence of the extended rate coefficients, those of the other targets are noticeably affected, especially in the rotational transitions (${<}50\%$ in the peak intensity). Future quantum dynamics calculations, which require a full-dimensional (12-D) PES of NH$_3$--H$_2$, are necessary to determine the complete set of rate coefficients for circumstellar NH$_3$ modelling.


\subsection{Dust continuum emission}
\label{sec:dust}

The thermal continuum emission from dust is related to the emission and absorption coefficients. In {\ratran}, the absorption coefficient is given by
\begin{equation}
\alpha^\text{dust}_{\nu} = \kappa_\nu \rho_\text{dust} = \kappa_\nu \cdot 2.4\,\text{u} \cdot (\text{d/g}) \cdot n_{\text{H}_2},
\end{equation}
where $\text{u}$ is the atomic mass unit, $\text{d/g}$ is the dust-to-gas mass ratio, and $n_{\text{H}_2}$ is the number density of molecular H$_2$ \citep{ratran}. The code assumes that the gas consists of 80\% H$_2$ and 20\% He by number. The emission coefficient can then be computed with the knowledge of the dust temperature, $T_\text{dust}$,
\begin{equation}
j^\text{dust}_{\nu} = \alpha^\text{dust}_{\nu} B_{\nu} \left( T_\text{dust} \right),
\end{equation}
where $B_{\nu}(T)$ is the Planck function.

We produce the dust temperature profiles of {\ik}, {\vy}, and {\irc} by modelling their spectral energy distributions (SEDs) with the continuum radiative transfer code DUSTY \citep{dustymanual}. We constrain the MIR fluxes of {\ik} by the IRAS\footnote{Infrared Astronomical Satellite \citep{iras1984}.} Low Resolution Spectra corrected by \citet{volk1989} and \citet{cohen1992} and those of {\vy} and {\irc} by the ISO\footnote{Infrared Space Observatory \citep{iso1996}.} SWS\footnote{Short Wavelength Spectrometer \citep{sws1996}.} data processed by \citet{sloan2003}. The FIR fluxes of these three stars are constrained by the ISO LWS\footnote{Long Wavelength Spectrometer \citep{lws1996}.} data processed by \citet{lloyd2003}.

We only modelled the SEDs with cool oxygen-rich interstellar silicates of which the optical constants were computed by \citet{ossenkopf1992}. The comparison of \citet{suh1999} shows that the opacity functions and optical constants of the cool silicates from \citet{ossenkopf1992} are consistent with those of other silicates in the literature. The optical constants of silicates mainly depend on the contents of magnesium and iron in the assumed chemical composition \citep{dorschner1995}. The dust emissivity function (and hence the dust opacity, $\kappa_{\nu}$) of cool silicates from \citet{ossenkopf1992} are readily available in the {\ratran} package, therefore allowing us to employ a dust temperature profile that describes the SED self-consistently. Our SED models are primarily based on that presented in \citet{maercker2008,maercker2016} for {\ik} and those in \citet{shenoy2016} for {\vy} and {\irc}. The derived dust temperature profiles of {\ik} and {\irc} agree very well with the optically thin model for dust temperature \citep{sopka1985},
\begin{equation}
T_\text{dust} (r) = \tstar \left( \frac{\rstar}{2r} \right)^{2/(4+p)}
\label{eq:tdust}
\end{equation}
where $p$ has a typical value of 1 \citep{olofsson2003}.

MIR images of {\oh} by \citet{lagadec2011} have revealed a bright, compact core and an extended halo around the star. The MIR-emitting circumstellar material is dusty \citep{jura2002,matsuura2006}. The MIR emission of {\oh} near the wavelengths of silicate bands extends to FWHM sizes of $2{\farcs}6 \times 4{\farcs}3$ at 9.7\,{\micron} and $5{\farcs}4 \times 9{\farcs}9$ at 18.1\,{\micron} \citep{lagadec2011}, covering most if not all of the NH$_3$-emitting region. \citet{jura2002} were able to explain the MIR emission from the compact core at 11.7\,{\micron} and 17.9\,{\micron} by an opaque, flared disc with an outer radius of $0{\farcs}25$ and outer temperature of 130\,K. This temperature is about 45\% of that expected from the optically thin dust temperature model as shown in Eq. (\ref{eq:tdust}). For simplicity, therefore, we scale Eq. (\ref{eq:tdust}) by a factor of 0.45 and adopt $p=1$ as the dust temperature profile of {\oh} in the NH$_3$-emitting region, which is beyond the opaque disc model of \citet{jura2002}.


\section{Results and discussion}
\label{sec:results}

\begin{figure*}[!htbp]
\centering
\includegraphics[trim=0.0cm 0.2cm 0.0cm 0.2cm, clip, width=\irtfwidth]{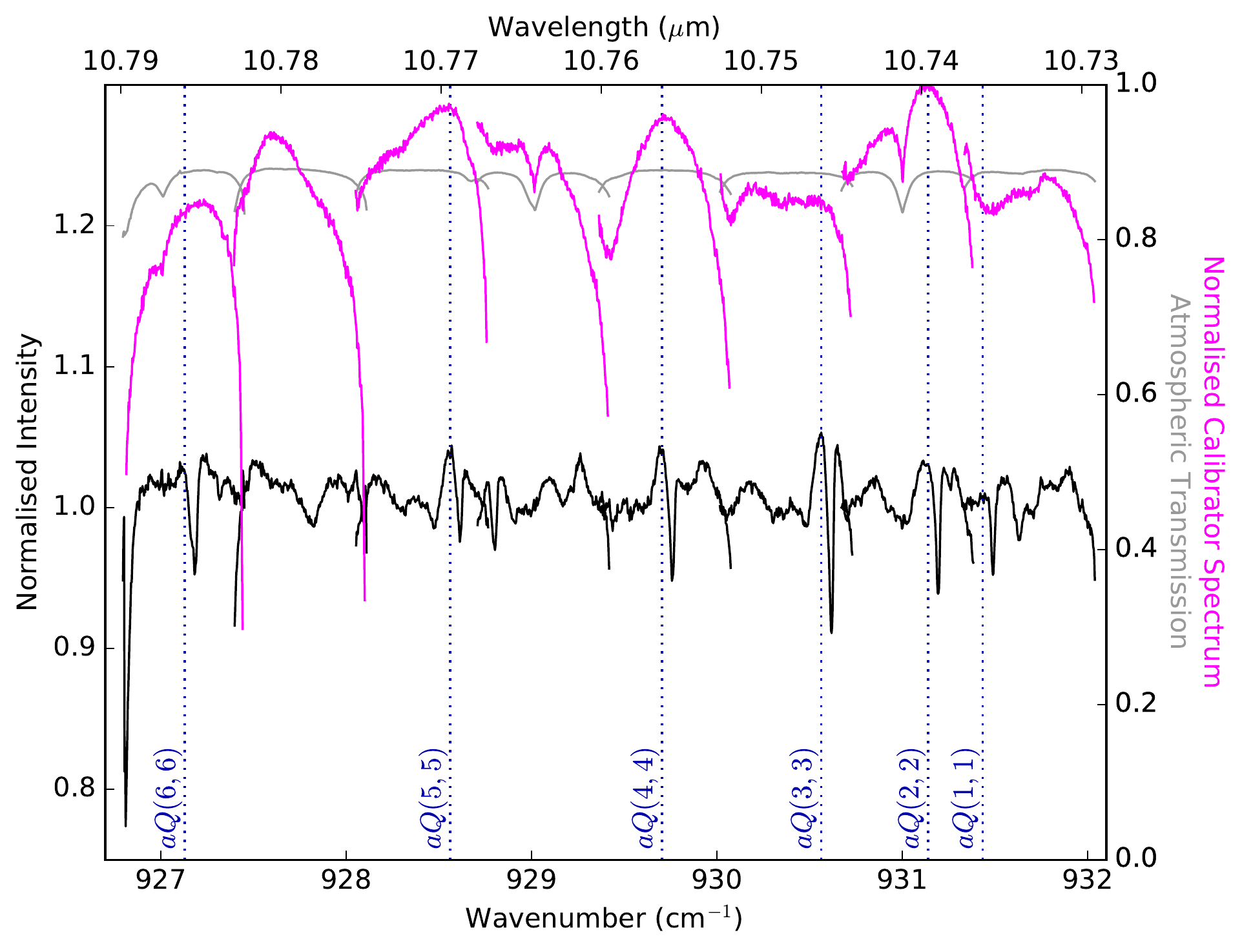}
\includegraphics[trim=0.0cm 0.2cm 0.0cm 0.2cm, clip, width=\irtfwidth]{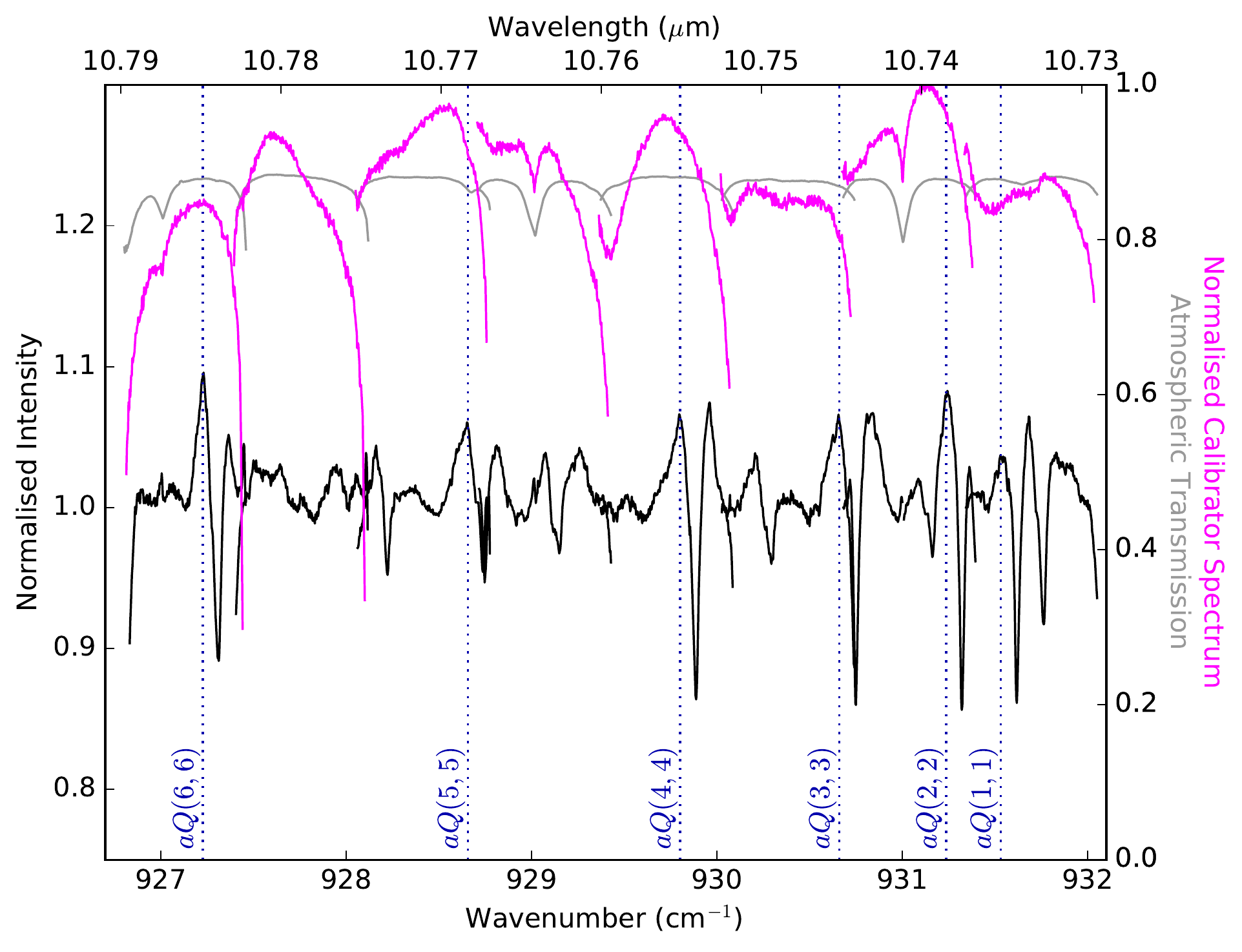} \\
\includegraphics[trim=0.0cm 0.2cm 0.0cm 0.2cm, clip, width=\irtfwidth]{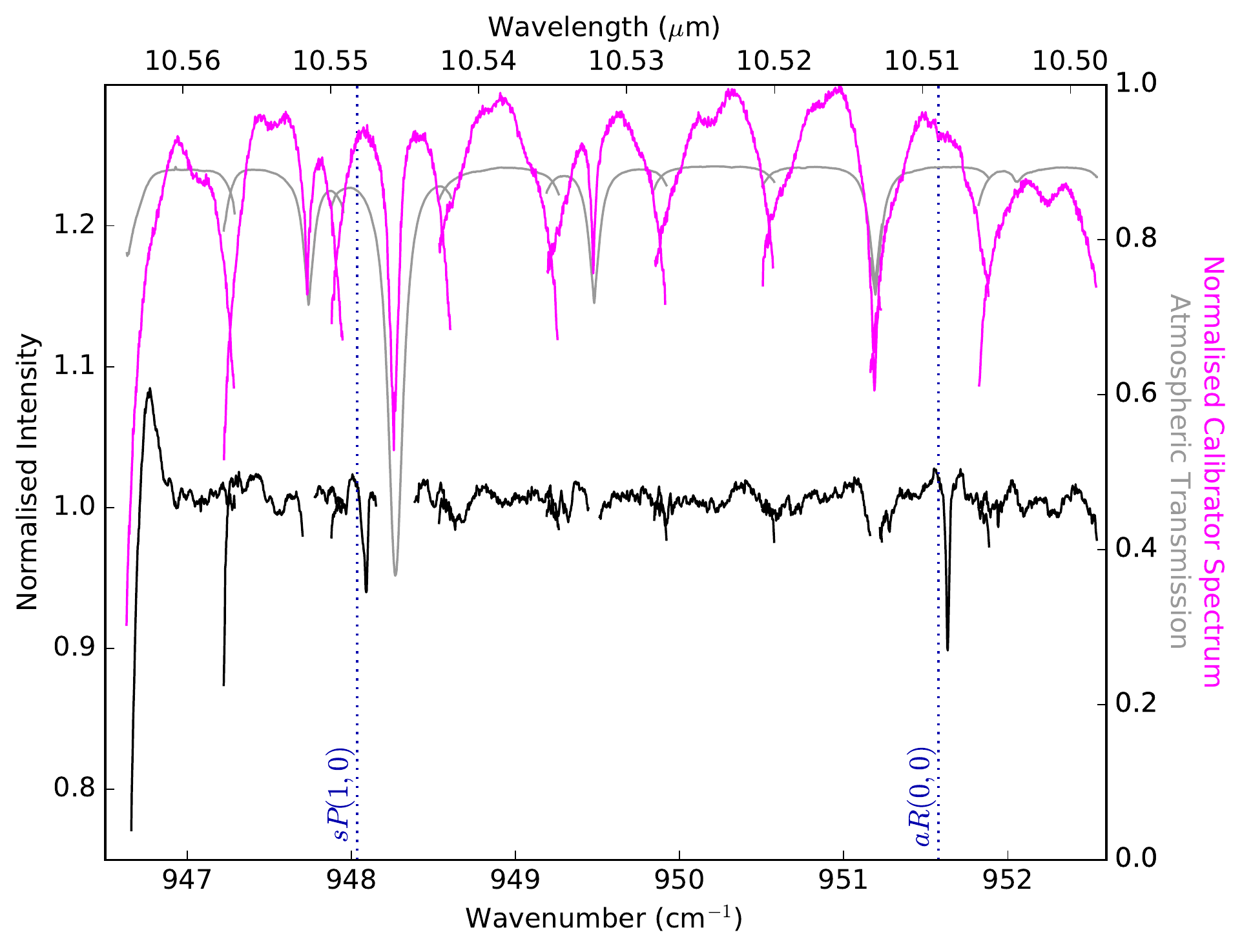}
\includegraphics[trim=0.0cm 0.2cm 0.0cm 0.2cm, clip, width=\irtfwidth]{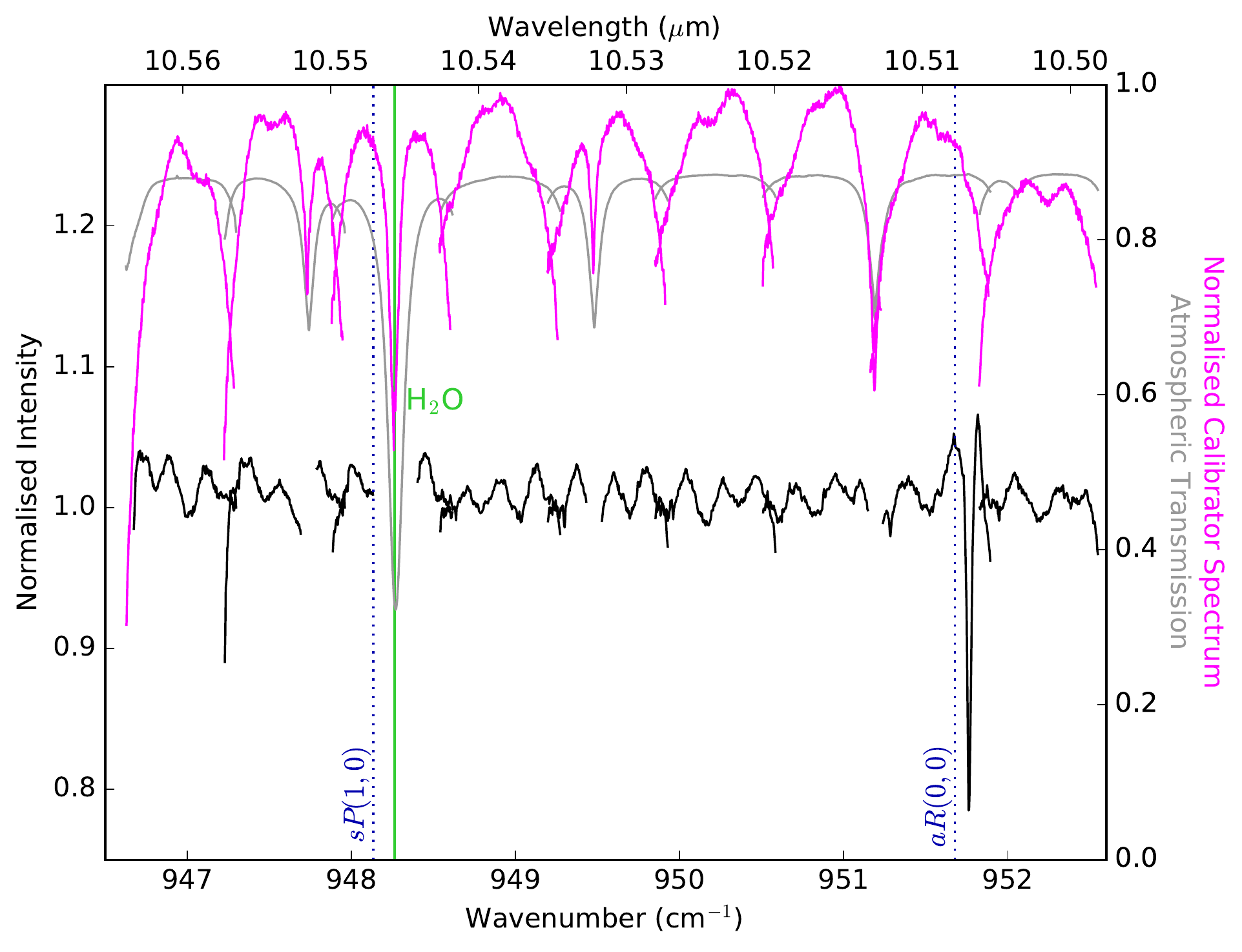} \\
\includegraphics[trim=0.0cm 0.2cm 0.0cm 0.2cm, clip, width=\irtfwidth]{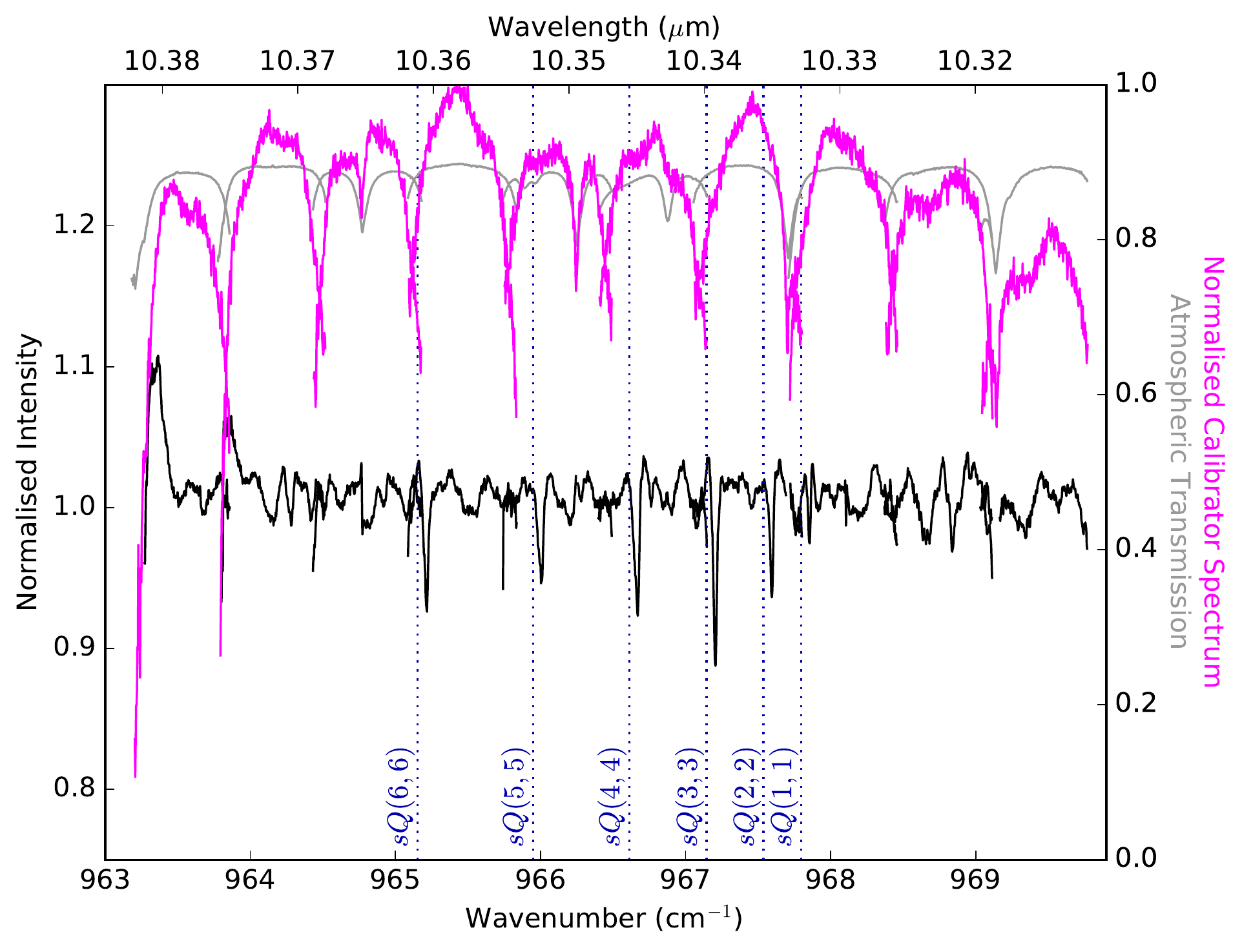}
\includegraphics[trim=0.0cm 0.2cm 0.0cm 0.2cm, clip, width=\irtfwidth]{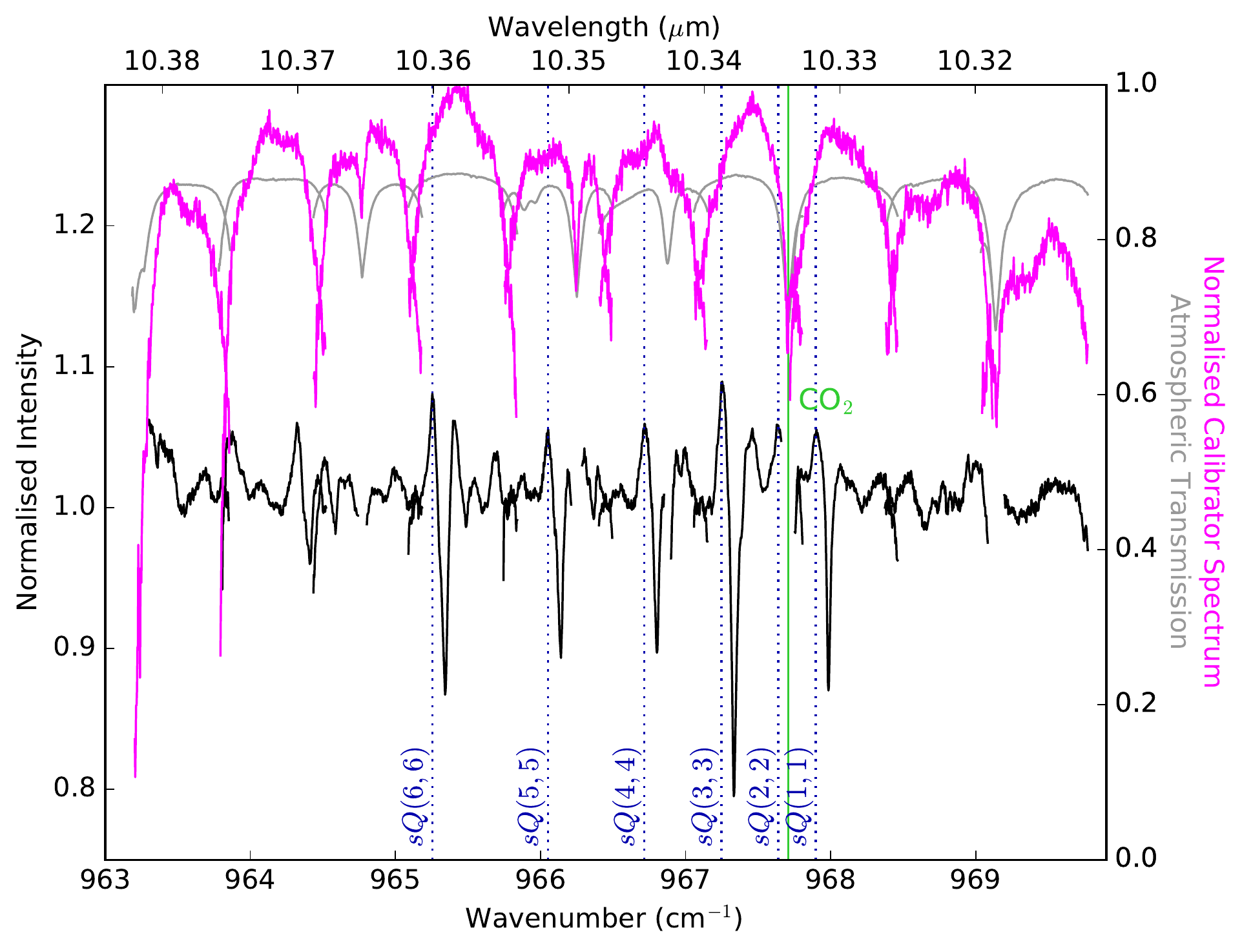}
\caption[]{NASA IRTF/TEXES spectra (in black) of {\ik} (\textit{left}) and {\vy} (\textit{right}) near the wavelengths (wavenumbers) of 10.7\,{\micron} (930\,{\percm}; \textit{top}), 10.5\,{\micron} (950\,{\percm}; \textit{middle}), and 10.3\,{\micron} (966\,{\percm}; \textit{bottom}). The wavenumber axes are in the topocentric frame of reference. The target spectra are normalised and zoomed. The normalised calibrator spectra are plotted in magenta and the fractional atmospheric transmissions are shown in grey. Only the target data with atmospheric transmission greater than 0.8 are plotted. The dark blue dotted lines in each spectrum indicate the positions of the labelled NH$_3$ rovibrational transitions in the stellar rest frame and the green solid lines indicate the telluric lines (in the observatory frame) that blend with the NH$_3$ lines of {\vy}. We assume the systemic velocities of {\ik} and {\vy} to be $34{\kms}$ and $22{\kms}$, respectively.}
\label{fig:irtf_all}
\end{figure*}

With the VLA, we detect all covered metastable ($J=K$) inversion transitions from all four targets, including $(J,K)=(1,1)$ to $(4,4)$ for {\ik}, $(1,1)$ to $(6,6)$ for {\vy} and {\oh}, and $(1,1)$ and $(2,2)$ for {\irc}. However, no non-metastable inversion emission from {\ik} is securely detected. Ground-state rotational line emission is also detected in all available {\hifi} spectra. These include up to $J=3$--$2$ towards {\ik}, {\vy}, and {\oh}; and up to $J=2$--$1$ towards {\irc}. We do not detect the 140-GHz rotational transition in the $\varv_2=1$ state towards {\ik} from the IRAM 30m observations. In the MIR spectra, all 14 rovibrational transitions from {\ik} and 12 unblended transitions from {\vy} are detected in absorption. Figure \ref{fig:irtf_all} shows the full IRTF spectra of both stars in three wavelength settings. There are no strong emission components in these spectra. We note that, however, the spectra are susceptible to imperfect telluric line removal and baseline subtraction as mentioned in Sect. \ref{sec:irtf}.

In Sects. \ref{sec:results_ik}--\ref{sec:results_irc}, we shall discuss the observational and modelling results for each target separately; in Sect. \ref{sec:results_chem}, we summarise the modelling results and discuss the possible origin of circumstellar ammonia. We made use of the \texttt{IPython} shell\footnote{\url{http://ipython.org/}} \citep[version 5.1.0;][]{ipython} and the Python libraries including \texttt{NumPy}\footnote{\url{http://www.numpy.org/}} \citep[version 1.11.2;][]{numpy}, \texttt{SciPy}\footnote{\url{http://www.scipy.org/}} \citep[version 0.18.1;][]{scipy}, \texttt{matplotlib}\footnote{\url{http://matplotlib.org/}} \citep[version 1.5.3;][]{matplotlib,matplotlib_v153}, and \texttt{Astropy}\footnote{\url{http://www.astropy.org/}} \citep[version 1.3;][]{astropy} in computing and plotting the results in this article.


\subsection{{\ik}}
\label{sec:results_ik}


\subsubsection{VLA observations of {\ik}}
\label{sec:vla_ik}

\begin{figure*}[!htbp]
\centering
\includegraphics[height=\jvlaheight]{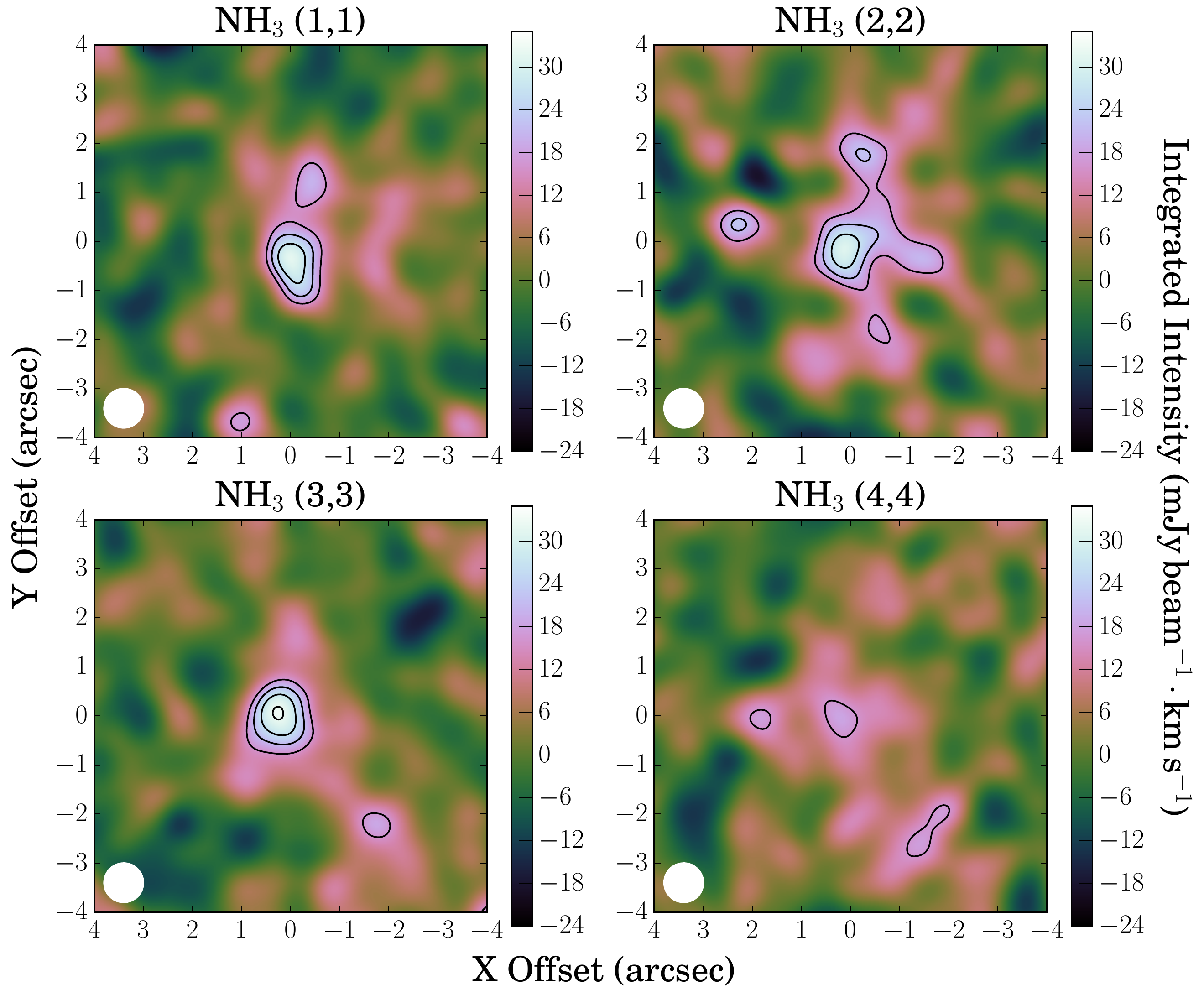}
\caption[]{Integrated intensity maps of the four lowest NH$_3$ inversion lines from {\ik} as observed with the VLA in B and C configurations over the LSR velocity range of $[13, 55]{\kms}$ (relative velocity between ${\pm}21{\kms}$ to the systemic). Horizontal and vertical axes represent the offsets (arcsec) relative to the fitted centre of the continuum in the directions of right ascension and declination, respectively. The circular restoring beam of $0{\farcs}83$ is indicated at the bottom-left corner in each map. The contour levels for the NH$_3$ lines are drawn at $3, 4, 5, 6\sigma$, where $\sigma = 5.4\intfluxunits$.}
\label{fig:vla_ik_moment}
\end{figure*}

\begin{figure*}[!htbp]
\centering
\includegraphics[trim=0.5cm 2.3cm 1.5cm 1.5cm, clip, width=\specwidth]{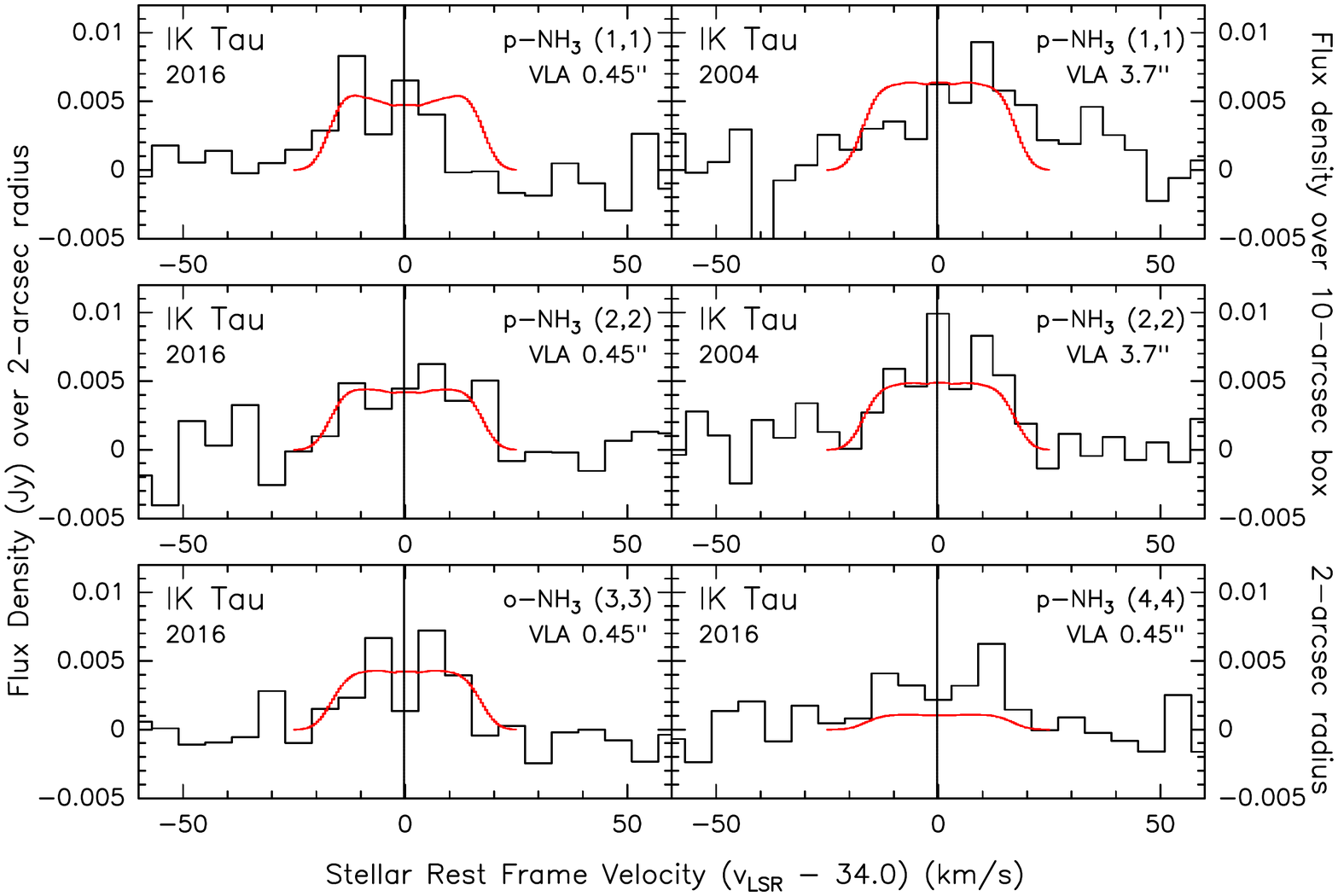}
\caption[]{NH$_3$ inversion line spectra of {\ik}. Black lines show the VLA spectra in flux density and red lines show the modelled spectra. NH$_3$ (1,1) and (2,2) have been observed in 2004 (\textit{top right} and \textit{middle right}) with D configuration and 2016 (\textit{top left} and \textit{middle left}) with both B and C configurations. The 2004 spectra were integrated over a $10{\arcsec} \times 10{\arcsec}$ box and the 2016 spectra were integrated over a $2{\arcsec}$-radius circle.}
\label{fig:model_ik_inv}
\end{figure*}

Figure \ref{fig:vla_ik_moment} shows the integrated intensity maps of all the four metastable inversion lines ($J=K$) over the LSR velocity range between $34 \pm 21{\kms}$. These images combined the data from our 2016 observations with the B and C arrays of the VLA. The native resolution was about $0{\farcs}45$ in FWHM and we smoothed the images to the final resolution of $0{\farcs}83$ for better visualisation. NH$_3$ emission is detected at rather low S/N ratio (${\lesssim}6$). The images from the D-array observations in 2004 do not resolve the NH$_3$ emission and are not presented. Figure \ref{fig:model_ik_inv} shows the flux density spectra, integrated over the regions specified in the vertical axes, of all VLA observations.

Comparing the NH$_3$ $(1,1)$ (top row) and $(2,2)$ (middle row) spectra taken at two epochs, the $(1,1)$ line emission in 2004 was predominantly in the redshifted velocities while that in 2016 was mainly blueshifted. This may indicate temporal variations in the NH$_3$ emission at certain LSR velocities, but is not conclusive because of the low S/N ratios of the spectra. The total fluxes of the 2004 spectra within a $10{\arcsec} \times 10{\arcsec}$ box are consistent with those of the 2016 spectra within a 2{\arcsec}-radius aperture centred at the stellar radio continuum. Hence, most of the NH$_3$ inversion line emission is confined within a radius of ${\sim}2{\arcsec}$. In the full-resolution images, the strongest emission is generally concentrated within ${\sim}0{\farcs}7$ from the star -- but not exactly at the continuum position -- and appear to be very clumpy. Individual channel maps show localised, outlying emission up to ${\sim}2{\arcsec}$ from the centre.


\subsubsection{{\hifi} observations of {\ik}}
\label{sec:hifi_ik}

\begin{figure*}[!htbp]
\centering
\includegraphics[trim=0.5cm 2.3cm 1.5cm 1.5cm, clip, width=\specwidth]{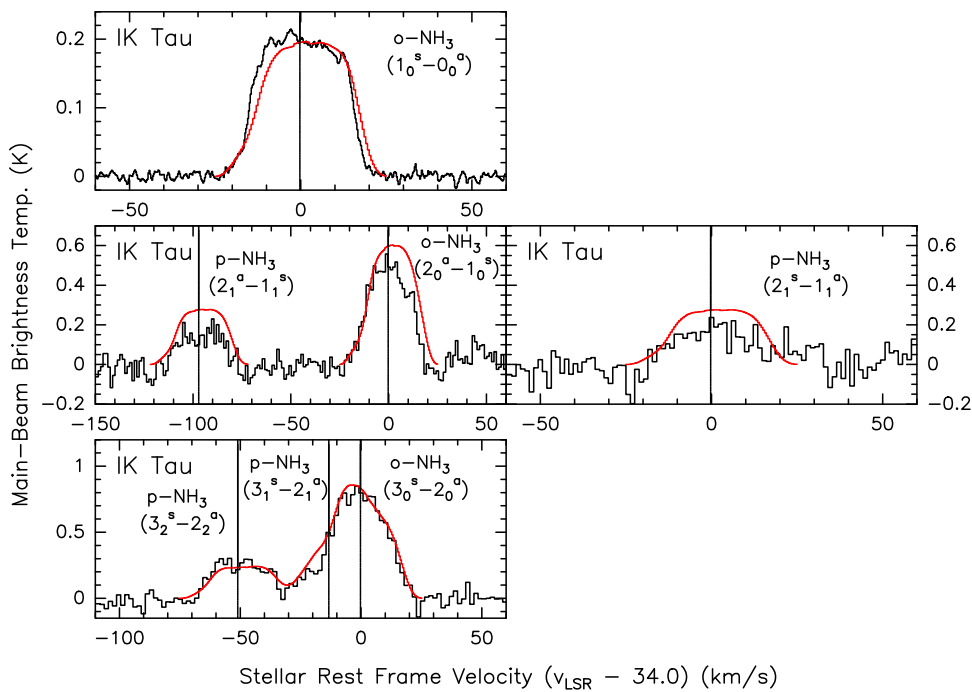}
\caption[]{NH$_3$ rotational line spectra of {\ik}. Black lines show the observed spectra from {\hifi} and red lines show the modelled spectra.}
\label{fig:model_ik_rot}
\end{figure*}

The {\hifi} spectra of {\ik} are shown in Fig. \ref{fig:model_ik_rot}. Except for the blended transitions $J_K=3_0$--$2_0$ and $3_1$--$2_1$, the rotational spectra appear to be generally symmetric with a weak blueshifted wing near the relative velocity between $-21$ and $-16{\kms}$ from the systemic. This weakness of the blueshifted wing may be a result of weak self-absorption. The central part of the profiles within the relative velocity of ${\pm}10{\kms}$ is relatively flat. Since the NH$_3$-emitting region is unresolved within the {\hifi} beams, the flat-topped spectra suggest that the emission only has a small or modest optical thickness. In the high-S/N spectrum of $1_0$--$0_0$, the central part also shows numerous small bumps, suggesting the presence of substructures of various kinematics in the NH$_3$-emitting regions. Since the line widths of NH$_3$ spectra resemble those from other major molecular species \citep[e.g. CO, HCN;][]{decin2010}, the bulk of NH$_3$ emission comes from the fully accelerated wind in the outer CSE with a maximum expansion velocity of $17.7{\kms}$ \citep{decin2010}.

\subsubsection{IRAM observations of {\ik}}
\label{sec:iram_ik}

We did not detect the rotational transition $\varv_2=1$ $J_K=2_1$--$1_1$ near 140\,GHz within the uncertainty of 0.61\,mK per 5.0{\kms}. If we assume the line-emitting region of this transition is also located in the fully accelerated wind with an expansion velocity of $\vexp=17.7{\kms}$, as for the ground-state rotational transitions, then the rms noise of the velocity-integrated spectrum would be ${\sim}0.24\,\text{mK}$ (over $35.4{\kms}$) in $T_\text{mb}$.

Our observations repeated the detection of PN near 141.0\,GHz \citep{debeck2013} and H$_2$CO near 150.5\,GHz \citep{vp2017} at higher S/N ratios. We also detected NO near 150.18 and 150.55\,GHz that were not reported in \citet{vp2017}, in which the higher NO transitions near 250.5\,GHz were detected. In the context of circumstellar environments and stellar merger remnants, NO has only been detected elsewhere towards {\irc} \citep{ql2013}, {\oh} \citep{vp2015}, and {\ck} \citep[Nova Vul 1670;][]{kaminski2017b} thus far.


\subsubsection{IRTF observations of {\ik}}
\label{sec:irtf_ik}

\begin{figure*}[!htbp]
\centering
\includegraphics[trim=0.5cm 2.3cm 1.5cm 1.5cm, clip, width=\specwidth]{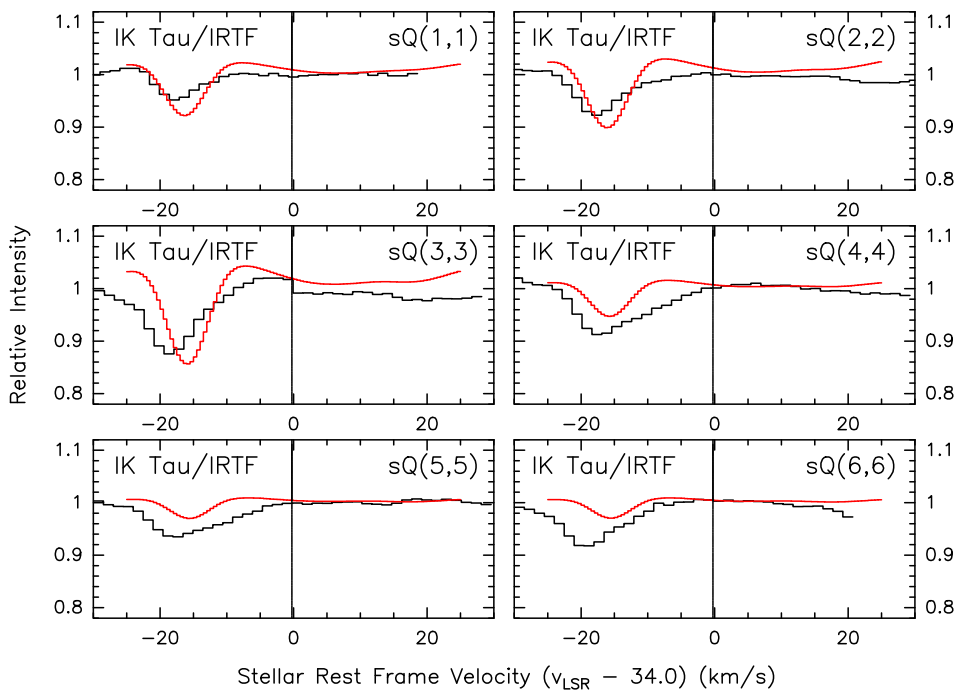}
\caption[]{$Q$-branch NH$_3$ rovibrational spectra of {\ik}. These transitions connects the lower energy levels of the ground-state inversion doublets and the upper levels in the vibrationally excited state. Black lines show the observed spectra from the NASA IRTF and red lines show the modelled spectra. Each modelled spectrum was divided by a smoothed version of itself as in the real data (see Sect. \ref{sec:irtf}) in order to induce the similar distortion in the line wings.}
\label{fig:model_ik_mir1}
\end{figure*}

\begin{figure*}[!htbp]
\centering
\includegraphics[trim=0.5cm 13cm 1.5cm 1.5cm, clip, width=\specwidth]{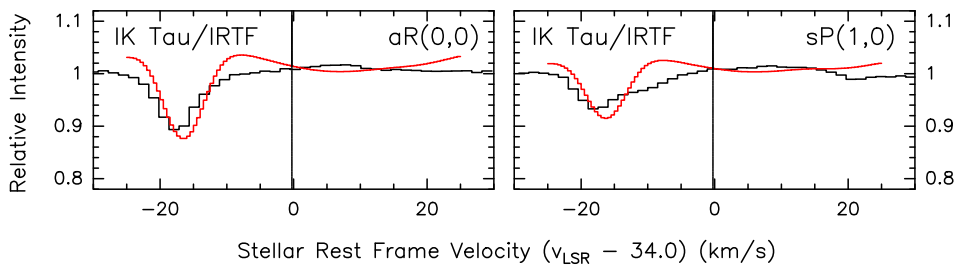}
\caption[]{NH$_3$ rovibrational spectra of {\ik} in the transitions $aR(0,0)$ and $sP(1,0)$. Black lines show the observed spectra from the NASA IRTF and red lines show the modelled spectra. Each modelled spectrum was divided by a smoothed version of itself as in the real data (see Sect. \ref{sec:irtf}) in order to induce the similar distortion in the line wings.}
\label{fig:model_ik_mir2}
\end{figure*}

\begin{figure*}[!htbp]
\centering
\includegraphics[trim=0.5cm 2.3cm 1.5cm 1.5cm, clip, width=\specwidth]{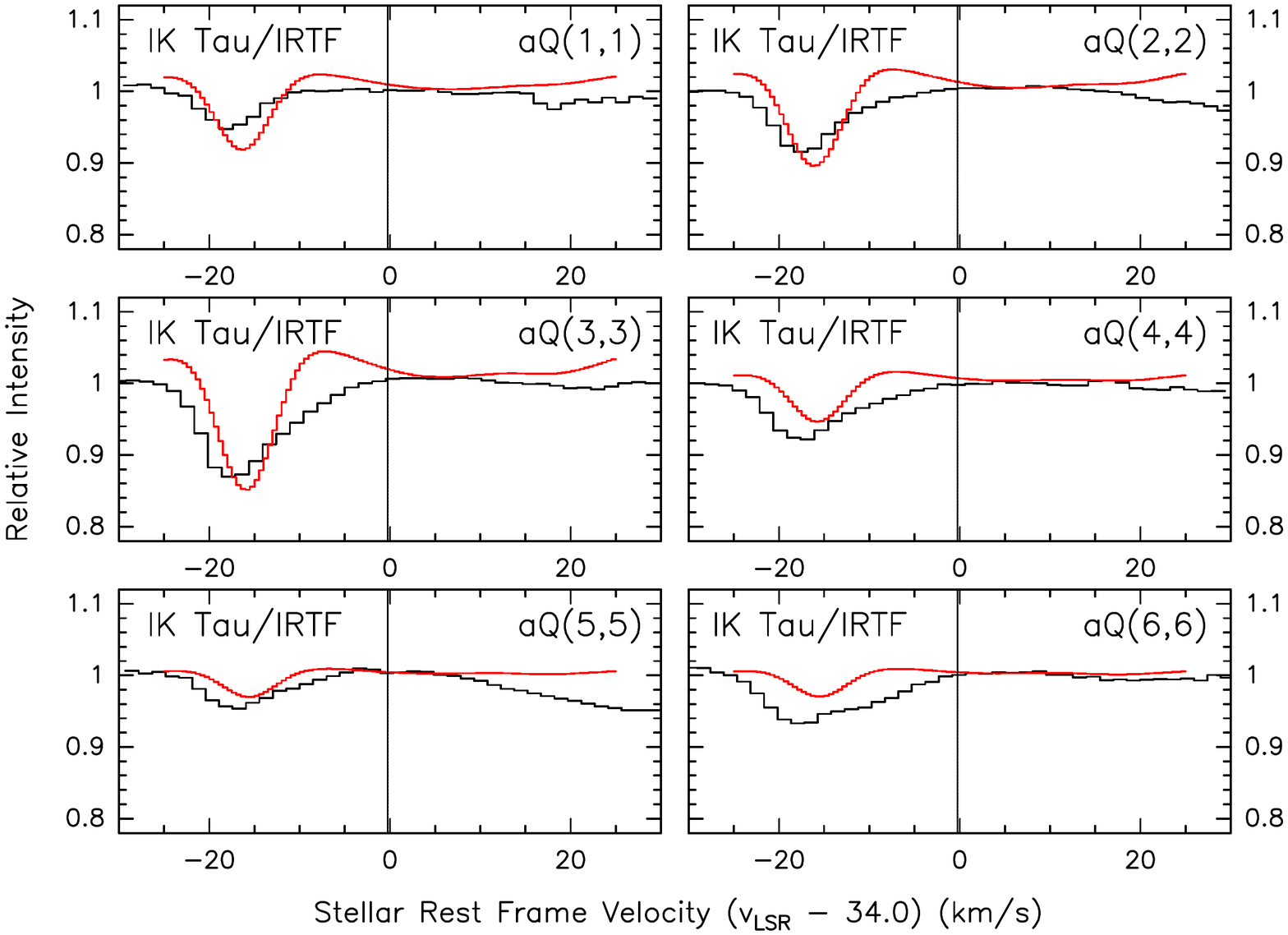}
\caption[]{$Q$-branch NH$_3$ rovibrational spectra of {\ik}. These transitions connects the upper energy levels of the ground-state inversion doublets and the lower levels in the vibrationally excited state. Black lines show the observed spectra from the NASA IRTF and red lines show the modelled spectra. Each modelled spectrum was divided by a smoothed version of itself as in the real data (see Sect. \ref{sec:irtf}) in order to induce the similar distortion in the line wings.}
\label{fig:model_ik_mir3}
\end{figure*}

Figure \ref{fig:irtf_all} shows the normalised IRTF spectra of {\ik} in three wavelength settings prior to baseline subtraction. All 14 targeted NH$_3$ rovibrational transitions are clearly detected in absorption. Figures \ref{fig:model_ik_mir1}--\ref{fig:model_ik_mir3} present the baseline-subtracted spectra in the stellar rest frame. The maximum absorption of all MIR transitions appear near the LSR velocities in the range of $16$--$17{\kms}$, which corresponds to the blueshifted velocity between $-18$ and $-17{\kms}$ and is consistent with the terminal expansion velocity of $17.7{\kms}$ as determined by \citet{decin2010}. Blueshifted absorption near terminal velocity is consistent with the scenario that the NH$_3$-absorbing gas is moving towards us in the fully-accelerated outflow in front of the background MIR continuum.

The MIR NH$_3$ absorption in {\ik} cannot come from the inner wind. First of all, our SED modelling (Sect. \ref{sec:dust}) shows that the optical depth of the circumstellar dust is about 1.3 at 11\,{\micron}. Continuum emission from dust grains would probably fill the line absorption near 10.5\,{\micron} had it originated beneath the dust shells. Furthermore, the visual phase of {\ik} in our IRTF observations is about 0.35, which is based on our recalculations of its visual light curve using modern photometric data (see Fig. \ref{fig:iktau_v_phase} in Appendix \ref{sec:app_iktau}). Mira variables at this phase, including {\ik} \citep{hinkle1997}, show CO second overtone ($\Delta \varv=3$) absorption, which sample the deep pulsating layers down to ${\sim}1\,\rstar$, close to the stellar systemic velocity or in the redshifted (infall) velocities \citep{lebzelter2002,nowotny2010,liljegren2017}. If NH$_3$ was produced in high abundance close to the stellar surface, we would see the absorption features in much redder (more positive) velocities; if it was abundant at the distance of a few stellar radii from the star, the observed velocity blueshifted by about $18{\kms}$ from the systemic would be too high for the gas in the inner wind of Mira variables \citep[e.g.][]{hofner2016,wong2016}. As a result, we only consider NH$_3$ in the dust-accelerated regions of the CSE in our subsequent modelling (Sect. \ref{sec:mdl_ik}) even though we cannot exclude the presence of NH$_3$ gas in the inner wind based on the available MIR spectra.

Among the same series of rotationally elastic transitions ($aQ$ or $sQ$), the absorption profiles are slightly broader in higher-$J$ transitions. For NH$_3$, the thermal component of the line broadening is below $1{\kms}$ when $T_{\text{NH}_3}$ is lower than $1000\,\text{K}$, so the observed line widths of several up to ${\sim}10{\kms}$ are mainly contributed by the microturbulent velocity of the circumstellar gas. This suggests that the higher-$J$ levels are more readily populated in more turbulent parts of the CSE.


\subsubsection{NH$_3$ model of {\ik}}
\label{sec:mdl_ik}

\citet{decin2010} presented detailed non-LTE radiative transfer analysis of molecular line emission from the CSE of {\ik}. In particular, the gas density and thermodynamic structure, including the gas kinetic temperature and expansion velocity, of the CSE have been constrained by modelling multiple CO and HCN spectra observed by single-dish telescopes. Therefore, we adopt the same density, kinetic temperature and expansion velocity profiles as in model 3 of \citet{decin2010}. The gas density of the CSE is given by the conservation of mass in a uniformly expanding outflow, 
\begin{equation}
n_{\text{H}_2}(r) = \frac{\mdot}{4\pi r^2 V(r)},
\label{eq:dens}
\end{equation}
where $\mdot = 8 \times 10^{-6}\mspy$ is the average gas mass-loss rate, $V(r)$ is the expansion velocity at radius $r$. We use the classical $\beta$-law to describe the radial velocity field, i.e.
\begin{equation}
\label{eq:velo}
V(r) = \vin + \left( \vexp - \vin \right) \left( 1-\frac{R_\text{wind}}{r} \right)^{\beta},
\end{equation}
where $\vin = 3{\kms}$ is the inner wind velocity, $\vexp = 17.7{\kms}$ is the terminal expansion velocity, $R_\text{wind} = 12\,\rstar$ is the radius from which wind acceleration begins, and $\beta=1$ \citep{decin2010}.

The gas kinetic temperature is described by two power laws which approximate the profile computed from energy balance equation \citep{decin2010},
\begin{equation}
T_\text{kin}(r) =
\begin{cases}
40\,\text{K} \cdot {r_{16}}^{-0.68} & \text{if $r \le 10^{16}\,\text{cm} = 667\,\rstar \approx 2{\farcs}5$} \\
40\,\text{K} \cdot {r_{16}}^{-0.8} & \text{if $r > 10^{16}\,\text{cm}$} \\
\end{cases},
\end{equation}
where
\begin{equation}
\displaystyle {r_{16}} = \frac{r}{10^{16}\,\text{cm}}.
\end{equation}

We parametrise the radial distribution of NH$_3$ abundance with a Gaussian function, i.e.
\begin{equation}
\displaystyle f(r) = f_0 \exp\left[-\left(\frac{r}{\refd}\right)^{2}\right] \quad \text{if $r > \rin$},
\label{eq:abun}
\end{equation}
where $f(r)$ and $f_0$ are the fractional abundance of NH$_3$ relative to H$_2$ at radius $r$ and at its maximum value, respectively; $\refd$ is the $e$-folding radius at which the NH$_3$ abundance drops to $1/e$ of the maximum value presumably due to photodissociation. We did not parametrise the formation of NH$_3$ in the inner CSE with a smoothly increasing function of radius. Instead, we introduce an inner radius, {\rin}, for the distribution of NH$_3$ at which its abundance increases sharply from the `default' value of $10^{-12}$ to the maximum. We assume the dust-to-gas mass ratio to be $\text{d/g}=0.002$, a value typical for {\ik} and also AGB stars in general \citep[e.g.][]{gobrecht2016}. In order to reproduce the broad IRTF rovibrational spectra (Sect. \ref{sec:irtf_ik}), we have to adopt a high Doppler-$b$ parameter of $\bdopp = 4\kms$.

We are able to fit the spectra with $\rin=75\,\rstar=0{\farcs}3$, $\refd=600\,\rstar=2{\farcs}3$, and $f_0=6 \times 10^{-7}$. Both choices of {\rin} and {\refd} are motivated by the inner and outer radii of the observed inversion line emission from the VLA images (Sect. \ref{sec:vla_ik}). We note, however, that a smaller {\refd} (${\sim}500\,\rstar=1{\farcs}9$) and a higher $f_0$ (${\sim}7 \times 10^{-7}$) can also reproduce fits of similar quality to the available spectra. On the other hand, a very small {\refd} would reduce the line intensity ratios between higher and lower rotational transitions because the lower transitions come from the outer part of the NH$_3$ distribution.

\begin{figure*}[!htbp]
\centering
\includegraphics[height=\profheight]{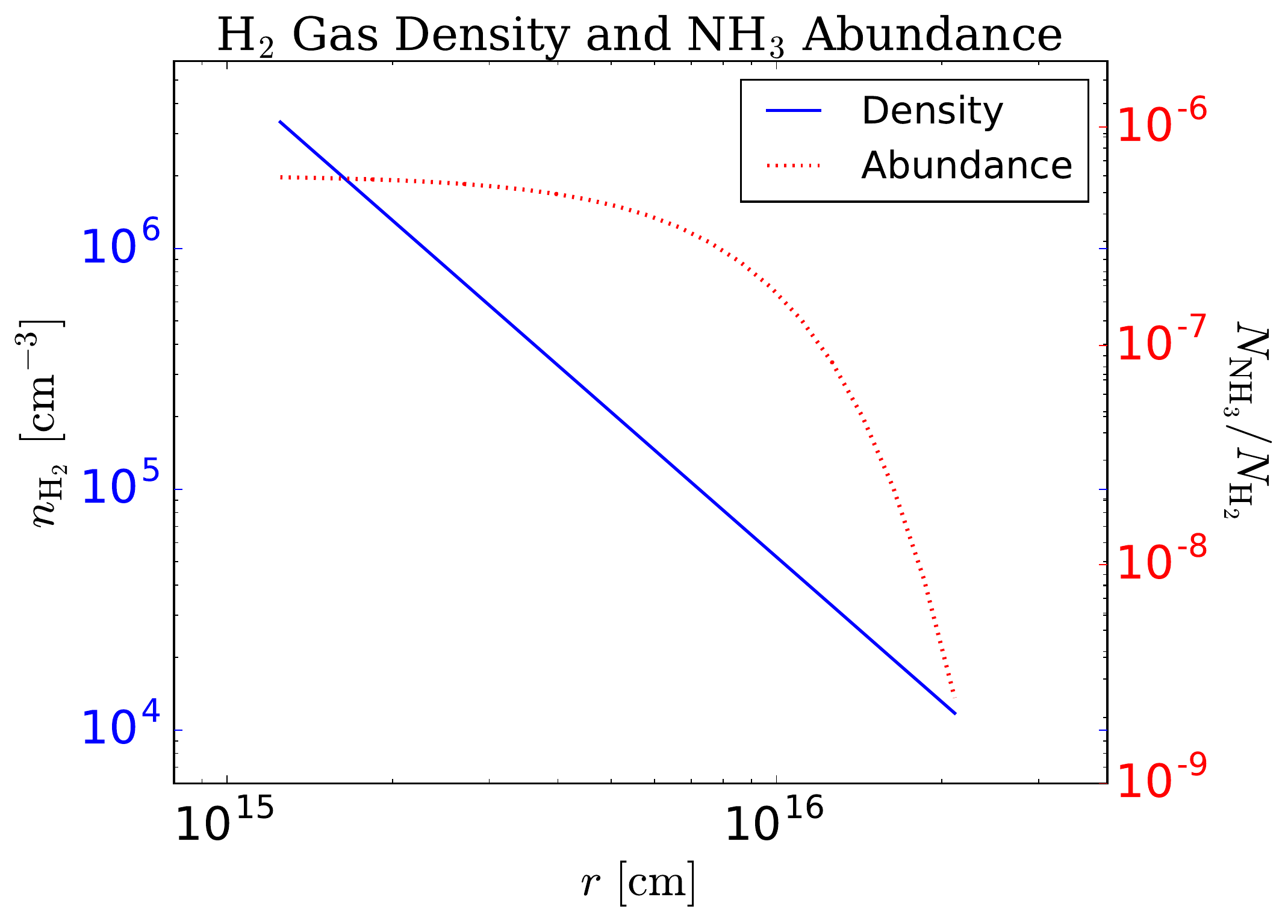}
\includegraphics[height=\profheight]{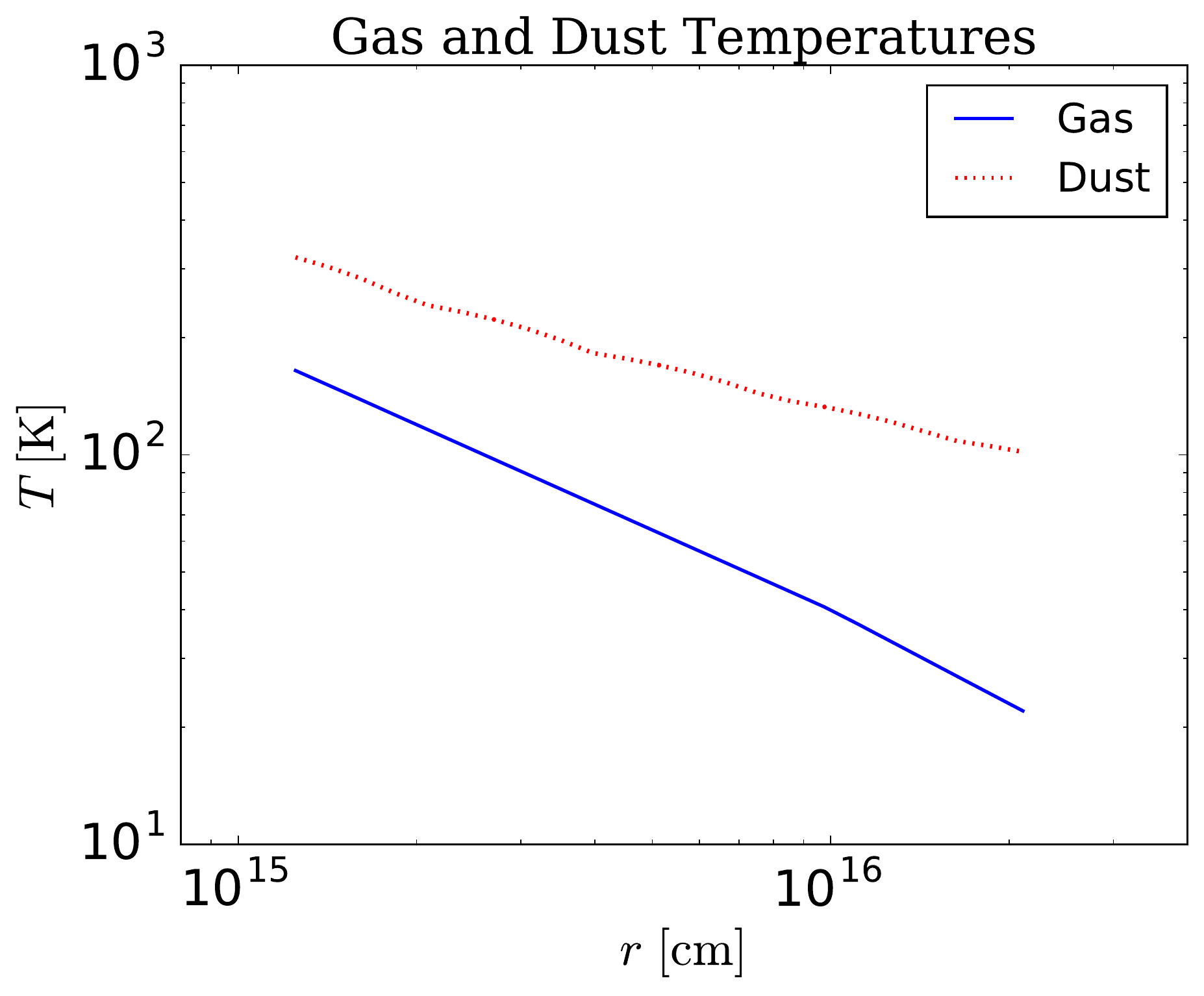}
\includegraphics[height=\profheight]{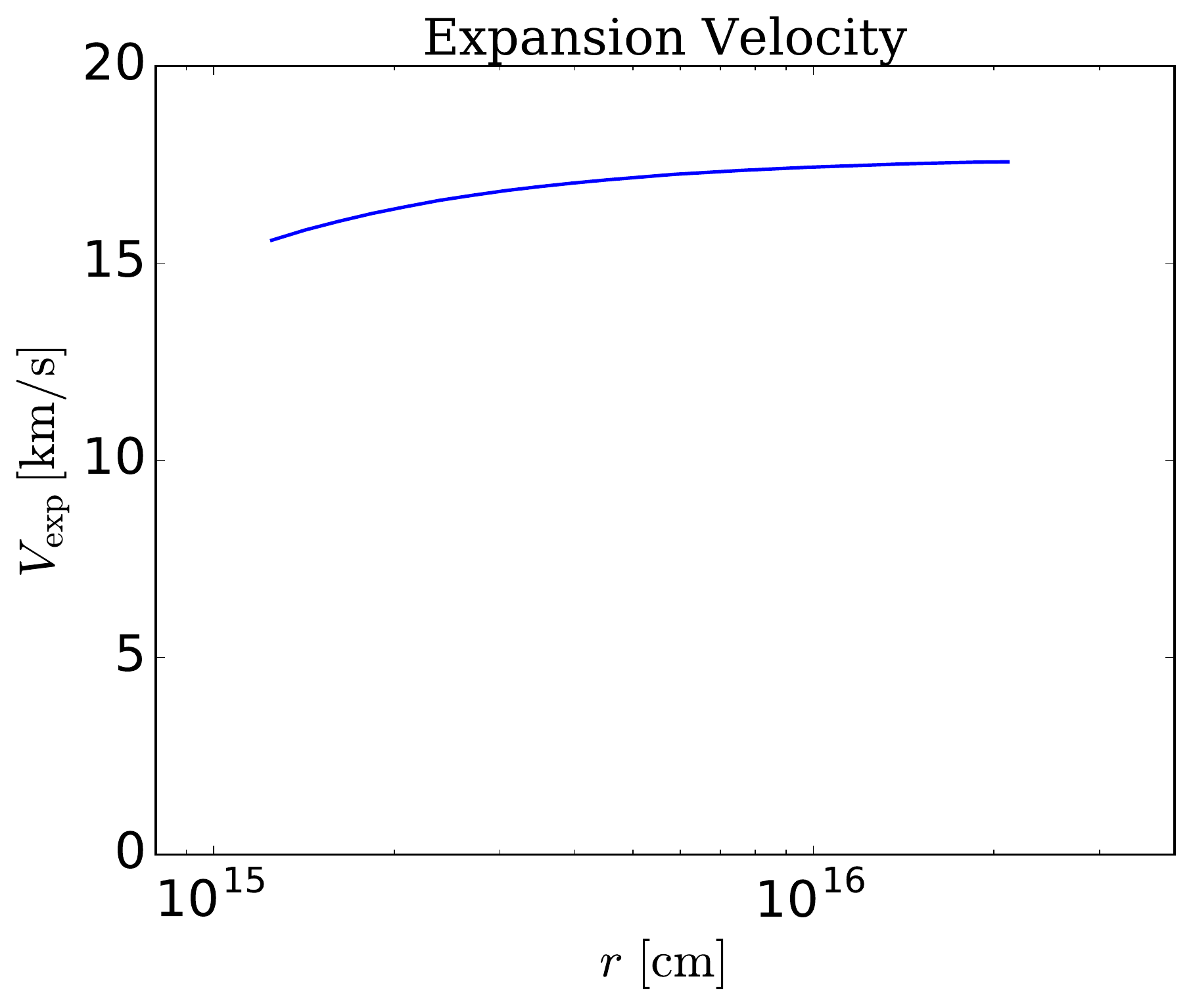}
\caption[]{The model for {\ik}. \textit{Left}: Input gas density and NH$_3$ abundance profiles. Only the radii with abundance ${>}10^{-10}$ are plotted. \textit{Middle}: Input gas and dust temperature profiles. \textit{Right}: Input expansion velocity profile.}
\label{fig:model_ik_prof}
\end{figure*}

The inner radius is not tightly constrained because the small volume of the line-emitting region only contributes to a relatively small fraction of the emission. Although the predicted intensities of the rotational transitions within the $\varv_2=1$ state (near 140.1 and 466.2\,GHz) are sensitive to the adopted value of {\rin} in our test models, they are very weak compared to the noise levels achievable by (sub)millimetre telescopes with the typical amount of observing time. In the presented model of {\ik}, the peak brightness temperature of the $\varv_2=1$ $J_K=2_1$--$1_1$ transition is ${\sim}0.03\,\text{mK}$, much lower than the rms noise of our observation (Sect. \ref{sec:iram_ik}). High rotational transitions from $J=4$--$3$ (${\sim}125$\,{\micron}; 2.4\,THz) up to $J=9$--$8$ (${\sim}57$\,{\micron}; 5.3\,THz) were covered by the {\pacs}\footnote{The Photodetector Array Camera and Spectrometer \citep{poglitsch2010}.} instrument. Most of them were not detected except for $J=4$--$3$. We did not include the PACS data\footnote{{\herschel} OBSIDs: 1342203679, 1342203680, 1342203681, 1342238942.} in our detailed modelling because the spectral resolutions are too coarse for detailed analysis. The predicted total flux of the blended $J=4$--$3$ lines by our model is consistent, within 30\%, with the PACS spectra. In this study, we can only conclude that the inner boundary of the NH$_3$ distribution is ${\lesssim}100\,\rstar$. Figure \ref{fig:model_ik_prof} shows the radial profiles of the input density, abundance, dust and gas temperatures, and expansion velocity.

\subsection{{\vy}}
\label{sec:results_vy}


\subsubsection{VLA observations of {\vy}}
\label{sec:vla_vy}

\begin{figure*}[!htbp]
\centering
\includegraphics[height=\jvlaheight]{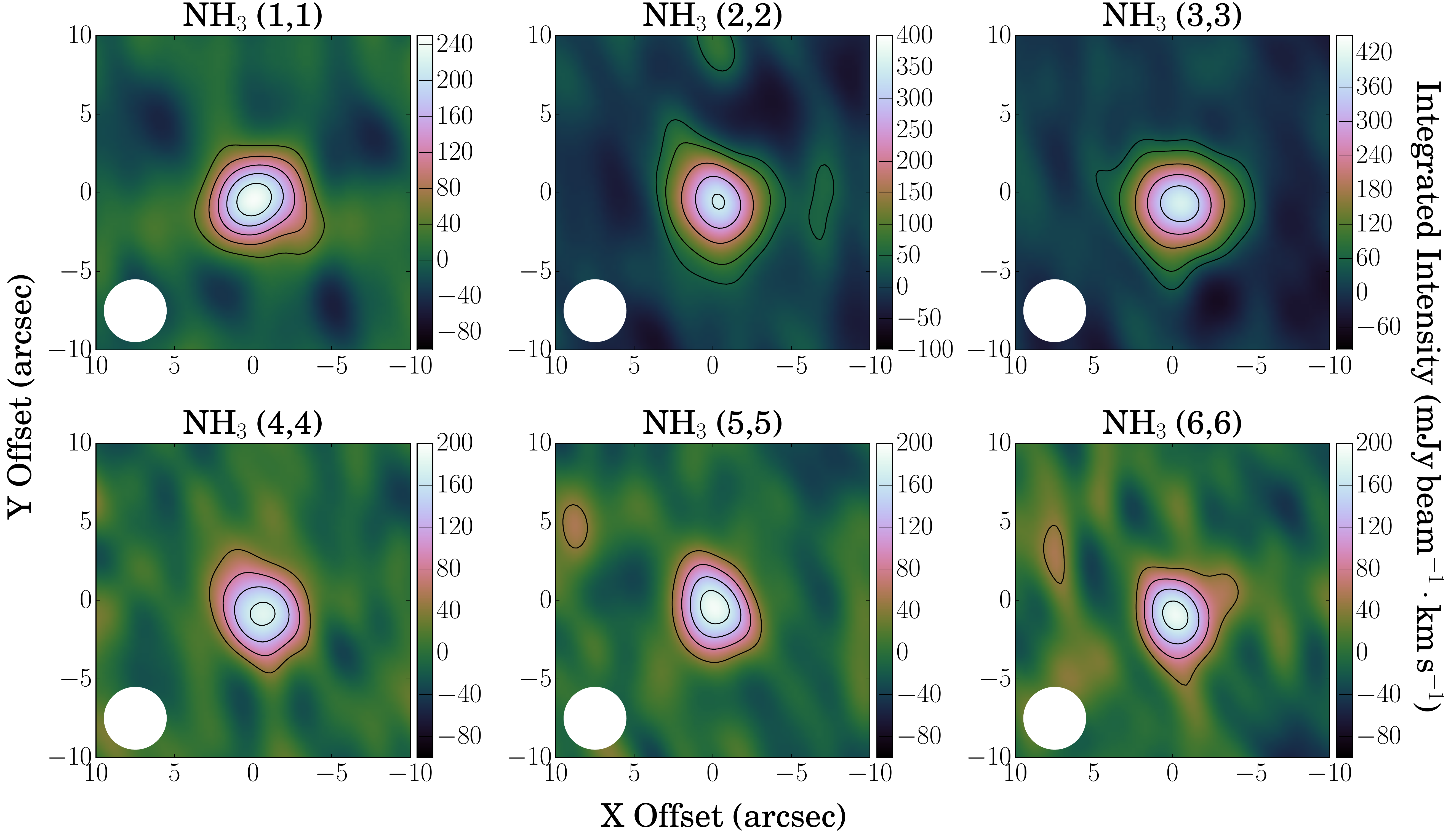}
\caption[]{Integrated intensity maps of the six lowest NH$_3$ inversion lines from {\vy} as observed with the VLA over the LSR velocity range of $[-17, 85]{\kms}$ (relative velocity between ${\pm}51{\kms}$). Horizontal and vertical axes represent the offsets (arcsec) relative to the fitted centre of the continuum in the directions of right ascension and declination, respectively. The circular restoring beam of $4{\farcs}0$ is indicated at the bottom-left corner in each map. The contour levels for the NH$_3$ lines are drawn at: $3, 6, 9, 12, 15\sigma$ for $(1,1)$; $3, 6, 12, 18, 24\sigma$ for $(2,2)$ and $(3,3)$; $3, 6, 9, 12\sigma$ for $(4,4)$ to $(6,6)$, where $\sigma = 14.2\intfluxunits$.}
\label{fig:vla_vy_moment}
\end{figure*}

\begin{figure*}[!htbp]
\centering
\includegraphics[trim=0.5cm 2.3cm 1.5cm 1.5cm, clip, width=\specwidth]{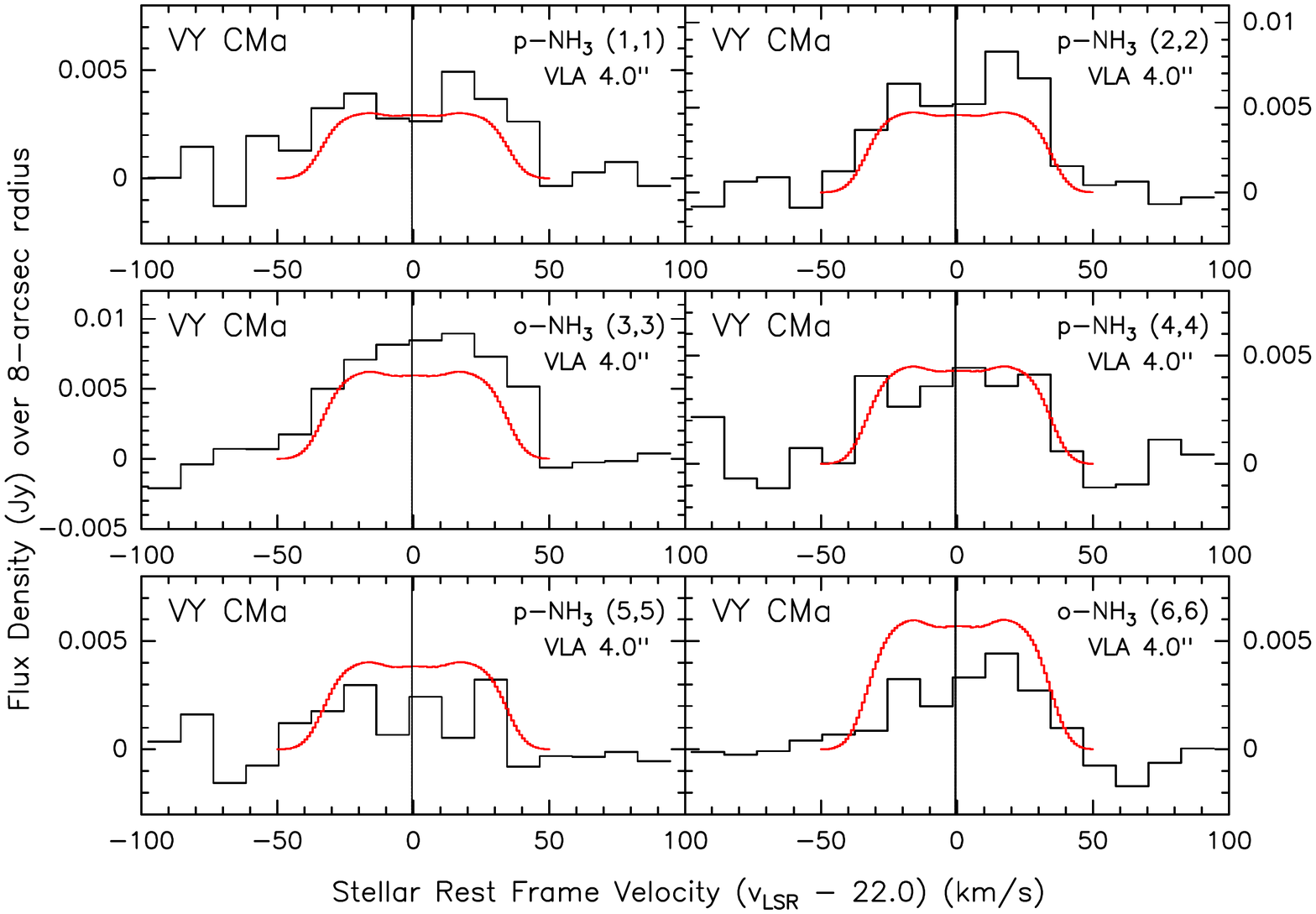}
\caption[]{NH$_3$ inversion line spectra of {\vy}. Black lines show the VLA spectra in flux density and red lines show the modelled spectra. All transitions were observed in 2013 with D configuration. The spectra were integrated over an $8{\arcsec}$-radius circle.}
\label{fig:model_vy_inv}
\end{figure*}

The radio inversion line emission from the CSE of {\vy} is not clearly resolved by the VLA beam of ${\sim}4\arcsec$. Figures \ref{fig:vla_vy_moment} and \ref{fig:model_vy_inv} shows the integrated intensity maps and flux density spectra, respectively, of all transitions covered in our VLA observation. The strongest emission is detected in NH$_3$ (3,3) and (2,2), which also show somewhat extended emission up to ${\sim}4\arcsec$--$6\arcsec$ in the maps. Adopting the trigonometric parallax derived distance of 1.14\,kpc \citep[][see also \citealp{zhang2012}]{choi2008} and the stellar radius of $1420\,\rsun$ \citep{wittkowski2012}, NH$_3$ could maintain a moderate abundance out to ${\sim}700\,\rstar$ ($4\arcsec$).

A close inspection on the maps reveals that the emission peaks at ${\lesssim}1\arcsec$ to the south-west from the centre of the radio continuum, which was determined by fitting the images combining all line-free channels near the NH$_3$ spectra. This offset in peak molecular emission has also been observed under high angular resolution (${\sim}0{\farcs}9$) with the Submillimeter Array (SMA) in multiple other molecules, including dense gas tracers such as CS, SO, SO$_2$, SiO, and HCN \citep[Fig. 8 of][]{kaminski2013}. \citet{kaminski2013} noted that the spatial location of the peak emission may be associated with the `Southwest Clump' (SW Clump) as seen in the infrared images from the {\hubble} Space Telescope (HST) \citep{smith2001,humphreys2007}. Since SW Clump was only seen in the infrared filter but not in any visible filters, \citet{humphreys2007} suggested that the feature is very dusty and obscures all the visible light. However, submillimetre continuum emission from SW Clump has never been detected \citep{kaminski2013,ogorman2015}, which cannot be explained by typical circumstellar grains with a small spectral emissivity index \citep{ogorman2015}. From K\,{\sc i} spectra near 7700\,{\AA}, \citet{humphreys2007} found that SW Clump is likely moving away from us with a small line-of-sight velocity of ${\sim}2.5{\kms}$. Molecules from SW Clump, however, do not always emit from the same redshifted velocity. For example, the H$_2$S and CS emission appears over a wide range of velocities \citep[Sect. 3.1.4 of][]{kaminski2013}; TiO$_2$ appears to emit in blueshifted velocities only \citep[Fig. 2 of][]{debeck2015}; NaCl emission peaks at the redshifted velocity of $3{\kms}$ \citep[Sect. 3.3 of][]{decin2016}, similar to the velocity determined from the K\,{\sc i} spectra. In our VLA images, we see the south-west offset of NH$_3$ emission from much higher redshifted velocities between ${\sim}9$ and $30{\kms}$ relative to our adopted systemic LSR velocity ($22{\kms}$). The peak of the south-west emission is seen near $20$--$25{\kms}$. If the peak NH$_3$ emission is indeed associated with the infrared SW Clump, then NH$_3$ may trace a distinct kinematic component in this possibly dusty and dense feature.

The spectra show relatively flat profiles, with two weakly distinguishable peaks near the LSR velocities of about $-5$--$0\kms$ and $40$--$45\kms$. These features correspond to the relative velocities of roughly $-25\kms$ and $20\kms$, respectively. As we will show in the next subsection, the NH$_3$ $J_K=1_0(\text{s})$--$0_0(\text{a})$ rotational transition, but not the higher ones, also shows similar peaks in its high S/N spectrum. Furthermore, high-spectral resolution spectra of SO and SO$_2$ also show prominent peaks near the same velocities of the NH$_3$ features \citep[e.g.][]{tenenbaum2010,fu2012,adande2013,kaminski2013}.


\subsubsection{{\hifi} observations of {\vy}}
\label{sec:hifi_vy}

\begin{figure*}[!htbp]
\centering
\includegraphics[trim=0.5cm 2.3cm 1.5cm 1.5cm, clip, width=\specwidth]{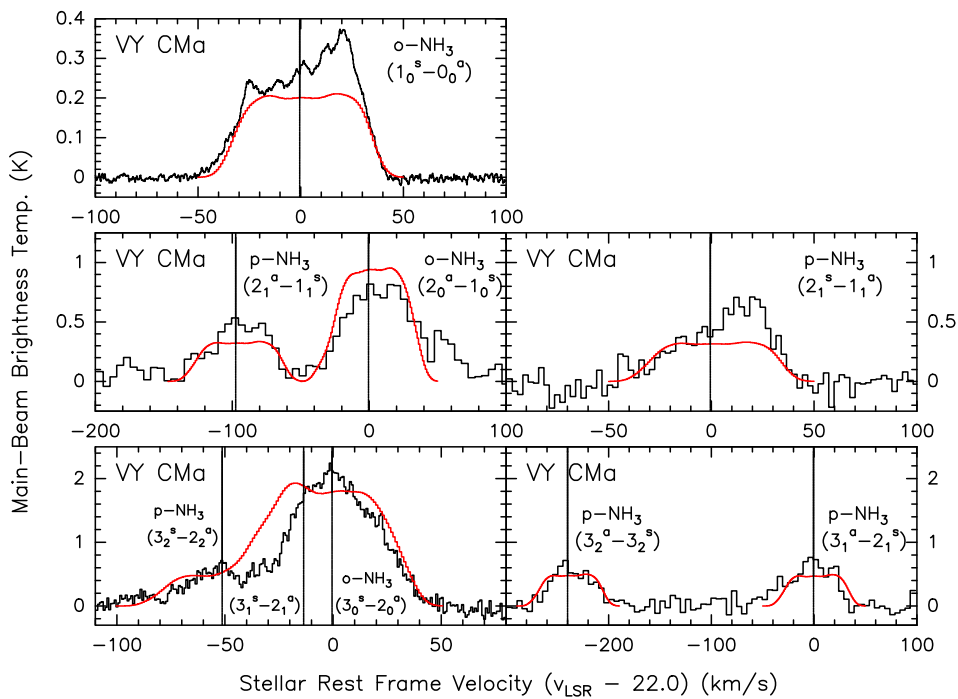}
\caption[]{NH$_3$ rotational line spectra of {\vy}. Black lines show the observed spectra from {\hifi} and red lines show the modelled spectra.}
\label{fig:model_vy_rot}
\end{figure*}

Figure \ref{fig:model_vy_rot} shows all rotational spectra of {\vy}. The NH$_3$ rotational transitions $J_K=1_0(\text{s})$--$0_0(\text{a})$ and $J_K=3_K(\text{s})$--$2_K(\text{a})$ ($K=0,1,2$) observed under the HIFISTARS programme have already been published by \citet{menten2010} and \citet{alcolea2013}. Other transitions were covered in the spectral line survey of the proposal ID `OT1{\_}jcernich{\_}5' (PI: J. Cernicharo). {\vy} is the only target in this study with complete coverage of ground-state rotational transitions up to $J_\text{up}=3$.

The full width (at zero power) of the NH$_3$ $1_0(\text{s})$--$0_0(\text{a})$ and $2_0(\text{a})$--$1_0(\text{s})$ spectra spans the relative velocity range of almost ${\pm}50{\kms}$. The expansion velocity of the CSE of {\vy} is about $35{\kms}$ \citep{decin2006}; and therefore the expected total line width should not exceed $70{\kms}$ if the NH$_3$ emission arises from a single, uniformly expanding component of the circumstellar wind. Huge line widths have also been observed in other abundant molecules \citep[e.g. CO, SiO, H$_2$O, HCN, CN, SO, and SO$_2$;][]{tenenbaum2010,tenenbaum2010cat,alcolea2013,kaminski2013}. Some of the optically thin transitions even show multiple peaks in the spectra \citep[e.g. Fig. 7 of][]{kaminski2013}, suggesting the presence of multiple kinematic components moving at various line-of-sight velocities. 

As already discussed by \citet[][their Sect. 3.3.1]{alcolea2013}, the line profiles of $1_0(\text{s})$--$0_0(\text{a})$ and $3_0(\text{s})$--$2_0(\text{a})$ are distinctly different, with the former showing significantly stronger emission in the redshifted velocities and the latter showing only one main component at the centre. With the new spectra, which show profiles that are somewhat intermediate between the two extremes, we find that the line shapes mainly depend on the relative intensities between the central component and the redshifted component near $+20$--$21{\kms}$ relative to the systemic velocity. The redshifted component is stronger than the central in lower transitions, such as $1_0(\text{s})$--$0_0(\text{a})$ and $2_1(\text{s})$--$1_1(\text{a})$, but becomes weaker and less noticeable in higher transitions. The peak velocity of this redshifted component is also consistent with that of the south-west offset structure as seen in our VLA images, especially in NH$_3$ (2,2) and (3,3). Therefore, at least part of the enhanced redshifted emission may originate from the local NH$_3$-emitting region to the south-west of the star.

The $1_0(\text{s})$--$0_0(\text{a})$ spectrum, which has the highest S/N ratio, shows five peaks at the relative velocities of approximately $-25$, $-11$, $1$, $13$, and $21{\kms}$. Similar spectra showing multiple emission peaks are also seen in CO, SO, and SO$_2$ \citep[e.g.][]{muller2007,ziurys2007,tenenbaum2010,fu2012,adande2013,kaminski2013}. The most prominent peaks in these molecules occur at LSR velocities of $-7$, $20$--$25$, and $42{\kms}$, corresponding to the relative velocities of $-29$, $-2$--$+3$, and $+20{\kms}$, respectively. High-angular resolution SMA images and position-velocity diagrams of these molecules show complex geometry and kinematics of the emission; accordingly, dense and localised structures were introduced to explain individual velocity components \citep{muller2007,fu2012,kaminski2013}. Except for the high-spectral resolution spectrum of SO $J_N=7_6$--$6_5$, which also shows a peak near the relative velocity of $12{\kms}$ \citep[Fig. 6 in][]{tenenbaum2010}, no other molecules seem to exhibit the same peaks at intermediate relative velocities ($-11$ and $13\kms$) as in NH$_3$.

The NH$_3$ inversion (Sect. \ref{sec:vla_vy}) and rotational lines show spectral features at similar velocities to the SO and SO$_2$ lines, suggesting that all these molecules may share similar kinematics and spatial distribution. High-angular resolution SMA images show that SO and SO$_2$ emission is extended by about $2$--$3{\arcsec}$ \citep[Fig. 8 of][]{kaminski2013}. Hence, the strongest NH$_3$ emission is likely concentrated within a radius of ${\sim}2{\arcsec}$.

High rotational transitions including $J=4$--$3$ (${\sim}125$\,{\micron}) and, possibly, $J=6$--$5$ (${\sim}84$\,{\micron}) have been detected by \citet{royer2010} with {\pacs}. However, line confusion from other molecular species (e.g. H$_2$O) and the low spectral resolution of PACS spectra prevented detailed analysis of the NH$_3$ lines \citep[Fig. 1 of][]{royer2010}.


\subsubsection{IRTF observations of {\vy}}
\label{sec:irtf_vy}

\begin{figure*}[!htbp]
\centering
\includegraphics[trim=0.5cm 2.3cm 1.5cm 1.5cm, clip, width=\specwidth]{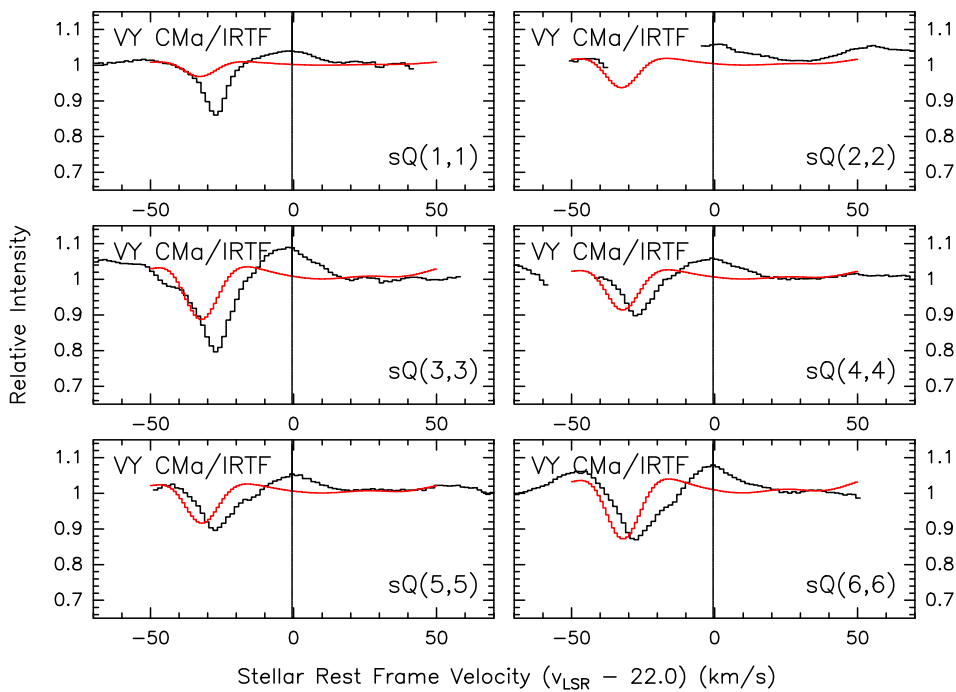}
\caption[]{$Q$-branch NH$_3$ rovibrational spectra of {\vy}. These transitions connects the lower energy levels of the ground-state inversion doublets and the upper levels in the vibrationally excited state. Black lines show the observed spectra from the NASA IRTF and red lines show the modelled spectra. Each modelled spectrum was divided by a smoothed version of itself as in the real data (see Sect. \ref{sec:irtf}) in order to induce the similar distortion in the line wings. The transition $sQ(2,2)$ is contaminated by a telluric CO$_2$ absorption line and therefore a large part of its absorption profile is discarded.}
\label{fig:model_vy_mir1}
\end{figure*}

\begin{figure*}[!htbp]
\centering
\includegraphics[trim=0.5cm 7.7cm 1.5cm 1.5cm, clip, width=\specwidth]{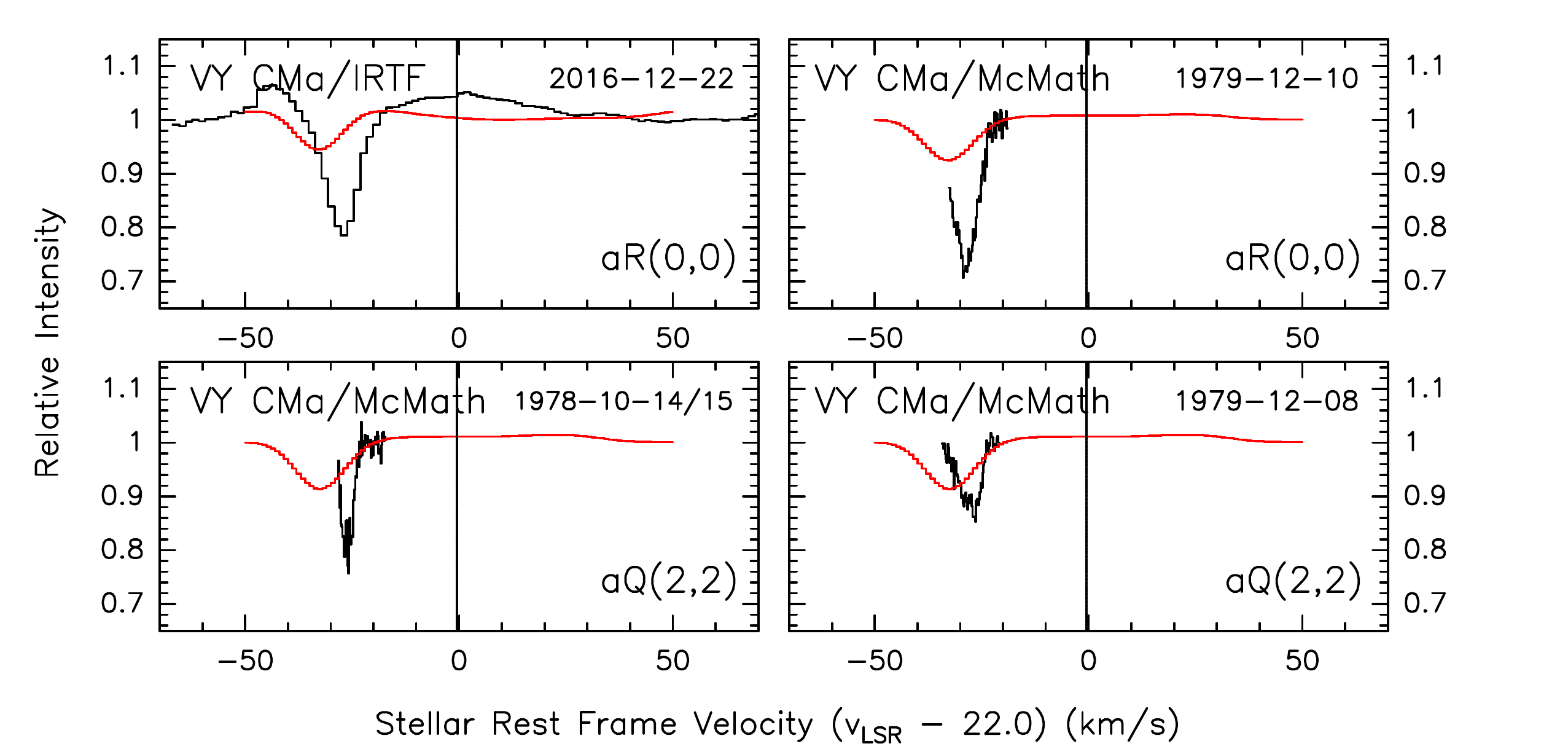}
\caption[]{NH$_3$ rovibrational spectra of {\vy} in the transitions $aR(0,0)$ (\textit{top}) and $aQ(2,2)$ (\textit{bottom}). The black line in the \textit{top-left} panel shows the observed spectrum from the NASA IRTF while those in other panels show the old spectra from the McMath Solar Telescope, as reproduced from Fig. 1 of \citet[][{\aasperm}]{mclaren1980}. Red lines show the modelled spectra in all panels. The modelled spectrum in the \textit{top-left} panel was divided by a smoothed version of itself as in the real data (see Sect. \ref{sec:irtf}) in order to induce the similar distortion in the line wings.}
\label{fig:model_vy_mir2}
\end{figure*}

\begin{figure*}[!htbp]
\centering
\includegraphics[trim=0.5cm 2.3cm 1.5cm 1.5cm, clip, width=\specwidth]{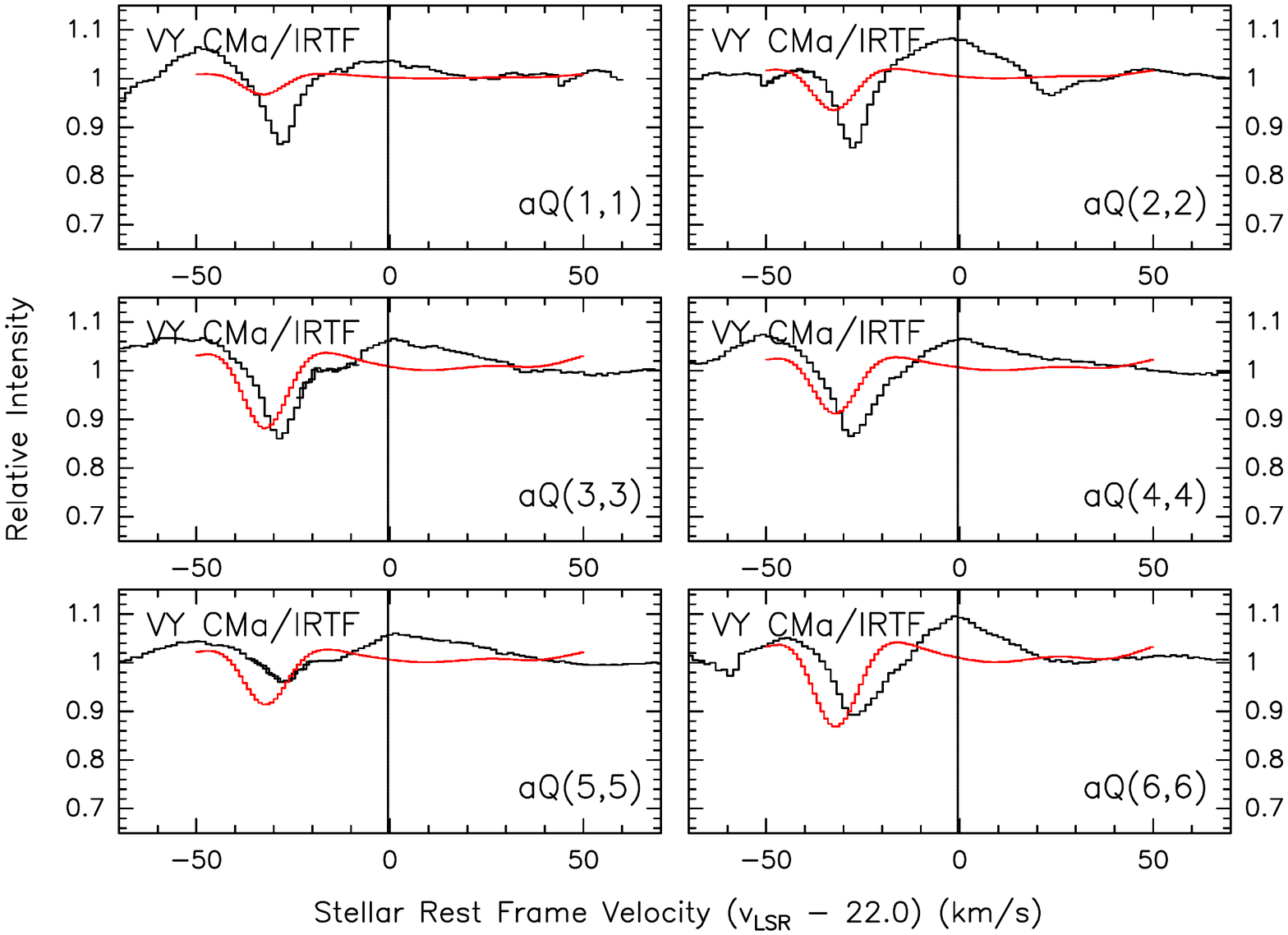}
\caption[]{$Q$-branch NH$_3$ rovibrational spectra of {\vy}. These transitions connects the upper energy levels of the ground-state inversion doublets and the lower levels in the vibrationally excited state. Black lines show the observed spectra from the NASA IRTF and red lines show the modelled spectra. Each modelled spectrum was divided by a smoothed version of itself as in the real data (see Sect. \ref{sec:irtf}) in order to induce the similar distortion in the line wings.}
\label{fig:model_vy_mir3}
\end{figure*}

Figure \ref{fig:irtf_all} shows the normalised IRTF spectra of {\vy} prior to baseline subtraction. Except for the two transitions blended by telluric lines, all other 12 targeted NH$_3$ rovibrational transitions are detected in absorption. The normalised spectra in the stellar rest frame are shown in Figs. \ref{fig:model_vy_mir1}--\ref{fig:model_vy_mir3} along with the modelling results. Figure \ref{fig:model_vy_mir2} additionally shows the old $aR(0,0)$ and $aQ(2,2)$ spectra observed with the McMath Solar Telescope, which are reproduced from Fig. 1 of \citet{mclaren1980}.

The absorption minima of the MIR transitions appear in the LSR velocities between $-7$ and $-4{\kms}$, which correspond to the blueshifted velocities of $26$--$29\kms$. The positions of our $aR(0,0)$ and $aQ(2,2)$ absorption profiles are consistent with the old heterodyne spectra taken by \citet{mclaren1980}, \citet{goldhaber1988}, and \citet{monnier2000s} within the spectral resolution of the TEXES instrument (${\sim}3{\kms}$). The velocities of the absorption minima of these MIR spectra are consistent with that of the prominent blueshifted emission peak in the $1_0(\text{s})$--$0_0(\text{a})$ spectrum (Fig. \ref{fig:model_vy_rot}). Assuming the expansion velocity of $35{\kms}$ \citep{decin2006}, the bulk of the NH$_3$-absorbing gas may come from the wind acceleration zone or from a distinct component (along the line of sight through the slit) of which the kinematics is not represented by the presumed uniform expansion.

Similar to {\ik}, we see slightly more broadening of the line profiles towards the less blueshifted wing (i.e. higher LSR velocities) in higher-$J$ transitions. This trend may be indicative of the formation of NH$_3$ in the accelerating wind where the hotter gas expands at a lower velocity than the outer, cooler gas. In their data of better velocity resolution \citep[${\sim}1\kms$;][]{monnier2000i}, \citet{monnier2000s} found a slight shift in the core of absorption to a lower blueshifted velocity in the higher transition and interpreted -- with additional support from the fitting to the visibility data -- that NH$_3$ probably formed near the outer end of the accelerating part of the CSE, i.e. at the radius of ${\sim}70\,\rstar$ (460\,AU)\footnote{The NH$_3$ formation radius was reported as $40\,{\rstar}'$ in \citet{monnier2000s}, where ${\rstar}' = 0{\farcs}4 = 69.1 \rstar$ \citep{monnier2000c}.}.


\subsubsection{NH$_3$ model of {\vy}}
\label{sec:mdl_vy}

We model the NH$_3$ emission from {\vy} in the similar manner as for {\ik} (Sect. \ref{sec:mdl_ik}). In our 1-D model, we assume that the bulk of the NH$_3$-emitting material comes from the fully accelerated wind at the expansion velocity of $35{\kms}$ with a relatively large turbulence width so as to explain the broad rotational lines (Sect. \ref{sec:hifi_vy}). However, this velocity does not agree with that of the MIR line absorption, which are blueshifted by ${\lesssim}30\kms$ (Sect. \ref{sec:irtf_vy}). In addition, our model would not reproduce individual velocity components in the observed spectra. The CSE of {\vy} is particularly complex and inhomogeneous and our simple model is far from being realistic in describing the kinematics of the NH$_3$-carrying gas.

The gas density of the NH$_3$-emitting CSE is given by Eq. (\ref{eq:dens}) with $\mdot \approx 2 \times 10^{-4}\mspy$ in order to fit the line intensity ratios of the central components of the rotational transitions. The velocity is described by Eq. (\ref{eq:velo}) with $\vin = 4{\kms}$, $\vexp = 35{\kms}$, $R_\text{wind} = 10\,\rstar$, and $\beta=0.5$ \citep[similar to][]{decin2006}. The gas temperature is described by a power law,
\begin{equation}
T_\text{kin}(r) = \tstar \left(\frac{r}{\rstar}\right)^{-0.5},
\end{equation}
where $\tstar = 3490\,\text{K}$ and $\rstar = 1420\,\rsun = 9.88 \times 10^{13}\,\text{cm}$ are both adopted from \citet{wittkowski2012}.

\begin{figure*}[!htbp]
\centering
\includegraphics[height=\profheight]{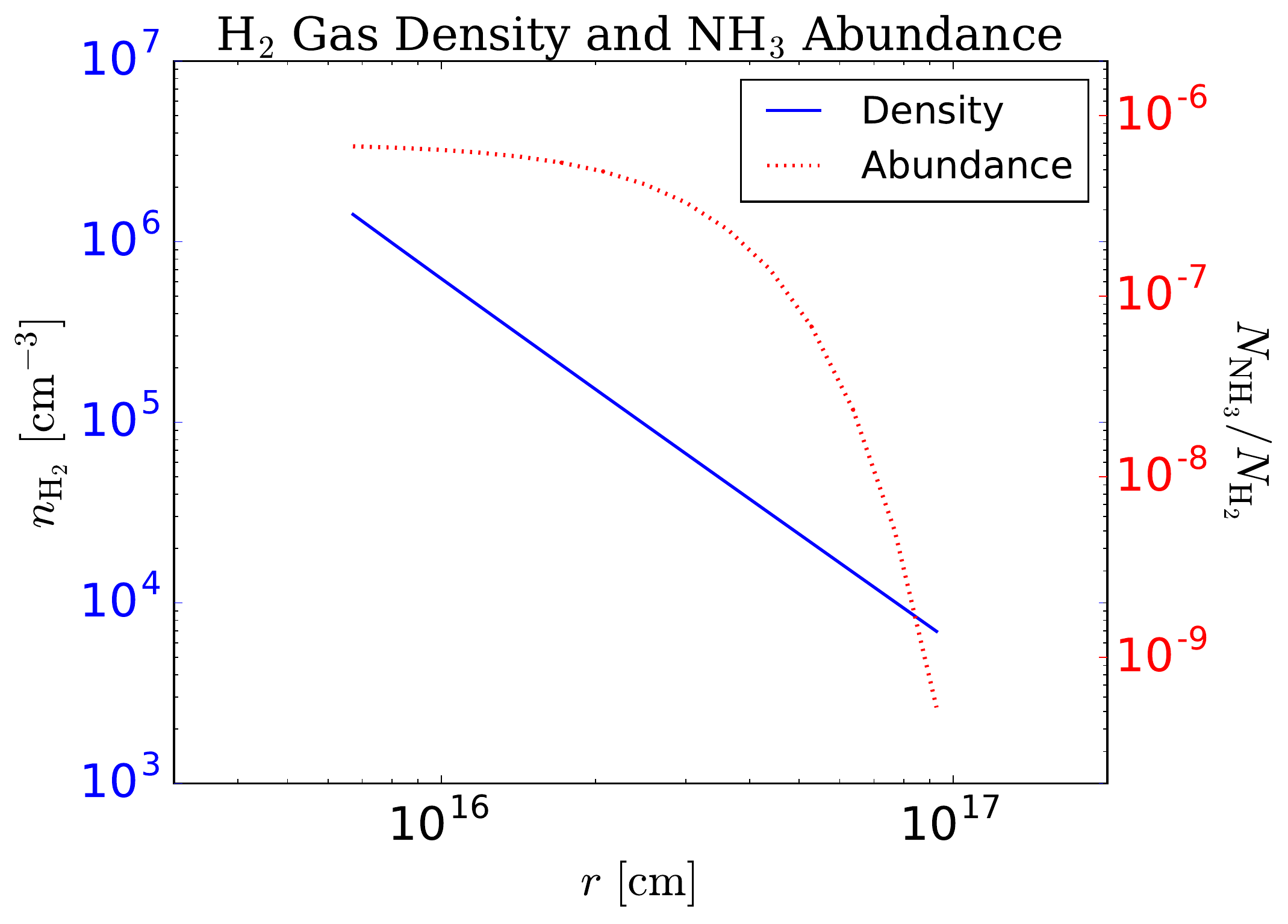}
\includegraphics[height=\profheight]{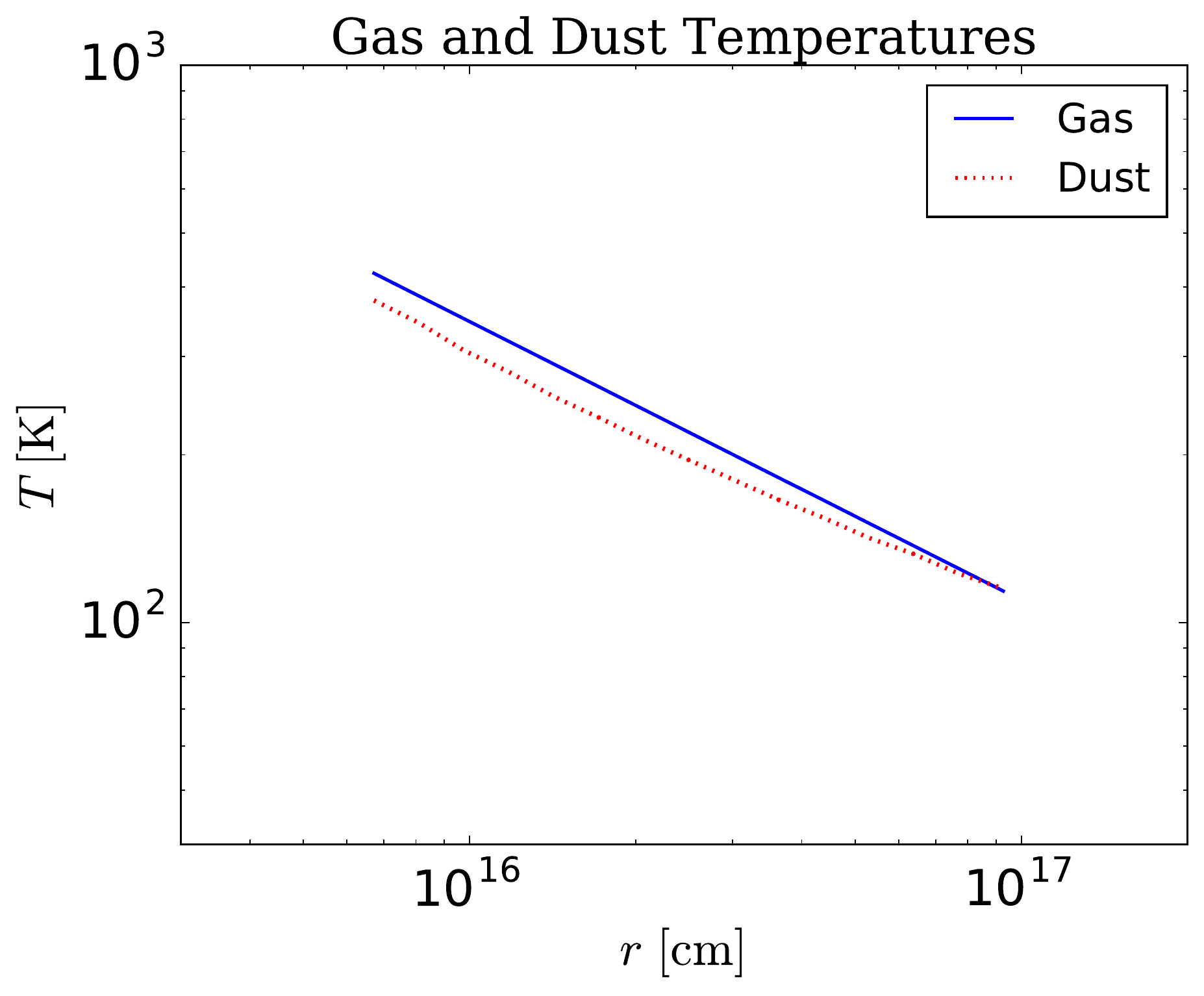}
\includegraphics[height=\profheight]{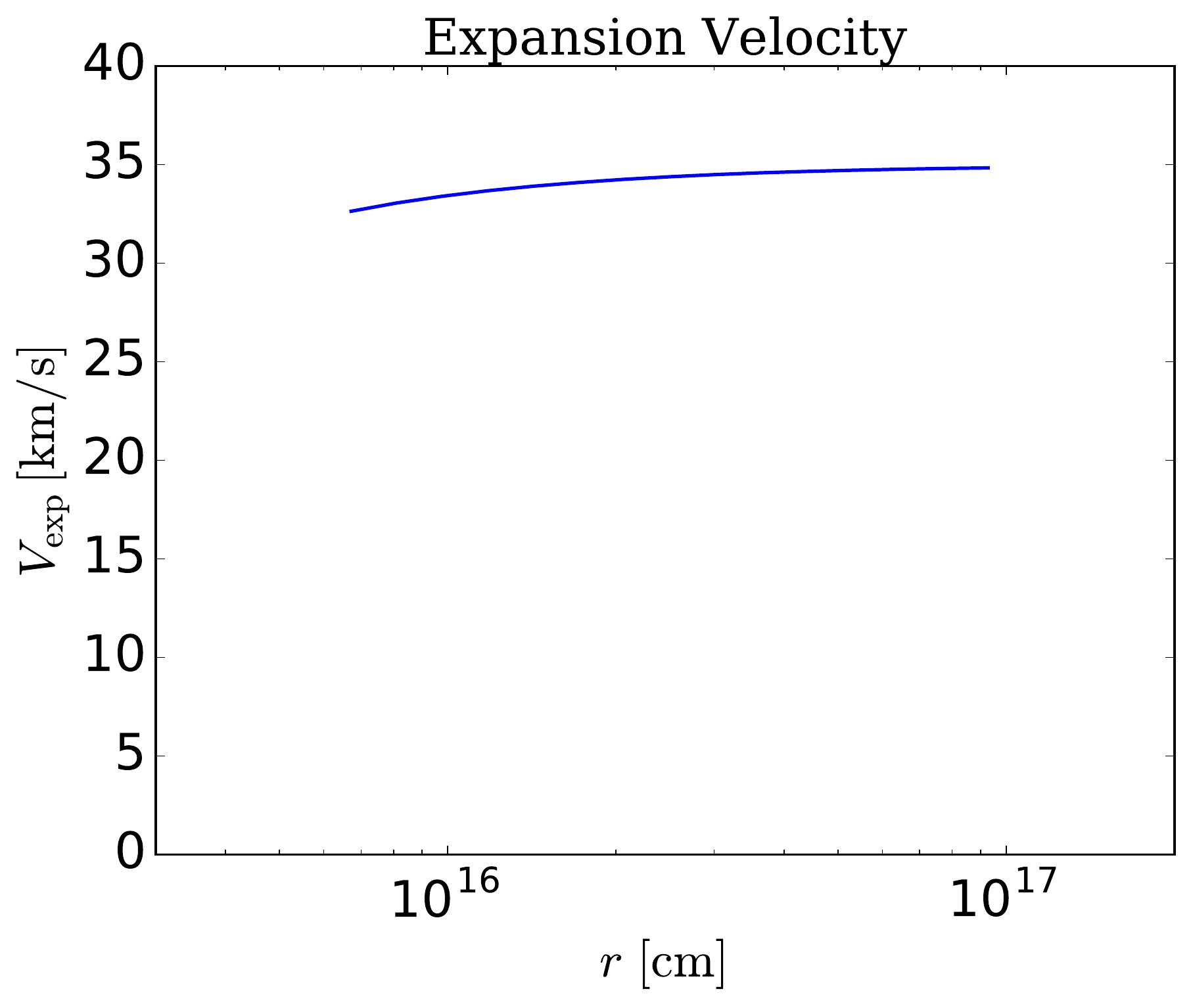}
\caption[]{The model for {\vy}. \textit{Left}: Input gas density and NH$_3$ abundance profiles. Only the radii with abundance ${>}10^{-10}$ are plotted. \textit{Middle}: Input gas and dust temperature profiles. \textit{Right}: Input expansion velocity profile.}
\label{fig:model_vy_prof}
\end{figure*}

We assume a typical d/g of 0.002. Assuming that $\vexp = 35{\kms}$, we need a very high Doppler-$b$ parameter, of ${\sim}7.5\kms$, to explain the broad line profiles as seen in almost all observed spectra (except the high-$J$ rotational lines and a few rovibrational lines). We used the same description of NH$_3$ abundance as in Eq. (\ref{eq:abun}) and found that $\rin = 60\,\rstar = 0{\farcs}35$, $\refd = 350\,\rstar = 2\arcsec$, and $f_0 = 7 \times 10^{-7}$. The NH$_3$ inner radius is mainly constrained by the relative intensity ratios of the rovibrational transitions. If we adopted a smaller {\rin}, the higher-$J$ absorption profiles would become much stronger than the lower ones. The profiles of gas density, abundance, gas and dust temperatures, and outflow velocity are plotted in Fig. \ref{fig:model_vy_prof}. 

In general, the model is able to fit the central velocity component of most rotational transitions. However, there is a large excess of predicted flux in the spectrum of the blended $3$--$2$ transitions (lower-left panel of Fig. \ref{fig:model_vy_rot}). In our model, we did not consider radiative transfer between the blended NH$_3$ transitions. The modelled line flux at each velocity is simply the sum of fluxes contributed from the blending lines. For the inversion transitions (Fig. \ref{fig:model_vy_inv}), our model tends to overestimate the flux densities of higher transitions. Increasing the adopted {\rin} could reduce the discrepancies, but would also underestimate the line absorption in high-$J$ rovibrational transitions.

Better fitting to the MIR absorption profiles (Figs. \ref{fig:model_vy_mir1}--\ref{fig:model_vy_mir3}) would have been obtained had we adopted an expansion velocity of ${\approx}30\kms$, which was also the value adopted by \citet[][Sect. VII.1.2.B]{goldhaber1988} to fit the MIR lines; however, such velocity cannot explain the broad rotational and inversion spectra with our single-component model (e.g. Figs. \ref{fig:model_vy_inv} and \ref{fig:model_vy_rot}) and is not consistent with the commonly adopted terminal expansion velocity as derived from more abundant molecules \citep[$35$--$45{\kms}$; e.g.][]{decin2006,muller2007,ziurys2007}. The origin of MIR NH$_3$ absorption may be close to the end of the wind acceleration as already suggested by \citet{monnier2000s}.


\subsection{{\oh}}
\label{sec:results_oh}

{\oh} (also known as Rotten Egg Nebula or Calabash Nebula) is a bipolar pre-planetary nebula (PPN) hosting the Mira-type AGB star QX Pup. This object is believed to be evolving into a planetary nebula \citep{sc2004}. In this study, we assume a distance of 1.54\,kpc \citep{choi2012} and a systemic LSR velocity of $34{\kms}$ \citep{alcolea2001}. The average estimates of its luminosity and effective temperature are, respectively, ${\sim}10^{4}\,\lsun$ and ${\sim}2300\,\text{K}$\citep{cohen1981,kastner1992,sc2002}. Using the Stefan--Boltzmann law, ${\lstar} \propto {\rstar}^2 {\tstar}^4$, the stellar radius is estimated to be ${\sim}4.4 \times 10^{13}\,\text{cm} = 2.9\,\text{AU}$. High-resolution images of molecular emission of the nebula reveal a very clumpy bipolar outflow with a clear axis of symmetry at the position angle (PA) of about $21\degr$ and a plane-of-sky-projected, global velocity gradient ($\nabla V$) of about $6.5\velgradunits$ from the north-east to the south-west \citep[e.g.][]{sc2000,alcolea2001}. The fact that the CO emission is very compact (${\lesssim}4{\arcsec}$) within each velocity channel of width $6.5{\kms}$ suggests that the molecular outflow is highly collimated \citep{alcolea2001}. Most of the mass (${\sim}70\%$) as derived from molecular emission concentrates at the central velocity component, which is known as `I3', between the LSR velocities of $+10$ and $+55{\kms}$. This component spatially corresponds to the central molecular clump within a radius of ${\lesssim}4\arcsec$ from the continuum centre \citep{alcolea1996,alcolea2001}. The circumstellar material is thought to have been suddenly accelerated about 770 years ago, resulting in a Hubble-like outflow with a constant velocity gradient, $\nabla V$ \citep{alcolea2001}.

At the core of clump I3, the velocity gradient of SO is opposite in direction (i.e. velocity increasing from the south to the north) to the global $\nabla V$ as traced by CO, HCO$^{+}$, and H$^{13}$CN within a radius of about $2\arcsec$ and up to a velocity of about $10{\kms}$ relative to our adopted systemic velocity \citep{sc2000}. In the position-velocity (PV) diagram of SO along the axis of symmetry \citep[Fig. 6 in][]{sc2000}, there are two regions of enhanced SO emission: one peaks at about $1\arcsec$ to the north-east and ${\sim}5{\kms}$ redshifted to the systemic and the other peaks around the zero offset and ${\sim}5{\kms}$ blueshifted to the systemic. \citet{sc2000} suggested that the peculiar velocity gradient of the core SO emission could indicate the presence of a disc-like structure that is possibly expanding.


\subsubsection{VLA observation of {\oh}}
\label{sec:vla_oh}

\begin{figure*}[!htbp]
\centering
\includegraphics[height=\jvlaheight]{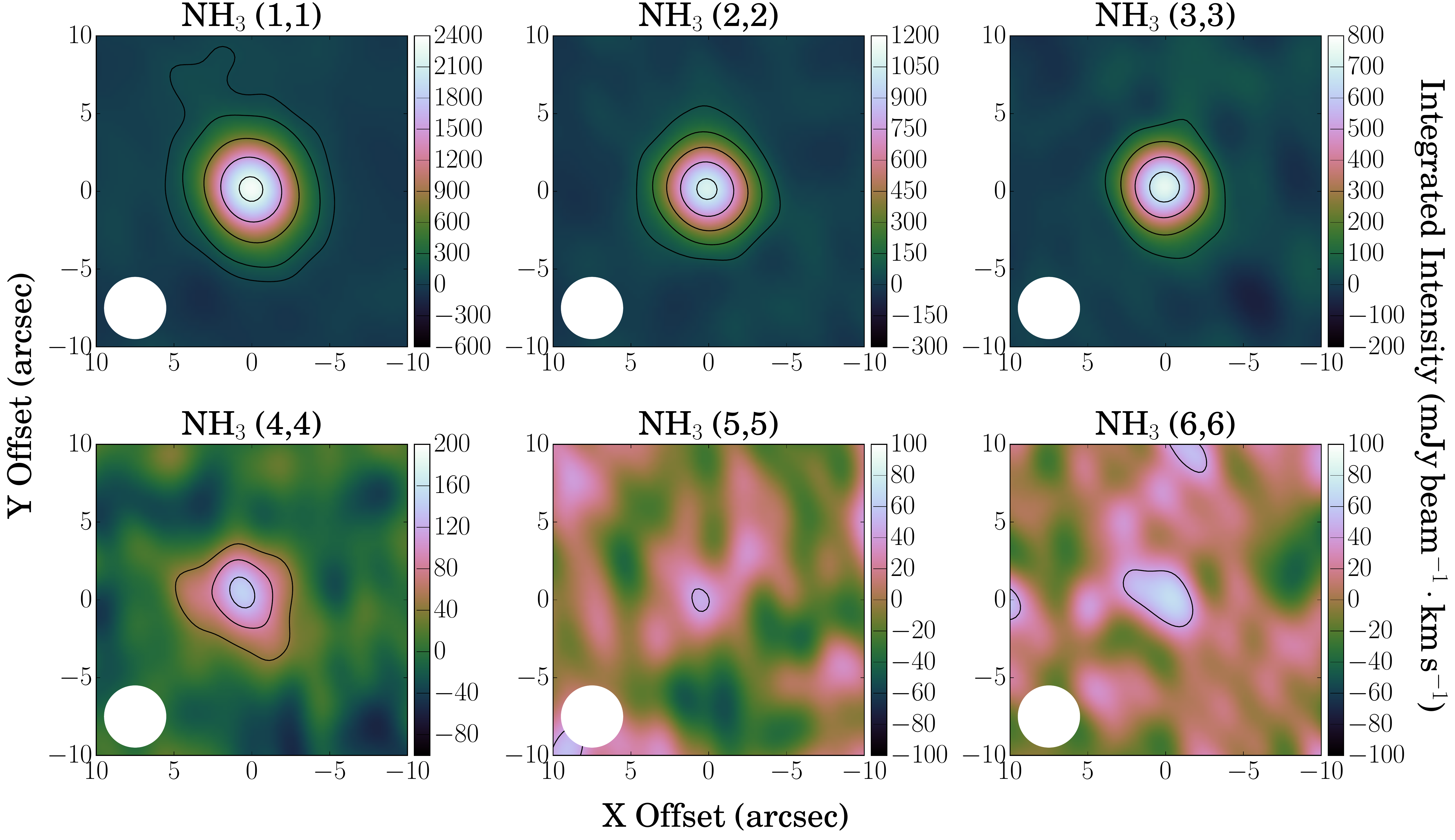}
\caption[]{Integrated intensity maps of the six lowest NH$_3$ inversion lines from {\oh} as observed with the VLA over the LSR velocity range of $[-20, 88]{\kms}$ (relative velocity between ${\pm}54{\kms}$). Horizontal and vertical axes represent the offsets (arcsec) relative to the fitted centre of the continuum in the directions of right ascension and declination, respectively. The circular restoring beam of $4{\farcs}0$ is indicated at the bottom-left corner in each map. The contour levels for the NH$_3$ lines are drawn at: $5, 15, 50, 100, 150\sigma$ for $(1,1)$; $5, 15, 30, 50, 70\sigma$ for $(2,2)$; $5, 15, 30, 45\sigma$ for $(3,3)$; and $3, 6, 9\sigma$ for $(4,4)$ to $(6,6)$, where $\sigma = 14.6\intfluxunits$.}
\label{fig:vla_oh_moment}
\end{figure*}

Our VLA observations of {\oh} detected all the six covered metastable inversion lines. In each $3{\kms}$-wide velocity channel, the NH$_3$ emission is not resolved by the $4\arcsec$-beam. This is consistent with the compact channel-by-channel CO emission as observed by \citet{alcolea2001} assuming that the NH$_3$ emission also originates from the collimated bipolar outflow or the central clump I3. Figure \ref{fig:vla_oh_moment} shows the integrated intensity maps of all the inversion lines over the LSR velocity range between $34 \pm 54{\kms}$. The NH$_3$ (1,1) transition shows the most extended spatial distribution of typically ${\sim}6\arcsec$ (up to ${\sim}9\arcsec$ along the north-eastern part of the outflow) and the widest velocity range of ${\pm}50{\kms}$ relative to the systemic velocity. All other transitions show emission well within the above limits. Therefore, the vast majority of the ammonia molecules are situated in clump I3 as identified by \citet{alcolea1996,alcolea2001}.

\begin{figure*}[!htbp]
\centering
\includegraphics[height=\jvlaheight]{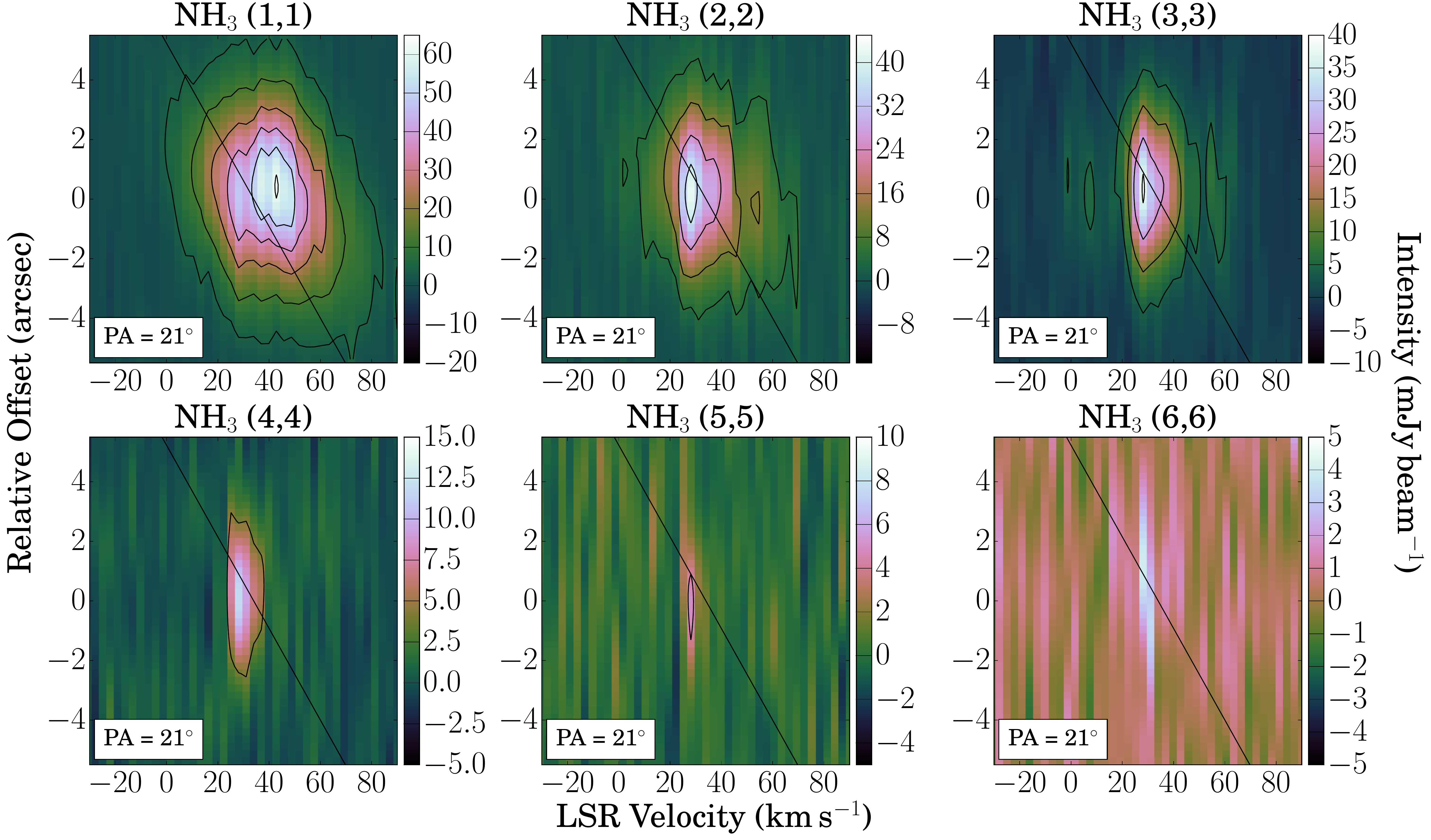}
\caption[]{Position-velocity (PV) diagrams of the six lowest NH$_3$ lines as observed with the VLA cutting along the axis of symmetry of {\oh}. Vertical axes represent offsets (arcsec) in the north-east direction at the position angle of $21\degr$ and horizontal axes represent the LSR velocity ({\kms}). In each PV diagram, the global velocity gradient of $\nabla V = 6.5\velgradunits$ is indicated by the straight line passing through the continuum centre (i.e. zero offset) and the systemic LSR velocity ($34{\kms}$) of {\oh}.}
\label{fig:vla_oh_pv}
\end{figure*}

Although the emission is not spatially resolved by channels, we see a systematic shift of the emission centroid from the north-east to the south-west with increasing LSR velocity. In Fig. \ref{fig:vla_oh_pv}, we show the PV diagrams of all six NH$_3$ transitions extracted along the symmetry axis at $\text{PA}=21\degr$. A straight line of $\nabla V = 6.5\velgradunits$ is also drawn in each diagram to indicate the global velocity gradient as traced by CO \citep{alcolea2001}. NH$_3$ (1,1) is the only transition that somehow follows the global $\nabla V$ in its extended (${\gtrsim}2\arcsec$) emission. Within $2\arcsec$, however, the intensity of the PV diagram is dominated by a component of enhanced emission at around $0\arcsec$--$1\arcsec$ north-east and with the redshifted velocities of $0$--$15{\kms}$. Even more intriguing is the fact that other inversion lines, from (2,2) to (5,5) (and perhaps also (6,6)), show an enhancement near zero offset and in blushifted velocities; but none of them show the distinct redshifted feature at the same position-velocity coordinates in the diagram of (1,1).

The redshifted and blueshifted components of the NH$_3$-emitting region probably have different kinetic temperatures. Assuming that all inversion lines are optically thin, we can compute the rotational temperatures using the line ratios of different line combinations by the following formula, which is Eq. (1) of \citet{menten1995}:
\begin{equation}
T_{JJ'} = \frac{ \displaystyle \frac{E_\text{low} (J, K)}{k} - \frac{E_\text{low} (J', K')}{k} }{ \displaystyle \ln \left[ \frac{\int T_\text{mb} (J,K) \,\text{d}{\varv}}{\int T_\text{mb} (J',K') \,\text{d}{\varv}} \frac{J(J+1)}{J'(J'+1)} \frac{K'^2(2J'+1)}{K^2(2J+1)} \frac{{\varg}_\text{op} (K')}{{\varg}_\text{op} (K)} \right] }
\end{equation}
where $E_\text{low} (J, K)/k$ (in Kelvin) is the lower energy level and $\int T_\text{mb} (J,K) \,\text{d}{\varv}$ is the velocity-integrated main-beam brightness temperature of the inversion line $(J,K)$. The weighting ${\varg}_\text{op} (K)$ equals $2$ for ortho-NH$_3$ ($K=3,6,\ldots$) and equals $1$ for para-NH$_3$ ($K=1,2,4,5,\ldots$). We compute the rotational temperatures for the redshifted and blueshifted components separately. From the binned spectra presented in the next subsection, we get the main-beam brightness temperature of the two components from the relative velocity ranges of $[7.5, 13.5]{\kms}$ (where (1,1) peaks) and $[-10.5, -4.5]{\kms}$ (where other lines peak). The rotational temperature $T_{12}$ for the redshifted component is 20\,K and that of the blueshifted component is 37\,K. The upper limits of $T_{14}$ and $T_{24}$ for the redshifted component (${<}41$\,K and ${<}58$\,K) are also consistently smaller that the derived values for the blueshifted (54\,K and 63\,K). We do not show the rotational temperatures involving both ortho- and para-NH$_3$ lines, which are of limited meaning because the conversion between the two species is extremely slow and is not primarily determined by the present temperature \citep[Sect. \ref{sec:mol}, see also][]{ho1979}.

As already predicted by \citet{menten1995} based on single-dish inversion spectra of NH$_3$, the (2,2) and (3,3) lines probably exclusively trace the warm material that is close to the central star, while the (1,1) line (among other molecules) traces the cooler part of the circumstellar gas and the high-velocity outflow. From our analysis of the VLA data, we are able to distinguish the spatial and kinematic components that emit at different inversion lines. NH$_3$ (1,1) emission predominantly comes from the cool material in the bulk of the clump I3, the inner part of the bipolar outflow up to ${\sim}6\arcsec$--$9\arcsec$, and the redshifted component of enhanced emission, which is possibly part of a disc-like structure, as traced by the inner SO emission. On the other hand, most if not all emission of the higher transitions originates from the warm, blueshifted component of the inner SO emission. This component is spatially close to the centre of the radio continuum.

\begin{figure*}[!htbp]
\centering
\includegraphics[trim=0.5cm 2.3cm 1.5cm 1.5cm, clip, width=\specwidth]{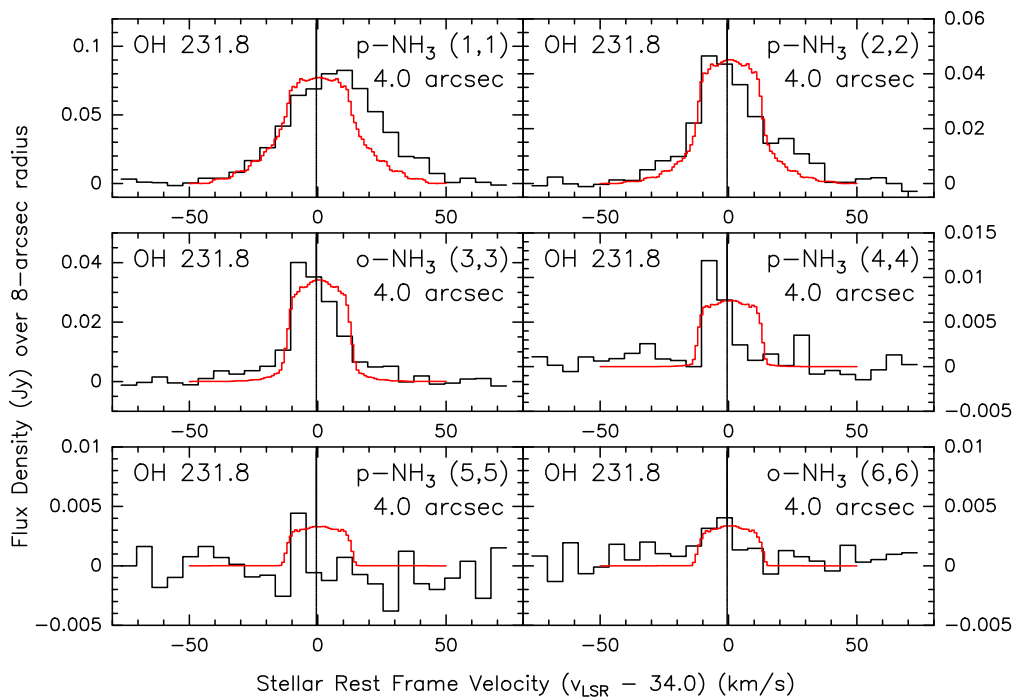}
\caption[]{NH$_3$ inversion line spectra of {\oh}. Black lines show the VLA spectra in flux density and red lines show the modelled spectra. All transitions were observed in 2013 with D configuration. The spectra were integrated over an $8{\arcsec}$-radius circle.}
\label{fig:model_oh_inv}
\end{figure*}


\subsubsection{{\hifi} observations of {\oh}}
\label{sec:hifi_oh}

The NH$_3$ $J_K=1_0$--$0_0$ spectrum of {\oh} clearly shows two components: a strong, narrow component within ${\pm}20$--$25{\kms}$ relative to the systemic velocity and a weaker, broad component that is only apparent in the redshifted velocity range between ${\sim}18{\kms}$ and ${\sim}45{\kms}$. Both components are asymmetric with signs of self-absorption in blueshifted velocities. Due to poor S/N ratios, higher transitions only show the narrow component but not the broad one. The relative velocity range of the narrow component is consistent with that of the clump I3 \citep{alcolea1996}, therefore suggesting that most of the rotational line emission originates from this clump within the radius of ${\sim}4\arcsec$.

\begin{figure*}[!htbp]
\centering
\includegraphics[trim=0.5cm 2.3cm 1.5cm 1.5cm, clip, width=\specwidth]{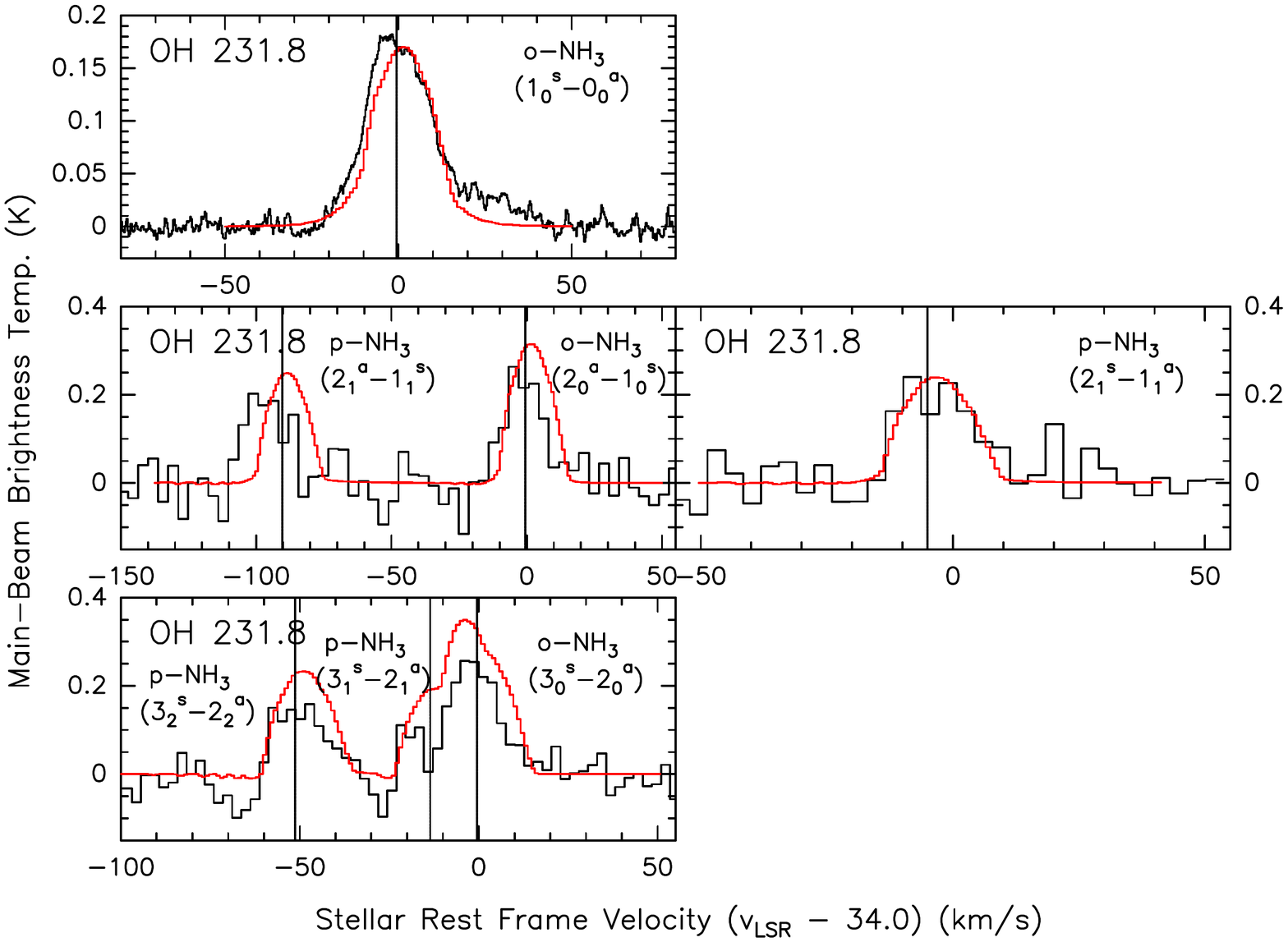}
\caption[]{NH$_3$ rotational line spectra of {\oh}. Black lines show the observed spectra from {\hifi} and red lines show the modelled spectra.}
\label{fig:model_oh_rot}
\end{figure*}


\subsubsection{NH$_3$ model of {\oh}}
\label{sec:mdl_oh}

From the above discussion, the NH$_3$ distribution can be distinguished into at least two parts: (1) a central core containing a possible disc-like structure and (2) an outer outflow following a Hubble-like expansion with a constant velocity gradient. Since the VLA images do not resolve the structure of the possible disc and our code is only one-dimensional, we can only assume spherical symmetry and uniform distribution of the molecule (without any substructures). We adopt the expansion velocity profile as
\begin{equation}
\label{eq:oh_vel}
{\vexp}(r) =
\begin{cases}
13.0{\kms} & \text{if $ r \le 1{\farcs}5 = 787\,\rstar$} \\
(8.9\velgradunits) \times r & \text{if $r > 1{\farcs}5$} \\
\end{cases},
\end{equation}

\begin{figure*}[!htbp]
\centering
\includegraphics[height=\profheight]{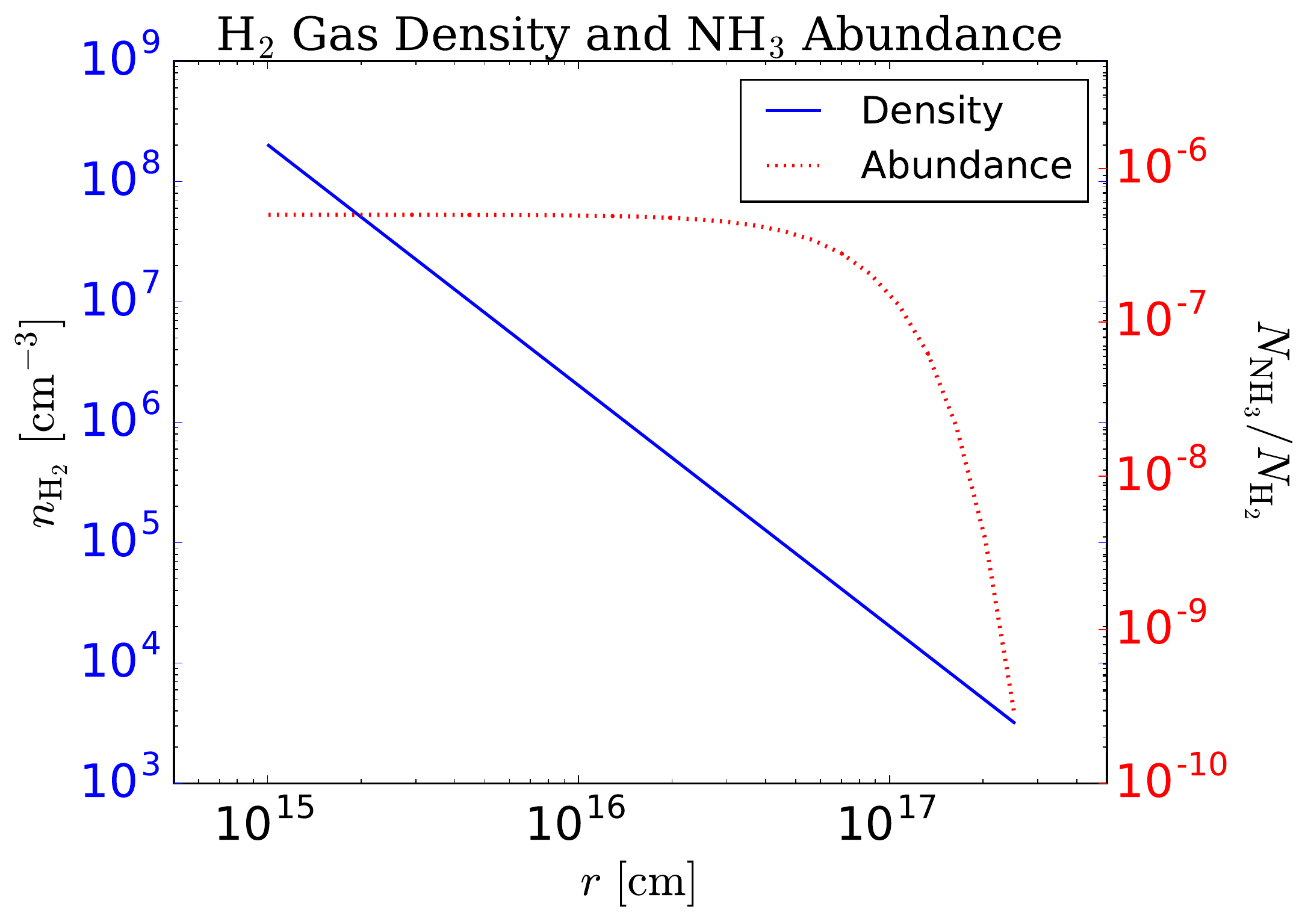}
\includegraphics[height=\profheight]{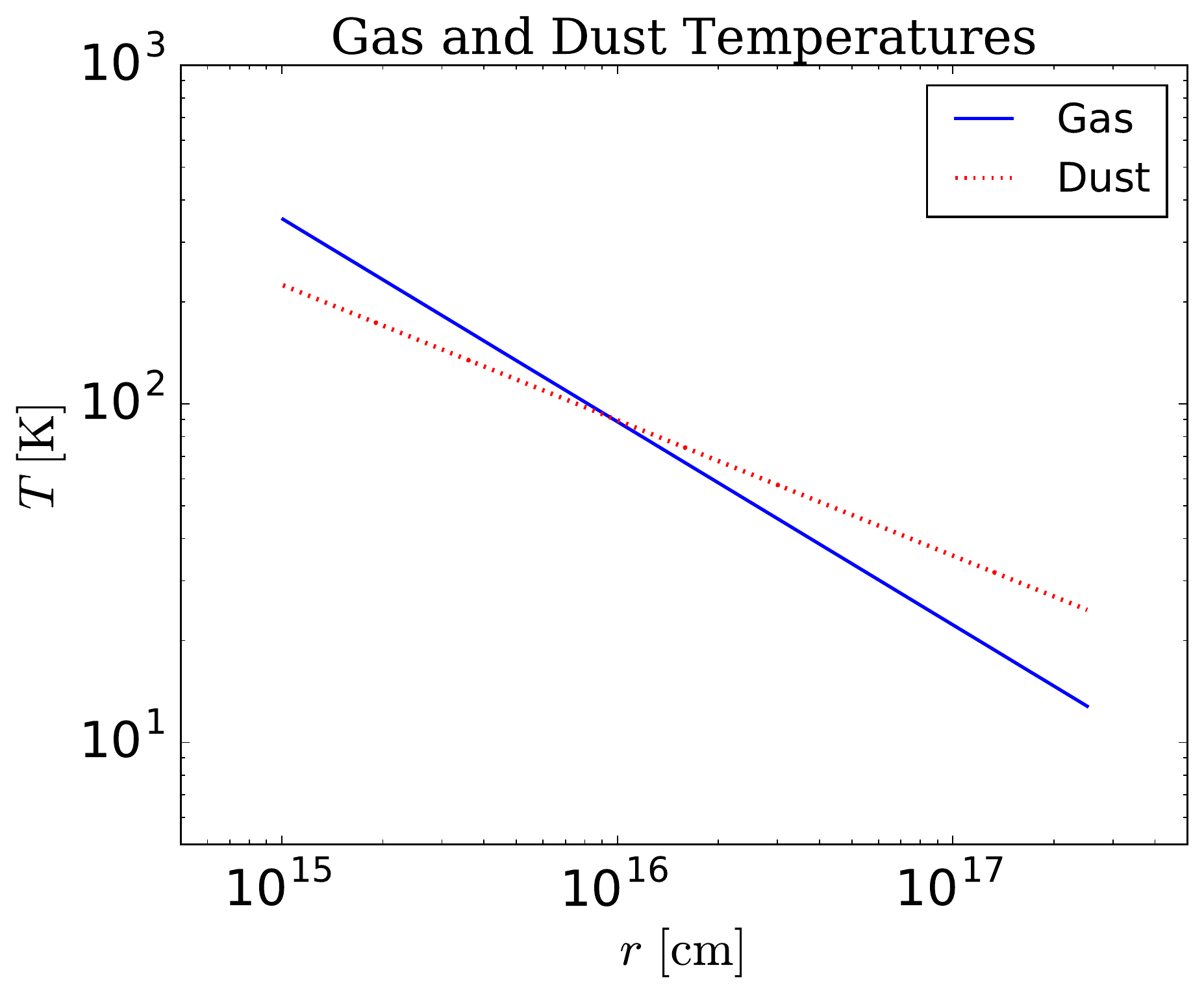}
\includegraphics[height=\profheight]{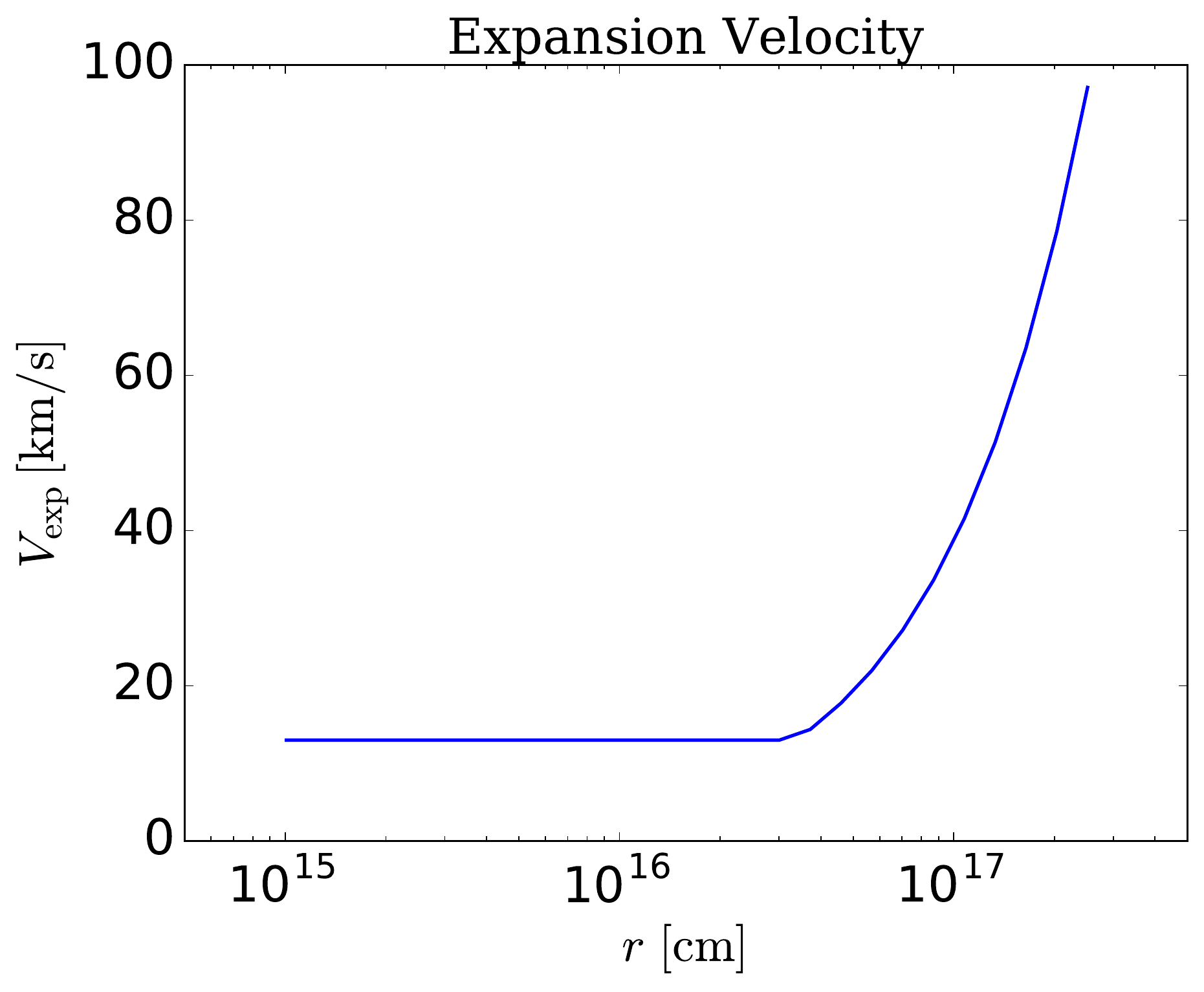}
\caption[]{The model for {\oh}. \textit{Left}: Input gas density and NH$_3$ abundance profiles. Only the radii with abundance ${>}10^{-10}$ are plotted. \textit{Middle}: Input gas and dust temperature profiles. \textit{Right}: Input expansion velocity profile.}
\label{fig:model_oh_prof}
\end{figure*}

The gas density distribution of the nebula is not known. We use again Eq. (\ref{eq:dens}) to describe the density by assuming the conservation of mass. As in the models of \citet{sc2015} and \citet{vp2015}, we adopt a constant mass-loss rate of $\mdot = 1 \times 10^{-4}\mspy$ and a characteristic velocity of ${\langle \vexp \rangle} = 20{\kms}$. However, we found that such density is too low to reproduce the observed intensities of the rotational lines. We have to introduce a density enhancement factor of 3.5 to fit all the spectra. We use the similar abundance profile as in Eq. (\ref{eq:abun}) with $\rin = 20\,\rstar$, $\refd = 2100\,\rstar \approx 4\arcsec$, and $f_0 = 5 \times 10^{-7}$. The gas temperature in our model is described by a simple power-law
\begin{equation}
T_\text{kin}(r) = 2300\,\text{K} \left(\frac{r}{\rstar}\right)^{-0.6}.
\end{equation} Figure \ref{fig:model_oh_prof} shows the radial profiles of our input parameters. 

Figs. \ref{fig:model_oh_inv} and \ref{fig:model_oh_rot} show the VLA and {\hifi} spectra, respectively. In general, our model is able to predict the observed emission from all available spectra. However, the fitting to the redshifted part of the spectra is unsatisfactory, especially in the $(J,K)=(1,1)$ and $J_K=1_0$--$0_0$ lines. As discussed in Sect. \ref{sec:vla_oh}, we speculate that some of the redshifted emission comes from a cool component that is shifted to the north-east of the outflow and predominantly emits in (1,1). Our 1-D model assumes the density of the outflow at those velocities to be ${\sim}10^4$--$10^{5}\,{\percc}$, while the critical density of the $1_0$--$0_0$ transition is about $1 \times 10^7\,{\percc}$; therefore, this component is likely to be significantly denser than the ambient gas.

\subsection{{\irc}}
\label{sec:results_irc}


\subsubsection{VLA observation of {\irc}}
\label{sec:vla_irc}

\begin{figure*}[!htbp]
\centering
\includegraphics[trim=1.5cm 3.5cm 1.5cm 3.5cm, clip, width=\aipswidth]{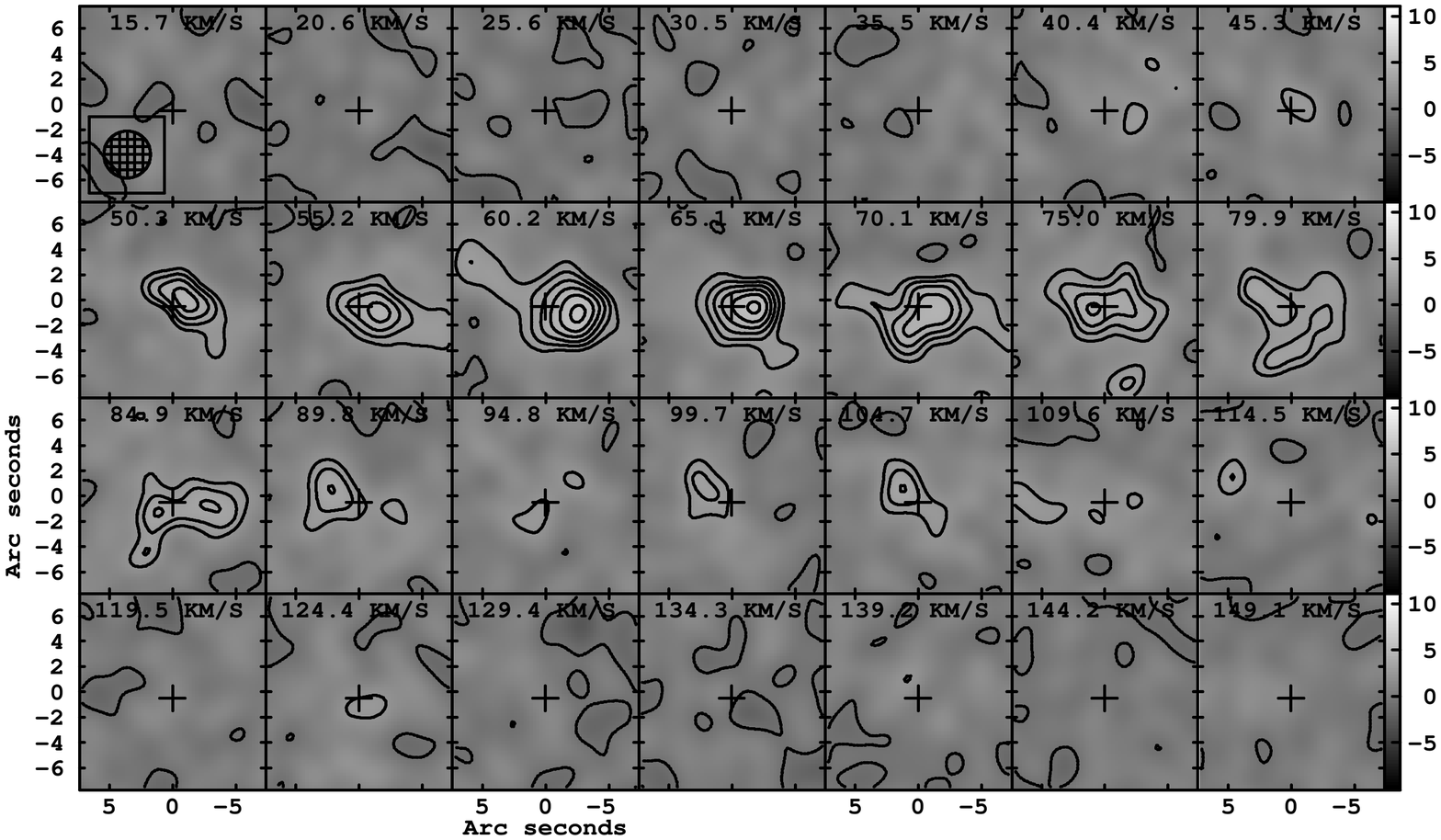}\\
\includegraphics[trim=1.5cm 4.0cm 1.5cm 4.0cm, clip, width=\aipswidth]{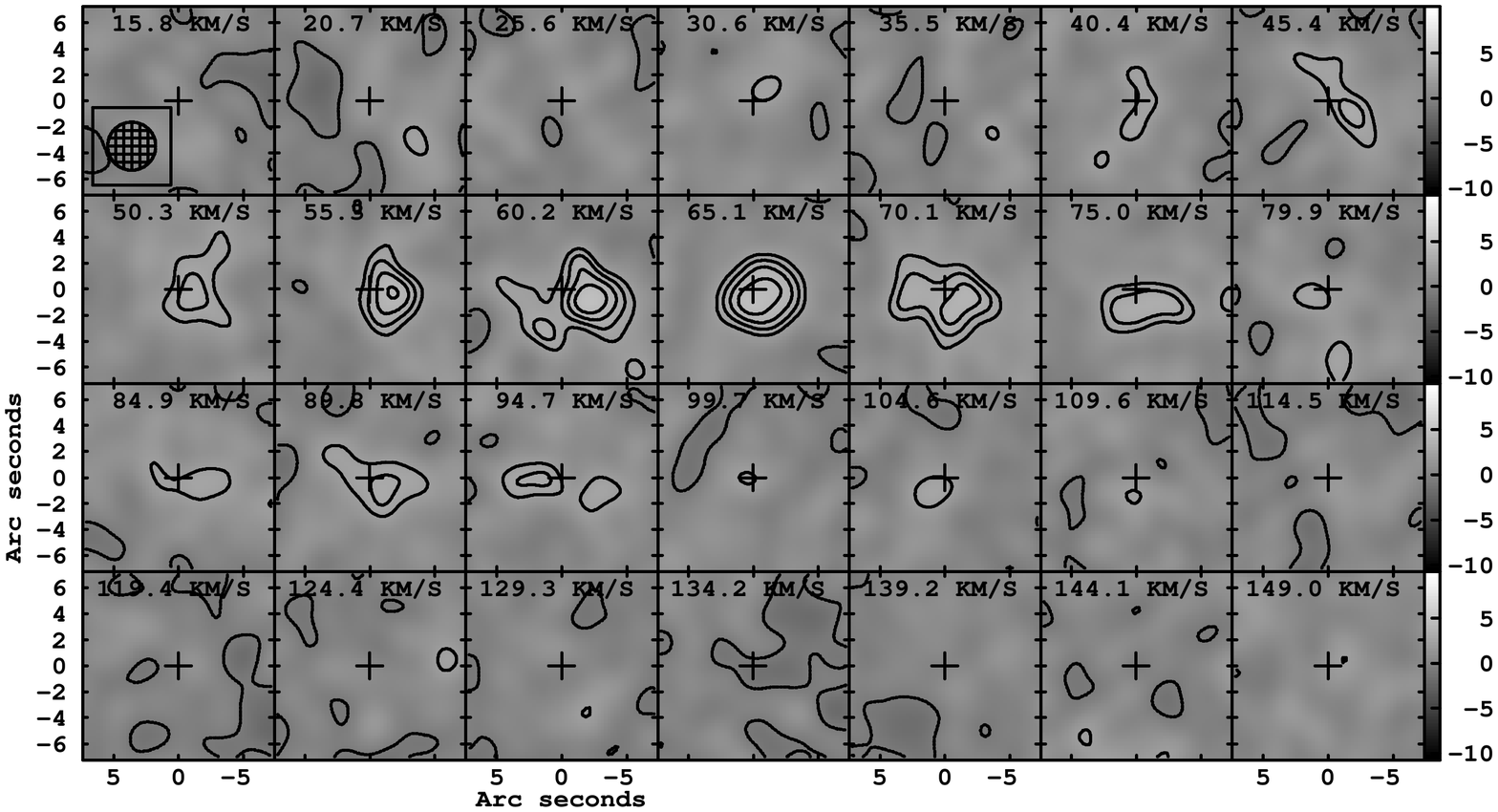}
\caption[]{VLA images of NH$_3$ (1,1) (\textit{top}) and (2,2) (\textit{bottom}) emission from {\irc}. In each panel, the LSR velocity is shown and the position of {\irc} is marked by a cross. The circular restoring beam with a FWHM of $3{\farcs}7$ is indicated at the bottom left corner of the first panel. The contour levels for the NH$_3$ lines are drawn at: $3, 4, 5, 6, 7, 8\sigma$, where $\sigma = 0.69\fluxunits$ for $(1,1)$ and $\sigma = 0.68\fluxunits$ for $(2,2)$.}
\label{fig:vla_irc}
\end{figure*}

\begin{figure*}[!htbp]
\centering
\includegraphics[trim=0.5cm 2.3cm 1.5cm 1.5cm, clip, width=\specwidth]{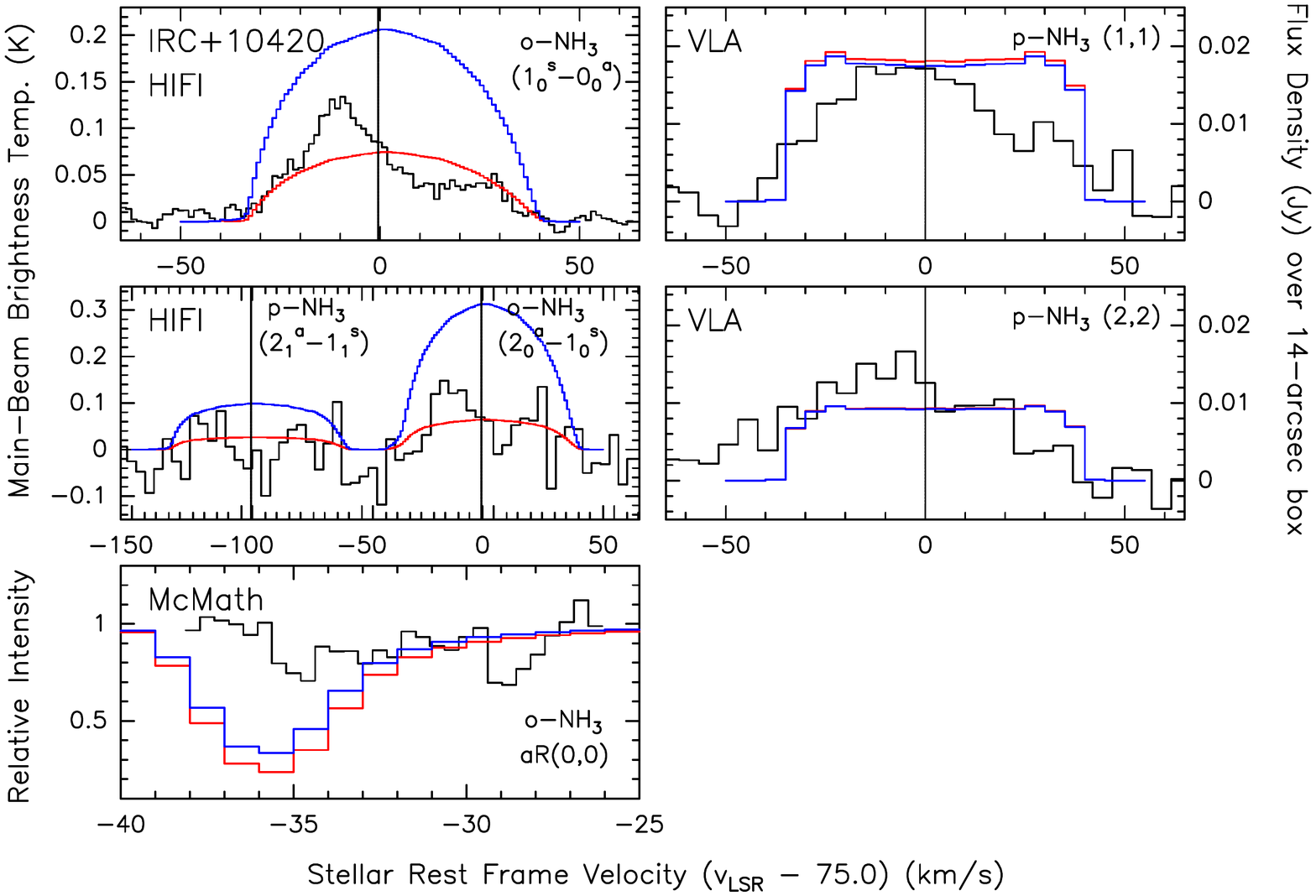}
\caption[]{NH$_3$ spectra of {\irc}. Black lines show the observed spectra; red and blue lines show the modelled spectra from the first model (low density, high abundance) assuming a dust-to-gas mass ratio of 0.001 and 0.005, respectively. The \textit{top-left} and \textit{middle-left} spectra are {\hifi} spectra of {\irc}. The two spectra on the \textit{right} are VLA spectra showing the flux density integrated over a $14\arcsec \times 14\arcsec$ box centred at {\irc}, i.e. the same region as shown in Fig. \ref{fig:vla_irc}. The black spectrum in the \textit{bottom} panel shows the old $aR(0,0)$ spectrum from the McMath Solar Telescope, as reproduced from Fig. 3 of \citet[][{\aasperm}]{mclaren1980}.}
\label{fig:model_irc_cc07_spec}
\end{figure*}

Figure \ref{fig:vla_irc} shows the images of NH$_3$ (1,1) and (2,2) emission from {\irc} as observed with the VLA in 2004, before major upgrade of the interferometer. We binned the image cube every $4.9{\kms}$ to produce the channel maps. We adopt the systemic velocity to be $\vlsr = 75{\kms}$ \citep{cc2007}. The two right panels in Fig. \ref{fig:model_irc_cc07_spec} show the observed spectra of the flux density over the same $14\arcsec$-box as displayed in the VLA images (Fig. \ref{fig:vla_irc}) together with the synthesised spectra as discussed in Sect. \ref{sec:mdl_irc}.

The NH$_3$ emission appears to be resolved by the restoring beam of $\text{FWHM} = 3{\farcs}7$ and appears to be distributed primarily along the east-west directions. The radius of $3\sigma$ detection extends up to about $7{\arcsec}$ for (1,1) and $5{\arcsec}$ for (2,2). In general, the spatial distributions of the emission from NH$_3$ (1,1) and (2,2) are very similar, with the (1,1) line being slightly stronger than (2,2). Most of the NH$_3$ emission is offset to the west/south-west in the $\vlsr$ range of ${\sim}40$--$70{\kms}$, while the emission is shifted to the east in extreme redshifted velocities near $95$--$105{\kms}$. Similar trend of the velocity gradient is also seen in the rotational emission of $^{13}$CO and SO as observed with the SMA by \citet{dvt2009}.

A close, channel-by-channel comparison of the interferometric images finds that the spatial distribution of the NH$_3$ inversion emission, especially that in the $(2,2)$ line, is strikingly similar to that of rotational emission in SO $J_K=6_5$--$5_4$ \citep[Fig. 6 of][]{dvt2009}. Both molecules appear to trace the similar clumpy structures in the CSE at various velocities, suggesting that both molecules share the similar kinematics and abundance distribution. The chemical models of \citet{willacy1997} for O-rich AGB stars suggest that SO is produced when S, S$^{+}$, or HS is produced from ion-neutral reactions or photodissociation and reacts with OH or O. Therefore, most of the SO molecules form in the gas-phase chemistry of the CSE. If the spatial distribution of NH$_3$ is similar to that of SO, then NH$_3$ has to be produced in the CSE rather than the inner wind as presumed in existing chemical models for lower-mass stars. The apparent lack of centrally peaked emission in our VLA images is consistent with such scenario.

There is a bright component emitting in both NH$_3$ lines at the radius of ${\sim}2\arcsec$ to the west of the star. This component is the most prominent in the $\vlsr$ range of $50$--$75{\kms}$, particularly near $65{\kms}$, i.e. a blueshifted velocity of $10{\kms}$. The high-resolution CO images of {\irc} also show a similar bright component between $60$ and $75{\kms}$, elongated along a PA of $75\degr$ \citep[Fig. 6 of][]{cc2007}. The peaks of the CO-enhanced component are about $4\arcsec$ and $1\arcsec$ south-west of the star in the CO $J=1$--$0$ and $J=2$--$1$ transitions, respectively. At $\vlsr = 65{\kms}$, the brightest CO emission arises from an offset of ${\sim}2\arcsec$.


\subsubsection{{\hifi} observations of {\irc}}
\label{sec:hifi_irc}

{\hifi} spectra of {\irc} are only available for the $J_K=1_0(\text{s})$--$0_0(\text{a})$, $2_0(\text{a})$--$1_0(\text{s})$, and $2_1(\text{a})$--$1_1(\text{s})$ transitions. The spectra are shown in the top-left and middle-left panels of Fig. \ref{fig:model_irc_cc07_spec}. The $1_0$--$0_0$ line has the highest S/N ratio and clearly shows asymmetry. The brightness of this line -- and perhaps also that of $2_0$--$1_0$ -- peaks at $\vlsr = 65{\kms}$, which is the same `peak velocity' as in the VLA data. Therefore, this spectral feature likely corresponds to the bright clump at ${\sim}2\arcsec$ west of the star in the VLA images. Additionally, there is a weaker, broad spectral component spanning up to the expansion velocity of ${\sim}37{\kms}$ \citep[e.g.][]{cc2007}. 

By comparing the NH$_3$ $1_0$--$0_0$ spectrum with other {\hifi} spectra of {\irc} in \citet{teyssier2012}, the NH$_3$ spectral features resemble those of the ortho-H$_2$O and perhaps OH transitions, especially o-H$_2$O $J_{K_a,K_c}=3_{2,1}$--$3_{1,2}$ and $3_{1,2}$--$2_{2,1}$. \citet{teyssier2012} suggested that both water and ammonia abundances may be enhanced by shock-induced chemistry in the CSE of {\irc}.


\subsubsection{Old McMath observations of {\irc}}
\label{sec:mcmath_irc}

The bottom panel of Fig. \ref{fig:model_irc_cc07_spec} shows the MIR $aR(0,0)$ spectrum reproduced from Fig. 3 of \citet{mclaren1980}. In this spectrum taken in 1979, there are two absorption components at the LSR velocities of ${\sim}40{\kms}$ and ${\sim}46{\kms}$, which correspond to the relative velocities of approximately $-35{\kms}$ and $-29{\kms}$, respectively, to the systemic velocity. While the more-blueshifted component probably arose from the gas moving towards us at the terminal expansion velocity (as in the cases of {\ik} and {\vy}), the intermediate-velocity component is more intriguing. Owing to the coincidence of their velocities, \citet{mclaren1980} associated this NH$_3$ absorption component with the 1612\,MHz OH maser spots imaged by \citet{benson1979}. \citet{nedoluha1992} interpreted the 1612\,MHz OH masers as coming from a shell of radius ${\sim}8000$--$9000$\,AU (${\sim}1{\farcs}5$), which was slightly smaller than the MIR continuum of {\irc} as imaged by \citet{shenoy2016}. Hence, the NH$_3$ component at $\vlsr \sim 46{\kms}$ could have come from the expanding shell at some offset from the line of sight towards the star, thus causing it to absorb the extended background continuum at a projected line-of-sight velocity.


\subsubsection{NH$_3$ model of {\irc}}
\label{sec:mdl_irc}

{\irc} is a yellow hypergiant with an extremely high and variable mass-loss rate \citep[${\sim}10^{-4}$--$10^{-3}\mspy$, e.g.][]{shenoy2016}. The CSE of the {\irc} is expanding at a maximum speed of about $37{\kms}$ \citep{cc2007}, which is the value we adopt as the constant expansion velocity of the NH$_3$-carrying gas. The gas density and temperature of the CSE have been studied in detail by modelling multiple CO rotational transitions \citep{teyssier2012}. The CO emission can be well explained by two distinct mass-loss episodes with a period very low global mass loss in between \citep{cc2007,dvt2009}.

In our NH$_3$ models, however, we found that the gas density profiles derived from such a mass-loss history are unable to explain the NH$_3$ emission, especially near the radii of 2{\arcsec}--4{\arcsec} where there is a low-density `gap' separating the two dominant gas shells. A higher density in the inter-shell region is required in our model. We assume that the density profile of this region corresponds to a uniformly expanding shell with an equivalent mass-loss rate being the average of the inner and outer shells of the model of \citet{cc2007}, i.e.
\begin{equation}
n_{\text{H}_2}(r) =
\begin{cases}
\displaystyle 9.37 \times 10^3\,{\percc} \cdot {r_{17}}^{-2} & \text{if $0.25 \le r_{17} \le 1.24$} \\
\displaystyle 7.46 \times 10^3\,{\percc} \cdot {r_{17}}^{-2} & \text{if $1.24 < r_{17} \le 2.20$} \\
\displaystyle 5.55 \times 10^3\,{\percc} \cdot {r_{17}}^{-2} & \text{if $2.20 < r_{17} \le 5.20$} \\
\end{cases},
\end{equation}
where
\begin{equation}
\displaystyle {r_{17}} = \frac{r}{10^{17}\,\text{cm}}.
\end{equation}
The gas temperature profile is based on that derived by \citet{teyssier2012}. We assume the gas temperature in the inter-shell region to follow the same power-law as the outer shell, i.e.
\begin{equation}
T_\text{kin}(r) =
\begin{cases}
\displaystyle 170\,\text{K} \cdot {r_{17}}^{-1.2} & \text{if $0.25 \le r_{17} \le 1.24$} \\
\displaystyle 100\,\text{K} \cdot {r_{17}}^{-0.8} & \text{if $1.24  <  r_{17} \le 5.20$} \\
\end{cases}.
\end{equation}

We found that infrared pumping by dust grains has significant influence on the intensity of the rotational transitions. Assuming a typical dust-to-gas mass ratio of $\text{d/g}=0.005$ \citep{oudmaijer1996}, the modelled rotational line intensities are significantly higher than the observed. We had to adopt a lower d/g in order to reduce the contribution of radiative excitation from dust. The following NH$_3$ abundance distribution is found to be able to reproduce the radial intensity profiles of NH$_3$(1,1) and NH$_3$(2,2) at the peak velocity ($\vlsr = 65{\kms}$): 
\begin{equation}
{\ammabun} =
\begin{cases}
\displaystyle 1.0 \times 10^{-6} & \text{if $0.25 \le r_{17} \le 1.24$} \\
\displaystyle 2.5 \times 10^{-6} & \text{if $1.24  <  r_{17} \le 2.20$} \\
\displaystyle 1.0 \times 10^{-6} & \text{if $2.20  <  r_{17} \le 5.20$} \\
\end{cases}.
\end{equation}

Figure \ref{fig:model_irc_cc07_spec} shows our model fitting to all available spectra with two different d/g ratios based on the above model. Our modelled $aR(0,0)$ spectrum does not reproduce the absorption component at the intermediate velocity near $-29{\kms}$ because our model does not consider an extended MIR continuum; therefore, line absorption can only occur along the central line of sight and hence at the blueshifted expansion velocity. The radio inversion lines are not affected by the adopted d/g ratio. Our model predicts the inversion lines to be optically thin and show flat-topped spectra. The observed VLA profiles, however, are far from being flat. The synthesised absorption profile at the terminal blueshifted velocity is much deeper than the observed, so is the emission profile in NH$_3$(1,1). Since our model adopts the oversimplified assumption of a smooth and uniform distribution of NH$_3$ in the CSE, the discrepancies are probably due to the non-uniform distribution of the molecule, particularly in the outer, cooler part of the CSE which dominates NH$_3$(1,1) emission and $aR(0,0)$ absorption.

\begin{figure*}[!htbp]
\centering
\includegraphics[height=\ircparamh]{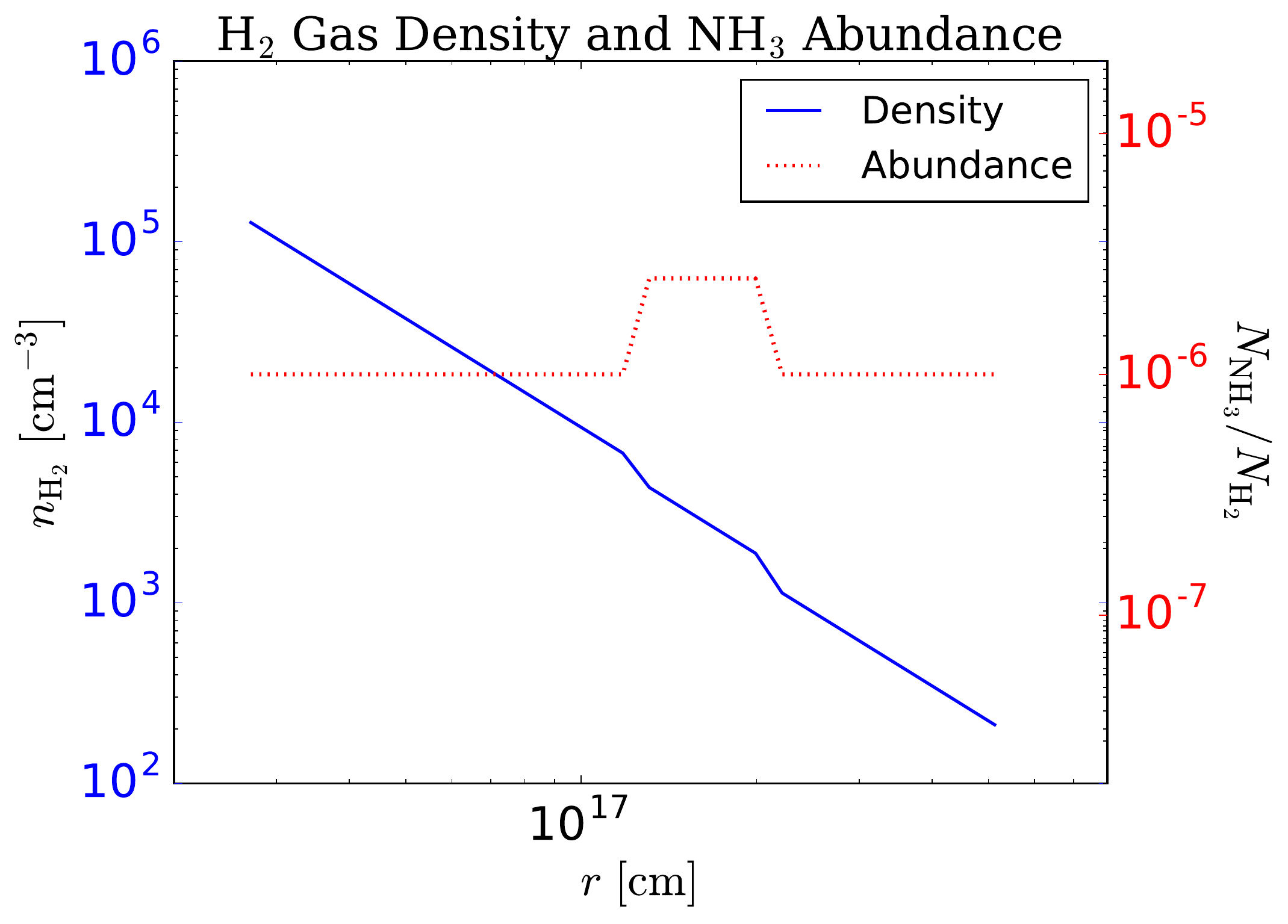}
\includegraphics[height=\ircparamh]{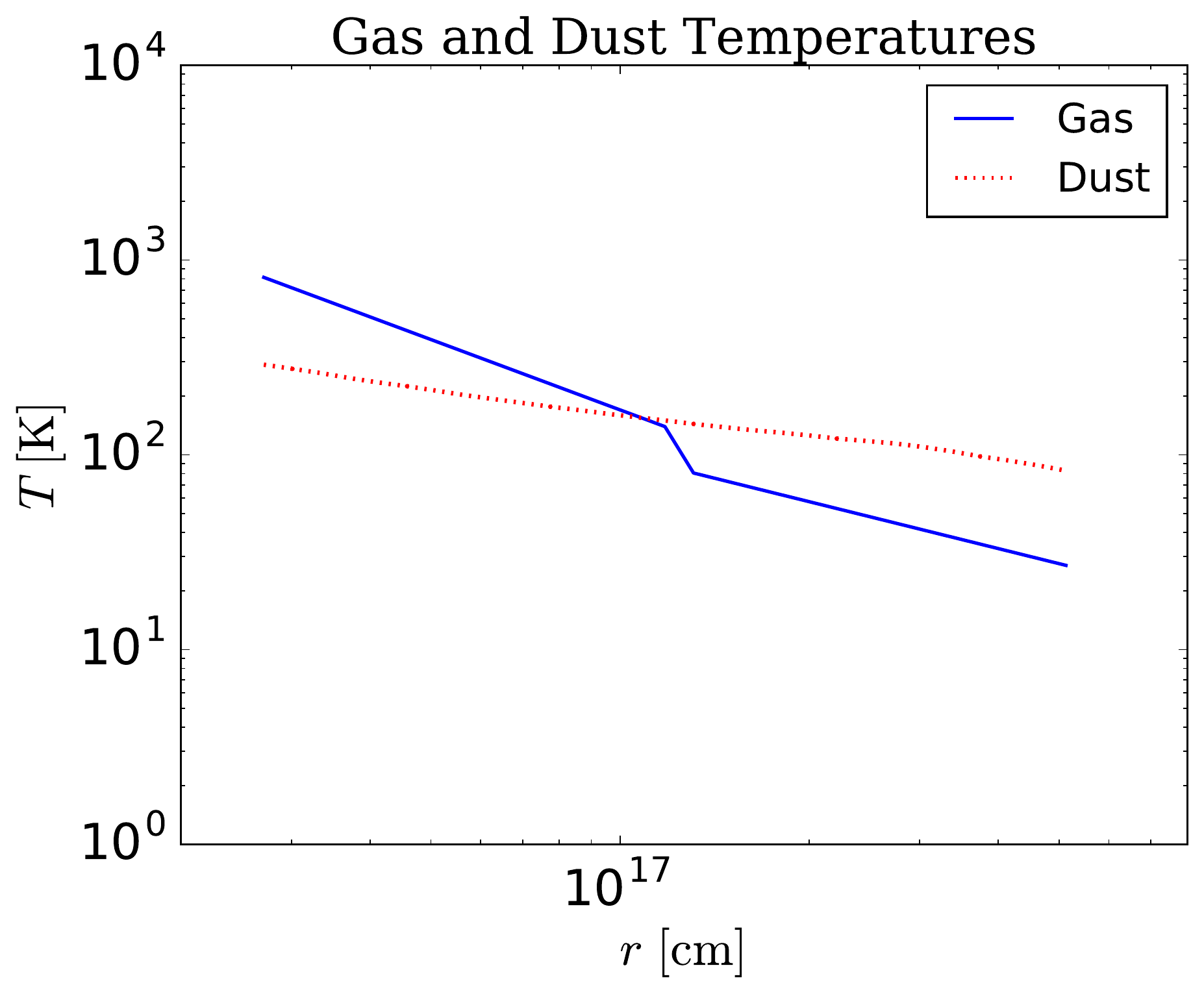}\\
\includegraphics[height=\ircheight]{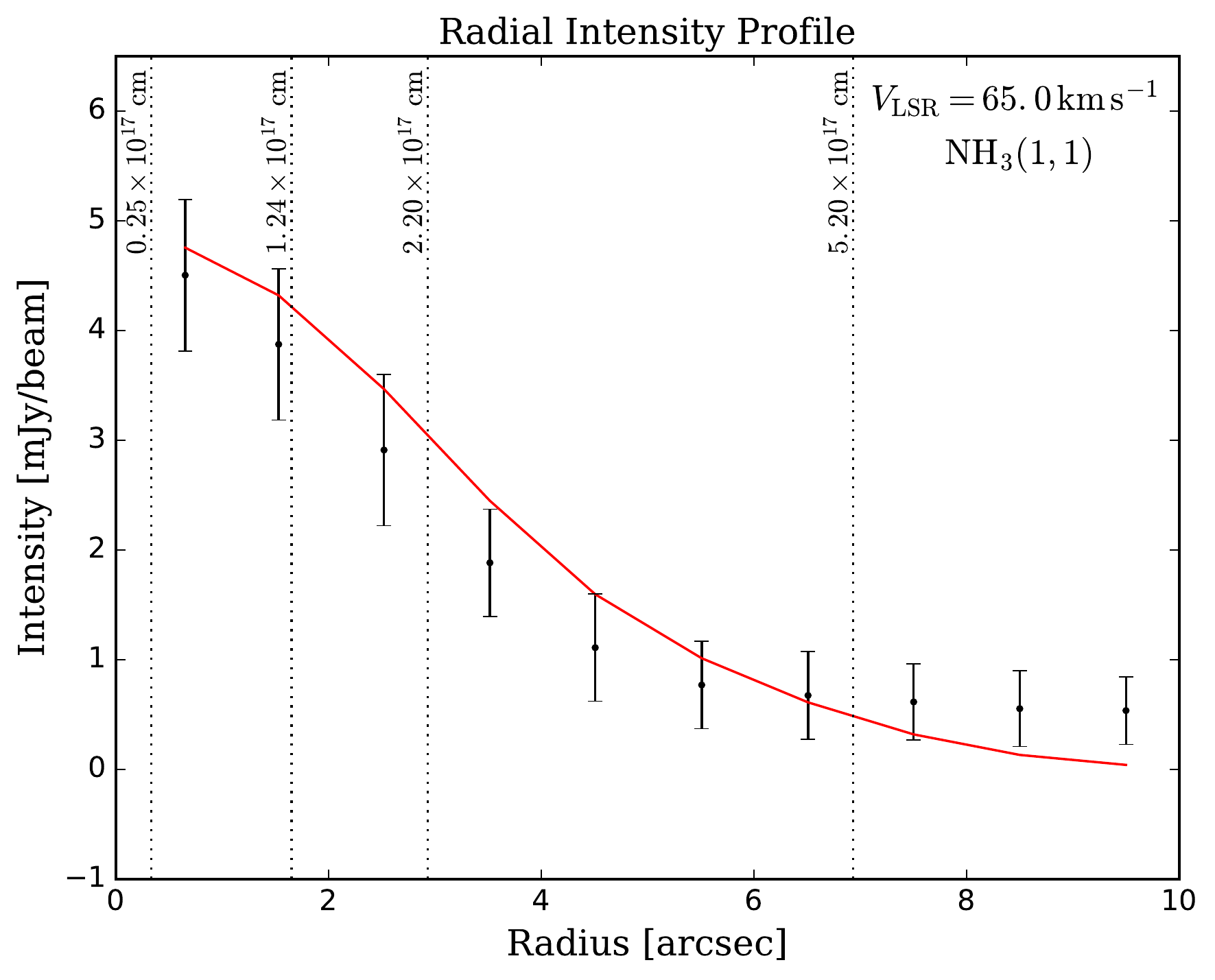}
\includegraphics[height=\ircheight]{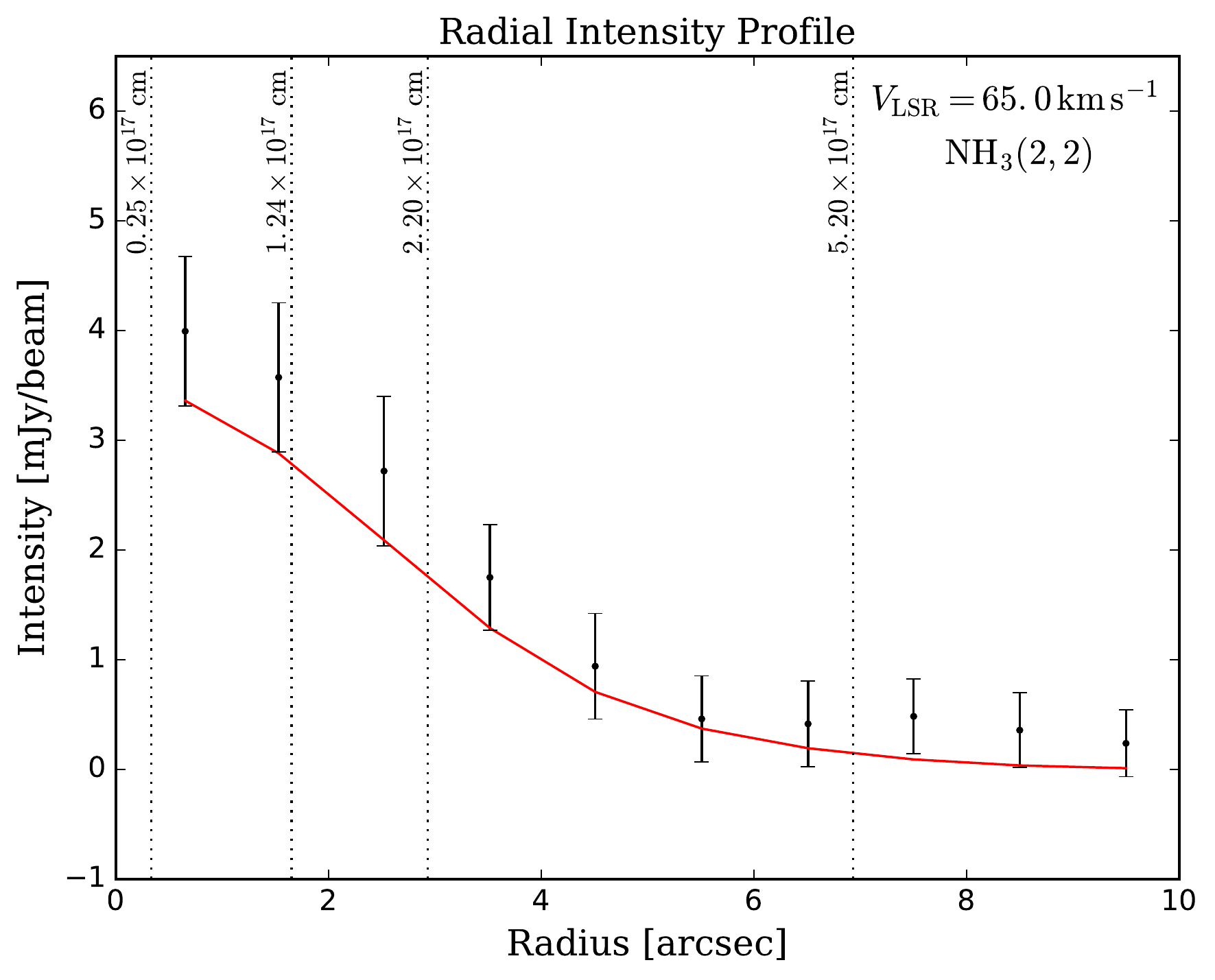}\\
\includegraphics[height=\ircheight]{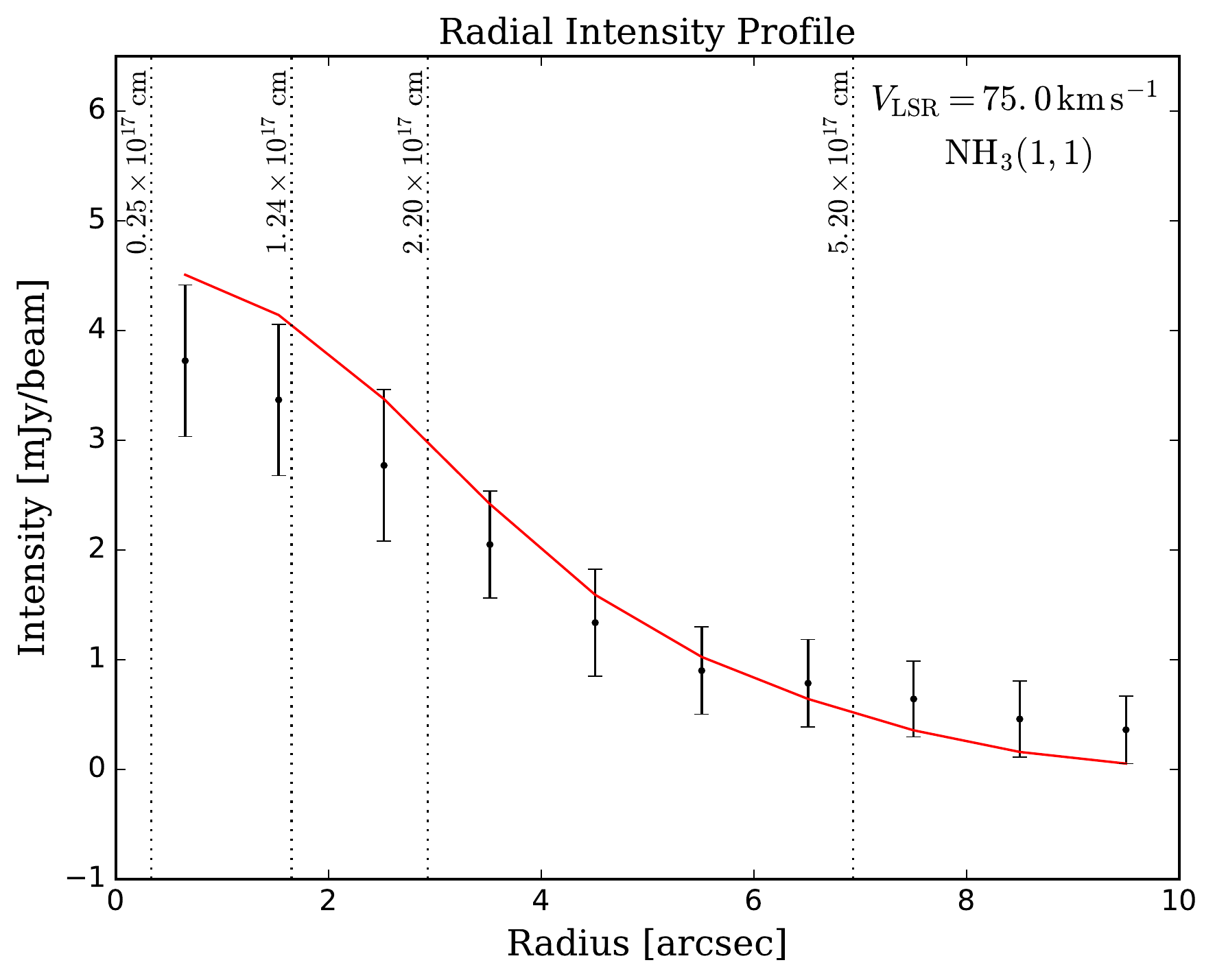}
\includegraphics[height=\ircheight]{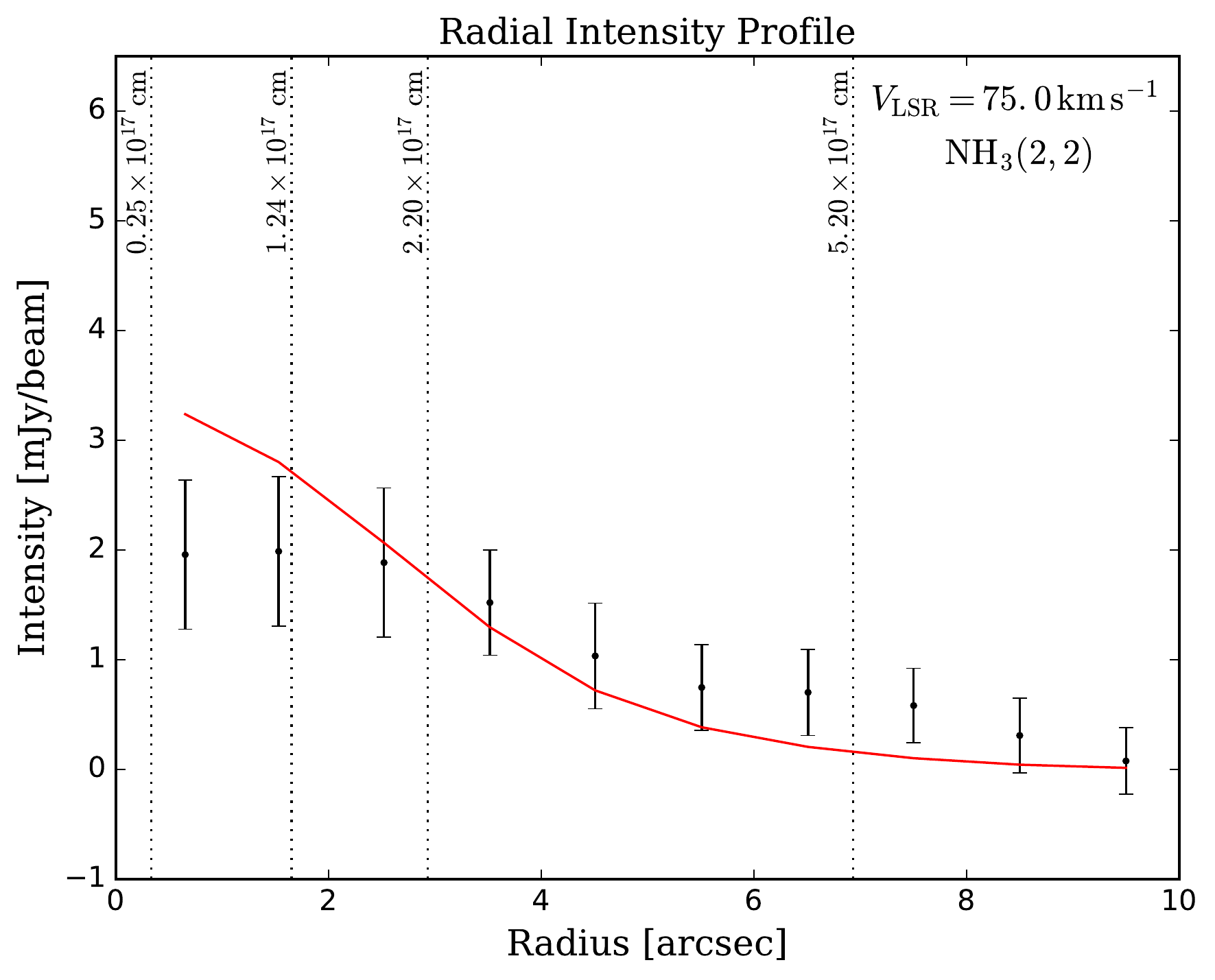}
\caption[]{The first model for {\irc} with low gas density and high NH$_3$ abundance. \textit{Top left}: Input gas density and NH$_3$ abundance profiles. Only the radii with abundance ${>}10^{-10}$ are plotted. \textit{Top right}: Input gas and dust temperature profiles. \textit{Middle}: The radial intensity profiles of NH$_3$(1,1) (\textit{left}) and NH$_3$(2,2) (\textit{right}) at the velocity channel $\vlsr = 65{\kms}$, which show the strongest emission in both rotational and inversion transitions. \textit{Bottom}: The radial intensity profiles of NH$_3$(1,1) (\textit{left}) and NH$_3$(2,2) (\textit{right}) at the velocity channel $\vlsr = 75{\kms}$, which is the systemic velocity of {\irc}. In the radial intensity profiles, black points indicate the binned data points from the VLA images and red curves show the modelled profiles.}
\label{fig:model_irc_cc07_prof}
\end{figure*}

The input parameters, together with the resultant radial intensity profiles at $\vlsr = 65{\kms}$ and at the systemic velocity ($\vlsr = 75{\kms}$), are plotted in Fig. \ref{fig:model_irc_cc07_prof}. At $\vlsr = 65{\kms}$, while this model is able to fit the radial intensity profiles within uncertainties, it tends to underestimate the line intensity ratio of NH$_3$(2,2)/NH$_3$(1,1). The model cannot produce a satisfactory fit to the radial intensity profiles at the systemic velocity.

The NH$_3$ abundance of ${\sim}10^{-6}$ is significantly higher, by a factor ranging from 5 to 12.5, than that derived for the inversion lines by \citet{menten2010}. In their model, they used a much higher density which corresponds to a constant mass-loss rate of $2 \times 10^{-3}\mspy$ and is approximately seven times higher than our assumed density. Similar models requiring high gas density to explain molecular line emission have also been reported by \citet{wong2013} and \citet{ql2016} for the SiO molecule. The modelling exercises of \citet{wong2013} found that it was impossible to explain the SiO emission, in particular the radial intensity profiles, with the model of two disconnected gas shells from \citet{cc2007}. Indeed, a higher gas density throughout the entire SiO-emitting region is required \citep{wong2013,ql2016}. Therefore, it is speculated that the SiO emission originates from localised clumpy structures, possibly created by shocks, that are not represented by the global mass-loss history as traced by CO emission \citep[][Dinh-V.-Trung et al., in prep.]{wong2013}.

Assuming that the dense gas associated with SiO emission is distinct from the diffuse gas producing the bulk of the CO emission, \citet{wong2013} introduced two models of density and abundance stratification both different from those based on CO. The `three-zone model' retains the shell boundaries in the model of \citet{cc2007}, but adopts an at least four times higher gas density throughout the modelled region \citep[Sect. 5.4 of][]{wong2013}; whereas the `two-zone model' consists of two bordering zones with a higher SiO abundance in the inner one and has the gas density following a single power-law at a shallower gradient than a uniformly expanding envelope \citep[Sect. 6.4 of][]{wong2013}. The gas density of the two-zone model is generally higher than the CO-based density from \citet{cc2007} except at the innermost radii (${\lesssim} 5 \times 10^{16}\,\text{cm}$).

We consider the gas density from the two-zone model as the high-end estimate for {\irc}'s CSE and derive the minimum NH$_3$ abundance from the optically thin inversion lines. Our adopted gas density profile is slightly modified from that of \citet{wong2013} and is described as
\begin{equation}
\displaystyle n_{\text{H}_2}(r) = 2.30 \times 10^4\,{\percc} \cdot {r_{17}}^{-0.7} \quad \text{if} \quad 0.25 \le r_{17} \le 6.00.
\end{equation}
By fitting the radial intensity profiles at the systemic velocity, we obtain the following NH$_3$ abundance distribution:
\begin{equation}
{\ammabun} =
\begin{cases}
\displaystyle 2.5 \times 10^{-7} & \text{if $0.25 \le r_{17} \le 2.35$} \\
\displaystyle 4.0 \times 10^{-8} & \text{if $2.35  <  r_{17} \le 6.00$} \\
\end{cases}.
\end{equation}

\begin{figure*}[!htbp]
\centering
\includegraphics[trim=0.5cm 2.3cm 1.5cm 1.5cm, clip, width=\specwidth]{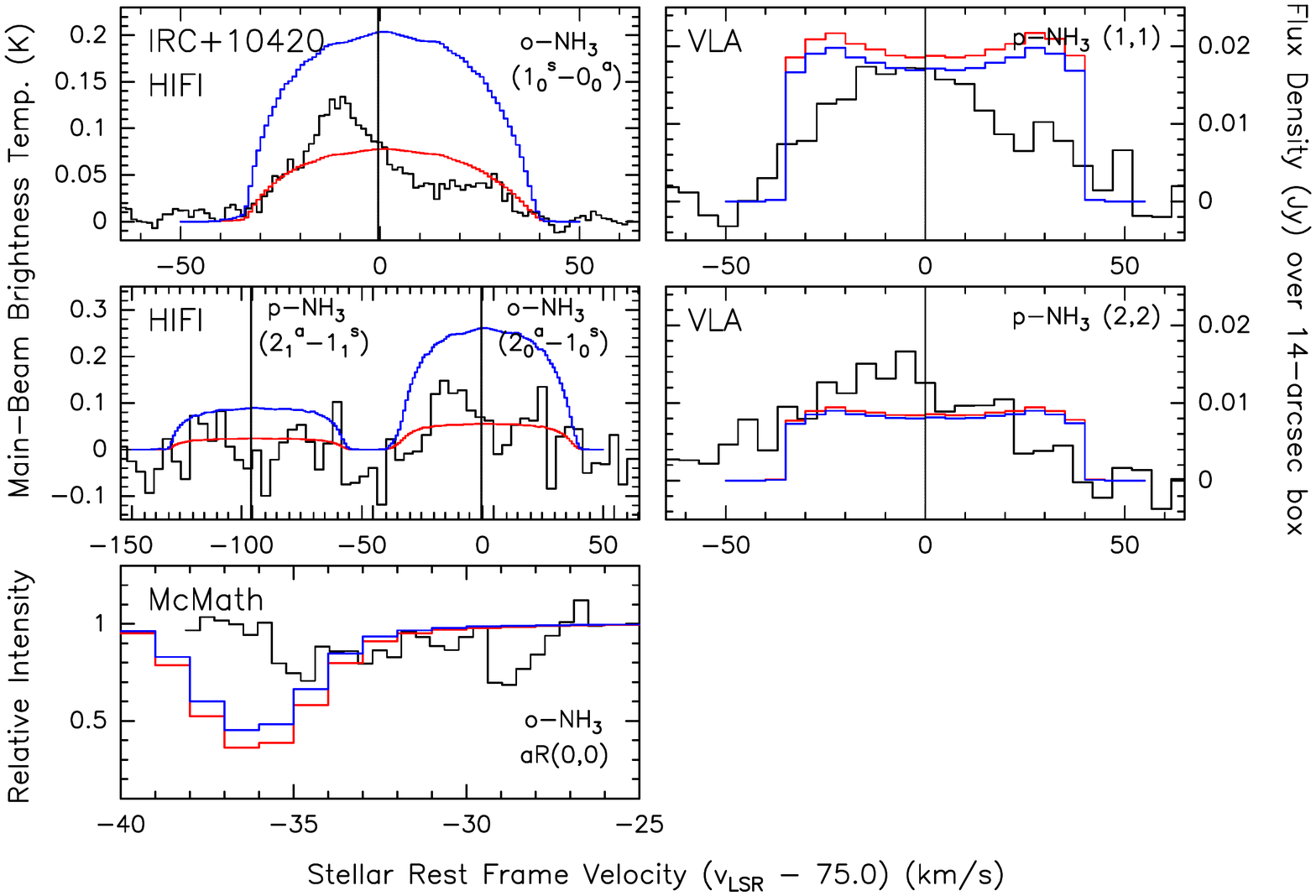}
\caption[]{NH$_3$ spectra of {\irc}. Black lines show the observed spectra; red and blue lines show the modelled spectra from the second model (high density, low abundance) assuming a dust-to-gas mass ratio of 0.001 and 0.005, respectively. The \textit{top-left} and \textit{middle-left} spectra are {\hifi} spectra of {\irc}. The two spectra on the \textit{right} are VLA spectra showing the flux density integrated over a $14\arcsec \times 14\arcsec$ box centred at {\irc}, i.e. the same region as shown in Fig. \ref{fig:vla_irc}. The black spectrum in the \textit{bottom} panel shows the old $aR(0,0)$ spectrum from the McMath Solar Telescope, as reproduced from Fig. 3 of \citet[][{\aasperm}]{mclaren1980}.}
\label{fig:model_irc_w13_spec}
\end{figure*}

\begin{figure*}[!htbp]
\centering
\includegraphics[height=\ircparamh]{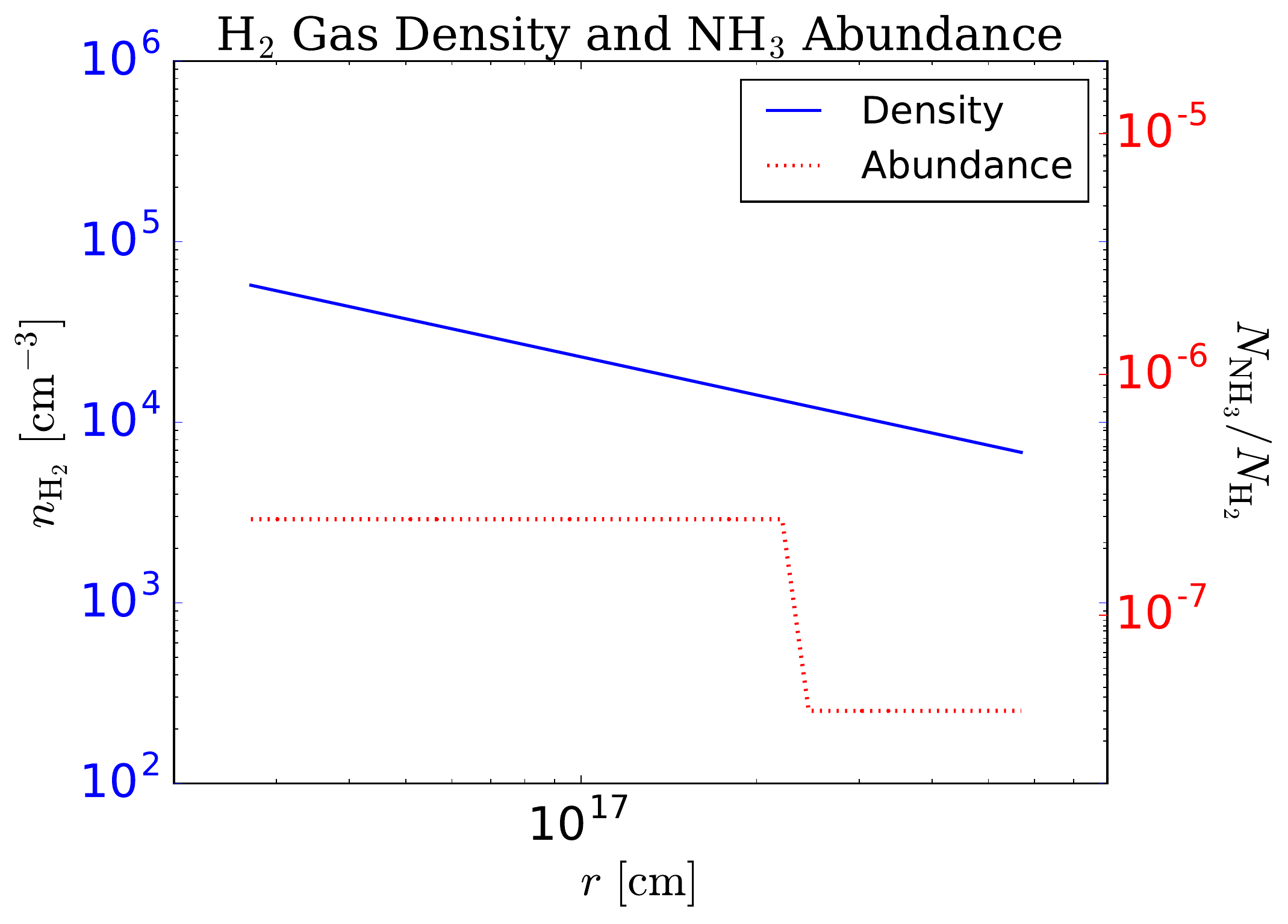}
\includegraphics[height=\ircparamh]{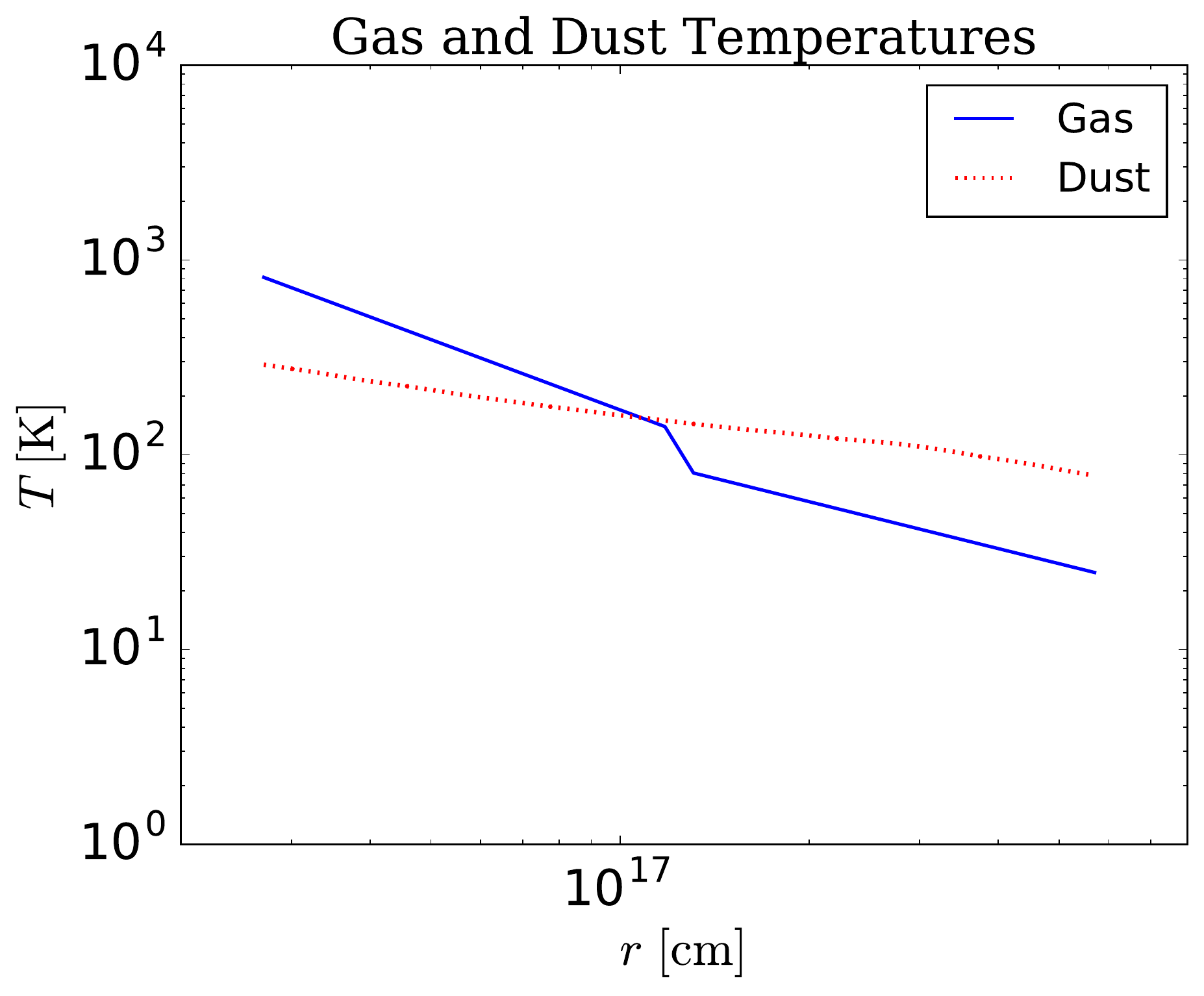}\\
\includegraphics[height=\ircheight]{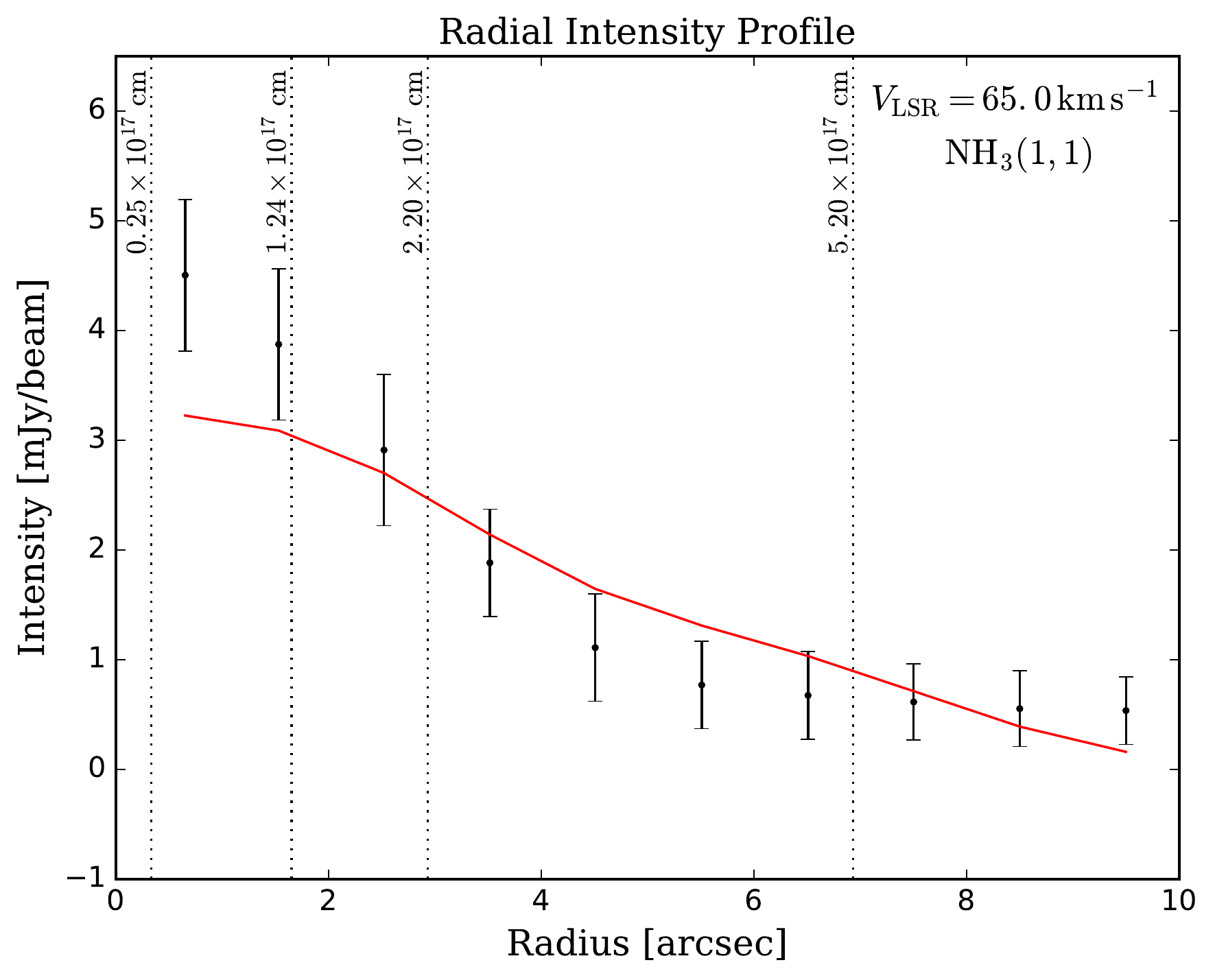}
\includegraphics[height=\ircheight]{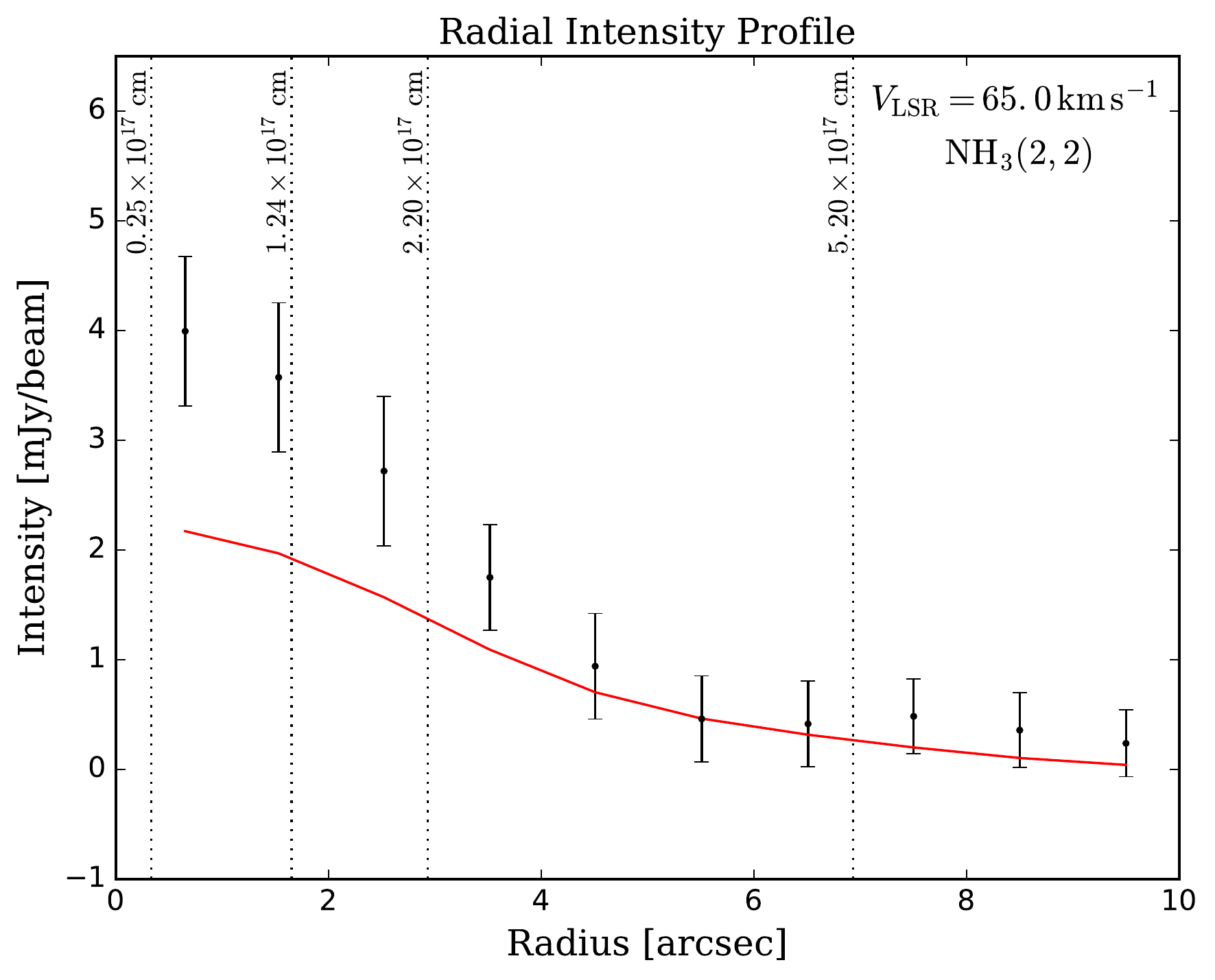}\\
\includegraphics[height=\ircheight]{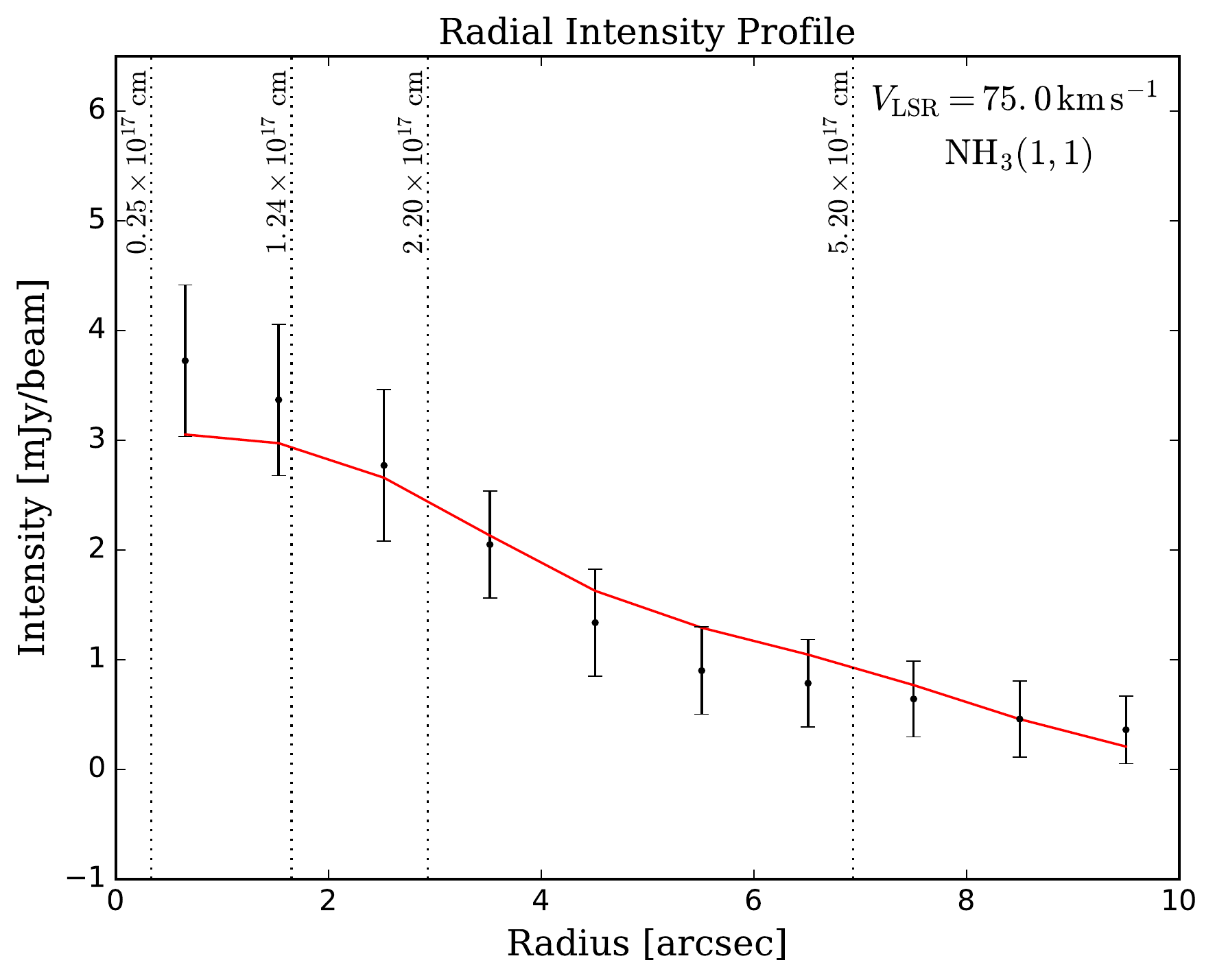}
\includegraphics[height=\ircheight]{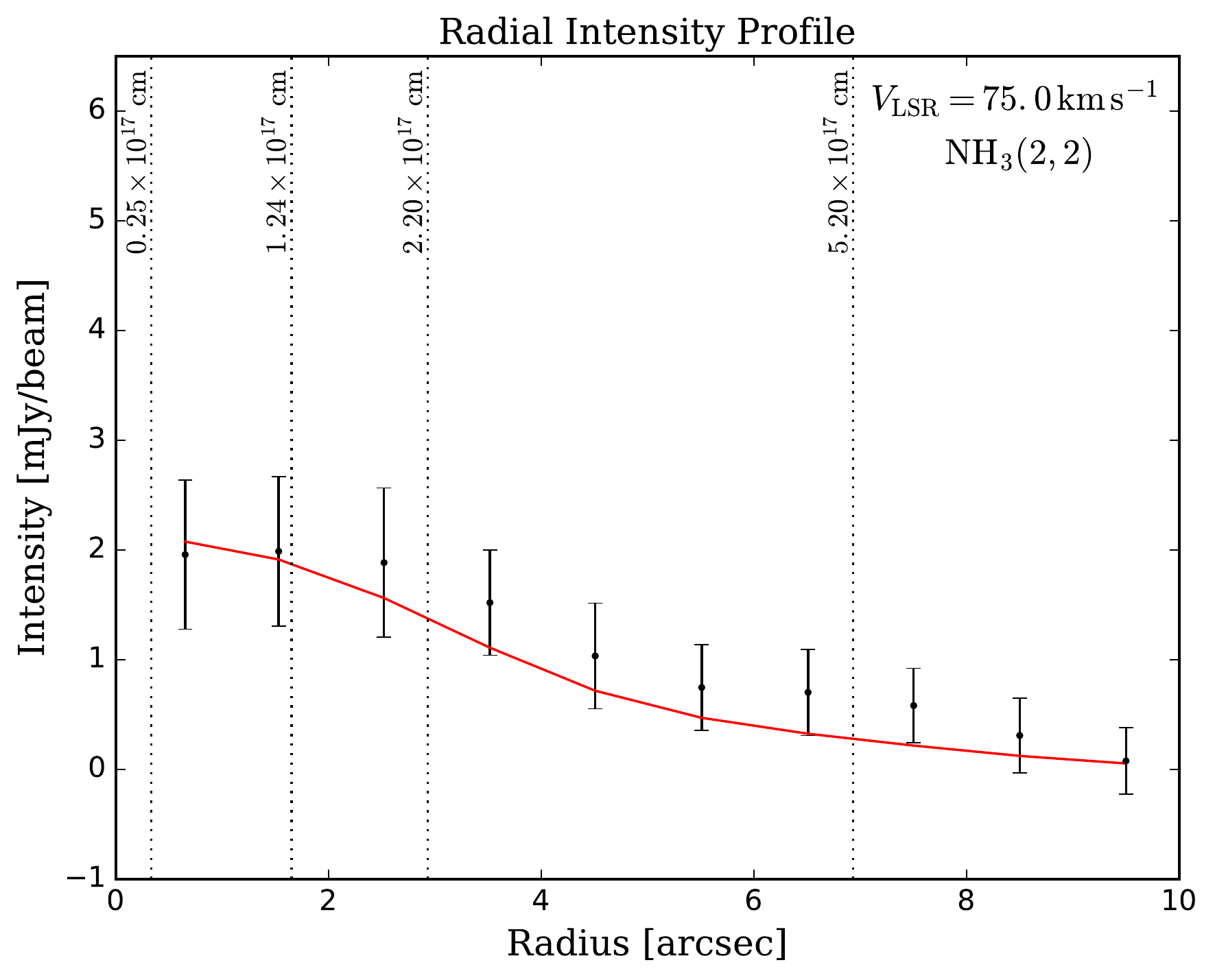}
\caption[]{The second model for {\irc} with high gas density and low NH$_3$ abundance. \textit{Top left}: Input gas density and NH$_3$ abundance profiles. Only the radii with abundance ${>}10^{-10}$ are plotted. \textit{Top right}: Input gas and dust temperature profiles. \textit{Middle}: The radial intensity profiles of NH$_3$(1,1) (\textit{left}) and NH$_3$(2,2) (\textit{right}) at the velocity channel $\vlsr = 65{\kms}$, which show the strongest emission in both rotational and inversion transitions. \textit{Bottom}: The radial intensity profiles of NH$_3$(1,1) (\textit{left}) and NH$_3$(2,2) (\textit{right}) at the velocity channel $\vlsr = 75{\kms}$, which is the systemic velocity of {\irc}. In the radial intensity profiles, black points indicate the binned data points from the VLA images and red curves show the modelled profiles.}
\label{fig:model_irc_w13_prof}
\end{figure*}

Figure \ref{fig:model_irc_w13_spec} shows our model fitting with two different d/g ratios to all available spectra and Fig. \ref{fig:model_irc_w13_prof} shows the plots of the input parameters and the resultant radial intensity profiles. The quality of the fitting and the effects of the d/g ratio are similar to our first model, but the NH$_3$ abundance has been reduced by at least one order of magnitude. Because of the ambiguity of the gas density and abundance, we are unable to conclusively determine the NH$_3$ abundance in the CSE of {\irc}, which is likely in the range ${\sim}10^{-7}$--$10^{-6}$. In any case, the density of NH$_3$-emitting gas cannot be the same as that of the ambient medium as traced by CO for a considerable range of radii; we therefore speculate that NH$_3$ emission probably also comes from the dense, localised clumps throughout {\irc}'s CSE.

Similar to {\ik}, {\pacs} observations\footnote{{\herschel} OBSIDs: 1342208886, 1342208887.} also covered higher rotational transitions from $J=3$--$2$ up to $J=9$--$8$. The only detected features are the $J=3$--$2$ lines. In both of our models, the total flux of the blended lines is within the range of the predicted fluxes spanned by the two adopted d/g ratios.

\subsection{Ammonia chemistry}
\label{sec:results_chem}

Table \ref{tab:models} summarises the adopted parameters and results of our NH$_3$ models. In general, the NH$_3$ abundance required to explain the spectral line data is of the order of $10^{-7}$, except for {\irc} where this value is a lower limit due to the ambiguous gas density of the CSE. The abundances in {\ik} and {\vy} are a few times lower than those obtained by \citet{menten2010}, in which MIR radiative pumping of the molecule to the $\varv_2=1$ state was not considered. The decrease in the inferred abundance upon the inclusion of infrared pumping has also been reported by \citet{schmidt2016} for the carbon star {\cw}, which exhibits a peak abundance of about $3 \times 10^{-8}$.

\begin{table*}[!htbp]
\caption{Parameters of our NH$_3$ models.}
\label{tab:models}
\centering
\begin{tabular}{cccccccccc}
\hline\hline
Target & $d$ & {\rstar} & {\tstar} & $V_\text{sys}$ & {\vexp} & {\bdopp} & {\rin} & {\refd} & $f_0$ \\
       & (pc) & (AU, mas) & (K) & ({\nkms}) & ({\nkms}) & ({\nkms}) & ({\rstar}) & ({\rstar}, arcsec) & ({\ammabun}) \\ \hline
{\ik}  &  265\tm{a} & 1.00, 3.8\tm{a}  & 2200\tm{a} & 34\tm{b} & 17.7\tm{a} & 4.0 &  75 &  600, $2{\farcs}3$ & $6 \times 10^{-7}$ \\
{\vy}  & 1140\tm{c} & 6.60, 5.8\tm{d}  & 3490\tm{d} & 22\tm{e} & 35\tm{f}   & 7.5 &  60 &  350, $2{\farcs}0$ & $7 \times 10^{-7}$ \\
{\oh}  & 1540\tm{g} & 2.93, 1.9\tm{h}  & 2300\tm{i} & 34\tm{j} & ---\tm{k}  & 1.0 &  20 & 2100, $4{\arcsec}$ & $5 \times 10^{-7}$ \\
{\irc} & 5000\tm{l} & 1.75, 0.35\tm{m} & 8250\tm{n} & 75\tm{p} & 37\tm{p}   & 1.8 & 955 & ${\sim}20\,000$, $7{\arcsec}$--$8{\arcsec}$ & ${\gtrsim}2.5 \times 10^{-7}$ \\
\hline
\end{tabular}
\tablefoot{
The columns are (from \textit{left} to \textit{right}): target, distance ($d$; pc), stellar radius ({\rstar}; AU or mas), stellar temperature ({\tstar}; K), systemic velocity in LSR ($V_\text{sys}$; {\nkms}), terminal expansion velocity ({\vexp}; {\nkms}), Doppler-$b$ parameter ({\bdopp}; {\nkms}), inner radius of NH$_3$ distribution ({\rin}; {\rstar}), $e$-folding radius of NH$_3$ ({\refd}; {\rstar} or arcsec), and the peak NH$_3$ abundance ($f_0$; relative to H$_2$). \\
\tablefoottext{a}{\citet{decin2010}.}
\tablefoottext{b}{\citet{vp2017}.}
\tablefoottext{c}{\citet{choi2008}.}
\tablefoottext{d}{\citet{wittkowski2012}.}
\tablefoottext{e}{\citet{alcolea2013}.}
\tablefoottext{f}{\citet{decin2006}.}
\tablefoottext{g}{\citet{choi2012}.}
\tablefoottext{h}{From Stefan--Boltzmann law (Sect. \ref{sec:results_oh}).}
\tablefoottext{i}{\citet{sc2002}.}
\tablefoottext{j}{\citet{sc2015,vp2015}.}
\tablefoottext{k}{Equation (\ref{eq:oh_vel}).}
\tablefoottext{l}{\citet{jones1993}.}
\tablefoottext{m}{\citet{oudmaijer2013}.}
\tablefoottext{n}{\citet{shenoy2016}.}
\tablefoottext{p}{\citet{cc2007}.}
}
\end{table*}

The inner boundaries of the NH$_3$ abundance distribution are ${<}100\,\rstar$ for {\ik}, {\vy}, and {\oh}. The values of {\rin} are chosen to reproduce the observed line ratios and depend heavily on our assumptions of the gas density and temperature profiles. The MIR profiles of {\ik}, {\vy}, and {\irc} appear near the terminal expansion velocities, suggesting that the MIR-absorbing NH$_3$ is in the outer, accelerated part of the CSEs. 

None of the four targets in this study shows a significant amount of NH$_3$ at radii close to the stellar photosphere, where pulsation-driven shocks dominate the chemistry. Spatially-resolved images of {\ik} and {\irc} show NH$_3$-emitting clumps that are offset from the star; unresolved images of {\vy} and {\oh} also show slight departure of the peak emission from the stellar continuum. Furthermore, had NH$_3$ been produced at smaller radii where both density and temperature are very high, emission from higher-$J$ transitions would have been much stronger than the lower ones and the inversion line profiles would have shown narrow features at the outflow or infall velocities during the time of observations. Hence, we conclude that NH$_3$ do not form near the stellar photosphere or in the pulsation-driven shocks in our targets even thought tight constraints of {\rin} could not be determined from available data\footnote{(Sub)millimetre rotational transitions within the $\varv_2=1$ state near 140\,GHz and 466\,GHz from would be detected in emission towards {\vy} with on-source time of ${\sim}10$ hours with the IRAM 30m Telescope or the APEX 12-metre telescope (Atacama Pathfinder EXperiment) if NH$_3$ reached a high abundance near the stellar photosphere. Given a velocity law, the predicted line profiles of these transitions are sensitive to the inner radius. However, no such observations exist to our knowledge.}.

Earlier detection of NH$_3$ MIR absorption from O-rich AGB stars includes the Mira variables {\omi} and, less likely, {\rleo} (see Table \ref{tab:obs_mir}). \citet{betz1985} reported large velocity variation in the absorption profiles observed towards {\omi} in 1981--1982. At some epochs, NH$_3$ even exhibited redshifted absorption, which is indicative of infall and has never been seen in the NH$_3$ spectra of any other evolved stars. The NH$_3$ absorption up to $20\%$ of the continuum intensity was very strong. In their unpublished spectra, the velocity of the absorption relative to the systemic could reach ${\sim}30{\kms}$ at its utmost \citep{betz1985}. Modern dynamic atmosphere models of Mira variables predict that such strongly-varying and high-velocity motions occur in the vicinity of the infrared photosphere \citep[${\sim}1\,\rstar$; e.g.][]{ireland2008,ireland2011,freytag2017}. Chemical model suggests that a strong shock of velocity ${\gtrsim}30{\kms}$ at $1.0\,\rstar$ could produce a considerable amount of NH$_3$ \citep{gobrecht2016}. Hence, the MIR lines detected towards {\omi} in 1981--1982 likely came from the pulsating wind. However, NH$_3$ detection in {\omi} is only sporadic. Our IRTF/TEXES observations in 2016 did not detect NH$_3$ absorption above ${\sim}5\%$ of the continuum intensity; the {\hifi} observation\footnote{{\herschel} OBSID: 1342200904.} in 2010 also did not find any emission in the rotational transition $J=1$--$0$. In any case, {\omi} does not show evidence of high NH$_3$ abundance in the outer CSE and, therefore, it has a very different ammonia chemistry from the targets of this study. The main difference between {\omi} and {\ik} is the mass-loss rate (and hence the gas density), which is much lower in {\omi} than in {\ik} by a factor of ${\sim}30$ \citep{ryde2001,decin2010}. High gas density may be one of the criteria of efficient NH$_3$ production.

The photodissociation models of \citet{decin2010nat} and \citet{agundez2010} suggest that a significant amount of NH$_3$ could be produced if the CSE were clumpy enough to allow deep penetration of interstellar UV photons. Assuming that one-fourth of the solid angle of the incoming UV flux was unattenuated, \citet{agundez2010} predicted a high peak NH$_3$ abundance of nearly $10^{-7}$ in an O-rich star with the mass-loss rate of {\ik} (${\sim}10^{-5}\mspy$). This abundance is only a few times lower than our derived value. On the other hand, another consequence of the deep UV photodissociation is the rapid formation of methane (CH$_4$) due to the availability of free carbon atoms. Photochemical models suggest that a substantial amount of CH$_4$ would lead to the formation of excess methanol (CH$_3$OH) and ethynyl radical (C$_2$H) \citep{charnley1995}; however, the later two molecules have never been detected towards `typical' O-rich AGB stars\footnote{CH$_3$OH has only been detected (recently) towards two evolved stars and one evolved star candidate thus far, including the post-AGB object HD\,101584 \citep{olofsson2017}, the YHG {\irc} at marginal S/N ratios \citep{ql2016}, and the peculiar object IRAS\,19312+1950 \citep{nakashima2015}, which is located in a complex region that contains dense interstellar gas \citep[see also][]{nakashima2016,cordiner2016}.} like {\ik} \citep[e.g.][]{latter1996a,latter1996b,charnley1997,marvel2005,gomez2014}.

Enhanced nitrogen production by stellar nucleosynthesis has been proposed to explain the nitrogen-rich chemistry of some of our targets (Sect. \ref{sec:intro}). However, nitrogen enhancement alone cannot be the solution of the NH$_3$ abundance problem. Hot bottom burning (HBB) in intermediate-mass AGB stars ($M=4$--$8\,\msun$) can enhance the surface nitrogen abundance at the end of the AGB evolution by up to an order of magnitude \citep{karakas2014,ventura2017}. In massive stars ($M \gtrsim 8\,\msun$), nitrogen is primarily produced in the CN branch of the CNO cycles \citep{henry2000} and the surface nitrogen abundance in A-type supergiants could increase by a factor of a few ($<10$) compared to the solar value due to partial mixing of the CN-cycled gas \citep{venn1995}. Optical spectroscopy of the massive hypergiant {\irc} also found a similar nitrogen enhancement factor to A-type supergiants \citep{klochkova1997}. The enhancement of nitrogen by a factor of ${\lesssim}10$ is insufficient to explain the nitrogen chemistry of our targets. For example, nitrogen in {\irc} has to be enriched by at least 20 times of the solar abundance in order to explain the NO emission \citep{ql2013}; the abundance of N-bearing molecules could not be explained even if nitrogen was enhanced by 40 times in the chemical models of \citet{vp2015} and \citet{sc2015}. The line surveys of \citet{tenenbaum2010} and \citet{kaminski2013} did not find nitrogen-rich chemistry in {\vy}, even though the supergiant shows the brightest NH$_3$ rotational emission among our targets. On the other hand, the low-mass AGB star {\ik} is not expected to be enriched in elemental nitrogen abundance by HBB or CNO cycles, yet it shows signs of N-rich chemistry \citep[Sect. \ref{sec:iram_ik} and][]{vp2017}. Furthermore, our observations do not find significant production of NH$_3$ near the stellar photosphere. The effects of nitrogen enrichment on the NH$_3$ abundance are therefore insignificant.

Our modelling shows that NH$_3$ emission probably arises from denser and more turbulent gas than the ambient medium. We find that the turbulence parameters of the NH$_3$-emitting gas in {\ik} and {\vy} have to be significantly higher than the typical values in the bulk of the circumstellar gas so as to fit the broad emission wings in the rotational transitions; our model of {\oh} suggests that a density enhancement of about three to four times is required to fit the spectral lines (Sect. \ref{sec:mdl_oh}); the strongest NH$_3$ emission of {\vy} and {\oh} comes from localised spatial-kinematic features (in the outer CSE) that could be associated with molecules such as SO and SO$_2$ (Sects. \ref{sec:vla_vy}, \ref{sec:hifi_vy}, and \ref{sec:vla_oh}). We note that the abundance of sulphur oxides is yet another mystery in circumstellar chemistry of O-rich stars \citep[e.g.][]{yamamura1999,fu2012,adande2013,danilovich2016,vp2017}. Our results may be indicative of some violent processes, such as shocks, in the formation of NH$_3$. A similar interpretation has also been suggested to explain the high abundance and clump distributions of NH$_3$ observed towards the carbon-rich PPNs {\egg} (Egg Nebula) and {\crl} \citep{mp1992a,mp1993,mp1995,tb1996,dvt2009nh3}.

While circumstellar shocks may play a role in producing the excess amount of NH$_3$, the formation mechanism of the molecule in such a scenario is still unknown. In particular, shocks cannot explain the source of free nitrogen atoms. The bond dissociation energy of N$_2$ exceeds 110\,000\,K and is much higher than that of NH$_3$ \citep[by about 60\,000\,K;][]{bde1970} because of the $\text{N}{\equiv}\text{N}$ triple bond. Strong shocks are likely to be present near the stellar photosphere, but our observations and chemical models do not find much NH$_3$ in those radii. In addition, the non-detection of emission in the non-metastable ($J>K$) inversion transitions or rotational transitions within the vibrationally excited state from {\ik} argues against the presence of very dense and hot NH$_3$-emitting gas in the CSE, which would be expected if NH$_3$ forms in shocked gas right after collisional dissociation of the parent species N$_2$.

{\irc} is a peculiar target with a known history of episodic mass loss. The NH$_3$ emission from this star is very far away from the centre of the CSE compared to other targets of this study. Because of the resemblance of NH$_3$'s kinematics and radial intensity distribution to other molecules that are associated with photochemistry \citep[e.g. SO;][]{willacy1997} or shock chemistry \citep[e.g. SiO;][Dinh-V.-Trung et al., in prep.]{wong2013}, the formation of NH$_3$ in {\irc} is probably linked to photodissociation or shocks, or both. It is unclear whether the internal UV radiation field from the star is strong enough to dissociate N-bearing parent species in the inner envelope. The CSE of {\irc} is highly obscured by dust and no X-rays or UV emission was detected \citep{debecker2014}. We note that the lack of central NH$_3$ emission has also been observed in {\egg}. \citet{dvt2009nh3} found that NH$_3$ in {\egg} mainly traces warm molecular gas in four distinct spatial-kinematic components/lobes, which are associated with shocked gas as traced by CO and near-infrared H$_2$ emission; therefore, the authors suggested that the NH$_3$ abundance in the nebula could be enhanced by shocks.


\section{Conclusions and prospects}
\label{sec:conclsions}

We presented multi-wavelength observations of NH$_3$ emission or absorption in the circumstellar environments of four carefully selected oxygen-rich post-main sequence stars, the AGB star {\ik}, the red supergiant {\vy}, the pre-planetary nebula {\oh}, and the yellow hypergiant {\irc}. These targets were selected because of their diversity in evolutionary stages, previous detection of NH$_3$, and the availability of radiative transfer and chemical models. We detected emission in multiple inversion transitions in the radio $K$ band (${\sim}1.3$\,cm) with the VLA and ground-state rotational transitions in the submillimetre wavelengths with {\hifi} from all four targets. We did not detect the rotational transition in the first vibrationally excited state ($\varv_2=1$) near 140\,GHz (2.14\,mm) from {\ik}. In addition, we detected multiple NH$_3$ rovibrational absorption lines from {\ik} and {\vy} in the MIR $N$ band (${\sim}10$\,\micron) with the TEXES instrument at IRTF. 

The VLA images show that NH$_3$ in {\ik} forms quite close to the star at radii within $100\,\rstar$. In {\vy}, the angular resolution is not enough to clearly resolve the distribution of the gas; however, there are some indications that the strongest NH$_3$ emission is slightly displaced from the stellar continuum by ${\lesssim}1\arcsec$ to the south-west and is apparently associated with the SW Clump. The NH$_3$ emission from {\oh} predominantly arises from the central clump, I3. Some fraction of NH$_3$ in {\oh} resides in the bipolar outflow and a distinct kinematic feature (possibly a disc-like structure).

{\hifi} spectra show broad emission lines in all four targets. In {\ik}, {\vy}, and {\irc}, the line widths are close to or slightly exceed the terminal expansion velocities of the respective targets; therefore, the bulk of the line-emitting gas locate in the fully accelerated CSEs. IRTF spectra show blueshifted absorption close to ({\ik}) or smaller than ({\vy}) the expansion velocity. Hence, the MIR-absorbing gas in {\vy} is probably near the end of the wind acceleration zone.

We performed radiative transfer modelling on all available spectra. MIR excitation of the molecule to the first vibrationally excited state ($\varv_2=1$) is included. The NH$_3$ abundance is generally of the order of $10^{-7}$, which is a few times lower than previous results without MIR pumping \citep[e.g.][]{menten2010}. Comparing our results with those of {\cw} \citep{schmidt2016}, the NH$_3$ abundance in O-rich CSEs is more than 10 times higher than in C-rich CSEs. Both studies find that the inclusion of radiative pumping of NH$_3$ by the MIR continuum from the star and dust shells is important to correctly estimate the molecular abundance: the revised abundances in both environments are significantly lower than previously thought.

We speculate that the high NH$_3$ abundance in circumstellar environments is possibly associated with shock-induced chemistry and photodissociation. However, detailed theoretical models including both shock- and photo-chemistry are not available. Furthermore, the origins of free nitrogen atoms or ions cannot be identified. Free nitrogen is required to trigger successive hydrogenation which leads to the formation of NH$_3$ \citep[e.g.][]{willacy1997,decin2010nat}.

Future imaging of the NH$_3$ inversion line emission with the VLA at sub-arcsecond resolutions would be useful to constrain the inner radius and the spatial-kinematic distribution of the NH$_3$ emission, especially for the strong emission detected in {\oh} and {\irc}. In addition, simultaneous coverage of SO$_2$ transitions in the radio $K$ band would allow us to verify the association of NH$_3$ with SO$_2$ as seen in some targets. The envisioned next generation Very Large Array \citep[ngVLA;][]{ngvla2015,ngvla2017} will be useful in searching for non-metastable ($J>K$) inversion lines, which exclusively trace the densest part (${\gtrsim}10^{8}\,\percc$) of the NH$_3$-carrying gas \citep{sweitzer1979,ho1983}. In the (sub)millimetre wavelengths, it is also important to compare the spatial distribution and kinematics of NH$_3$ with other species associated with shocks and photodissociation (such as CS, SO, SO$_2$) and species in the nitrogen network (such as NO, N$_2$H$^+$) under high angular resolutions. Shocked-excited H$_2$ emission can be observed in the near-infrared to verify the speculation of circumstellar shocks.

The rovibrational absorption profiles provide important constraints on the radii of NH$_3$ molecules by comparing the line absorption velocity and the CSE expansion velocity. There are indications that NH$_3$ forms close to the end of the wind acceleration zone ({\vy}) or exhibits temporal variations in abundance and absorption velocity ({\omi}). Additional MIR $N$-band observations of other massive stars and Mira variables (such as {\irc}, {\vx}, {\rleo}) at high spectral resolution ($R \sim 10^{5}$) are necessary to fully understand the formation of the molecule in evolved stars.

Above $\varv_2=1$, the next lowest vibrationally excited levels are $\varv_2=2$ ($s$-state only) at $1597.47\,{\percm}$ and $\varv_4=1$ (both states) at $1626.28$ and $1627.38\,{\percm}$ \citep[e.g.][]{yu2010}. These vibrational states are connected to the ground state ($0$ and $0.79\,{\percm}$) by MIR photons of wavelengths between 6.1--6.3\,{\micron}. In this study, we assumed that these energy levels are not significantly populated by the MIR radiation of the star and dust. Verification of the assumption by modelling is extremely computationally expensive because the number of transitions increases exponentially. Observations with MIR instruments near 6\,{\micron}, such as TEXES at ground-based telescopes or EXES\footnote{Echelon-Cross-Echelle Spectrograph \citep{richter2006,richter2010}.} aboard the airborne telescope of SOFIA\footnote{Stratospheric Observatory For Infrared Astronomy \citep{young2012,temi2014}.}, would be useful to examine the extent of vibrational excitation of NH$_3$.


\begin{acknowledgements}
We thank the anonymous referee for the useful comments. 
We thank M.~R.~Schmidt for kindly providing the molecular data files of ortho- and para-NH$_3$. 
We are grateful to A.~Faure for the collisional rate coefficients of ortho- and para-NH$_3$ with ortho-H$_2$, para-H$_2$, and H \citep{bouhafs2017}; and for the useful comments of the extrapolation methods of the rate coefficients. 
We are indebted to D.~Gobrecht for the insightful discussions of the chemical models of IK~Tau \citep{gobrecht2016}. \\
We would like to express our gratitude to M.~J.~Richter for the assistance in the preparation of the IRTF proposal and data analysis. 
We appreciate the generosity of R.~A.~McLaren and A.~L.~Betz for giving us the consent and IOP Publishing Ltd. for granting us the permission to reproduce the NH$_3$ spectra as shown in Figs. 1 and 3 of \citet{mclaren1980}. \\
This research has made use of the \href{http://vizier.u-strasbg.fr/viz-bin/VizieR/}{VizieR catalogue access tool}, CDS, Strasbourg, France. The original description of the VizieR service was published in \citet{vizier}. 
We thank \citet{vp2016} for the millimetre spectra of {\ik} and \citet{tenenbaum2010cat} for the table of emission lines in {\vy} deposited in the VizieR database. \\
This research has made use of NASA's \href{http://adsabs.harvard.edu/abstract_service.html}{Astrophysics Data System Bibliographic Services}. 
This research has made use of the \href{http://simbad.u-strasbg.fr/simbad/}{SIMBAD database}, operated at CDS, Strasbourg, France. 
We acknowledge with thanks the variable star observations from the \href{https://www.aavso.org/}{AAVSO International Database} contributed by observers worldwide and used in this research. \\
The {\herschel} spacecraft was designed, built, tested, and launched under a contract to ESA managed by the {\herschel}/{\planck} Project team by an industrial consortium under the overall responsibility of the prime contractor Thales Alenia Space (Cannes), and including Astrium (Friedrichshafen) responsible for the payload module and for system testing at spacecraft level, Thales Alenia Space (Turin) responsible for the service module, and Astrium (Toulouse) responsible for the telescope, with in excess of a hundred subcontractors. \\
HIFI has been designed and built by a consortium of institutes and university departments from across Europe, Canada and the United States under the leadership of SRON Netherlands Institute for Space Research, Groningen, The Netherlands and with major contributions from Germany, France and the US. Consortium members are: Canada: CSA, U.Waterloo; France: CESR, LAB, LERMA, IRAM; Germany: KOSMA, MPIfR, MPS; Ireland, NUI Maynooth; Italy: ASI, IFSI-INAF, Osservatorio Astrofisico di Arcetri-INAF; Netherlands: SRON, TUD; Poland: CAMK, CBK; Spain: Observatorio Astron{\'o}mico Nacional (IGN), Centro de Astrobiolog{\'i}a (CSIC-INTA). Sweden: Chalmers University of Technology - MC2, RSS \& GARD; Onsala Space Observatory; Swedish National Space Board, Stockholm University - Stockholm Observatory; Switzerland: ETH Zurich, FHNW; USA: Caltech, JPL, NHSC. \\
HCSS / HSpot / HIPE is a joint development (are joint developments) by the Herschel Science Ground Segment Consortium, consisting of ESA, the NASA Herschel Science Center, and the HIFI, PACS and SPIRE consortia. \\
This research has made use of the \href{https://irsa.ipac.caltech.edu/frontpage/}{NASA/IPAC Infrared Science Archive}, which is operated by the Jet Propulsion Laboratory, California Institute of Technology, under contract with the National Aeronautics and Space Administration. 
The version of the ISO data presented in this article correspond to the \href{https://www.cosmos.esa.int/web/iso/highly-processed-data-products}{Highly Processed Data Product (HPDP)} sets called `Uniformly processed LWS L01 spectra' by C. Lloyd, M. Lerate, and T. Grundy; and `A uniform database of SWS $2.4$--$45.4$ micron spectra' by G.~C. Sloan et al., available for public use in the \href{https://www.cosmos.esa.int/web/iso/access-the-archive}{ISO Data Archive}. \\
This research is partly based on observations at Kitt Peak National Observatory, National Optical Astronomy Observatory (NOAO), which is operated by the Association of Universities for Research in Astronomy (AURA) under cooperative agreement with the National Science Foundation. \\
This research made use of Astropy, a community-developed core Python package for Astronomy \citep{astropy}. PyFITS is a product of the Space Telescope Science Institute, which is operated by AURA for NASA. \\
K.~T. Wong was supported for this research through a stipend from the International Max Planck Research School (IMPRS) for Astronomy and Astrophysics at the Universities of Bonn and Cologne and also by the Bonn-Cologne Graduate School of Physics and Astronomy (BCGS). 
\end{acknowledgements}


\bibliographystyle{aa}
\bibliography{aa31873_arXiv}

\begin{thebibliography}{242}
\expandafter\ifx\csname natexlab\endcsname\relax\def\natexlab#1{#1}\fi

\bibitem[{{Adande} {et~al.}(2013){Adande}, {Edwards}, \& {Ziurys}}]{adande2013}
{Adande}, G.~R., {Edwards}, J.~L., \& {Ziurys}, L.~M. 2013, \apj, 778, 22

\bibitem[{{Ag{\'u}ndez} {et~al.}(2010){Ag{\'u}ndez}, {Cernicharo}, \&
  {Gu{\'e}lin}}]{agundez2010}
{Ag{\'u}ndez}, M., {Cernicharo}, J., \& {Gu{\'e}lin}, M. 2010, \apjl, 724, L133

\bibitem[{{Alcolea} {et~al.}(2013){Alcolea}, {Bujarrabal}, {Planesas},
  {Teyssier}, {Cernicharo}, {De Beck}, {Decin}, {Dominik}, {Justtanont}, {de
  Koter}, {Marston}, {Melnick}, {Menten}, {Neufeld}, {Olofsson}, {Schmidt},
  {Sch{\"o}ier}, {Szczerba}, \& {Waters}}]{alcolea2013}
{Alcolea}, J., {Bujarrabal}, V., {Planesas}, P., {et~al.} 2013, \aap, 559, A93

\bibitem[{{Alcolea} {et~al.}(1996){Alcolea}, {Bujarrabal}, \& {Sanchez
  Contreras}}]{alcolea1996}
{Alcolea}, J., {Bujarrabal}, V., \& {Sanchez Contreras}, C. 1996, \aap, 312,
  560

\bibitem[{{Alcolea} {et~al.}(2001){Alcolea}, {Bujarrabal}, {S{\'a}nchez
  Contreras}, {Neri}, \& {Zweigle}}]{alcolea2001}
{Alcolea}, J., {Bujarrabal}, V., {S{\'a}nchez Contreras}, C., {Neri}, R., \&
  {Zweigle}, J. 2001, \aap, 373, 932

\bibitem[{{Arai} {et~al.}(2016){Arai}, {Nagai}, {Fujita}, {Nakai}, {Seta},
  {Yamauchi}, {Kaneko}, {Hagiwara}, {Mamyoda}, {Miyamoto}, {Horie}, {Ishii},
  {Koide}, {Ogino}, {Maruyama}, {Hirai}, {Oshiro}, {Nagai}, {Akiyama},
  {Konakawa}, {Nonogawa}, {Salak}, {Terabe}, {Nihonmatsu}, \&
  {Funahashi}}]{arai2016}
{Arai}, H., {Nagai}, M., {Fujita}, S., {et~al.} 2016, \pasj, 68, 76

\bibitem[{{Astropy Collaboration} {et~al.}(2013){Astropy Collaboration},
  {Robitaille}, {Tollerud}, {Greenfield}, {Droettboom}, {Bray}, {Aldcroft},
  {Davis}, {Ginsburg}, {Price-Whelan}, {Kerzendorf}, {Conley}, {Crighton},
  {Barbary}, {Muna}, {Ferguson}, {Grollier}, {Parikh}, {Nair}, {Unther},
  {Deil}, {Woillez}, {Conseil}, {Kramer}, {Turner}, {Singer}, {Fox}, {Weaver},
  {Zabalza}, {Edwards}, {Azalee Bostroem}, {Burke}, {Casey}, {Crawford},
  {Dencheva}, {Ely}, {Jenness}, {Labrie}, {Lim}, {Pierfederici}, {Pontzen},
  {Ptak}, {Refsdal}, {Servillat}, \& {Streicher}}]{astropy}
{Astropy Collaboration}, {Robitaille}, T.~P., {Tollerud}, E.~J., {et~al.} 2013,
  \aap, 558, A33

\bibitem[{{Balan{\c c}a} \& {Dayou}(2017)}]{balanca2017}
{Balan{\c c}a}, C. \& {Dayou}, F. 2017, \mnras, 469, 1673

\bibitem[{{Bell} {et~al.}(1980){Bell}, {Kwok}, \& {Feldman}}]{bell1980}
{Bell}, M.~B., {Kwok}, S., \& {Feldman}, P.~A. 1980, in \iaus, Vol.~87,
  Interstellar Molecules, ed. B.~H. {Andrew}, 495--496

\bibitem[{{Belov} {et~al.}(1998){Belov}, {Urban}, \& {Winnewisser}}]{belov1998}
{Belov}, S.~P., {Urban}, {\v S}., \& {Winnewisser}, G. 1998, \jmsp, 189, 1

\bibitem[{{Benson} {et~al.}(1979){Benson}, {Mutel}, {Fix}, \&
  {Claussen}}]{benson1979}
{Benson}, J.~M., {Mutel}, R.~L., {Fix}, J.~D., \& {Claussen}, M.~J. 1979,
  \apjl, 229, L87

\bibitem[{{Betz}(1975)}]{betz1975}
{Betz}, A. 1975, \ssr, 17, 677

\bibitem[{{Betz}(1996)}]{betz1996}
{Betz}, A.~L. 1996, in Amazing Light, ed. R.~Y. {Chiao}, 73--78

\bibitem[{{Betz} \& {Goldhaber}(1985)}]{betz1985}
{Betz}, A.~L. \& {Goldhaber}, D.~M. 1985, in \assl, Vol. 117, Mass Loss from
  Red Giants, ed. M.~{Morris} \& B.~{Zuckerman}, 83

\bibitem[{{Betz} \& {McLaren}(1980)}]{betz1980}
{Betz}, A.~L. \& {McLaren}, R.~A. 1980, in \iaus, Vol.~87, Interstellar
  Molecules, ed. B.~H. {Andrew}, 503--507

\bibitem[{{Betz} {et~al.}(1979){Betz}, {McLaren}, \& {Spears}}]{betz1979}
{Betz}, A.~L., {McLaren}, R.~A., \& {Spears}, D.~L. 1979, \apjl, 229, L97

\bibitem[{{Bouhafs} {et~al.}(2017){Bouhafs}, {Rist}, {Daniel}, {Dumouchel},
  {Lique}, {Wiesenfeld}, \& {Faure}}]{bouhafs2017}
{Bouhafs}, N., {Rist}, C., {Daniel}, F., {et~al.} 2017, \mnras, 470, 2204

\bibitem[{{Bujarrabal} {et~al.}(2012){Bujarrabal}, {Alcolea}, {Soria-Ruiz},
  {Planesas}, {Teyssier}, {Cernicharo}, {Decin}, {Dominik}, {Justtanont}, {de
  Koter}, {Marston}, {Melnick}, {Menten}, {Neufeld}, {Olofsson}, {Schmidt},
  {Sch{\"o}ier}, {Szczerba}, \& {Waters}}]{bujarrabal2012}
{Bujarrabal}, V., {Alcolea}, J., {Soria-Ruiz}, R., {et~al.} 2012, \aap, 537, A8

\bibitem[{{Bujarrabal} {et~al.}(2011){Bujarrabal}, {Hifistars}, \& {Success
  Teams}}]{bujarrabal2011}
{Bujarrabal}, V., {Hifistars}, \& {Success Teams}. 2011, in \aspc, Vol. 445,
  Why Galaxies Care about AGB Stars II: Shining Examples and Common
  Inhabitants, ed. F.~{Kerschbaum}, T.~{Lebzelter}, \& R.~F. {Wing}, 577

\bibitem[{{Bunker}(1979)}]{bunkerbook1979}
{Bunker}, P.~R. 1979, {Microwave Symmetry and Spectroscopy} (New York: Academic
  Press, Inc.)

\bibitem[{{Cannon}(1967)}]{cannon1967}
{Cannon}, R.~D. 1967, The Observatory, 87, 231

\bibitem[{{Carter} {et~al.}(2012){Carter}, {Lazareff}, {Maier}, {Chenu},
  {Fontana}, {Bortolotti}, {Boucher}, {Navarrini}, {Blanchet}, {Greve}, {John},
  {Kramer}, {Morel}, {Navarro}, {Pe{\~n}alver}, {Schuster}, \&
  {Thum}}]{carter2012}
{Carter}, M., {Lazareff}, B., {Maier}, D., {et~al.} 2012, \aap, 538, A89

\bibitem[{{Castro-Carrizo} {et~al.}(2007){Castro-Carrizo}, {Quintana-Lacaci},
  {Bujarrabal}, {Neri}, \& {Alcolea}}]{cc2007}
{Castro-Carrizo}, A., {Quintana-Lacaci}, G., {Bujarrabal}, V., {Neri}, R., \&
  {Alcolea}, J. 2007, \aap, 465, 457

\bibitem[{{Castro-Carrizo} {et~al.}(2010){Castro-Carrizo}, {Quintana-Lacaci},
  {Neri}, {Bujarrabal}, {Sch{\"o}ier}, {Winters}, {Olofsson}, {Lindqvist},
  {Alcolea}, {Lucas}, \& {Grewing}}]{cc2010}
{Castro-Carrizo}, A., {Quintana-Lacaci}, G., {Neri}, R., {et~al.} 2010, \aap,
  523, A59

\bibitem[{{Cazzoli} {et~al.}(2009){Cazzoli}, {Dore}, \&
  {Puzzarini}}]{cazzoli2009}
{Cazzoli}, G., {Dore}, L., \& {Puzzarini}, C. 2009, \aap, 507, 1707

\bibitem[{{Charnley} \& {Latter}(1997)}]{charnley1997}
{Charnley}, S.~B. \& {Latter}, W.~B. 1997, \mnras, 287, 538

\bibitem[{{Charnley} {et~al.}(1995){Charnley}, {Tielens}, \&
  {Kress}}]{charnley1995}
{Charnley}, S.~B., {Tielens}, A.~G.~G.~M., \& {Kress}, M.~E. 1995, \mnras, 274,
  L53

\bibitem[{{Chen} {et~al.}(2006){Chen}, {Pearson}, {Pickett}, {Matsuura}, \&
  {Blake}}]{chen2006}
{Chen}, P., {Pearson}, J.~C., {Pickett}, H.~M., {Matsuura}, S., \& {Blake},
  G.~A. 2006, \jmsp, 236, 116

\bibitem[{{Cheung} {et~al.}(1973){Cheung}, {Chui}, {Matsakis}, {Townes}, \&
  {Yngvesson}}]{cheung1973}
{Cheung}, A.~C., {Chui}, M.~F., {Matsakis}, D., {Townes}, C.~H., \&
  {Yngvesson}, K.~S. 1973, \apjl, 186, L73

\bibitem[{{Cheung} {et~al.}(1969){Cheung}, {Rank}, {Townes}, {Knowles}, \&
  {Sullivan}}]{cheung1969}
{Cheung}, A.~C., {Rank}, D.~M., {Townes}, C.~H., {Knowles}, S.~H., \&
  {Sullivan}, III, W.~T. 1969, \apjl, 157, L13

\bibitem[{{Cheung} {et~al.}(1968){Cheung}, {Rank}, {Townes}, {Thornton}, \&
  {Welch}}]{cheung1968}
{Cheung}, A.~C., {Rank}, D.~M., {Townes}, C.~H., {Thornton}, D.~D., \& {Welch},
  W.~J. 1968, \prl, 21, 1701

\bibitem[{{Choi} {et~al.}(2012){Choi}, {Brunthaler}, {Menten}, \&
  {Reid}}]{choi2012}
{Choi}, Y.~K., {Brunthaler}, A., {Menten}, K.~M., \& {Reid}, M.~J. 2012, in
  \iaus, Vol. 287, Cosmic Masers - from OH to H0, ed. R.~S. {Booth}, W.~H.~T.
  {Vlemmings}, \& E.~M.~L. {Humphreys}, 407--410

\bibitem[{{Choi} {et~al.}(2008){Choi}, {Hirota}, {Honma}, {Kobayashi},
  {Bushimata}, {Imai}, {Iwadate}, {Jike}, {Kameno}, {Kameya}, {Kamohara},
  {Kan-Ya}, {Kawaguchi}, {Kijima}, {Kim}, {Kuji}, {Kurayama}, {Manabe},
  {Maruyama}, {Matsui}, {Matsumoto}, {Miyaji}, {Nagayama}, {Nakagawa},
  {Nakamura}, {Oh}, {Omodaka}, {Oyama}, {Sakai}, {Sasao}, {Sato}, {Sato},
  {Shibata}, {Tamura}, {Tsushima}, \& {Yamashita}}]{choi2008}
{Choi}, Y.~K., {Hirota}, T., {Honma}, M., {et~al.} 2008, \pasj, 60, 1007

\bibitem[{{Clegg} {et~al.}(1996){Clegg}, {Ade}, {Armand}, {Baluteau}, {Barlow},
  {Buckley}, {Berges}, {Burgdorf}, {Caux}, {Ceccarelli}, {Cerulli}, {Church},
  {Cotin}, {Cox}, {Cruvellier}, {Culhane}, {Davis}, {di Giorgio}, {Diplock},
  {Drummond}, {Emery}, {Ewart}, {Fischer}, {Furniss}, {Glencross},
  {Greenhouse}, {Griffin}, {Gry}, {Harwood}, {Hazell}, {Joubert}, {King},
  {Lim}, {Liseau}, {Long}, {Lorenzetti}, {Molinari}, {Murray}, {Naylor},
  {Nisini}, {Norman}, {Omont}, {Orfei}, {Patrick}, {Pequignot}, {Pouliquen},
  {Price}, {Nguyen-Q-Rieu}, {Rogers}, {Robinson}, {Saisse}, {Saraceno},
  {Serra}, {Sidher}, {Smith}, {Smith}, {Spinoglio}, {Swinyard}, {Texier},
  {Towlson}, {Trams}, {Unger}, \& {White}}]{lws1996}
{Clegg}, P.~E., {Ade}, P.~A.~R., {Armand}, C., {et~al.} 1996, \aap, 315, L38

\bibitem[{{Cohen}(1981)}]{cohen1981}
{Cohen}, M. 1981, \pasp, 93, 288

\bibitem[{{Cohen} {et~al.}(1992){Cohen}, {Witteborn}, {Carbon}, {Augason},
  {Wooden}, {Bregman}, \& {Goorvitch}}]{cohen1992}
{Cohen}, M., {Witteborn}, F.~C., {Carbon}, D.~F., {et~al.} 1992, \aj, 104, 2045

\bibitem[{{Cordiner} {et~al.}(2016){Cordiner}, {Boogert}, {Charnley},
  {Justtanont}, {Cox}, {Smith}, {Tielens}, {Wirstr{\"o}m}, {Milam}, \&
  {Keane}}]{cordiner2016}
{Cordiner}, M.~A., {Boogert}, A.~C.~A., {Charnley}, S.~B., {et~al.} 2016, \apj,
  828, 51

\bibitem[{{Danby} {et~al.}(1988){Danby}, {Flower}, {Valiron}, {Schilke}, \&
  {Walmsley}}]{danby1988}
{Danby}, G., {Flower}, D.~R., {Valiron}, P., {Schilke}, P., \& {Walmsley},
  C.~M. 1988, \mnras, 235, 229

\bibitem[{{Danilovich} {et~al.}(2014){Danilovich}, {Bergman}, {Justtanont},
  {Lombaert}, {Maercker}, {Olofsson}, {Ramstedt}, \& {Royer}}]{danilovich2014}
{Danilovich}, T., {Bergman}, P., {Justtanont}, K., {et~al.} 2014, \aap, 569,
  A76

\bibitem[{{Danilovich} {et~al.}(2016){Danilovich}, {De Beck}, {Black},
  {Olofsson}, \& {Justtanont}}]{danilovich2016}
{Danilovich}, T., {De Beck}, E., {Black}, J.~H., {Olofsson}, H., \&
  {Justtanont}, K. 2016, \aap, 588, A119

\bibitem[{{Darwent}(1970)}]{bde1970}
{Darwent}, B.~d. 1970, {National Standard Reference Data Series}, 31, {DOI:
  \href{https://doi.org/10.6028/NBS.NSRDS.31}{10.6028/NBS.NSRDS.31}}

\bibitem[{{De Beck} {et~al.}(2013){De Beck}, {Kami{\'n}ski}, {Patel}, {Young},
  {Gottlieb}, {Menten}, \& {Decin}}]{debeck2013}
{De Beck}, E., {Kami{\'n}ski}, T., {Patel}, N.~A., {et~al.} 2013, \aap, 558,
  A132

\bibitem[{{De Beck} {et~al.}(2015){De Beck}, {Vlemmings}, {Muller}, {Black},
  {O'Gorman}, {Richards}, {Baudry}, {Maercker}, {Decin}, \&
  {Humphreys}}]{debeck2015}
{De Beck}, E., {Vlemmings}, W., {Muller}, S., {et~al.} 2015, \aap, 580, A36

\bibitem[{{De Becker} {et~al.}(2014){De Becker}, {Hutsem{\'e}kers}, \&
  {Gosset}}]{debecker2014}
{De Becker}, M., {Hutsem{\'e}kers}, D., \& {Gosset}, E. 2014, \na, 29, 75

\bibitem[{{de Graauw} {et~al.}(1996){de Graauw}, {Haser}, {Beintema},
  {Roelfsema}, {van Agthoven}, {Barl}, {Bauer}, {Bekenkamp}, {Boonstra},
  {Boxhoorn}, {Cote}, {de Groene}, {van Dijkhuizen}, {Drapatz}, {Evers},
  {Feuchtgruber}, {Frericks}, {Genzel}, {Haerendel}, {Heras}, {van der Hucht},
  {van der Hulst}, {Huygen}, {Jacobs}, {Jakob}, {Kamperman}, {Katterloher},
  {Kester}, {Kunze}, {Kussendrager}, {Lahuis}, {Lamers}, {Leech}, {van der
  Lei}, {van der Linden}, {Luinge}, {Lutz}, {Melzner}, {Morris}, {van Nguyen},
  {Ploeger}, {Price}, {Salama}, {Schaeidt}, {Sijm}, {Smoorenburg}, {Spakman},
  {Spoon}, {Steinmayer}, {Stoecker}, {Valentijn}, {Vandenbussche}, {Visser},
  {Waelkens}, {Waters}, {Wensink}, {Wesselius}, {Wiezorrek}, {Wieprecht},
  {Wijnbergen}, {Wildeman}, \& {Young}}]{sws1996}
{de Graauw}, T., {Haser}, L.~N., {Beintema}, D.~A., {et~al.} 1996, \aap, 315,
  L49

\bibitem[{{de Graauw} {et~al.}(2010){de Graauw}, {Helmich}, {Phillips},
  {Stutzki}, {Caux}, {Whyborn}, {Dieleman}, {Roelfsema}, {Aarts}, {Assendorp},
  {Bachiller}, {Baechtold}, {Barcia}, {Beintema}, {Belitsky}, {Benz}, {Bieber},
  {Boogert}, {Borys}, {Bumble}, {Ca{\"i}s}, {Caris}, {Cerulli-Irelli},
  {Chattopadhyay}, {Cherednichenko}, {Ciechanowicz}, {Coeur-Joly}, {Comito},
  {Cros}, {de Jonge}, {de Lange}, {Delforges}, {Delorme}, {den Boggende},
  {Desbat}, {Diez-Gonz{\'a}lez}, {di Giorgio}, {Dubbeldam}, {Edwards},
  {Eggens}, {Erickson}, {Evers}, {Fich}, {Finn}, {Franke}, {Gaier}, {Gal},
  {Gao}, {Gallego}, {Gauffre}, {Gill}, {Glenz}, {Golstein}, {Goulooze},
  {Gunsing}, {G{\"u}sten}, {Hartogh}, {Hatch}, {Higgins}, {Honingh}, {Huisman},
  {Jackson}, {Jacobs}, {Jacobs}, {Jarchow}, {Javadi}, {Jellema}, {Justen},
  {Karpov}, {Kasemann}, {Kawamura}, {Keizer}, {Kester}, {Klapwijk}, {Klein},
  {Kollberg}, {Kooi}, {Kooiman}, {Kopf}, {Krause}, {Krieg}, {Kramer},
  {Kruizenga}, {Kuhn}, {Laauwen}, {Lai}, {Larsson}, {Leduc}, {Leinz}, {Lin},
  {Liseau}, {Liu}, {Loose}, {L{\'o}pez-Fernandez}, {Lord}, {Luinge}, {Marston},
  {Mart{\'{\i}}n-Pintado}, {Maestrini}, {Maiwald}, {McCoey}, {Mehdi}, {Megej},
  {Melchior}, {Meinsma}, {Merkel}, {Michalska}, {Monstein}, {Moratschke},
  {Morris}, {Muller}, {Murphy}, {Naber}, {Natale}, {Nowosielski}, {Nuzzolo},
  {Olberg}, {Olbrich}, {Orfei}, {Orleanski}, {Ossenkopf}, {Peacock}, {Pearson},
  {Peron}, {Phillip-May}, {Piazzo}, {Planesas}, {Rataj}, {Ravera}, {Risacher},
  {Salez}, {Samoska}, {Saraceno}, {Schieder}, {Schlecht}, {Schl{\"o}der},
  {Schm{\"u}lling}, {Schultz}, {Schuster}, {Siebertz}, {Smit}, {Szczerba},
  {Shipman}, {Steinmetz}, {Stern}, {Stokroos}, {Teipen}, {Teyssier}, {Tils},
  {Trappe}, {van Baaren}, {van Leeuwen}, {van de Stadt}, {Visser}, {Wildeman},
  {Wafelbakker}, {Ward}, {Wesselius}, {Wild}, {Wulff}, {Wunsch}, {Tielens},
  {Zaal}, {Zirath}, {Zmuidzinas}, \& {Zwart}}]{deGraauw2010}
{de Graauw}, T., {Helmich}, F.~P., {Phillips}, T.~G., {et~al.} 2010, \aap, 518,
  L6

\bibitem[{{Decin} {et~al.}(2010{\natexlab{a}}){Decin}, {Ag{\'u}ndez}, {Barlow},
  {Daniel}, {Cernicharo}, {Lombaert}, {De Beck}, {Royer}, {Vandenbussche},
  {Wesson}, {Polehampton}, {Blommaert}, {De Meester}, {Exter}, {Feuchtgruber},
  {Gear}, {Gomez}, {Groenewegen}, {Gu{\'e}lin}, {Hargrave}, {Huygen}, {Imhof},
  {Ivison}, {Jean}, {Kahane}, {Kerschbaum}, {Leeks}, {Lim}, {Matsuura},
  {Olofsson}, {Posch}, {Regibo}, {Savini}, {Sibthorpe}, {Swinyard}, {Yates}, \&
  {Waelkens}}]{decin2010nat}
{Decin}, L., {Ag{\'u}ndez}, M., {Barlow}, M.~J., {et~al.} 2010{\natexlab{a}},
  \nat, 467, 64

\bibitem[{{Decin} {et~al.}(2006){Decin}, {Hony}, {de Koter}, {Justtanont},
  {Tielens}, \& {Waters}}]{decin2006}
{Decin}, L., {Hony}, S., {de Koter}, A., {et~al.} 2006, \aap, 456, 549

\bibitem[{{Decin} {et~al.}(2010{\natexlab{b}}){Decin}, {Justtanont}, {De Beck},
  \& et~al.}]{decin2010}
{Decin}, L., {Justtanont}, K., {De Beck}, E., \& et~al. 2010{\natexlab{b}},
  \aap, 521, L4

\bibitem[{{Decin} {et~al.}(2016){Decin}, {Richards}, {Millar}, {Baudry}, {De
  Beck}, {Homan}, {Smith}, {Van de Sande}, \& {Walsh}}]{decin2016}
{Decin}, L., {Richards}, A.~M.~S., {Millar}, T.~J., {et~al.} 2016, \aap, 592,
  A76

\bibitem[{{Deguchi} {et~al.}(1986){Deguchi}, {Claussen}, \&
  {Goldsmith}}]{deguchi1986}
{Deguchi}, S., {Claussen}, M.~J., \& {Goldsmith}, P.~F. 1986, \apj, 303, 810

\bibitem[{{Deguchi} \& {Goldsmith}(1985)}]{deguchi1985}
{Deguchi}, S. \& {Goldsmith}, P.~F. 1985, \nat, 317, 336

\bibitem[{{Dinh-V.-Trung} {et~al.}(2009{\natexlab{a}}){Dinh-V.-Trung}, {Chiu},
  \& {Lim}}]{dvt2009nh3}
{Dinh-V.-Trung}, {Chiu}, P.~J., \& {Lim}, J. 2009{\natexlab{a}}, \apj, 700, 86

\bibitem[{{Dinh-V.-Trung} {et~al.}(2009{\natexlab{b}}){Dinh-V.-Trung},
  {Muller}, {Lim}, {Kwok}, \& {Muthu}}]{dvt2009}
{Dinh-V.-Trung}, {Muller}, S., {Lim}, J., {Kwok}, S., \& {Muthu}, C.
  2009{\natexlab{b}}, \apj, 697, 409

\bibitem[{{Dolan}(1965)}]{dolan1965}
{Dolan}, J.~F. 1965, \apj, 142, 1621

\bibitem[{{Dorschner} {et~al.}(1995){Dorschner}, {Begemann}, {Henning},
  {Jaeger}, \& {Mutschke}}]{dorschner1995}
{Dorschner}, J., {Begemann}, B., {Henning}, T., {Jaeger}, C., \& {Mutschke}, H.
  1995, \aap, 300, 503

\bibitem[{{Duari} {et~al.}(1999){Duari}, {Cherchneff}, \&
  {Willacy}}]{duari1999}
{Duari}, D., {Cherchneff}, I., \& {Willacy}, K. 1999, \aap, 341, L47

\bibitem[{{Faure} {et~al.}(2013){Faure}, {Hily-Blant}, {Le Gal}, {Rist}, \&
  {Pineau des For{\^e}ts}}]{faure2013}
{Faure}, A., {Hily-Blant}, P., {Le Gal}, R., {Rist}, C., \& {Pineau des
  For{\^e}ts}, G. 2013, \apjl, 770, L2

\bibitem[{{Faure} \& {Josselin}(2008)}]{faure2008}
{Faure}, A. \& {Josselin}, E. 2008, \aap, 492, 257

\bibitem[{{Flower} \& {Watt}(1984)}]{flower1984}
{Flower}, D.~R. \& {Watt}, G.~D. 1984, \mnras, 209, 25

\bibitem[{{Fonfr{\'{\i}}a} {et~al.}(2017){Fonfr{\'{\i}}a}, {Hinkle},
  {Cernicharo}, {Richter}, {Ag{\'u}ndez}, \& {Wallace}}]{fonfria2017}
{Fonfr{\'{\i}}a}, J.~P., {Hinkle}, K.~H., {Cernicharo}, J., {et~al.} 2017,
  \apj, 835, 196

\bibitem[{{Freytag} {et~al.}(2017){Freytag}, {Liljegren}, \&
  {H{\"o}fner}}]{freytag2017}
{Freytag}, B., {Liljegren}, S., \& {H{\"o}fner}, S. 2017, \aap, 600, A137

\bibitem[{{Fu} {et~al.}(2012){Fu}, {Moullet}, {Patel}, {Biersteker}, {Derose},
  \& {Young}}]{fu2012}
{Fu}, R.~R., {Moullet}, A., {Patel}, N.~A., {et~al.} 2012, \apj, 746, 42

\bibitem[{{Fujita}(1966)}]{fujita1966}
{Fujita}, Y. 1966, \va, 7, 71

\bibitem[{{Gobrecht} {et~al.}(2016){Gobrecht}, {Cherchneff}, {Sarangi},
  {Plane}, \& {Bromley}}]{gobrecht2016}
{Gobrecht}, D., {Cherchneff}, I., {Sarangi}, A., {Plane}, J.~M.~C., \&
  {Bromley}, S.~T. 2016, \aap, 585, A6

\bibitem[{{Goldhaber}(1988)}]{goldhaber1988}
{Goldhaber}, D.~M. 1988, {PhD Dissertation}, University of California at
  Berkeley, {available via subscription at
  \url{https://search.proquest.com/docview/303689762}}

\bibitem[{{G{\'o}mez} {et~al.}(2014){G{\'o}mez}, {Uscanga}, {Su{\'a}rez},
  {Rizzo}, \& {de Gregorio-Monsalvo}}]{gomez2014}
{G{\'o}mez}, J.~F., {Uscanga}, L., {Su{\'a}rez}, O., {Rizzo}, J.~R., \& {de
  Gregorio-Monsalvo}, I. 2014, \rmxaa, 50, 137

\bibitem[{{Gong} {et~al.}(2015){Gong}, {Henkel}, {Spezzano}, {Thorwirth},
  {Menten}, {Wyrowski}, {Mao}, \& {Klein}}]{gong2015}
{Gong}, Y., {Henkel}, C., {Spezzano}, S., {et~al.} 2015, \aap, 574, A56

\bibitem[{{Green}(1976)}]{green1976}
{Green}, S. 1976, \jcp, 64, 3463

\bibitem[{{Hale} {et~al.}(1997){Hale}, {Bester}, {Danchi}, {Hoss}, {Lipman},
  {Monnier}, {Tuthill}, {Townes}, {Johnson}, {Lopez}, \& {Geballe}}]{hale1997}
{Hale}, D.~D.~S., {Bester}, M., {Danchi}, W.~C., {et~al.} 1997, \apj, 490, 407

\bibitem[{{Harju} {et~al.}(1993){Harju}, {Walmsley}, \&
  {Wouterloot}}]{harju1993}
{Harju}, J., {Walmsley}, C.~M., \& {Wouterloot}, J.~G.~A. 1993, \aaps, 98, 51

\bibitem[{{Hasegawa} {et~al.}(2006){Hasegawa}, {Kwok}, {Koning}, {Volk},
  {Justtanont}, {Olofsson}, {Sch{\"o}ier}, {Sandqvist}, {Hjalmarson}, {Olberg},
  {Winnberg}, {Nyman}, \& {Frisk}}]{hasegawa2006}
{Hasegawa}, T.~I., {Kwok}, S., {Koning}, N., {et~al.} 2006, \apj, 637, 791

\bibitem[{{Henry} {et~al.}(2000){Henry}, {Edmunds}, \&
  {K{\"o}ppen}}]{henry2000}
{Henry}, R.~B.~C., {Edmunds}, M.~G., \& {K{\"o}ppen}, J. 2000, \apj, 541, 660

\bibitem[{{Herzberg}(1945)}]{herzbergbook1945}
{Herzberg}, G. 1945, {Molecular Spectra and Molecular Structure. II: Infrared
  and Raman Spectra of Polyatomic Molecules, 13th print, 1968} (Princeton: D.
  Van Nostrand Company, Inc.)

\bibitem[{{Hinkle} {et~al.}(1997){Hinkle}, {Lebzelter}, \&
  {Scharlach}}]{hinkle1997}
{Hinkle}, K.~H., {Lebzelter}, T., \& {Scharlach}, W.~W.~G. 1997, \aj, 114, 2686

\bibitem[{{Ho} {et~al.}(1979){Ho}, {Barrett}, {Myers}, {Matsakis}, {Chui},
  {Townes}, {Cheung}, \& {Yngvesson}}]{ho1979}
{Ho}, P.~T.~P., {Barrett}, A.~H., {Myers}, P.~C., {et~al.} 1979, \apj, 234, 912

\bibitem[{{Ho} \& {Townes}(1983)}]{ho1983}
{Ho}, P.~T.~P. \& {Townes}, C.~H. 1983, \araa, 21, 239

\bibitem[{{H{\"o}fner} {et~al.}(2016){H{\"o}fner}, {Bladh}, {Aringer}, \&
  {Ahuja}}]{hofner2016}
{H{\"o}fner}, S., {Bladh}, S., {Aringer}, B., \& {Ahuja}, R. 2016, \aap, 594,
  A108

\bibitem[{{Hogerheijde} \& {van der Tak}(2000)}]{ratran}
{Hogerheijde}, M.~R. \& {van der Tak}, F.~F.~S. 2000, \aap, 362, 697

\bibitem[{{Holler}(1999)}]{holler1999}
{Holler}, C. 1999, {\textit{Diplomarbeit} [Master's Thesis]},
  Ludwig-Maximilians-Universit\"{a}t M\"{u}nchen, quoted in
  \citet{monnier2000s}

\bibitem[{{Howard} \& {Bright Wilson}(1934)}]{howard1934}
{Howard}, J.~B. \& {Bright Wilson}, Jr., E. 1934, \jcp, 2, 630

\bibitem[{{Humphreys} {et~al.}(2007){Humphreys}, {Helton}, \&
  {Jones}}]{humphreys2007}
{Humphreys}, R.~M., {Helton}, L.~A., \& {Jones}, T.~J. 2007, \aj, 133, 2716

\bibitem[{Hunter(2007)}]{matplotlib}
Hunter, J.~D. 2007, Computing In Science \& Engineering, 9, 90

\bibitem[{{Ireland} {et~al.}(2008){Ireland}, {Scholz}, \& {Wood}}]{ireland2008}
{Ireland}, M.~J., {Scholz}, M., \& {Wood}, P.~R. 2008, \mnras, 391, 1994

\bibitem[{{Ireland} {et~al.}(2011){Ireland}, {Scholz}, \& {Wood}}]{ireland2011}
{Ireland}, M.~J., {Scholz}, M., \& {Wood}, P.~R. 2011, \mnras, 418, 114

\bibitem[{{Isella} {et~al.}(2015){Isella}, {Hull}, {Moullet},
  {Galv{\'a}n-Madrid}, {Johnstone}, {Ricci}, {Tobin}, {Testi}, {Beltran},
  {Lazio}, {Siemion}, {Liu}, {Du}, {{\"O}berg}, {Bergin}, {Caselli}, {Bourke},
  {Carilli}, {Perez}, {Butler}, {de Pater}, {Qi}, {Hofstadter}, {Moreno},
  {Alexander}, {Williams}, {Goldsmith}, {Wyatt}, {Loinard}, {Di Francesco},
  {Wilner}, {Schilke}, {Ginsburg}, {S{\'a}nchez-Monge}, {Zhang}, \&
  {Beuther}}]{ngvla2015}
{Isella}, A., {Hull}, C.~L.~H., {Moullet}, A., {et~al.} 2015, Next Generation
  Very Large Array Memos Series [\eprint[arXiv]{1510.06444}], available at
  \url{https://library.nrao.edu/public/memos/ngvla/NGVLA_06.pdf}

\bibitem[{{Ivezi{\'c}} {et~al.}(1999){Ivezi{\'c}}, {Nenkova}, \&
  {Elitzur}}]{dustymanual}
{Ivezi{\'c}}, {\v Z}., {Nenkova}, M., \& {Elitzur}, M. 1999, ArXiv e-prints
  [\eprint{astro-ph/9910475}], {University of Kentucky Internal Report,
  accessible at \url{http://www.pa.uky.edu/~moshe/dusty/}}

\bibitem[{{Jacquinet-Husson} {et~al.}(2016){Jacquinet-Husson}, {Armante},
  {Scott}, {Ch{\'e}din}, {Cr{\'e}peau}, {Boutammine}, {Bouhdaoui},
  {Crevoisier}, {Capelle}, {Boonne}, {Poulet-Crovisier}, {Barbe}, {Chris
  Benner}, {Boudon}, {Brown}, {Buldyreva}, {Campargue}, {Coudert}, {Devi},
  {Down}, {Drouin}, {Fayt}, {Fittschen}, {Flaud}, {Gamache}, {Harrison},
  {Hill}, {Hodnebrog}, {Hu}, {Jacquemart}, {Jolly}, {Jim{\'e}nez},
  {Lavrentieva}, {Liu}, {Lodi}, {Lyulin}, {Massie}, {Mikhailenko},
  {M{\"u}ller}, {Naumenko}, {Nikitin}, {Nielsen}, {Orphal}, {Perevalov},
  {Perrin}, {Polovtseva}, {Predoi-Cross}, {Rotger}, {Ruth}, {Yu}, {Sung},
  {Tashkun}, {Tennyson}, {Tyuterev}, {Vander Auwera}, {Voronin}, \&
  {Makie}}]{geisa2016}
{Jacquinet-Husson}, N., {Armante}, R., {Scott}, N.~A., {et~al.} 2016, \jmsp,
  327, 31

\bibitem[{{Johnson} \& {Sauval}(1982)}]{johnson1982}
{Johnson}, H.~R. \& {Sauval}, A.~J. 1982, \aaps, 49, 77

\bibitem[{Jones {et~al.}(2001)Jones, Oliphant, P., \& et~al.}]{scipy}
Jones, E., Oliphant, T., P., P., \& et~al. 2001, {SciPy}: Open source
  scientific tools for {Python}, \url{https://www.scipy.org/}

\bibitem[{{Jones} {et~al.}(1993){Jones}, {Humphreys}, {Gehrz}, {Lawrence},
  {Zickgraf}, {Moseley}, {Casey}, {Glaccum}, {Koch}, {Pina}, {Jones}, {Venn},
  {Stahl}, \& {Starrfield}}]{jones1993}
{Jones}, T.~J., {Humphreys}, R.~M., {Gehrz}, R.~D., {et~al.} 1993, \apj, 411,
  323

\bibitem[{{Jura} {et~al.}(2002){Jura}, {Chen}, \& {Plavchan}}]{jura2002}
{Jura}, M., {Chen}, C., \& {Plavchan}, P. 2002, \apj, 574, 963

\bibitem[{{Justtanont} {et~al.}(2012){Justtanont}, {Khouri}, {Maercker},
  {Alcolea}, {Decin}, {Olofsson}, {Sch{\"o}ier}, {Bujarrabal}, {Marston},
  {Teyssier}, {Cernicharo}, {Dominik}, {de Koter}, {Melnick}, {Menten},
  {Neufeld}, {Planesas}, {Schmidt}, {Szczerba}, \& {Waters}}]{justtanont2012}
{Justtanont}, K., {Khouri}, T., {Maercker}, M., {et~al.} 2012, \aap, 537, A144

\bibitem[{{Kami{\'n}ski} {et~al.}(2013){Kami{\'n}ski}, {Gottlieb}, {Young},
  {Menten}, \& {Patel}}]{kaminski2013}
{Kami{\'n}ski}, T., {Gottlieb}, C.~A., {Young}, K.~H., {Menten}, K.~M., \&
  {Patel}, N.~A. 2013, \apjs, 209, 38

\bibitem[{{Kami{\'n}ski} {et~al.}(2015){Kami{\'n}ski}, {Menten}, {Tylenda},
  {Hajduk}, {Patel}, \& {Kraus}}]{kaminski2015}
{Kami{\'n}ski}, T., {Menten}, K.~M., {Tylenda}, R., {et~al.} 2015, \nat, 520,
  322

\bibitem[{{Kami{\'n}ski} {et~al.}(2017){Kami{\'n}ski}, {Menten}, {Tylenda},
  {Karakas}, {Belloche}, \& {Patel}}]{kaminski2017b}
{Kami{\'n}ski}, T., {Menten}, K.~M., {Tylenda}, R., {et~al.} 2017, \aap, {in
  press} [\eprint[arXiv]{1708.02261}], {DOI:
  \href{https://doi.org/10.1051/0004-6361/201731287}{10.1051/0004-6361/201731287}}

\bibitem[{{Karakas} \& {Lattanzio}(2014)}]{karakas2014}
{Karakas}, A.~I. \& {Lattanzio}, J.~C. 2014, \pasa, 31, e030

\bibitem[{{Kastner} {et~al.}(1992){Kastner}, {Weintraub}, {Zuckerman},
  {Becklin}, {McLean}, \& {Gatley}}]{kastner1992}
{Kastner}, J.~H., {Weintraub}, D.~A., {Zuckerman}, B., {et~al.} 1992, \apj,
  398, 552

\bibitem[{{Keady} \& {Ridgway}(1993)}]{keady1993}
{Keady}, J.~J. \& {Ridgway}, S.~T. 1993, \apj, 406, 199

\bibitem[{{Kessler} {et~al.}(1996){Kessler}, {Steinz}, {Anderegg}, {Clavel},
  {Drechsel}, {Estaria}, {Faelker}, {Riedinger}, {Robson}, {Taylor}, \&
  {Xim{\'e}nez de Ferr{\'a}n}}]{iso1996}
{Kessler}, M.~F., {Steinz}, J.~A., {Anderegg}, M.~E., {et~al.} 1996, \aap, 315,
  L27

\bibitem[{{Klochkova} {et~al.}(1997){Klochkova}, {Chentsov}, \&
  {Panchuk}}]{klochkova1997}
{Klochkova}, V.~G., {Chentsov}, E.~L., \& {Panchuk}, V.~E. 1997, \mnras, 292,
  19

\bibitem[{{Kramer} {et~al.}(2013){Kramer}, {Pe{\~n}alver}, \&
  {Greve}}]{kramer2013}
{Kramer}, C., {Pe{\~n}alver}, J., \& {Greve}, A. 2013, {Improvement of the IRAM
  30m telescope beam pattern}, {v8.2, available at
  \url{http://www.iram.es/IRAMES/mainWiki/CalibrationPapers}}

\bibitem[{{Kukolich}(1967)}]{kukolich1967}
{Kukolich}, S.~G. 1967, \phrv, 156, 83

\bibitem[{{Kukolich} \& {Wofsy}(1970)}]{kukolich1970}
{Kukolich}, S.~G. \& {Wofsy}, S.~C. 1970, \jcp, 52, 5477

\bibitem[{{Kwok} {et~al.}(1981){Kwok}, {Bell}, \& {Feldman}}]{kwok1981}
{Kwok}, S., {Bell}, M.~B., \& {Feldman}, P.~A. 1981, \apj, 247, 125

\bibitem[{{Lacy} {et~al.}(2002){Lacy}, {Richter}, {Greathouse}, {Jaffe}, \&
  {Zhu}}]{lacy2002}
{Lacy}, J.~H., {Richter}, M.~J., {Greathouse}, T.~K., {Jaffe}, D.~T., \& {Zhu},
  Q. 2002, \pasp, 114, 153

\bibitem[{{Lagadec} {et~al.}(2011){Lagadec}, {Verhoelst}, {M{\'e}karnia},
  {Su{\'a}eez}, {Zijlstra}, {Bendjoya}, {Szczerba}, {Chesneau}, {van Winckel},
  {Barlow}, {Matsuura}, {Bowey}, {Lorenz-Martins}, \& {Gledhill}}]{lagadec2011}
{Lagadec}, E., {Verhoelst}, T., {M{\'e}karnia}, D., {et~al.} 2011, \mnras, 417,
  32

\bibitem[{{Latter} \& {Charnley}(1996{\natexlab{a}})}]{latter1996a}
{Latter}, W.~B. \& {Charnley}, S.~B. 1996{\natexlab{a}}, \apjl, 463, L37

\bibitem[{{Latter} \& {Charnley}(1996{\natexlab{b}})}]{latter1996b}
{Latter}, W.~B. \& {Charnley}, S.~B. 1996{\natexlab{b}}, \apjl, 465, L81

\bibitem[{{Le Bertre}(1993)}]{lebertre1993}
{Le Bertre}, T. 1993, \aaps, 97, 729

\bibitem[{{Lebzelter} \& {Hinkle}(2002)}]{lebzelter2002}
{Lebzelter}, T. \& {Hinkle}, K.~H. 2002, in \aspc, Vol. 259, IAU Colloq. 185:
  Radial and Nonradial Pulsationsn as Probes of Stellar Physics, ed.
  C.~{Aerts}, T.~R. {Bedding}, \& J.~{Christensen-Dalsgaard}, 556

\bibitem[{{Li} \& {Guo}(2014)}]{liguo2014}
{Li}, J. \& {Guo}, H. 2014, \pccp, 16, 6753

\bibitem[{{Li} {et~al.}(2013){Li}, {Heays}, {Visser}, {Ubachs}, {Lewis},
  {Gibson}, \& {van Dishoeck}}]{li2013}
{Li}, X., {Heays}, A.~N., {Visser}, R., {et~al.} 2013, \aap, 555, A14

\bibitem[{{Li} {et~al.}(2016){Li}, {Millar}, {Heays}, {Walsh}, {van Dishoeck},
  \& {Cherchneff}}]{li2016}
{Li}, X., {Millar}, T.~J., {Heays}, A.~N., {et~al.} 2016, \aap, 588, A4

\bibitem[{{Liljegren} {et~al.}(2017){Liljegren}, {H{\"o}fner}, {Eriksson}, \&
  {Nowotny}}]{liljegren2017}
{Liljegren}, S., {H{\"o}fner}, S., {Eriksson}, K., \& {Nowotny}, W. 2017, \aap,
  {in press} [\eprint[arXiv]{1706.08332}], {DOI:
  \href{https://doi.org/10.1051/0004-6361/201731137}{10.1051/0004-6361/201731137}}

\bibitem[{{Lloyd} {et~al.}(2013){Lloyd}, {Lerate}, \& {Grundy}}]{lloyd2003}
{Lloyd}, C., {Lerate}, M.~R., \& {Grundy}, T.~W. 2013, {The LWS LO1 Pipeline,
  Version 1}, {available at
  \url{http://ida.esac.esa.int:8080/hpdp/technical_reports/technote34.html}}

\bibitem[{{Lomb}(1976)}]{lomb1976}
{Lomb}, N.~R. 1976, \apss, 39, 447

\bibitem[{{Longuet-Higgins}(1963)}]{longuethiggins1963}
{Longuet-Higgins}, H.~C. 1963, \molph, 6, 445

\bibitem[{{Lyubimkov} {et~al.}(2011){Lyubimkov}, {Lambert}, {Korotin},
  {Poklad}, {Rachkovskaya}, \& {Rostopchin}}]{lyubimkov2011}
{Lyubimkov}, L.~S., {Lambert}, D.~L., {Korotin}, S.~A., {et~al.} 2011, \mnras,
  410, 1774

\bibitem[{{Madhusudhan} {et~al.}(2016){Madhusudhan}, {Ag{\'u}ndez}, {Moses}, \&
  {Hu}}]{madhusudhan2016}
{Madhusudhan}, N., {Ag{\'u}ndez}, M., {Moses}, J.~I., \& {Hu}, Y. 2016, \ssr,
  205, 285

\bibitem[{{Maercker} {et~al.}(2016){Maercker}, {Danilovich}, {Olofsson}, {De
  Beck}, {Justtanont}, {Lombaert}, \& {Royer}}]{maercker2016}
{Maercker}, M., {Danilovich}, T., {Olofsson}, H., {et~al.} 2016, \aap, 591, A44

\bibitem[{{Maercker} {et~al.}(2008){Maercker}, {Sch{\"o}ier}, {Olofsson},
  {Bergman}, \& {Ramstedt}}]{maercker2008}
{Maercker}, M., {Sch{\"o}ier}, F.~L., {Olofsson}, H., {Bergman}, P., \&
  {Ramstedt}, S. 2008, \aap, 479, 779

\bibitem[{{Maret} {et~al.}(2009){Maret}, {Faure}, {Scifoni}, \&
  {Wiesenfeld}}]{maret2009}
{Maret}, S., {Faure}, A., {Scifoni}, E., \& {Wiesenfeld}, L. 2009, \mnras, 399,
  425

\bibitem[{{Martin-Pintado} \& {Bachiller}(1992{\natexlab{a}})}]{mp1992a}
{Martin-Pintado}, J. \& {Bachiller}, R. 1992{\natexlab{a}}, \apjl, 391, L93

\bibitem[{{Martin-Pintado} \& {Bachiller}(1992{\natexlab{b}})}]{mp1992b}
{Martin-Pintado}, J. \& {Bachiller}, R. 1992{\natexlab{b}}, \apjl, 401, L47

\bibitem[{{Martin-Pintado} {et~al.}(1993){Martin-Pintado}, {Gaume},
  {Bachiller}, \& {Johnson}}]{mp1993}
{Martin-Pintado}, J., {Gaume}, R., {Bachiller}, R., \& {Johnson}, K. 1993,
  \apj, 419, 725

\bibitem[{{Martin-Pintado} {et~al.}(1995){Martin-Pintado}, {Gaume}, {Johnston},
  \& {Bachiller}}]{mp1995}
{Martin-Pintado}, J., {Gaume}, R.~A., {Johnston}, K.~J., \& {Bachiller}, R.
  1995, \apj, 446, 687

\bibitem[{{Marvel}(2005)}]{marvel2005}
{Marvel}, K.~B. 2005, \aj, 130, 261

\bibitem[{{Matplotlib Developers}(2016)}]{matplotlib_v153}
{Matplotlib Developers}. 2016, matplotlib: v1.5.3, {DOI:
  \href{https://doi.org/10.5281/zenodo.61948}{10.5281/zenodo.61948}}

\bibitem[{{Matsuura} {et~al.}(2006){Matsuura}, {Chesneau}, {Zijlstra}, {Jaffe},
  {Waters}, {Yates}, {Lagadec}, {Gledhill}, {Etoka}, \&
  {Richards}}]{matsuura2006}
{Matsuura}, M., {Chesneau}, O., {Zijlstra}, A.~A., {et~al.} 2006, \apjl, 646,
  L123

\bibitem[{{Mauersberger} {et~al.}(1988){Mauersberger}, {Henkel}, \&
  {Wilson}}]{mauersberger1988}
{Mauersberger}, R., {Henkel}, C., \& {Wilson}, T.~L. 1988, \aap, 205, 235

\bibitem[{{McCabe} {et~al.}(1979){McCabe}, {Smith}, \& {Clegg}}]{mccabe1979}
{McCabe}, E.~M., {Smith}, R.~C., \& {Clegg}, R.~E.~S. 1979, \nat, 281, 263

\bibitem[{{McCollom}(2013)}]{mccollom2013}
{McCollom}, T.~M. 2013, \areps, 41, 207

\bibitem[{{McLaren} \& {Betz}(1980)}]{mclaren1980}
{McLaren}, R.~A. \& {Betz}, A.~L. 1980, \apjl, 240, L159

\bibitem[{{McMullin} {et~al.}(2007){McMullin}, {Waters}, {Schiebel}, {Young},
  \& {Golap}}]{casa}
{McMullin}, J.~P., {Waters}, B., {Schiebel}, D., {Young}, W., \& {Golap}, K.
  2007, in \aspc, Vol. 376, Astronomical Data Analysis Software and Systems
  XVI, ed. R.~A. {Shaw}, F.~{Hill}, \& D.~J. {Bell}, 127

\bibitem[{{Menten} \& {Alcolea}(1995)}]{menten1995}
{Menten}, K.~M. \& {Alcolea}, J. 1995, \apj, 448, 416

\bibitem[{{Menten} {et~al.}(2010){Menten}, {Wyrowski}, {Alcolea}, \&
  et~al.}]{menten2010}
{Menten}, K.~M., {Wyrowski}, F., {Alcolea}, J., \& et~al. 2010, \aap, 521, L7

\bibitem[{{Mills} \& {Morris}(2013)}]{mills2013}
{Mills}, E.~A.~C. \& {Morris}, M.~R. 2013, \apj, 772, 105

\bibitem[{{Monnier} {et~al.}(2000{\natexlab{a}}){Monnier}, {Danchi}, {Hale},
  {Lipman}, {Tuthill}, \& {Townes}}]{monnier2000c}
{Monnier}, J.~D., {Danchi}, W.~C., {Hale}, D.~S., {et~al.} 2000{\natexlab{a}},
  \apj, 543, 861

\bibitem[{{Monnier} {et~al.}(2000{\natexlab{b}}){Monnier}, {Danchi}, {Hale},
  {Tuthill}, \& {Townes}}]{monnier2000s}
{Monnier}, J.~D., {Danchi}, W.~C., {Hale}, D.~S., {Tuthill}, P.~G., \&
  {Townes}, C.~H. 2000{\natexlab{b}}, \apj, 543, 868

\bibitem[{{Monnier} {et~al.}(2000{\natexlab{c}}){Monnier}, {Fitelson},
  {Danchi}, \& {Townes}}]{monnier2000i}
{Monnier}, J.~D., {Fitelson}, W., {Danchi}, W.~C., \& {Townes}, C.~H.
  2000{\natexlab{c}}, \apjs, 129, 421

\bibitem[{{Morris} {et~al.}(1987){Morris}, {Guilloteau}, {Lucas}, \&
  {Omont}}]{morris1987}
{Morris}, M., {Guilloteau}, S., {Lucas}, R., \& {Omont}, A. 1987, \apj, 321,
  888

\bibitem[{{Mortier} {et~al.}(2015{\natexlab{a}}){Mortier}, {Faria}, {Correia},
  {Santerne}, \& {Santos}}]{mortier2015}
{Mortier}, A., {Faria}, J.~P., {Correia}, C.~M., {Santerne}, A., \& {Santos},
  N.~C. 2015{\natexlab{a}}, \aap, 573, A101

\bibitem[{{Mortier} {et~al.}(2015{\natexlab{b}}){Mortier}, {Faria}, {Correia},
  {Santerne}, \& {Santos}}]{bgls2015}
{Mortier}, A., {Faria}, J.~P., {Correia}, C.~M., {Santerne}, A., \& {Santos},
  N.~C. 2015{\natexlab{b}}, {BGLS: A Bayesian formalism for the generalised
  Lomb-Scargle periodogram}, Astrophysics Source Code Library

\bibitem[{{Mueller} {et~al.}(2014){Mueller}, {Jellema}, {Olberg}, {Moreno}, \&
  {Teyssier}}]{mueller2014}
{Mueller}, M., {Jellema}, W., {Olberg}, M., {Moreno}, R., \& {Teyssier}, D.
  2014, {The HIFI Beam: Release 1, Release Note for Astronomers}, {Tech. Rep.
  HIFI-ICC-RP-2014-001, v1.1, HIFI-ICC, available at
  \url{http://herschel.esac.esa.int/twiki/pub/Public/HifiCalibrationWeb/HifiBeamReleaseNote_Sep2014.pdf}}

\bibitem[{{Muller} {et~al.}(2007){Muller}, {Dinh-V-Trung}, {Lim}, {Hirano},
  {Muthu}, \& {Kwok}}]{muller2007}
{Muller}, S., {Dinh-V-Trung}, {Lim}, J., {et~al.} 2007, \apj, 656, 1109

\bibitem[{{Nakashima} {et~al.}(2016){Nakashima}, {Ladeyschikov}, {Sobolev},
  {Zhang}, {Hsia}, \& {Yung}}]{nakashima2016}
{Nakashima}, J.-i., {Ladeyschikov}, D.~A., {Sobolev}, A.~M., {et~al.} 2016,
  \apj, 825, 16

\bibitem[{{Nakashima} {et~al.}(2015){Nakashima}, {Sobolev}, {Salii}, {Zhang},
  {Yung}, \& {Deguchi}}]{nakashima2015}
{Nakashima}, J.-i., {Sobolev}, A.~M., {Salii}, S.~V., {et~al.} 2015, \pasj, 67,
  95

\bibitem[{{Nedoluha} \& {Bowers}(1992)}]{nedoluha1992}
{Nedoluha}, G.~E. \& {Bowers}, P.~F. 1992, \apj, 392, 249

\bibitem[{{Nercessian} {et~al.}(1989){Nercessian}, {Omont}, {Benayoun}, \&
  {Guilloteau}}]{nercessian1989}
{Nercessian}, E., {Omont}, A., {Benayoun}, J.~J., \& {Guilloteau}, S. 1989,
  \aap, 210, 225

\bibitem[{{Neugebauer} {et~al.}(1984){Neugebauer}, {Habing}, {van Duinen},
  {Aumann}, {Baud}, {Beichman}, {Beintema}, {Boggess}, {Clegg}, {de Jong},
  {Emerson}, {Gautier}, {Gillett}, {Harris}, {Hauser}, {Houck}, {Jennings},
  {Low}, {Marsden}, {Miley}, {Olnon}, {Pottasch}, {Raimond}, {Rowan-Robinson},
  {Soifer}, {Walker}, {Wesselius}, \& {Young}}]{iras1984}
{Neugebauer}, G., {Habing}, H.~J., {van Duinen}, R., {et~al.} 1984, \apjl, 278,
  L1

\bibitem[{{Nguyen-Q-Rieu} {et~al.}(1984){Nguyen-Q-Rieu}, {Graham}, \&
  {Bujarrabal}}]{nqr1984}
{Nguyen-Q-Rieu}, {Graham}, D., \& {Bujarrabal}, V. 1984, \aap, 138, L5

\bibitem[{{ngVLA Science Advisory Council}(2017)}]{ngvla2017}
{ngVLA Science Advisory Council}. 2017, Next Generation Very Large Array Memos
  Series, available at
  \url{https://library.nrao.edu/public/memos/ngvla/NGVLA_19.pdf}

\bibitem[{{Nowotny} {et~al.}(2010){Nowotny}, {H{\"o}fner}, \&
  {Aringer}}]{nowotny2010}
{Nowotny}, W., {H{\"o}fner}, S., \& {Aringer}, B. 2010, \aap, 514, A35

\bibitem[{{Ochsenbein} {et~al.}(2000){Ochsenbein}, {Bauer}, \&
  {Marcout}}]{vizier}
{Ochsenbein}, F., {Bauer}, P., \& {Marcout}, J. 2000, \aaps, 143, 23

\bibitem[{{O'Gorman} {et~al.}(2015){O'Gorman}, {Vlemmings}, {Richards},
  {Baudry}, {De Beck}, {Decin}, {Harper}, {Humphreys}, {Kervella}, {Khouri}, \&
  {Muller}}]{ogorman2015}
{O'Gorman}, E., {Vlemmings}, W., {Richards}, A.~M.~S., {et~al.} 2015, \aap,
  573, L1

\bibitem[{{Oka}(1976)}]{oka1976}
{Oka}, T. 1976, in Molecular Spectroscopy: Modern Research, Volume II, ed.
  K.~N. {Rao} (New York: Academic Press, Inc.), 229--253

\bibitem[{{Oka} {et~al.}(1971){Oka}, {Shimizu}, {Shimizu}, \&
  {Watson}}]{oka1971}
{Oka}, T., {Shimizu}, F.~O., {Shimizu}, T., \& {Watson}, J.~K.~G. 1971, \apjl,
  165, L15

\bibitem[{{Olofsson}(2003)}]{olofsson2003}
{Olofsson}, H. 2003, in Asymptotic giant branch stars, ed. H.~J. {Habing} \&
  H.~{Olofsson} (New York: Springer Science+Business Media), 325--410

\bibitem[{{Olofsson} {et~al.}(2017){Olofsson}, {Vlemmings}, {Bergman},
  {Humphreys}, {Lindqvist}, {Maercker}, {Nyman}, {Ramstedt}, \&
  {Tafoya}}]{olofsson2017}
{Olofsson}, H., {Vlemmings}, W.~H.~T., {Bergman}, P., {et~al.} 2017, \aap, 603,
  L2

\bibitem[{{Ossenkopf} {et~al.}(1992){Ossenkopf}, {Henning}, \&
  {Mathis}}]{ossenkopf1992}
{Ossenkopf}, V., {Henning}, T., \& {Mathis}, J.~S. 1992, \aap, 261, 567

\bibitem[{{Ott}(2010)}]{ott2010}
{Ott}, S. 2010, in \aspc, Vol. 434, Astronomical Data Analysis Software and
  Systems XIX, ed. Y.~{Mizumoto}, K.-I. {Morita}, \& M.~{Ohishi}, 139

\bibitem[{{Oudmaijer} \& {de Wit}(2013)}]{oudmaijer2013}
{Oudmaijer}, R.~D. \& {de Wit}, W.~J. 2013, \aap, 551, A69

\bibitem[{{Oudmaijer} {et~al.}(1996){Oudmaijer}, {Groenewegen}, {Matthews},
  {Blommaert}, \& {Sahu}}]{oudmaijer1996}
{Oudmaijer}, R.~D., {Groenewegen}, M.~A.~T., {Matthews}, H.~E., {Blommaert},
  J.~A.~D.~L., \& {Sahu}, K.~C. 1996, \mnras, 280, 1062

\bibitem[{P{\'e}rez \& Granger(2007)}]{ipython}
P{\'e}rez, F. \& Granger, B.~E. 2007, \mcse, 9, 21, \url{https://ipython.org/}

\bibitem[{{Perley} \& {Butler}(2013)}]{perley2013}
{Perley}, R.~A. \& {Butler}, B.~J. 2013, \apjs, 204, 19

\bibitem[{{Pety}(2005)}]{gildas}
{Pety}, J. 2005, in SF2A-2005: Semaine de l'Astrophysique Francaise, ed.
  F.~{Casoli}, T.~{Contini}, J.~M. {Hameury}, \& L.~{Pagani}, 721

\bibitem[{{Pickett} {et~al.}(1998){Pickett}, {Poynter}, {Cohen}, {Delitsky},
  {Pearson}, \& {M{\"u}ller}}]{jpl1998}
{Pickett}, H.~M., {Poynter}, R.~L., {Cohen}, E.~A., {et~al.} 1998, \jqsrt, 60,
  883

\bibitem[{{Pilbratt} {et~al.}(2010){Pilbratt}, {Riedinger}, {Passvogel},
  {Crone}, {Doyle}, {Gageur}, {Heras}, {Jewell}, {Metcalfe}, {Ott}, \&
  {Schmidt}}]{pilbratt2010}
{Pilbratt}, G.~L., {Riedinger}, J.~R., {Passvogel}, T., {et~al.} 2010, \aap,
  518, L1

\bibitem[{{Pillai} {et~al.}(2006){Pillai}, {Wyrowski}, {Carey}, \&
  {Menten}}]{pillai2006}
{Pillai}, T., {Wyrowski}, F., {Carey}, S.~J., \& {Menten}, K.~M. 2006, \aap,
  450, 569

\bibitem[{{Poglitsch} {et~al.}(2010){Poglitsch}, {Waelkens}, {Geis},
  {Feuchtgruber}, {Vandenbussche}, {Rodriguez}, {Krause}, {Renotte}, {van
  Hoof}, {Saraceno}, {Cepa}, {Kerschbaum}, {Agn{\`e}se}, {Ali}, {Altieri},
  {Andreani}, {Augueres}, {Balog}, {Barl}, {Bauer}, {Belbachir}, {Benedettini},
  {Billot}, {Boulade}, {Bischof}, {Blommaert}, {Callut}, {Cara}, {Cerulli},
  {Cesarsky}, {Contursi}, {Creten}, {De Meester}, {Doublier}, {Doumayrou},
  {Duband}, {Exter}, {Genzel}, {Gillis}, {Gr{\"o}zinger}, {Henning},
  {Herreros}, {Huygen}, {Inguscio}, {Jakob}, {Jamar}, {Jean}, {de Jong},
  {Katterloher}, {Kiss}, {Klaas}, {Lemke}, {Lutz}, {Madden}, {Marquet},
  {Martignac}, {Mazy}, {Merken}, {Montfort}, {Morbidelli}, {M{\"u}ller},
  {Nielbock}, {Okumura}, {Orfei}, {Ottensamer}, {Pezzuto}, {Popesso},
  {Putzeys}, {Regibo}, {Reveret}, {Royer}, {Sauvage}, {Schreiber}, {Stegmaier},
  {Schmitt}, {Schubert}, {Sturm}, {Thiel}, {Tofani}, {Vavrek}, {Wetzstein},
  {Wieprecht}, \& {Wiezorrek}}]{poglitsch2010}
{Poglitsch}, A., {Waelkens}, C., {Geis}, N., {et~al.} 2010, \aap, 518, L2

\bibitem[{{Poynter} \& {Kakar}(1975)}]{poynter1975}
{Poynter}, R.~L. \& {Kakar}, R.~K. 1975, \apjs, 29, 87

\bibitem[{{Poynter} \& {Margolis}(1983)}]{poynter1983}
{Poynter}, R.~L. \& {Margolis}, J.~S. 1983, \molph, 48, 401

\bibitem[{{Quintana-Lacaci} {et~al.}(2013){Quintana-Lacaci}, {Ag{\'u}ndez},
  {Cernicharo}, {Bujarrabal}, {S{\'a}nchez Contreras}, {Castro-Carrizo}, \&
  {Alcolea}}]{ql2013}
{Quintana-Lacaci}, G., {Ag{\'u}ndez}, M., {Cernicharo}, J., {et~al.} 2013,
  \aap, 560, L2

\bibitem[{{Quintana-Lacaci} {et~al.}(2016){Quintana-Lacaci}, {Ag{\'u}ndez},
  {Cernicharo}, {Bujarrabal}, {S{\'a}nchez Contreras}, {Castro-Carrizo}, \&
  {Alcolea}}]{ql2016}
{Quintana-Lacaci}, G., {Ag{\'u}ndez}, M., {Cernicharo}, J., {et~al.} 2016,
  \aap, 592, A51

\bibitem[{{Richter} {et~al.}(2010){Richter}, {Ennico}, {McKelvey}, \&
  {Seifahrt}}]{richter2010}
{Richter}, M.~J., {Ennico}, K.~A., {McKelvey}, M.~E., \& {Seifahrt}, A. 2010,
  in \procspie, Vol. 7735, Ground-based and Airborne Instrumentation for
  Astronomy III, 77356Q

\bibitem[{{Richter} {et~al.}(2006){Richter}, {Lacy}, {Jaffe}, {Mar}, {Goertz},
  {Moller}, {Strong}, \& {Greathouse}}]{richter2006}
{Richter}, M.~J., {Lacy}, J.~H., {Jaffe}, D.~T., {et~al.} 2006, in \procspie,
  Vol. 6269, Society of Photo-Optical Instrumentation Engineers (SPIE)
  Conference Series, 62691H

\bibitem[{{Rizzo} {et~al.}(2014){Rizzo}, {Palau}, {Jim{\'e}nez-Esteban}, \&
  {Henkel}}]{rizzo2014}
{Rizzo}, J.~R., {Palau}, A., {Jim{\'e}nez-Esteban}, F., \& {Henkel}, C. 2014,
  \aap, 564, A21

\bibitem[{{Roelfsema} {et~al.}(2012){Roelfsema}, {Helmich}, {Teyssier},
  {Ossenkopf}, {Morris}, {Olberg}, {Shipman}, {Risacher}, {Akyilmaz},
  {Assendorp}, {Avruch}, {Beintema}, {Biver}, {Boogert}, {Borys}, {Braine},
  {Caris}, {Caux}, {Cernicharo}, {Coeur-Joly}, {Comito}, {de Lange},
  {Delforge}, {Dieleman}, {Dubbeldam}, {de Graauw}, {Edwards}, {Fich},
  {Flederus}, {Gal}, {di Giorgio}, {Herpin}, {Higgins}, {Hoac}, {Huisman},
  {Jarchow}, {Jellema}, {de Jonge}, {Kester}, {Klein}, {Kooi}, {Kramer},
  {Laauwen}, {Larsson}, {Leinz}, {Lord}, {Lorenzani}, {Luinge}, {Marston},
  {Mart{\'{\i}}n-Pintado}, {McCoey}, {Melchior}, {Michalska}, {Moreno},
  {M{\"u}ller}, {Nowosielski}, {Okada}, {Orlea{\'n}ski}, {Phillips}, {Pearson},
  {Rabois}, {Ravera}, {Rector}, {Rengel}, {Sagawa}, {Salomons},
  {S{\'a}nchez-Su{\'a}rez}, {Schieder}, {Schl{\"o}der}, {Schm{\"u}lling},
  {Soldati}, {Stutzki}, {Thomas}, {Tielens}, {Vastel}, {Wildeman}, {Xie},
  {Xilouris}, {Wafelbakker}, {Whyborn}, {Zaal}, {Bell}, {Bjerkeli}, {De Beck},
  {Cavali{\'e}}, {Crockett}, {Hily-Blant}, {Kama}, {Kaminski}, {Lefl{\'o}ch},
  {Lombaert}, {de Luca}, {Makai}, {Marseille}, {Nagy}, {Pacheco}, {van der
  Wiel}, {Wang}, \& {Y{\i}ld{\i}z}}]{roelfsema2012}
{Roelfsema}, P.~R., {Helmich}, F.~P., {Teyssier}, D., {et~al.} 2012, \aap, 537,
  A17

\bibitem[{{Rohatgi} \& {Stanojevic}(2016)}]{webplotdigitizer}
{Rohatgi}, A. \& {Stanojevic}, Z. 2016, {WebPlotDigitizer: Version 3.10 of
  WebPlotDigitizer}, {DOI:
  \href{https://doi.org/10.5281/zenodo.51157}{10.5281/zenodo.51157}}

\bibitem[{{Rothermel} {et~al.}(1986){Rothermel}, {Schrey}, {Kaufl}, \&
  {Drapatz}}]{rothermel1986}
{Rothermel}, H., {Schrey}, U., {Kaufl}, H.~U., \& {Drapatz}, S. 1986, in \iaus,
  Vol. 118, Instrumentation and Research Programmes for Small Telescopes, ed.
  J.~B. {Hearnshaw} \& P.~L. {Cottrell}, 431

\bibitem[{{Roueff} \& {Lique}(2013)}]{roueff2013}
{Roueff}, E. \& {Lique}, F. 2013, \chrv, 113, 8906

\bibitem[{{Royer} {et~al.}(2010){Royer}, {Decin}, {Wesson}, {Barlow},
  {Polehampton}, {Matsuura}, {Ag{\'u}ndez}, {Blommaert}, {Cernicharo}, {Cohen},
  {Daniel}, {Degroote}, {De Meester}, {Exter}, {Feuchtgruber}, {Gear}, {Gomez},
  {Groenewegen}, {Hargrave}, {Huygen}, {Imhof}, {Ivison}, {Jean}, {Kerschbaum},
  {Leeks}, {Lim}, {Lombaert}, {Olofsson}, {Posch}, {Regibo}, {Savini},
  {Sibthorpe}, {Swinyard}, {Vandenbussche}, {Waelkens}, {Witherick}, \&
  {Yates}}]{royer2010}
{Royer}, P., {Decin}, L., {Wesson}, R., {et~al.} 2010, \aap, 518, L145

\bibitem[{{Ryde} \& {Sch{\"o}ier}(2001)}]{ryde2001}
{Ryde}, N. \& {Sch{\"o}ier}, F.~L. 2001, \apj, 547, 384

\bibitem[{{Salinas} {et~al.}(2016){Salinas}, {Hogerheijde}, {Bergin},
  {Cleeves}, {Brinch}, {Blake}, {Lis}, {Melnick}, {Pani{\'c}}, {Pearson},
  {Kristensen}, {Y{\i}ld{\i}z}, \& {van Dishoeck}}]{salinas2016}
{Salinas}, V.~N., {Hogerheijde}, M.~R., {Bergin}, E.~A., {et~al.} 2016, \aap,
  591, A122

\bibitem[{{S{\'a}nchez Contreras} {et~al.}(2000){S{\'a}nchez Contreras},
  {Bujarrabal}, {Neri}, \& {Alcolea}}]{sc2000}
{S{\'a}nchez Contreras}, C., {Bujarrabal}, V., {Neri}, R., \& {Alcolea}, J.
  2000, \aap, 357, 651

\bibitem[{{S{\'a}nchez Contreras} {et~al.}(2002){S{\'a}nchez Contreras},
  {Desmurs}, {Bujarrabal}, {Alcolea}, \& {Colomer}}]{sc2002}
{S{\'a}nchez Contreras}, C., {Desmurs}, J.~F., {Bujarrabal}, V., {Alcolea}, J.,
  \& {Colomer}, F. 2002, \aap, 385, L1

\bibitem[{{S{\'a}nchez Contreras} {et~al.}(2004){S{\'a}nchez Contreras}, {Gil
  de Paz}, \& {Sahai}}]{sc2004}
{S{\'a}nchez Contreras}, C., {Gil de Paz}, A., \& {Sahai}, R. 2004, \apj, 616,
  519

\bibitem[{{S{\'a}nchez Contreras} {et~al.}(2015){S{\'a}nchez Contreras},
  {Velilla Prieto}, {Ag{\'u}ndez}, {Cernicharo}, {Quintana-Lacaci},
  {Bujarrabal}, {Alcolea}, {Goicoechea}, {Herpin}, {Menten}, \&
  {Wyrowski}}]{sc2015}
{S{\'a}nchez Contreras}, C., {Velilla Prieto}, L., {Ag{\'u}ndez}, M., {et~al.}
  2015, \aap, 577, A52

\bibitem[{{Sault} {et~al.}(1995){Sault}, {Teuben}, \& {Wright}}]{miriad}
{Sault}, R.~J., {Teuben}, P.~J., \& {Wright}, M.~C.~H. 1995, in \aspc, Vol.~77,
  Astronomical Data Analysis Software and Systems IV, ed. R.~A. {Shaw}, H.~E.
  {Payne}, \& J.~J.~E. {Hayes}, 433

\bibitem[{{Savitzky} \& {Golay}(1964)}]{savitzkygolay1964}
{Savitzky}, A. \& {Golay}, M.~J.~E. 1964, \anac, 36, 1627

\bibitem[{{Scargle}(1982)}]{scargle1982}
{Scargle}, J.~D. 1982, \apj, 263, 835

\bibitem[{{Schilke} {et~al.}(1992){Schilke}, {Guesten}, {Schulz}, {Serabyn}, \&
  {Walmsley}}]{schilke1992}
{Schilke}, P., {Guesten}, R., {Schulz}, A., {Serabyn}, E., \& {Walmsley}, C.~M.
  1992, \aap, 261, L5

\bibitem[{{Schilke} {et~al.}(1990){Schilke}, {Walmsley}, {Wilson}, \&
  {Mauersberger}}]{schilke1990}
{Schilke}, P., {Walmsley}, C.~M., {Wilson}, T.~L., \& {Mauersberger}, R. 1990,
  \aap, 227, 220

\bibitem[{{Schmidt} {et~al.}(2016){Schmidt}, {He}, {Szczerba}, {Bujarrabal},
  {Alcolea}, {Cernicharo}, {Decin}, {Justtanont}, {Teyssier}, {Menten},
  {Neufeld}, {Olofsson}, {Planesas}, {Marston}, {Sobolev}, {de Koter}, \&
  {Sch{\"o}ier}}]{schmidt2016}
{Schmidt}, M.~R., {He}, J.~H., {Szczerba}, R., {et~al.} 2016, \aap, 592, A131

\bibitem[{{Schrey}(1986)}]{schrey1986}
{Schrey}, U. 1986, {\textit{Dissertation} [PhD Dissertation]}, Technische
  Universität München, {MPE Report 198, translated abstract is available at
  \url{https://ntrl.ntis.gov/NTRL/dashboard/searchResults/titleDetail/N8715040.xhtml}}

\bibitem[{{Schrey} {et~al.}(1985){Schrey}, {Rothermel}, {Kaeufei}, \&
  {Drapatz}}]{schrey1985}
{Schrey}, U., {Rothermel}, H., {Kaeufei}, H.~U., \& {Drapatz}, S. 1985, in
  Millimeter and Submillimeter Wave Radio Astronomy, 213--216

\bibitem[{{Shenoy} {et~al.}(2016){Shenoy}, {Humphreys}, {Jones}, {Marengo},
  {Gehrz}, {Helton}, {Hoffmann}, {Skemer}, \& {Hinz}}]{shenoy2016}
{Shenoy}, D., {Humphreys}, R.~M., {Jones}, T.~J., {et~al.} 2016, \aj, 151, 51

\bibitem[{{Sloan} {et~al.}(2003){Sloan}, {Kraemer}, {Price}, \&
  {Shipman}}]{sloan2003}
{Sloan}, G.~C., {Kraemer}, K.~E., {Price}, S.~D., \& {Shipman}, R.~F. 2003,
  \apjs, 147, 379

\bibitem[{{Smith} {et~al.}(2001){Smith}, {Humphreys}, {Davidson}, {Gehrz},
  {Schuster}, \& {Krautter}}]{smith2001}
{Smith}, N., {Humphreys}, R.~M., {Davidson}, K., {et~al.} 2001, \aj, 121, 1111

\bibitem[{{Sopka} {et~al.}(1985){Sopka}, {Hildebrand}, {Jaffe}, {Gatley},
  {Roellig}, {Werner}, {Jura}, \& {Zuckerman}}]{sopka1985}
{Sopka}, R.~J., {Hildebrand}, R., {Jaffe}, D.~T., {et~al.} 1985, \apj, 294, 242

\bibitem[{{Suh}(1999)}]{suh1999}
{Suh}, K.-W. 1999, \mnras, 304, 389

\bibitem[{{Sweitzer} {et~al.}(1979){Sweitzer}, {Palmer}, {Morris}, {Turner}, \&
  {Zuckerman}}]{sweitzer1979}
{Sweitzer}, J.~S., {Palmer}, P., {Morris}, M., {Turner}, B.~E., \& {Zuckerman},
  B. 1979, \apj, 227, 415

\bibitem[{{Temi} {et~al.}(2014){Temi}, {Marcum}, {Young}, {Adams}, {Adams},
  {Andersson}, {Becklin}, {Boogert}, {Brewster}, {Burgh}, {Cobleigh}, {Culp},
  {De Buizer}, {Dunham}, {Engfer}, {Ediss}, {Fujieh}, {Grashuis}, {Gross},
  {Harmon}, {Helton}, {Hoffman}, {Homan}, {H{\"u}twohl}, {Jakob}, {Jensen},
  {Kaminski}, {Kozarsky}, {Krabbe}, {Klein}, {Lammen}, {Lampater}, {Latter},
  {Le}, {McKown}, {Melchiorri}, {Meyer}, {Miles}, {Miller}, {Miller}, {Moore},
  {Nickison}, {Opshaug}, {Pf{\"u}eller}, {Radomski}, {Rasmussen}, {Reach},
  {Reinacher}, {Roellig}, {Sandell}, {Sankrit}, {Savage}, {Shenoy},
  {Schonfeld}, {Shuping}, {Smith}, {Talebi}, {Teufel}, {Tseng}, {Vacca},
  {Vaillancourt}, {Van Cleve}, {Wiedemann}, {Wolf}, {Zavala}, {Zeile}, {Zell},
  \& {Zinnecker}}]{temi2014}
{Temi}, P., {Marcum}, P.~M., {Young}, E., {et~al.} 2014, \apjs, 212, 24

\bibitem[{{Templeton} \& {Karovska}(2009)}]{templeton2009}
{Templeton}, M.~R. \& {Karovska}, M. 2009, \apj, 691, 1470

\bibitem[{{Tenenbaum} {et~al.}(2010{\natexlab{a}}){Tenenbaum}, {Dodd}, {Milam},
  {Woolf}, \& {Ziurys}}]{tenenbaum2010}
{Tenenbaum}, E.~D., {Dodd}, J.~L., {Milam}, S.~N., {Woolf}, N.~J., \& {Ziurys},
  L.~M. 2010{\natexlab{a}}, \apjs, 190, 348

\bibitem[{{Tenenbaum} {et~al.}(2010{\natexlab{b}}){Tenenbaum}, {Dodd}, {Milam},
  {Woolf}, \& {Ziurys}}]{tenenbaum2010cat}
{Tenenbaum}, E.~D., {Dodd}, J.~L., {Milam}, S.~N., {Woolf}, N.~J., \& {Ziurys},
  L.~M. 2010{\natexlab{b}}, VizieR Online Data Catalog: J/ApJS/190/348

\bibitem[{{Teyssier} {et~al.}(2012){Teyssier}, {Quintana-Lacaci}, {Marston},
  {Bujarrabal}, {Alcolea}, {Cernicharo}, {Decin}, {Dominik}, {Justtanont}, {de
  Koter}, {Melnick}, {Menten}, {Neufeld}, {Olofsson}, {Planesas}, {Schmidt},
  {Soria-Ruiz}, {Sch{\"o}ier}, {Szczerba}, \& {Waters}}]{teyssier2012}
{Teyssier}, D., {Quintana-Lacaci}, G., {Marston}, A.~P., {et~al.} 2012, \aap,
  545, A99

\bibitem[{{Townes} \& {Schawlow}(1955)}]{townesbook1955}
{Townes}, C.~H. \& {Schawlow}, A.~L. 1955, {Microwave Spectroscopy} (New York:
  Dover Publications, Inc.)

\bibitem[{{Truong-Bach} {et~al.}(1996){Truong-Bach}, {Graham}, \&
  {Nguyen-Q-Rieu}}]{tb1996}
{Truong-Bach}, {Graham}, D., \& {Nguyen-Q-Rieu}. 1996, \aap, 312, 565

\bibitem[{{Truong-Bach} {et~al.}(1988){Truong-Bach}, {Nguyen-Q-Rieu}, \&
  {Graham}}]{tb1988}
{Truong-Bach}, {Nguyen-Q-Rieu}, \& {Graham}, D. 1988, \aap, 199, 291

\bibitem[{{Tsuboi} {et~al.}(1958){Tsuboi}, {Shimanouchi}, \&
  {Mizushima}}]{tsuboi1958}
{Tsuboi}, M., {Shimanouchi}, T., \& {Mizushima}, S. 1958, \acspe, 13, 80

\bibitem[{{Tsuji}(1964)}]{tsuji1964}
{Tsuji}, T. 1964, \antok, 9, 1

\bibitem[{{Tsuji}(1973)}]{tsuji1973}
{Tsuji}, T. 1973, \aap, 23, 411

\bibitem[{{Urquhart} {et~al.}(2011){Urquhart}, {Morgan}, {Figura}, {Moore},
  {Lumsden}, {Hoare}, {Oudmaijer}, {Mottram}, {Davies}, \&
  {Dunham}}]{urquhart2011}
{Urquhart}, J.~S., {Morgan}, L.~K., {Figura}, C.~C., {et~al.} 2011, \mnras,
  418, 1689

\bibitem[{{van den Ancker} {et~al.}(2016){van den Ancker}, {Asmus}, {Hummel},
  {K{\"a}ufl}, {Kerber}, {Smette}, {Taylor}, {Tristram}, {Vinther}, \&
  {Wolff}}]{vdancker2016}
{van den Ancker}, M.~E., {Asmus}, D., {Hummel}, C., {et~al.} 2016, in
  \procspie, Vol. 9910, Society of Photo-Optical Instrumentation Engineers
  (SPIE) Conference Series, 99102T

\bibitem[{van~der Walt {et~al.}(2011)van~der Walt, Colbert, \&
  Varoquaux}]{numpy}
van~der Walt, S., Colbert, S.~C., \& Varoquaux, G. 2011, \mcse, 13, 22,
  \url{http://www.numpy.org/}

\bibitem[{{VanderPlas} {et~al.}(2012){VanderPlas}, {Connolly}, {Ivezic}, \&
  {Gray}}]{astroml2012}
{VanderPlas}, J., {Connolly}, A.~J., {Ivezic}, Z., \& {Gray}, A. 2012, in Proc.
  Conf. on Intelligent Data Understanding (CIDU), pp. 47-54, 2012., 47--54

\bibitem[{{VanderPlas} \& {Ivezi{\'c}}(2015)}]{vanderplas2015}
{VanderPlas}, J.~T. \& {Ivezi{\'c}}, {\v Z}. 2015, \apj, 812, 18

\bibitem[{{Velilla Prieto} {et~al.}(2015){Velilla Prieto}, {S{\'a}nchez
  Contreras}, {Cernicharo}, {Ag{\'u}ndez}, {Quintana-Lacaci}, {Alcolea},
  {Bujarrabal}, {Herpin}, {Menten}, \& {Wyrowski}}]{vp2015}
{Velilla Prieto}, L., {S{\'a}nchez Contreras}, C., {Cernicharo}, J., {et~al.}
  2015, \aap, 575, A84

\bibitem[{{Velilla Prieto} {et~al.}(2016){Velilla Prieto}, {S{\'a}nchez
  Contreras}, {Cernicharo}, {Ag{\'u}ndez}, {Quintana-Lacaci}, {Bujarrabal},
  {Alcolea}, {Balan{\c c}a}, {Herpin}, {Menten}, \& {Wyrowski}}]{vp2016}
{Velilla Prieto}, L., {S{\'a}nchez Contreras}, C., {Cernicharo}, J., {et~al.}
  2016, VizieR Online Data Catalog: J/A+A/597/A25

\bibitem[{{Velilla Prieto} {et~al.}(2017){Velilla Prieto}, {S{\'a}nchez
  Contreras}, {Cernicharo}, {Ag{\'u}ndez}, {Quintana-Lacaci}, {Bujarrabal},
  {Alcolea}, {Balan{\c c}a}, {Herpin}, {Menten}, \& {Wyrowski}}]{vp2017}
{Velilla Prieto}, L., {S{\'a}nchez Contreras}, C., {Cernicharo}, J., {et~al.}
  2017, \aap, 597, A25

\bibitem[{{Venn}(1995)}]{venn1995}
{Venn}, K.~A. 1995, \apj, 449, 839

\bibitem[{{Ventura} {et~al.}(2017){Ventura}, {Stanghellini}, {Dell'Agli}, \&
  {Garc{\'{\i}}a-Hern{\'a}ndez}}]{ventura2017}
{Ventura}, P., {Stanghellini}, L., {Dell'Agli}, F., \&
  {Garc{\'{\i}}a-Hern{\'a}ndez}, D.~A. 2017, \mnras, 471, 4648

\bibitem[{{Volk} \& {Cohen}(1989)}]{volk1989}
{Volk}, K. \& {Cohen}, M. 1989, \aj, 98, 1918

\bibitem[{{Watson}(1971)}]{watson1971}
{Watson}, J.~K.~G. 1971, \jmsp, 40, 536

\bibitem[{{Wieching}(2014)}]{kbandrx}
{Wieching}, G. 2014, in \effn, Vol.~5, Issue 1, ed. B.~{Hutawarakorn Kramer},
  {available at
  \url{https://issuu.com/effelsbergnewsletter/docs/effnews-jan14-final}}

\bibitem[{{Wienen} {et~al.}(2012){Wienen}, {Wyrowski}, {Schuller}, {Menten},
  {Walmsley}, {Bronfman}, \& {Motte}}]{wienen2012}
{Wienen}, M., {Wyrowski}, F., {Schuller}, F., {et~al.} 2012, \aap, 544, A146

\bibitem[{{Willacy} \& {Cherchneff}(1998)}]{willacy1998}
{Willacy}, K. \& {Cherchneff}, I. 1998, \aap, 330, 676

\bibitem[{{Willacy} \& {Millar}(1997)}]{willacy1997}
{Willacy}, K. \& {Millar}, T.~J. 1997, \aap, 324, 237

\bibitem[{{Wing} \& {Lockwood}(1973)}]{wing1973}
{Wing}, R.~F. \& {Lockwood}, G.~W. 1973, \apj, 184, 873

\bibitem[{{Winnewisser} {et~al.}(1996){Winnewisser}, {Belov}, {Klaus}, \&
  {Urban}}]{winnewisser1996}
{Winnewisser}, G., {Belov}, S.~P., {Klaus}, T., \& {Urban}, S. 1996,
  Zeitschrift Naturforschung Teil A, 51, 200

\bibitem[{{Wittkowski} {et~al.}(2012){Wittkowski}, {Hauschildt},
  {Arroyo-Torres}, \& {Marcaide}}]{wittkowski2012}
{Wittkowski}, M., {Hauschildt}, P.~H., {Arroyo-Torres}, B., \& {Marcaide},
  J.~M. 2012, \aap, 540, L12

\bibitem[{{Wong}(2013)}]{wong2013}
{Wong}, K.~T. 2013, {MPhil Thesis}, University of Hong Kong, {available at
  \url{http://hub.hku.hk/handle/10722/195970}}

\bibitem[{{Wong} {et~al.}(2016){Wong}, {Kami{\'n}ski}, {Menten}, \&
  {Wyrowski}}]{wong2016}
{Wong}, K.~T., {Kami{\'n}ski}, T., {Menten}, K.~M., \& {Wyrowski}, F. 2016,
  \aap, 590, A127

\bibitem[{{Yamamura} {et~al.}(1999){Yamamura}, {de Jong}, {Onaka}, {Cami}, \&
  {Waters}}]{yamamura1999}
{Yamamura}, I., {de Jong}, T., {Onaka}, T., {Cami}, J., \& {Waters},
  L.~B.~F.~M. 1999, \aap, 341, L9

\bibitem[{{Yang} {et~al.}(2011){Yang}, {Shutler}, \& {Grischkowsky}}]{yang2011}
{Yang}, Y., {Shutler}, A., \& {Grischkowsky}, D. 2011, \oexpr, 19, 8830

\bibitem[{{Young} {et~al.}(2012){Young}, {Becklin}, {Marcum}, {Roellig}, {De
  Buizer}, {Herter}, {G{\"u}sten}, {Dunham}, {Temi}, {Andersson}, {Backman},
  {Burgdorf}, {Caroff}, {Casey}, {Davidson}, {Erickson}, {Gehrz}, {Harper},
  {Harvey}, {Helton}, {Horner}, {Howard}, {Klein}, {Krabbe}, {McLean}, {Meyer},
  {Miles}, {Morris}, {Reach}, {Rho}, {Richter}, {Roeser}, {Sandell}, {Sankrit},
  {Savage}, {Smith}, {Shuping}, {Vacca}, {Vaillancourt}, {Wolf}, \&
  {Zinnecker}}]{young2012}
{Young}, E.~T., {Becklin}, E.~E., {Marcum}, P.~M., {et~al.} 2012, \apjl, 749,
  L17

\bibitem[{{Yu} {et~al.}(2010){Yu}, {Pearson}, {Drouin}, {Sung}, {Pirali},
  {Vervloet}, {Martin-Drumel}, {Endres}, {Shiraishi}, {Kobayashi}, \&
  {Matsushima}}]{yu2010}
{Yu}, S., {Pearson}, J.~C., {Drouin}, B.~J., {et~al.} 2010, \jcp, 133, 174317

\bibitem[{{Yurchenko} {et~al.}(2011){Yurchenko}, {Barber}, \&
  {Tennyson}}]{yurchenko2011}
{Yurchenko}, S.~N., {Barber}, R.~J., \& {Tennyson}, J. 2011, \mnras, 413, 1828

\bibitem[{{Zhang} {et~al.}(2012){Zhang}, {Reid}, {Menten}, \&
  {Zheng}}]{zhang2012}
{Zhang}, B., {Reid}, M.~J., {Menten}, K.~M., \& {Zheng}, X.~W. 2012, \apj, 744,
  23

\bibitem[{{Ziurys} {et~al.}(2007){Ziurys}, {Milam}, {Apponi}, \&
  {Woolf}}]{ziurys2007}
{Ziurys}, L.~M., {Milam}, S.~N., {Apponi}, A.~J., \& {Woolf}, N.~J. 2007, \nat,
  447, 1094

\end{thebibliography}
\addcontentsline{toc}{section}{References}


\begin{appendix}


\onecolumn


\section{VLA observations}
\label{sec:app_vla}

\subsection{Observing log}
\label{sec:app_log}

Table \ref{tab:obs_vla} lists all the VLA $K$-band observations that are relevant to this study and the covered NH$_3$ ground-state inversion lines.

\begin{table*}[!htbp]
\caption{VLA observing log.}
\label{tab:obs_vla}
\centering
\begin{tabular}{clccccc}
\hline\hline
Target & \hspace{63pt} Date & Config. & Project & Gain       & Flux       & Freq. coverage \\
       &                    &         &         & calibrator & calibrator & ($(J,K)$ or GHz) \\ \hline
{\ik}  & 2004 Jul 20                         & D & AM797   & J0409+1217 & 3C48 & $(1,1)$, $(2,2)$ \\
       & 2015 Oct 30, Nov 05, Jan 03         & D & 15B-167 & J0409+1217 & 3C48 & 21.12--23.14\tablefootmark{b} \\
       & 2016 Jan 29, Feb 06                 & C & 16A-011 & J0409+1217 & 3C48 & 22.12--24.18\tablefootmark{c} \\
       & 2016 May 29, Jun 17, Jun 19, Jul 01\tablefootmark{a} & B & 16A-011 & J0409+1217 & 3C48 & 22.12--24.18\tablefootmark{c} \\
\hline
{\irc} & 2004 Aug 09 & D   & AM797   & J1922+1530 & 3C286 & $(1,1)$, $(2,2)$ \\
\hline
{\oh}  & 2004 Jul 26 & D   & AM797   & J0735-1735 & 3C286 & $(1,1)$, $(2,2)$ \\
       &             &     &         & J0730-1141 &       &   \\
       & 2013 Apr 05 & D   & 13A-325 & J0748-1639 & 3C147 & 23.29--25.26\tablefootmark{d} \\
\hline
{\vy}  & 2013 Apr 05 & D   & 13A-325 & J0731-2341 & 3C147 & 23.29--25.26\tablefootmark{d} \\
       & 2016 Jan 10 & DnC & 15B-167 & J0731-2341 & 3C147 & 21.12--23.14\tablefootmark{b} \\
\hline
\end{tabular}
\tablefoot{
\tablefoottext{a}{This dataset is completely discarded because of poor phase stability and high rms noise. }
\tablefoottext{b}{Covered NH$_3$ lines include $(J,K)=(5,3)$, $(4,2)$, $(3,1)$, $(5,4)$, $(4,3)$, $(6,5)$, $(3,2)$, $(7,6)$, and $(2,1)$. }
\tablefoottext{c}{Covered NH$_3$ lines include $(J,K)=(3,1)$, $(5,4)$, $(4,3)$, $(6,5)$, $(3,2)$, $(7,6)$, $(2,1)$, $(1,1)$, $(2,2)$, $(3,3)$, and $(4,4)$. }
\tablefoottext{d}{Covered NH$_3$ lines include $(J,K)=(1,1)$, $(2,2)$, $(3,3)$, $(4,4)$, $(5,5)$, and $(6,6)$. }
}
\end{table*}


\subsection{Radio continuum}
\label{sec:app_cont}

Table \ref{tab:cont} lists the radio continuum measurements of our targets.

\begin{table*}[!htbp]
\caption{Radio continuum measurements.}
\label{tab:cont}
\centering
\begin{tabular}{clllcc}
\hline\hline
Target & Date & R.A.    & Decl.   & $\nu_\text{rep}$ & $F(\nu_\text{rep})$ \\
       &      & (J2000) & (J2000) &      (GHz)       & ($\mu$Jy)           \\ \hline
{\ik}  & 2015 Oct 30 (D1) & $03{^\text{h}}53{^\text{m}}28{\fs}924(9)$ & $+11{\degr}24{\arcmin}21{\farcs}5(1)$ & 22.13 & $350 \pm 31$ \\
       & 2015 Nov 05 (D2) & $03{^\text{h}}53{^\text{m}}28{\fs}915(5)$ & $+11{\degr}24{\arcmin}21{\farcs}50(9)$ & 22.13 & $271 \pm 22$ \\
       & 2016 Jan 03 (D3) & $03{^\text{h}}53{^\text{m}}28{\fs}926(2)$ & $+11{\degr}24{\arcmin}21{\farcs}45(5)$ & 22.13 & $250 \pm 11$ \\
       & 2016 Jan 29 (C1) & $03{^\text{h}}53{^\text{m}}28{\fs}9236(8)$ & $+11{\degr}24{\arcmin}21{\farcs}51(1)$ & 23.21 & $308 \pm 14$ \\
       & 2016 Feb 06 (C2) & $03{^\text{h}}53{^\text{m}}28{\fs}9230(6)$ & $+11{\degr}24{\arcmin}21{\farcs}52(1)$ & 23.21 & $278 \pm 13$ \\
       & 2016 May 29 (B1) & $03{^\text{h}}53{^\text{m}}28{\fs}9255(4)$ & $+11{\degr}24{\arcmin}21{\farcs}509(6)$ & 23.21 & $346 \pm 23$ \\
       & 2016 Jun 17 (B2) & $03{^\text{h}}53{^\text{m}}28{\fs}9244(2)$ & $+11{\degr}24{\arcmin}21{\farcs}493(6)$ & 23.21 & $361 \pm 21$ \\
       & 2016 Jun 19 (B3) & $03{^\text{h}}53{^\text{m}}28{\fs}9253(3)$ & $+11{\degr}24{\arcmin}21{\farcs}485(6)$ & 23.21 & $366 \pm 23$ \\
\hline
{\vy}  & 2013 Apr 05 & $07{^\text{h}}22{^\text{m}}58{\fs}3253(9)$ & $-25{\degr}46{\arcmin}02{\farcs}98(3)$ & 24.27 & $1404 \pm 13$ \\
\hline
{\oh}  & 2013 Apr 05 & $07{^\text{h}}42{^\text{m}}16{\fs}943(3)$ & $-14{\degr}42{\arcmin}49{\farcs}77(6)$ & 24.27 & $735 \pm 18$ \\
\hline
\end{tabular}
\tablefoot{
The columns are (from \textit{left} to \textit{right}): target, date of VLA observation, right ascension and declination of the centre of the fitted continuum, representative frequency ($\nu_\text{rep}$; GHz), and the continuum flux at $\nu_\text{rep}$ ($F(\nu_\text{rep})$; $\mu$Jy) Each VLA observation of {\ik} in 2015--2016 is labelled by the array configuration (`D', `C', or `B') and an ordinal number.
}
\end{table*}


\newpage

\section{{\hifi} observations}
\label{sec:app_hifi}

Table \ref{tab:obs_hifi} lists all the {\hifi} observations of NH$_3$ ground-state rotational lines used in this study.

\begin{table*}[!htbp]
\caption{Summary of {\hifi} observations.}
\label{tab:obs_hifi}
\centering
\begin{tabular}{cccccccc}
\hline\hline
Target & Transition & Frequency & HPBW\tablefootmark{a} & $\eta_\text{mb}$\tablefootmark{b} & {\herschel} & Observing & Date \\
       & $J_K(\text{sym})$ & (GHz) & (arcsec) & & OBSID & mode & \\ \hline
{\ik}  & $1_0(\text{s})$--$0_0(\text{a})$ & 572.5 & 37.0 & 0.619 & 1342190182 & spectral scan & 2010 Feb 04 \\
       &                                &       &      &       & 1342190192 & single point  & 2010 Feb 04 \\
       &                                &       &      &       & 1342190839 & spectral scan & 2010 Feb 19 \\
       &                                &       &      &       & 1342190840\tablefootmark{c} & spectral scan & 2010 Feb 19 \\
       &                                &       &      &       & 1342191594 & spectral scan & 2010 Mar 02 \\
       & $2_1(\text{s})$--$1_1(\text{a})$ & 1168  & 18.1 & 0.590 & 1342190215 & spectral scan & 2010 Feb 05 \\
       &                                &       &      &       & 1342190904 & spectral scan & 2010 Feb 21 \\
       & $2_K(\text{a})$--$1_K(\text{s})$ & ${\sim}1215$ & 17.5 & 0.589 & 1342190215 & spectral scan & 2010 Feb 05 \\
       &            $(K=0,1)$           &       &      &       & 1342190904 & spectral scan & 2010 Feb 21 \\
       &                                &       &      &       & 1342226021 & single point  & 2011 Aug 11 \\
       & $3_K(\text{s})$--$2_K(\text{a})$ & ${\sim}1764$ & 12.0 & 0.584 & 1342190807 & spectral scan & 2010 Feb 16 \\
       &           $(K=0,1,2)$          &       &      &       & 1342226032\tablefootmark{d} & single point  & 2011 Aug 11 \\
\hline
{\irc} & $1_0(\text{s})$--$0_0(\text{a})$ & 572.5 & 37.0 & 0.619 & 1342194526 & single point  & 2010 Apr 12 \\
       & $2_K(\text{a})$--$1_K(\text{s})$ & ${\sim}1215$ & 17.5 & 0.589 & 1342229948\tablefootmark{e} & single point  & 2011 Sept 30 \\
\hline
{\oh}  & $1_0(\text{s})$--$0_0(\text{a})$ & 572.5 & 37.0 & 0.619 & 1342194500 & single point  & 2010 Apr 11 \\
       &                                &       &      &       & 1342245270 & spectral scan & 2012 May 02--03 \\
       &                                &       &      &       & 1342269334 & single point  & 2013 Apr 04 \\
       & $2_1(\text{s})$--$1_1(\text{a})$ & 1168  & 18.1 & 0.590 & 1342244942 & spectral scan & 2012 Apr 25 \\
       & $2_K(\text{a})$--$1_K(\text{s})$ & ${\sim}1215$ & 17.5 & 0.589 & 1342244942 & spectral scan & 2012 Apr 25 \\
       & $3_K(\text{s})$--$2_K(\text{a})$ & ${\sim}1764$ & 12.0 & 0.584 & 1342244608 & single point  & 2012 Apr 20 \\
\hline
{\vy}  & $1_0(\text{s})$--$0_0(\text{a})$ & 572.5 & 37.0 & 0.619 & 1342192528 & single point  & 2010 Mar 21 \\
       &                                &       &      &       & 1342244486 & spectral scan & 2012 Apr 17 \\
       & $2_1(\text{s})$--$1_1(\text{a})$ & 1168  & 18.1 & 0.590 & 1342228611 & spectral scan & 2011 Sept 14 \\
       & $2_K(\text{a})$--$1_K(\text{s})$ & ${\sim}1215$ & 17.5 & 0.589 & 1342228611 & spectral scan & 2011 Sept 14 \\
       & $3_K(\text{s})$--$2_K(\text{a})$ & ${\sim}1764$ & 12.0 & 0.584 & 1342230402 & spectral scan & 2011 Oct 09 \\
       &                                &       &      &       & 1342230403\tablefootmark{e} & single point  & 2011 Oct 09 \\
       & $3_K(\text{a})$--$2_K(\text{s})$ & ${\sim}1810$ & 11.7 & 0.583 & 1342244537 & spectral scan & 2012 Apr 18 \\
\hline
\end{tabular}
\tablefoot{
\tablefoottext{a}{The half power beam width (HPBW) is computed by Eq. (3) of \citet{roelfsema2012}. }
\tablefoottext{b}{The main beam efficiency ($\eta_\text{mb}$) is computed by Eq. (1) of \citet{roelfsema2012} with the improved coefficients determined by \citet{mueller2014}. }
\tablefootmark{c}{Frequency switch mode was used instead of dual beam switch (DBS) mode. }
\tablefootmark{d}{Bad results in V-polarisation. }
\tablefootmark{e}{Bad results in H-polarisation. }
}
\end{table*}


\twocolumn


\section{Light curve and period of {\ik}}
\label{sec:app_iktau}

The most commonly quoted pulsation period of {\ik} is $470 \pm 5$ days, which was determined by curve-fitting the near-infrared (NIR) flux variation of the star at 1.04\,{\micron} \citep{wing1973}. \citet{wing1973} suggested that the luminosity phase of {\ik} to be represented by 
\begin{equation}
E = \frac{JD-2\,439\,440}{470},
\label{eq:phase}
\end{equation}
where the integral and fractional parts of $E$ are the number of complete cycles counting from the optical $V$-band maximum in 1966 Nov \citep{cannon1967} and the visual phase in the cycle, respectively. \citet{hale1997} also derived the same period of $470 \pm 5$ days from the (sparsely sampled) MIR fluxes of {\ik} at 11.15\,{\micron}. On the other hand, \citet{lebertre1993} determined a different period of $462^{+3}_{-1}$ days from Fourier analysis of the multi-band NIR (1.24--4.64\,{\micron}) light curves of {\ik}.

Since the periods of long-period pulsating variable stars are known to be stochastically variable \citep{templeton2009}, the period of {\ik} derived ${\gtrsim}20$ years ago may no longer be valid. To reduce the accumulated errors in the stellar phases, the time of {\ik}'s visual maximum in recent cycles is needed. Based on the photometric data from the American Association of Variable Star Observers (AAVSO) International Database, the relatively well-sampled optical $V$-band maxima of {\ik} occured in 2007--2009. The difference in periods of 8 days per cycle would lead to a phase difference of about 0.1 in our VLA observations of 2015--2016. In this appendix, we present new analysis the $V$-band light curve of {\ik} in order to improve the estimates of its pulsation period and phase zero date.

\begin{figure}[!htbp]
\centering
\includegraphics[width=0.49\textwidth]{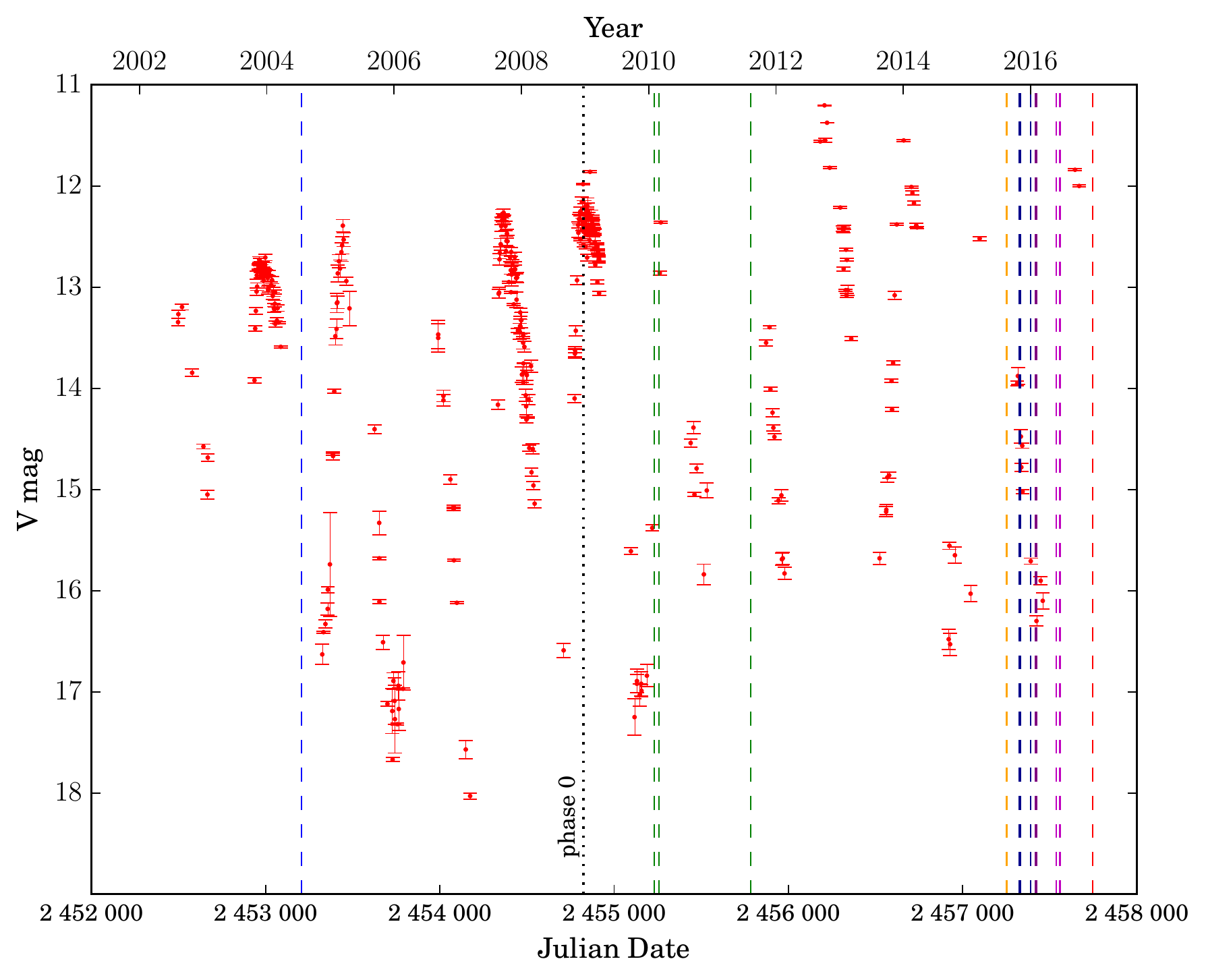}
\caption[]{The $V$-band light curve of {\ik} in 2002--2016. The red points are the observations from the AAVSO International Database. We only include the observations with $V$ magnitude uncertainties. The black vertical line indicates the date of $V$ maximum on JD 2\,454\,824 (2008 Dec 23) as determined from the smoothed light curve as displayed in Fig. \ref{fig:iktau_v_phase}. Other coloured vertical lines represent the dates of NH$_3$ observations with the VLA D-array (blue: 2004; dark blue: 2015--2016), C-array (purple), B-array (magenta), {\hifi} (green), IRAM 30m (orange), and the IRTF/TEXES instrument (red).}
\label{fig:iktau_v_date}
\end{figure}

From the AAVSO Database, we only select the $V$-band observations by the end of 2016 and exclude those without reported uncertainties. These criteria allow observations between 2002 Aug 10 and 2016 Oct 07. Figure \ref{fig:iktau_v_date} shows the $V$-band light curve of {\ik} and the corresponding time at which NH$_3$ observations were performed. We then carry out time-series analyses using the Lomb–Scargle (LS) periodogram \citep{lomb1976,scargle1982} in different Python implementations, including the standard LS periodogram in \texttt{Astropy} \citep[version 1.3;][]{astropy}, the generalised LS periodogram in \texttt{astroML} \citep{astroml2012,vanderplas2015}, and the Bayesian formalism of the generalised LS periodogram \citep[BGLS;][]{mortier2015,bgls2015}.

The output of BGLS is a probability distribution of period, which allows us to compute the standard deviation straightforwardly. In the \texttt{Astropy} implementation of the standard LS periodogram, we compute the uncertainty from the standard deviation of all the LS peak periods derived from a set of 10\,000 simulated light curves, assuming that the uncertainties of the nominal $V$-band magnitude obey Gaussian distribution. All the periodograms of the observed $V$-band light curve show a prominent peak corresponding to the period of 460.05 days. The standard deviations of the period from the BGLS probability and from simulations are 0.02 and 0.06 days, respectively. Therefore, we report the $V$-band period of {\ik} as $460.05 \pm 0.06$ days.

\begin{figure}[!htbp]
\centering
\includegraphics[width=0.49\textwidth]{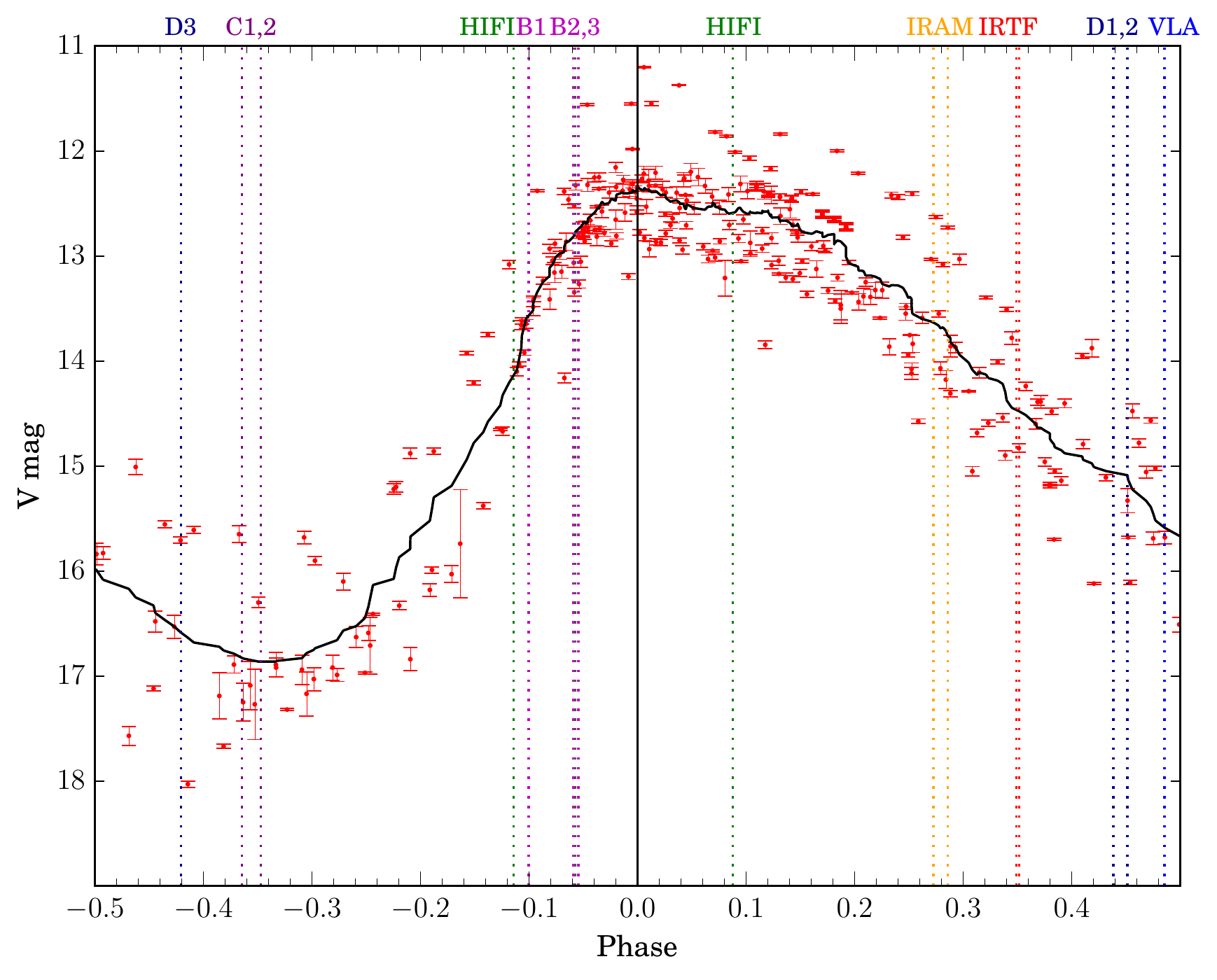}
\caption[]{The phased $V$-band light curve of {\ik} in 2002--2016. Red points are the observations from the AAVSO International Database and the black curve is the smoothed light curve with a Savitzky-Golay filter. The phases of relevant NH$_3$ observations are labelled by vertical lines and on the top axis, with the same colour scheme as Fig. \ref{fig:iktau_v_date}. The label `VLA' represents the 2004 VLA observation. The numerals after `D', `C', and `B' are the ordinal numbers representing the dates of VLA observations in 2015--2016 with the corresponding arrays (`B4' is discarded, see Tables \ref{tab:obs_vla} and \ref{tab:cont}).}
\label{fig:iktau_v_phase}
\end{figure}

We then phase the $V$-band light curve with the period of 460.05 days and smooth it with a Savitzky-Golay filter \citep{savitzkygolay1964}. We find that the phase zero date of JD 2\,454\,824 (2008 Dec 23) can produce an alignment of phase zero with the peak of the smoothed light curve. Figure \ref{fig:iktau_v_phase} shows the phased $V$-band data and the smoothed light curves. We search the phase zero date around the year 2009.0 because this peak is relatively well-sampled compared to more recent ones.

\end{appendix}


\end{document}